\documentclass[12pt]{article}
\renewcommand{\baselinestretch}{1.4}
\usepackage{fullpage,amsmath,amssymb,bm,url,mathrsfs}
\usepackage[protrusion=true,expansion=true]{microtype}
\usepackage{algorithmic}
\usepackage[ruled,linesnumbered]{algorithm2e}
\usepackage{amsthm}
\usepackage{hyperref}
\usepackage{makecell}
\usepackage{tablefootnote}
\usepackage{graphics,graphicx}
\usepackage{subcaption,caption,cleveref}
\hypersetup{
    colorlinks=true,
    linkcolor=blue,
    citecolor=blue,
    filecolor=magenta,
    urlcolor=cyan,
    pdftitle={Weak Factor Analysis via Temporally Weighted PCA},
    pdfauthor={Zhongyuan Lyu and Ming Yuan},
    hypertexnames=false,
}
\usepackage{multirow}
\usepackage[round,colon,authoryear]{natbib}
\usepackage{xcolor}
\usepackage{dsfont}
\usepackage{bbm}
\usepackage{enumitem}
\usepackage{blkarray}
\usepackage{amsfonts}
 \usepackage{booktabs,tabularx,makecell}
 \usepackage{amssymb} 

\def\frakD{{\mathfrak D}}

\def\frakP{{\mathfrak P}}

\def\EE{{\mathbb E}}

\def\II{{\mathbb I}}

\def\OO{{\mathbb O}}
\def\PP{{\mathbb P}}

\def\RR{{\mathbb R}}
\def\SS{{\mathbb S}}

\def\ZZ{{\mathbb Z}}

\def\a{{\boldsymbol a}}
\def\b{{\boldsymbol b}}

\def\e{{\boldsymbol e}}
\def\f{{\boldsymbol f}}
\def\g{{\boldsymbol g}}
\def\h{{\boldsymbol h}}

\def\l{{\boldsymbol l}}

\def\s{{\boldsymbol s}}

\def\u{{\boldsymbol u}}
\def\v{{\boldsymbol v}}
\def\w{{\boldsymbol w}}
\def\x{{\boldsymbol x}}
\def\y{{\boldsymbol y}}

\DeclareMathOperator*{\argmin}{arg\,min}

\def\calB{{\cal  B}}

\def\calE{{\cal  E}} 
\def\calF{{\cal  F}} 
\def\calG{{\cal  G}} 
\def\calH{{\cal  H}} 
\def\calI{{\cal  I}}

\def\calN{{\cal  N}}
 
\def\calP{{\cal  P}} 
 \def\scrQ{{\mathscr  Q}}
\def\calR{{\cal  R}} 
\def\calS{{\cal  S}}

\newcommand{\bfm}[1]{\ensuremath{{\boldsymbol{#1}}}}

   \def\bA{\bfm A}  
   \def\bB{\bfm B}  
   \def\bC{\bfm C}  
   \def\bD{\bfm D}  
\def\be{\bfm e}   \def\bE{\bfm E}  \def\EE{\mathbb{E}}
\def\bf{\bfm f}  \def\bF{\bfm F}  
   \def\bG{\bfm G}  
   \def\bH{\bfm H}  
   \def\bI{\bfm I}  \def\II{\mathbb{I}}

   \def\bL{\bfm L}  
   \def\bM{\bfm M}  
   \def\bN{\bfm N}  
   \def\bO{\bfm O}  \def\OO{\mathbb{O}}
   \def\bP{\bfm P}  \def\PP{\mathbb{P}}
   \def\bQ{\bfm Q}  
\def\br{\bfm r}   \def\bR{\bfm R}  \def\RR{\mathbb{R}}
   \def\bS{\bfm S}  \def\SS{\mathbb{S}}
     
   \def\bU{\bfm U}  
   \def\bV{\bfm V}  
   \def\bW{\bfm W}  
\def\bx{\bfm x}   \def\bX{\bfm X}  
   \def\bY{\bfm Y}  
   \def\bZ{\bfm Z}  \def\ZZ{\mathbb{Z}}
                  \def\bN{\bfm N}

  \def\bOmega{\bfm \Omega}
\def\bSigma{\bfm \Sigma}

\def\bLambda{\bfm \Lambda}
\def\bGamma{\bfm \Gamma}
\def\bgamma{{\boldsymbol{\gamma}}}
\def\beps{{\boldsymbol{\varepsilon}}}
\def\bDelta{\bfm \Delta}

\def\mPsi{\boldsymbol{\Psi}}
\def\mPhi{\boldsymbol{\Phi}}
\def\mXi{\boldsymbol{\Xi}}

\def\hat{\widehat}
\def\wt{\widetilde}

\def\beps{\boldsymbol{\varepsilon}}
\def\Tr{\textsf{Tr}}
\def\PTr{\textsf{PTr}}

\newcommand\inp[2]{\left\langle #1, #2 \right\rangle}
\newcommand\fro[1]{\left\| #1 \right\|_{\rm F}}
\newcommand\aop[1]{\left\|#1\right\|}
\newcommand\op[1]{\big\|#1\big\|}
\newcommand\bop[1]{\big\|#1\big\|}
\newcommand\infn[1]{\big\|#1\big\|_{2,\infty}}
\newcommand\binfn[1]{\big\|#1\big\|_{2,\infty}}
\newcommand\brac[1]{\left(#1\right)}
\newcommand\bbrac[1]{\big(#1\big)}
\newcommand\sqbrac[1]{\left[#1\right]}
\newcommand\ebrac[1]{\left\{#1\right\}}
\newcommand\ab[1]{\left|#1\right|}

\newtheorem{theorem}{Theorem}
\newtheorem{proposition}{Proposition}

\newtheorem{remark}{Remark}

\newtheorem{assumption}{Assumption}
\newtheorem{lemma}{Lemma}
\newtheorem{corollary}{Corollary}

\newtheorem{exmp}{Example}[section]

\SetKwInput{KwData}{Input}
\SetKwInput{KwResult}{Output}

\begin{document}
\title{Weak Factor Analysis via Temporally Weighted PCA\thanks{This research was supported in part by NSF Grant DMS-2052955. }}
\date{(October 19, 2026)}

\author{Zhongyuan Lyu\thanks{The University of Sydney Business School, Sydney, NSW 2008, Australia; Email: zhongyuan.lyu@sydney.edu.au} ~and Ming Yuan\thanks{Department of Statistics, Columbia University, New York, NY 10027, USA; Email: ming.yuan@columbia.edu}}
	
\maketitle

\begin{abstract}
Principal component analysis (PCA) is arguably the most widely used approach for large-dimensional factor analysis. While it is effective when the factors are sufficiently strong, it can be inconsistent when the factors are weak and/or the noise has complex dependence structure. We argue that the inconsistency often stems from bias and introduce a general approach to restore consistency. Specifically, we propose a general weighting scheme for PCA and {show that suitable weights can restore consistency for much weaker factors under the dependent-noise structures allowed by our model. Under the stronger signal-to-noise conditions needed for inference, we also establish asymptotic normality of the loading and factor estimators}. While {an oracle} weighting matrix may require knowledge about the factors and covariance of the idiosyncratic noise that are not known a priori, we develop an agnostic approach to adaptively choose from a large class of weighting matrices that can be viewed as PCA for weighted linear combinations of auto-covariances among the observations. Theoretical and numerical results demonstrate the merits of our methodology over the usual PCA and other recently developed techniques for large-dimensional approximate factor models.
\end{abstract}

\newpage
\section{Introduction}\label{sec:intro}
Large-dimensional factor analysis is an effective tool for extracting latent factors from noisy observations and is widely used in a variety of fields including economics, finance, and genomics among others. More specifically, consider the setting of panel data where observations, denoted by $\bX=(x_{it})_{1\le i\le N, 1\le t\le T}$, are obtained from $N$ units across $T$ time points. The factor model expresses $\bX$ as
\begin{align}\label{model:matrix-form}
	\bX=\bL\bF^\top +\bE,
\end{align}
where $\bF\in\RR^{T\times r}$ and $\bL\in\RR^{N\times r}$ represent the latent factors and factor loadings, respectively, and $\bE\in\RR^{N\times T}$ the centered idiosyncratic noise. Here $r$ is much smaller than both $N$ and $T$ so that the signal lies in a low-dimensional space, {and all formal results throughout the paper condition on a correctly specified \(r\).} In general, \eqref{model:matrix-form} is referred to as the \emph{approximate factor model} \citep[see, e.g.,][]{chamberlain1982arbitrage}. It is also called the \emph{strict factor model} if $E_{it}$'s are assumed to be uncorrelated \citep[see, e.g.,][]{anderson1956statistical}. It is clear that $\bL$ and $\bF$ are only identifiable up to scaling and rotation. It is thus customary to assume that $\bF^\top\bF/T \to_p \bI_r$ as $T\to\infty$.

Seminal works such as \cite{connor1986performance,connor1988risk,stock2002forecasting, bai2002determining, bai2003inferential} have established Principal Component Analysis (PCA) as the most popular approach for factor models \eqref{model:matrix-form}. See \cite{bai2008large} for a comprehensive review. In spite of the popularity and successes of PCA, its effectiveness for large-dimensional factor analysis should not be taken for granted. To fix ideas, let us focus on estimating the column space, denoted by $\bU$, of $\bL$. PCA estimator, $\hat{\bU}^{\rm PC}$, for $\bU$ can be viewed as the eigenspace of $\bX\bX^\top$. Intuitively, $\hat{\bU}^{\rm PC}$ can only be expected to perform well if the eigenspace, $\bU^{\rm PC}$, of $\EE[\bX\bX^\top]=\bL\bF^\top\bF\bL^\top + \EE[\bE\bE^\top]$ is identical or at least close to $\bU$. The former ($\bU^{\rm PC}=\bU$) is true, for example, if $E_{it}$'s are independent and identically distributed, whereas the latter ($\bU^{\rm PC}\approx\bU$) can hold, more generally, if the factors are strong in that the first term $\bL\bF^\top\bF\bL^\top$ dominates the second, $\EE[\bE\bE^\top]$. But what happens if the factors are not strong and $E_{it}$'s are cross-sectionally and temporally dependent? Are there situations where we can do better than the usual PCA for factor analysis? These are the questions we aim to address in this paper.

More specifically, we shall consider a weighted version of PCA: estimating $\bU$ by the eigenspace, $\hat{\bU}_\bQ$, of $\bX \bQ\bX^\top$ for an appropriately chosen weight matrix $\bQ$. The usual PCA can be viewed as a special case with $\bQ=\bI_T$. More generally, the weight matrix offers a mechanism to re-balance the signal $\bL\bF^\top\bQ\bF\bL^\top$ against the ``noise'' $\EE[\bE\bQ\bE^\top]$. Depending on the nature of the factors and noise, it is plausible that eigenspace, $\bU_{\bQ}$, of $\EE[\bX\bQ\bX^\top]$ can be close to $\bU$ while $\bU^{\rm PC}$ is not. In particular, taking $\bQ=T^{-1}\bF\bF^\top$ yields $\bL\bF^\top\bQ\bF\bL^\top = T^{-1}\bL(\bF^\top\bF)^2\bL^\top\to T\bL\bL^\top$. On the other hand, $\EE[\bE\bQ\bE^\top]=\textsf{PTr}_{T}\left\{\textsf{Cov}[\textsf{vec}(\bE)](\bI_N\otimes T^{-1}\bF\bF^\top)\right\}$ which is of the order $1$ and thus can be dominated by $\bL\bF^\top\bQ\bF\bL^\top$ even if $\EE[\bE\bE^\top]$ is not of smaller order than $\bL\bF^\top\bF\bL^\top$. Here $\textsf{PTr}_{T}$ stands for the partial trace taken over the time index, i.e., for any square matrix $\bA\in \RR^{pT\times qT}$, $\textsf{PTr}_{T}\ebrac{\bA}:=\sum_{t=1}^T\brac{\bI_p\otimes\e_t^\top}\bA \brac{\bI_q\otimes\e_t}$, where $\ebrac{\e_t}_{t=1}^T$ is the standard basis in $\RR^T$ and $\otimes$ denotes the Kronecker product. Of course this specific choice of the weight matrix is too idealistic and infeasible in practice since it requires knowing $\bF$ a priori. Nonetheless, there are numerous situations where simple and agnostic choices of $\bQ$ can lead to superior performance over PCA.

\begin{exmp}[Multivariate Time Series Model]
Consider the case when $\bE$ is temporally uncorrelated and $\bF$ follows a multivariate autoregressive model:
$$
\bf_t = \bA\bf_{t-1} + \beps_t,\qquad t=2,\ldots, T.
$$
Then by taking $\bQ=\sum_t(\be_t\be_{t-1}^\top+\be_{t-1}\be_t^\top)$, we get $\EE[\bE\bQ\bE^\top]=0$ yet $\bL\bF^\top\bQ\bF\bL^\top$ can be of the same order as $\bL\bF^\top\bF\bL^\top$.
\end{exmp}

\begin{exmp}[Multivariate Functional Data]
Consider the case where $\bE$ is temporally uncorrelated and each row of $\bF$ is a smooth function so that $F_{t,k}=g_k(t/T)$ for some smooth function $g_k: [0,1]\to \RR$. Let $\bQ$ be a banded matrix with ones on its first $B$ sub- and super-diagonals, and zeros otherwise. Then under mild conditions it can be shown that $\bL\bF^\top\bQ\bF\bL^\top$ is of the same order as $TB\bL\bL^\top$ yet $\EE[\bE\bQ\bE^\top]=0$.
\end{exmp}

See \Cref{sec:two-example} for further discussion about these two examples. To better understand the operating characteristics of weighted PCA, we shall derive non-asymptotic estimation error bounds and inferential theory for the estimated loading matrix and factors under general weighting schemes. These bounds pinpoint the effect of weighting and explain how it can provide an effective tool to address two of the most pressing challenges in large-dimensional factor analysis: weak factors and complex cross-sectional dependence.

Earlier studies for large-dimensional factor analysis have focused on strong factors in that $\bL^\top\bL/N^\alpha$ tends to a positive definite limit for some $\alpha\ge 1$. See \cite{bai2008large} for a survey. The shift toward weak factors can be traced back at least to \citet{onatski2012asymptotics}, who shows that $\alpha>0$ is necessary for the consistency of PCA. Subsequent advances establish consistency and asymptotic normality under weaker regimes \citep[e.g.,][]{uematsu2022estimation, bai2023approximate, choi2024high, fan2024can}.  In particular, \cite{choi2024high} established the consistency and asymptotic normality of PC estimator for $\alpha\in(0,1)$, and  \cite{fan2024can} further relaxed the factor strength  assumption for normality to logarithmic order.  These results indicate that PCA-based estimates remain consistent and are asymptotically normal for weaker factors, but under much more restrictive assumptions about the dependence among the entries of $\bE$ that can be relaxed when using weighted PCA. In addition, we shall see that weighted PCA can be consistent under suitable conditions even when $\bL^\top\bL\rightarrow 0$. 

As noted before, PCA-based estimates are expected to perform well when $\bE$ has independent and identically distributed entries. The potential limitations of PCA-based estimates when this assumption is violated have also come to light in the past few years. In particular, it has been observed that even if the entries of $\bE$ are independent but have different variances across units, the PCA-based estimates are suboptimal and can be much improved by the so-called HeteroPCA proposed in \cite{zhang2018heteroskedastic} that iteratively re-estimates the eigenspace of $\bX\bX^\top$ by imputing its diagonal entries in order to reduce the bias caused by the cross-sectional heteroskedasticity. See also \cite{yan2021inference, agterberg2022entrywise}. While HeteroPCA is designed specifically to address the cross-sectional heteroskedasticity, it remains unclear how to account for more general cross-sectional dependence. 

Another  related work is \citet{bai2013statistical} who proposes to estimate factors and loadings via a weighted least-squares criterion with the optimal weight $\bW=\bSigma_{\rm C}^{-1}$ under the assumption that $\bSigma_{\rm C}:=\textsf{Cov}(\bE_{\cdot,t})$ for $t\in[T]$. The performance of their estimators hinges upon this structural assumption and other conditions to allow for consistent estimation of $\bSigma_{\rm C}$. In fact, they show that even an estimate $\hat \bSigma_{\rm C}^{-1}$ that obeys $\op{\hat \bSigma_{\rm C}^{-1}- \bSigma_{\rm C}^{-1}}\overset{p}{\rightarrow} 0$ may not be sufficient for valid inference. In practice, when $\bSigma_{\rm C}$ cannot be reliably estimated, it is unclear how to proceed with estimation and inference for the factors and loadings within that framework.

Our approach provides a more generally applicable solution to the above challenges by leveraging the potential temporal dependence among the factors. In particular, we shall show that with appropriately chosen weight matrices, weighted PCA can significantly reduce the estimation error of the factors and their loadings. {Under the stronger conditions stated for inference, we also establish asymptotic normality under the cross-sectional and temporal dependence allowed by our model}.

Weighted PCA has been used for several distinct purposes in the literature. One line of work
uses low-rank or projective weights to suppress noise through projection, thereby improving
the signal-to-noise ratio for weak factors. Examples include projected PCA \cite{fan2016projected} and diversified
projection \citep{fan2022learning}. A second line uses
full-rank weights to accommodate heteroskedasticity or dependence in the idiosyncratic
component; see, e.g., \cite{choi2012efficient,bai2013statistical}. A third line uses supervised
or response-guided weights \citep{huang2022scaled}. {

Our paper is closest to the first line at the methodological level, but with a different scope: unlike the low-rank projective weights underlying projected PCA and diversified projection, our temporal weight \(\bQ\) is not required to be low rank, positive semidefinite, or a projection. To make the connection precise, suppose first that \(\bQ=\bW\bW^\top\). Then $\bX\bW=\bL(\bF^\top\bW)+\bE\bW$ and $\bX\bQ\bX^\top=(\bX\bW)(\bX\bW)^\top$. Two important conditions for the success of projected PCA are (i) \emph{relevance} \(\operatorname{rank}(\bW^\top\bF)=r\), and (ii) \emph{exogeneity} \(\EE(\bE\mid\bF,\bW)=\boldsymbol0\).
In this case, relevance is equivalent to \(\sigma_r(\bF^\top\bQ\bF)>0\) and exogeneity centers the projected error \(\bE\bW\), while concentration controls its finite-sample fluctuation. For a general symmetric temporal weight, we impose the analogous relevance requirement directly through positive definiteness and a quantitative lower bound on \(\bF^\top\bQ\bF\), and control the noise contribution through the projected perturbation in \eqref{eq:UQ-bound-gen}. More explicitly, after transposing the data matrix, the projective spectral matrix generated by a temporal proxy \(\bW\) is exactly \((\bX\bW)(\bX\bW)^\top=\bX\bW\bW^\top\bX^\top\), and hence corresponds to our construction with \(\bQ=\bW\bW^\top\). The complete procedures differ in how the proxy is constructed, but the spectral object is a member of the weighted-PCA family. WPCA is more general in rank and sign structure because it also covers high- or full-rank lag and autocovariance weights that reduce population noise bias.
}

The choice of the weighting matrix is clearly of great practical importance. Motivated by the examples before, we shall consider choosing the weighting matrix from the following broad class of Toeplitz matrices:
$$
\scrQ_T =\{\bQ=\textsf{Toeplitz}(\gamma_0,\gamma_1,\ldots, \gamma_{T-1}): \quad \gamma_0,\ldots,\gamma_{T-1}\ge0,\quad {\rm and}\quad \gamma_0+\cdots+\gamma_{T-1}=1\}.
$$
Here we use $\textsf{Toeplitz}(\gamma_0,\gamma_1,\ldots, \gamma_{T-1})$ to denote a  $T\times T$ symmetric matrix with  diagonals being $\gamma_0$ and $t$-th super- and sub-diagonals being $\gamma_t$ for $t\ge 1$.  Note that multiples of a weight matrix yield the same weighted PCA and thus the weighting matrices from both examples can be viewed as instances from $\scrQ_T$. Moreover, weight matrices from $\scrQ_T$ have an immediate statistical interpretation:
$$\bX \bQ\bX^\top = \gamma_0\sum_{t=1}^T \bx_t\bx_t^\top + \gamma_1\sum_{t=1}^{T-1} (\bx_{t-1}\bx_t^\top+\bx_t\bx_{t-1}^\top)+\cdots,$$
where $\bx_t$ is the $t$-th column vector of $\bX$. Thus, after centering, $\bX\bQ\bX^\top$ is proportional to a weighted linear combination of the covariance and auto-covariances of the column vectors of $\bX$. This connects our work with recent developments in multivariate time series literature to utilize higher order auto-covariances for identifying latent factors. See, e.g., \cite{lam2011estimation,lam2012factor,chang2018principal, chang2024autocovariance}. 

Instead of fixing a weighting matrix a priori, we propose a cross-validation (CV) procedure to adaptively choose a weighting matrix from $\scrQ_T$ as a practical finite-grid selection rule. Conceptually, our method shares a similar spirit to common techniques for matrix completion where missing entries are imputed to recover low-rank structure via spectral methods, nuclear-norm penalties, or EM-type algorithms  \citep[see, e.g.,][]{koltchinskii2011nuclear,xia2021statistical,mazumder2010spectral, chen2020noisy, choi2024matrix}. In our setting, no entry of $\bX$ is truly missing, yet the same logic of ``hiding information to gauge model fit'' underlies our CV scheme for choosing $\bQ$ from $\scrQ_T$. We temporarily withhold a uniformly chosen subset of entries, fit the weighted PCA on the retained data, and select the Toeplitz weights that best predict the held-out values. We shall refer to this procedure as \emph{adaptive weighted PCA} (AdaWPCA). 

In summary, our contributions are threefold. 
\begin{itemize}
	\item \emph{Methodology.} We propose weighted PCA (WPCA), utilizing a weighted sample covariance $\bX\bQ\bX^\top$ for factor model \eqref{model:matrix-form}, which accommodates general noise structures and weak factors.
	\item \emph{Theory.} We derive non-asymptotic estimation error bounds and inferential theory for our estimators $\bbrac{\hat\bL_\bQ,\hat\bF_\bQ}$. When factor structure is properly exploited via $\bQ$, the estimation error of $\hat\bL_\bQ$ can be significantly reduced. {We also establish asymptotic normality under the dependent-noise structures allowed by our model.} 
	\item  \emph{Adaptivity.} We propose a CV procedure for adaptively selecting  $\bQ\in\scrQ_T$ (AdaWPCA) {, establish a pairwise separation result over two specified subsets of the finite candidate class, and numerically evaluate its performance}. 
\end{itemize}

The rest of the paper is organized as follows. We shall introduce the general methodology for weighted PCA and study its performance in the next section. \Cref{sec:refined} develops estimation bounds and inferential theory for factors and loadings. In \Cref{sec:two-example}, we shall discuss Examples 1.1 and 1.2 in further detail. \Cref{sec:cv} focuses on weight matrices from $\scrQ_T$ and studies how to adaptively choose a weighting matrix by CV. Our methodological and theoretical developments are further complemented by numerical experiments, both simulated and real, in  \Cref{sec:num}.

\paragraph{Notation}  Let $\op{\cdot}_k$ denote either vector $k$-norm or the matrix norm induced by vector $k$-norm, and we omit the subscript when $k=2$ for brevity. We use $\OO_{p,r}$ to denote the collection of $p\times r$ matrices with orthonormal columns, and write $\OO_{p}:=\OO_{p,p}$. In addition, we use  $\fro{\cdot }$, $\infn{\cdot }$ and $\op{\cdot}_{\sf max}$ to denote the Frobenius norm,  $\ell_2\rightarrow\ell_\infty$ norm,  and entry-wise max norm, respectively. The operator $\textsf{vec}(\cdot)$ stacks the rows of a matrix. For two matrices $\bA$ and $\bB$ of the same dimension, the Hadamard product $\bA\circ\bB$ is  a matrix of the same dimension as the operands, with elements given by $[\bA\circ\bB]_{i,j}=A_{ij}B_{ij}$.  For two non-negative sequences $a_N$ and $b_N$, we write $a_N \lesssim b_N$ ($a_N \gtrsim b_N$) or $a_N=O(b_N)$ if there exists some universal constant $C>0$ independent of $N$ such
that $a_N \le Cb_N$ ($b_N \le  C a_N$); $a_N \asymp b_N$ if $a_N \lesssim b_N$ and $b_N \lesssim a_N$ hold simultaneously;
$a_N = o(b_N )$ or $b_N=\omega(a_N)$ or $a_N\ll b_N$ or $b_N\gg a_N$ if $b_N > 0$ and $a_N /b_N\rightarrow  0$. We use $C^k(D)$ to denote the space of all functions with $k$ continuous derivatives on a domain $D$. In addition, we shall write $n:=N\vee T$ for brevity. 

\section{Weighted PCA and Subspace Estimation}\label{sec:main}

In this section, we introduce our weighted PCA framework for recovering the factors and loadings in the approximate factor model. To fix ideas, we shall focus our discussion on random factors. But it is worth pointing out that our approach and analysis can also be readily applied to deterministic factors. We opt for random factors merely for brevity. 
Recall that $\bF=\sqbrac{\f_1,\cdots,\f_T}^\top \in\RR^{T\times r}$ , $\bL=\sqbrac{\l_1,\cdots,\l_N}^\top \in \RR^{N\times r}$. Denote by $\bM=\bL\bF^\top$ the low rank component.  Notice that $\bL$ and $\bF$ are only identifiable up to rotation and scale in terms of \eqref{model:matrix-form}. Following convention in the literature, we assume that
\begin{assumption}\label{assump:iden}
The factor matrix $\bF$ is independent of $\bE$ and $\sup_{t\in [T]}\op{\f_t}_{\psi_2}<\infty$. In addition, $\EE \f_t\f_t^\top =\bI_r$ for $t\in[T]$ and there exists some $\delta_F=o\brac{1}$ and $\eta_F=o(1)$ such that 
	\begin{align*}
		\PP\brac{\aop{\frac{1}{T}\sum_{t=1}^T\f_{t}\f^\top _{t}-\EE\brac{\frac{1}{T}\sum_{t=1}^T\f_{t}\f^\top _{t}}}\ge \delta_F}\le \eta_{F}.
	\end{align*}
Moreover, $\bL^\top \bL=\bLambda^2=\textsf{diag}\brac{\lambda_1^2,\cdots,\lambda_r^2}$, where $\lambda_1\ge \lambda_2\ge \cdots\ge \lambda_r>0$.
\end{assumption}  
For a deterministic factor matrix $\bF$, the first two sentences of Assumption~\ref{assump:iden} are instead understood as requiring $\max_{t\in[T]}\op{\f_t}\lesssim\sqrt{r+\log n}$ and $\op{T^{-1}\bF^\top\bF-\bI_r}\le\delta_F=o(1)$ with $\eta_F=0$.
Under \Cref{assump:iden}, $\bF^\top\bF/T\rightarrow_p\bI_r$, and the loadings $\bL^\top\bL$ grow in accordance with the factor strengths $\lambda_i^2$s. Similar assumption is commonly adopted as an identifiability condition \citep[see, e.g.,][]{fan2024can}. Throughout the paper, we allow $\lambda_i^2$s to vary from vanishing order to the order of $N$ which covers most practical scenarios.

The weighted PCA is simple: apply PCA to the weighted quadratic form \(\bX\bQ\bX^\top\), where \(\bQ\in\mathbb R^{T\times T}\) is a symmetric weight matrix normalized so that \(\|\bQ\|=1\), and then recover the loadings and factors from the resulting rank-\(r\) SVD. See \Cref{alg:main}. 

To gain intuition behind WPCA, recall that
\begin{align}\label{model:quadrac-form}
\bX\bQ\bX^\top=\bL\bF^\top\bQ\bF\bL^\top+\bN,
\end{align}
where $\bN:=\bM\bQ\bE^\top+\bE\bQ\bM^\top+\bE\bQ\bE^\top$. 
We shall restrict attention, without loss of generality, to symmetric weight matrices whose operator norm, i.e., the largest singular value, is $1$ unless otherwise indicated.

\begin{algorithm}[!tbp]
	\footnotesize
	\caption{Weighted PCA}\label{alg:main}
	\KwData{ Data matrix $\bX$, weight matrix $\bQ\in\RR^{T\times T}$,  rank $r$. }
	Compute  the leading-$r$ eigenvectors $\wt\bU_\bQ\in\RR^{N\times r}$  of $\bX\bQ\bX^\top $. 
	
	Compute  the rank-$r$ SVD $(\hat\bU_\bQ,\hat\bSigma_\bQ,\hat\bV_\bQ)$  of  $T^{-1/2}\wt\bU_\bQ\wt\bU_\bQ^\top \bX$.

	\KwResult{$\hat\bL_\bQ:=\hat\bU_\bQ\hat\bSigma_\bQ$ as the estimator of factor loadings, $\hat\bF_\bQ:=\sqrt{T}\hat\bV_\bQ$ as the estimator of factors.}
\end{algorithm}
Clearly the usual PCA amounts to the choice of $\bQ=\bI_T$. To better understand the benefit of a different weight matrix $\bQ$ in \eqref{model:quadrac-form}, denote $\brac{\bU,\bSigma,\bV}$ as the rank-$r$ SVD of $T^{-1/2}\bM$.
Provided that $\inf_{\zeta\ge0}\op{\bN-\zeta\bI_N}\le c\lambda_r^2\cdot\sigma_r\brac{\bF^\top\bQ\bF}$ for a sufficiently small universal constant $c>0$, following the explicit representation formula of the empirical spectral projectors from \cite{xia2021normal}, we obtain
\begin{align}\label{eq:UQ-bound-gen}
	\op{\hat\bU_\bQ\hat\bU_\bQ^\top -\bU\bU^\top }\lesssim\frac{\inf_{\zeta\in\RR}\op{\bU^\top \brac{\bN-\zeta\bI_N}\bU_\perp}}{\lambda_r^2\cdot \sigma_r\brac{\bF^\top\bQ\bF } }\wedge 1.
\end{align}
where the quantities $\sigma_r\brac{\bF^\top\bQ\bF }$ and $\inf_{\zeta\in\RR}\op{\bU^\top \brac{\bN-\zeta\bI_N}\bU_\perp}$ can be interpreted as the signal strength and the noise level, respectively, for estimating the singular subspace $\bU$. Note the bound above is sharp in that it cannot be improved in general. 
Thus the SNR condition (and hence the quality of $\hat\bU_{\bQ}$) depends on how well $\bQ$ aligns with both the factor matrix $\bF$ and the noise $\bE$. 

Equation \eqref{eq:UQ-bound-gen} explains the role of weighting in Example 1.1. First, since $\bU^\top \bI_N \bU_\perp = 0$, the scalar shift $\zeta \bI_N$ does not affect the projected perturbation term in
\eqref{eq:UQ-bound-gen}. Second, in the separable-noise case $\textsf{Cov}[\textsf{vec}(\bE)]=\bSigma_{\rm C}\otimes\bSigma_{\rm T}$, the
bias term below becomes
\[
\|\bU^\top \PTr_T\{\textsf{Cov}[\textsf{vec}(\bE)](\bI_N\otimes \bQ)\}\bU_\perp\|
=
|\Tr(\bSigma_{\rm T}\bQ)|\,\|\bU^\top\bSigma_C\bU_\perp\|.
\]
Hence, for ordinary PCA ($\bQ=\bI_T$), this term is typically of order
$|\Tr(\bSigma_{\rm T})|$, whereas in Example 1.1, where $\bSigma_{\rm T}=\bI_T$ and
$\bQ=\sum_{t=2}^T (\e_t \e_{t-1}^\top + \e_{t-1}\e_t^\top)$,
we have $\Tr(\bQ)=0$, so the leading bias vanishes. At the same time,
$\bF^\top \bQ\bF$ captures the lag-$1$ factor autocovariance and can remain of the same order as
$\bF^\top \bF$ under the certain dynamic structure. This is precisely the gain brought by $\bQ$ relative to the
naive choice $\bQ=\bI_T$. {For the estimation and inference results in \Cref{sec:main,sec:refined}, we assume that $\bF^\top\bQ\bF$ is positive definite.} We shall now consider this in more detail.

\subsection{Error Bounds for Subspace Estimation}
We begin with the general covariance structure. 

\begin{theorem}\label{thm:U-error-gen}
Suppose that \Cref{assump:iden} holds. {If
$\textsf{vec}\brac{\bE}=\bSigma_{\rm e}^{1/2}\g$ where $\g\in\RR^{NT}$ has independent entries with $\EE g_i=0$, $\EE g_i^2=1$ and $\|g_i\|_{\psi_2}\le K$,}
then
there exist some universal constants $c_0,C_0>0$ such that if $\lambda_r^2\cdot \sigma_r\brac{\bF^\top\bQ\bF }\ge C_0 \sqbrac{\textsf{GErr}_U(\bQ)\allowbreak+\inf_{\zeta\ge0}\op{\textsf{PTr}_{T}\ebrac{\bSigma_{\rm e}(\bI_N\otimes \bQ)}-\zeta\bI_N}}$, then with probability at least $1-\eta_F-O\brac{e^{-c_0N}}$,
	\begin{align*}
	\op{\hat\bU_\bQ\hat\bU_\bQ^\top -\bU\bU^\top }\lesssim \frac{\textsf{GErr}_{U}(\bQ)}{\lambda_r^2\cdot \sigma_r\brac{\bF^\top\bQ\bF }},
\end{align*} 
where {\small $\textsf{GErr}_{U}(\bQ):=K\lambda_1\sqrt{NT}\op{\bSigma_{\rm e}^{1/2}}+K^2\bbrac{\fro{\bQ}+\sqrt{N}}\sqrt{N}\op{\bSigma_{\rm e}}+\op{\bU^\top\textsf{PTr}_{T}\ebrac{\bSigma_{\rm e}(\bI_N\otimes \bQ)}\bU_\perp}$}.
\end{theorem}
\Cref{thm:U-error-gen} indicates the estimation error of $\hat{\bU}_{\bQ}$ is determined by the signal term $\lambda_{r}^{2}\cdot \sigma_r\brac{\bF^\top\bQ\bF }$ and  $\textsf{GErr}_{U}(\bQ)$. More specifically, the first two terms of 
$\textsf{GErr}_{U}(\bQ)$ can be understood as the first and second order terms of the variance, and the third term represents the bias. In particular, if the noise is isotropic, i.e., $\bSigma_{\rm e}=\sigma^{2}\bI_{NT}$, then the bias term vanishes. For general $\bSigma_{\rm e}$, the bias   $\op{\bU^\top\textsf{PTr}_{T}\ebrac{\bSigma_{\rm e}(\bI_N\otimes \bQ)}\bU_\perp}$ can be reduced depending on the alignment between $\bSigma_{\rm e}$ and $\bQ$.

For illustrative purposes, consider the case when $\bSigma_{\rm e}$ is decomposable in that  $\bSigma_{\rm e}=\bSigma_{\rm C}\otimes \bSigma_{\rm T}$ where  $\bSigma_{\rm C}\in\RR^{N\times N}$ and $\bSigma_{\rm T}\in \RR^{T\times T}$ are symmetric positive definite matrices. Formally, we write $\bE\sim \textsf{subG}_{K}(\bSigma_{\rm C},\bSigma_{\rm T})$ if $\bE=\bSigma_{\rm C}^{1/2} \bG \bSigma_{\rm T}^{1/2}$ where $\bG\in\mathbb R^{N\times T}$ has independent entries with
$\mathbb E G_{it}=0$, $\mathbb E G_{it}^2=1$, and $\|G_{it}\|_{\psi_2}\le K$. It is not hard to see that, for any $\bQ\in\RR^{T\times T}$, $\bU^\top\textsf{PTr}_{T}\ebrac{\bSigma_{\rm e}(\bI_N\otimes \bQ)}\bU_\perp
=\textsf{Tr}(\bSigma_{\rm T}\bQ)\bU^\top\bSigma_{\rm C}\bU_\perp$, and hence
\[
\op{\bU^\top\textsf{PTr}_{T}\ebrac{\bSigma_{\rm e}(\bI_N\otimes \bQ)}\bU_\perp}
=\ab{\textsf{Tr}(\bSigma_{\rm T}\bQ)}\op{\bU^\top\bSigma_{\rm C}\bU_\perp}.
\]  
In particular, the bias vanishes when $\textsf{Tr}(\bSigma_{\rm T}\bQ)=0$ or $\bSigma_{\rm C}$ is proportional to the identity. Following Theorem \ref{thm:U-error-gen}, we get

\begin{theorem}\label{thm:U-error}
	Suppose that \Cref{assump:iden} holds. {If $\bE\sim \textsf{subG}_{K}(\bSigma_{\rm C},\bSigma_{\rm T})$,} then there exist some universal constants $c_0,C_0>0$ such that if $\lambda_r^2\cdot \sigma_r\brac{\bF^\top\bQ\bF }\ge C_0 \sqbrac{\textsf{Err}_U(\bQ)\allowbreak+\inf_{\zeta\ge0}\op{\textsf{Tr}\brac{\bSigma_{\rm T}\bQ}\bSigma_{\rm C}-\zeta\bI_N}}$, then with probability at least $1-\eta_F-O\brac{e^{-c_0N}}$,
		\begin{align*}
			\op{\hat\bU_\bQ\hat\bU_\bQ^\top -\bU\bU^\top }\lesssim\frac{\textsf{Err}_{U}(\bQ)}{\lambda_r^2\cdot \sigma_r\brac{\bF^\top\bQ\bF }},
		\end{align*}
where $\textsf{Err}_{U}(\bQ):= K\lambda_1\sqrt{NT}
		\op{\bSigma_{\rm C}^{1/2}}\op{\bSigma_{\rm T}^{1/2}}+K^2\bbrac{\fro{\bQ}+\sqrt{N}}\sqrt{N}\op{\bSigma_{\rm C}}\op{\bSigma_{\rm T}}+\op{\bU^\top\bSigma_{\rm C}\bU_\perp}\ab{\textsf{Tr}\brac{\bSigma_{\rm T}\bQ}}$.
\end{theorem}

Let $\kappa:=\lambda_1/\lambda_r$. Consider the case when $\op{\bSigma_{\rm C}},\op{\bSigma_{\rm T}}, \op{\bU^\top\bSigma_{\rm C}\bU_\perp}, \kappa, K\asymp 1$. Then the error bound of \Cref{thm:U-error} can be simplified:
	\begin{align}\label{eq:U-error-simple}
		\op{\hat\bU_\bQ\hat\bU_\bQ^\top -\bU\bU^\top }\lesssim \frac{\sqrt{NT}}{\lambda_r\cdot \sigma_r\brac{\bF^\top\bQ\bF }}+\frac{\sqrt{nN}}{\lambda_r^2\cdot \sigma_r\brac{\bF^\top\bQ\bF } }+\frac{\ab{\textsf{Tr}\brac{\bSigma_{\rm T}\bQ}}}{\lambda_r^2\cdot \sigma_r\brac{\bF^\top\bQ\bF } }.
	\end{align}
In the usual PCA, $\bQ=\bI_T$, the last term dominates whenever $T\gg \lambda_r^2N$. On the other hand, with appropriately chosen $\bQ$, it is possible to reduce the last term and thus yield improved estimates.

We can also derive similar error bounds for $\hat{\bV}_{\bQ}$:
	\begin{theorem}\label{thm:V-error}
		Suppose that \Cref{assump:iden} holds and $T= o\brac{e^{cN}}$ for some universal constant $c>0$. {If $\bE\sim \textsf{subG}_{K}(\bSigma_{\rm C},\bSigma_{\rm T})$,} then there exist  some universal constants $c_0,C_0>0$ such that if $\lambda_r\sqrt{T}\ge C_0 \textsf{Err}_V(\bQ)$, then with probability at least $1-\eta_F-O\brac{e^{-c_0N}+e^{-c_0T}+n^{-9}}$,
			\begin{align*}
			\bop{\hat\bV_\bQ\hat\bV_\bQ^\top -\bV\bV^\top }\lesssim\frac{\textsf{Err}_{V}(\bQ)}{\lambda_r\sqrt{T}},
			\end{align*}
where $\textsf{Err}_V(\bQ):=K\brac{\sqrt{T}+\sqrt{n}\bop{\hat\bU_\bQ\hat\bU_\bQ^\top -\bU\bU^\top }}\bop{\bSigma_{\rm T}^{1/2}}\bop{\bSigma_{\rm C}^{1/2}}+\lambda_1\sqrt{T}\bop{\hat\bU_\bQ\hat\bU_\bQ^\top -\bU\bU^\top }$.
	\end{theorem} 
Different from the usual PC estimator, the estimation error of $\hat\bV_\bQ$ relies on that of $\hat\bU_{\bQ}$ as $\hat\bV_\bQ$ is obtained from the projected data $\wt\bU_\bQ\wt\bU_\bQ^\top \bX$. An improved first-stage estimate $\hat\bU_\bQ$ can carry through to the corresponding right-singular-space estimate. However, the bottleneck of estimating $\bV$, say,  in $\op{\cdot}$,  has a leading term of order $\lambda_r^{-1}\bop{\bSigma_{\rm T}^{1/2}}\bop{\bSigma_{\rm C}^{1/2}}$, hence we cannot  obtain an order improvement compared to the PC estimator. On the other hand, we also conduct numerical experiments to illustrate the superiority of our method in practice in \Cref{sec:num}.

\subsection{Error Bounds in $\ell_2\to\ell_\infty$ Norm }\label{subsec:two-to-inf-pertub}
To facilitate inferences about the factors and their loadings, it is often useful to derive error bounds in terms of $\ell_2\to\ell_\infty$ norm, i.e.,
$\infn{\hat\bU_\bQ\hat\bU_\bQ^\top -\bU\bU^\top }$ and $\infn{\hat\bV_\bQ\hat\bV_\bQ^\top -\bV\bV^\top }$. To this end, write
\begin{align}\label{eq:kappa-def}
	\kappa_{\rm C, 1}:=\bop{\bSigma_{\rm C}^{1/2}}_{1}\bop{\bSigma_{\rm C}^{-1/2}}_1, \quad\kappa_{\rm C, 2}:=\max_{i\in[N]}\bop{\bSigma_{\rm C}^{1/2}\e_i}\bop{\bSigma_{\rm C}^{-1/2}\e_i},\quad \kappa_{\rm T,1}:=\op{\bSigma_{\rm T}^{1/2}}_1\op{\bSigma_{\rm T}^{-1/2}}_1.
\end{align}
We shall also write $\rho:=\op{\bU^\top\bSigma_{\rm C}\bU_\perp}$ and  $\bar\rho:=\inf_{\zeta\in\RR}\op{\bSigma_{\rm C}-\zeta\bI_N}_{1}$.
Then we have
\begin{theorem}\label{thm:UV-error-inf}
	Suppose that Assumption \ref{assump:iden} holds and {$\bE\sim \textsf{subG}_{K}(\bSigma_{\rm C},\bSigma_{\rm T})$.} 
	There exist some universal constants $c_0,C_0>0$ such that if $\lambda_r^2\cdot \sigma_r\brac{\bF^\top\bQ\bF }\ge C_0 \overline{\textsf{Err}}_U(\bQ)$, then with probability at least $1-\eta_F-O\brac{e^{-c_0N}+n^{-9}}$,
		\begin{align*}
			\binfn{\hat\bU_\bQ\hat\bU_\bQ^\top -\bU\bU^\top }\lesssim \frac{\overline{\textsf{Err}}_U(\bQ) \infn{\bU}}{ \lambda_r^2\cdot \sigma_r\brac{\bF^\top\bQ\bF }},
		\end{align*}
where $\overline{\textsf{Err}}_U(\bQ):=\kappa_{\rm C, 1}\kappa_{\rm C, 2}\Big(\bar\rho\ab{\textsf{Tr}\brac{\bSigma_{\rm T}\bQ}}+K\binfn{\bU}^{-1}\lambda_1\sqrt{Tr}\log n\op{\bSigma_{\rm C}^{1/2}}_1\op{\bSigma_{\rm T}^{1/2}}
        + K^2\sqrt{nNr\log n}\allowbreak\op{\bSigma_{\rm C}^{1/2}}^2_{1}\op{\bSigma^{1/2}_{\rm T}}^2\Big)$. 
Moreover, there exist some universal constants $c_1, C_1>0$ such that if the signal condition of \Cref{thm:U-error} holds, $T\ge C_1\log n$, $T=o\brac{e^{c_1N}}$ and $\lambda_r\sqrt{T}>C_1 \overline{\textsf{Err}}_V(\bQ)$, then with probability at least $1-\eta_F-O\brac{n^{-9}}$,
\begin{align*}
 \infn{\hat\bV_\bQ\hat\bV_\bQ^\top-\bV\bV^\top}
 \lesssim\frac{\overline{\textsf{Err}}_V(\bQ)\infn{\bV}}{\lambda_r\sqrt{T}},
\end{align*}
where {\small $\overline{\textsf{Err}}_V(\bQ):=\lambda_1\sqrt{T}\op{\hat\bU_\bQ\hat\bU_\bQ^\top-\bU\bU^\top}+K\brac{\sqrt{T}+\sqrt{n}\op{\hat\bU_\bQ\hat\bU_\bQ^\top-\bU\bU^\top}}\op{\bSigma_{\rm T}^{1/2}}\op{\bSigma_{\rm C}^{1/2}}+\allowbreak K\infn{\bV}^{-1}\sqrt{r+\log n}\op{\bSigma_{\rm T}^{1/2}}_1\op{\bSigma_{\rm C}^{1/2}}+\allowbreak \sqbrac{\lambda_r\sigma_r(\bF^\top\bQ\bF)}^{-1}K^2\kappa \kappa_{\rm T,1}N\sqrt T\allowbreak\op{\bSigma_{\rm T}^{1/2}\bQ\bSigma_{\rm T}^{1/2}}_1\op{\bSigma_{\rm C}^{1/2}}^2$}.
\end{theorem}

To fix ideas, consider in particular the scenario where $\infn{\bU}\lesssim\sqrt{r/N}$, $\kappa_{\rm C, 1},\kappa_{\rm C, 2},K\asymp 1$, and $\sigma_r\brac{\bF^\top\bQ\bF }\gtrsim T$. Then the error bound of \Cref{thm:UV-error-inf} becomes
	{\small
\begin{align}\label{eq:U-infn-error-simple}
		\infn{\hat\bU_\bQ\hat\bU_\bQ^\top -\bU\bU^\top }\lesssim \frac{\op{\bSigma_{\rm C}^{1/2}}_1\op{\bSigma_{\rm T}^{1/2}}}{\lambda_r}\sqrt{\frac{\log^2 n}{T}}+\frac{\op{\bSigma_{\rm C}^{1/2}}^2_{1}\op{\bSigma^{1/2}_{\rm T}}^2}{\lambda_r^2 }\frac{\sqrt{n\log n}}{T}+\frac{\bar\rho\ab{\textsf{Tr}\brac{\bSigma_{\rm T}\bQ}}}{\lambda_r^2T\sqrt{N} },
	\end{align}
}
under the signal-to-noise requirement:
	\begin{align}\label{eq:U-infn-simple-cond}
		\lambda_r\gtrsim \op{\bSigma_{\rm C}^{1/2}}_1\op{\bSigma_{\rm T}^{1/2}}\brac{\frac{n}{T}}^{1/4}\brac{\frac{N}{T}}^{1/4}\log n+\frac{\bar\rho^{1/2} \ab{\textsf{Tr}\brac{\bSigma_{\rm T}\bQ}}^{1/2}}{T^{1/2}}.
	\end{align}
Similar to before, the last term of the error bound \Cref{eq:U-infn-error-simple} is the bias term induced by the heteroskedasticity of $\bSigma_{\rm C}$. When $T\lesssim N$,  it is dominated by the first two terms under the SNR condition \eqref{eq:U-infn-simple-cond}. It is also helpful to compare with some of the earlier results. 

\cite{fan2024can} studied the PC estimator for weak factor model with the additional assumption  that $\bSigma_{\rm T}=\bI_T$. In particular, under the above setting, their Lemma 4 and Lemma 7 therein entail that,  with probability at least $1-O\brac{n^{-2}}$,
	\begin{align}\label{eq:U-PCA-infn-error-simple}
		\infn{\hat\bU^{\rm PC}(\hat\bU^{\rm PC})^\top -\bU\bU^\top }\lesssim \frac{\op{\bSigma^{1/2}_{\rm C}}_{1}}{\lambda_r}\sqrt{\frac{\log^2 n}{T}}+\frac{\op{\bSigma^{1/2}_{\rm C}}^2_{1}}{\lambda_r^2}\frac{n\log n}{T\sqrt{N}},
	\end{align}
	provided that 
	\begin{align}\label{eq:U-PCA-infn-simple-cond}
		\lambda_r\gtrsim\op{\bSigma_{\rm C}^{1/2}}_1\brac{\frac{n}{T}}^{1/2}\log^{1/2} n.
	\end{align}
It is not hard to see that this coincides with our bound \eqref{eq:U-PCA-infn-error-simple} by taking $\bQ=\bI_T$. There is, however, a subtle difference between their SNR requirement and ours: the two are equivalent only when $T\lesssim N$, and \eqref{eq:U-PCA-infn-simple-cond} is more stringent  when $T\gg N$.

We can also compare the bound \eqref{eq:U-PCA-infn-error-simple} with those for HeteroPCA where $\bSigma_{\rm C}$ is assumed to be diagonal. To facilitate comparison, we consider $\bSigma_{\rm T}=\textsf{diag}\bbrac{\sigma_{{\rm T},1}^2,\cdots, \sigma_{{\rm T},T}^2}$ and $\bSigma_{\rm C}=\textsf{diag}\bbrac{\sigma_{{\rm C},1}^2,\cdots, \sigma_{{\rm C},N}^2}$. Denote by $\sigma_{\sf max}^2:=\bop{\bSigma_{\rm C}^{1/2}}\bop{\bSigma_{\rm T}^{1/2}}$ and $\hat\bU^{\rm HPC}$ the HeteroPCA estimate of the left singular space. Then the results from \cite{yan2021inference} and \cite{agterberg2022entrywise} imply that with high probability,
		\begin{align}\label{eq:U-heteroPCA-infn-error-simple}
			&\infn{\hat\bU^{\rm HPC}(\hat\bU^{\rm HPC})^\top-\bU\bU^\top }\lesssim \frac{\sigma_{\sf max}^2}{\lambda_r}\sqrt{\frac{\log n}{T}}+\frac{\sigma_{\sf max}}{\lambda_r^2}\sqrt{\frac{1}{T}}\log n,
		\end{align}
		provided that $\lambda_r\gtrsim  \brac{\dfrac{n}{T}}^{1/4}\brac{\dfrac{N}{T}}^{1/4}\sigma_{\sf max}\log^{1/2}n$.
Note that our bound \eqref{eq:U-infn-error-simple} and \eqref{eq:U-heteroPCA-infn-error-simple} coincide under this particular setting. Yet our proposed weighted PCA approach is more generally applicable and allows for both cross-sectional and temporal dependence beyond heteroskedasticity. Indeed, our numerical experiments in \Cref{sec:num} suggest that the HeteroPCA might deteriorate significantly when there is cross-sectional dependence, whereas the weighted PCA remains unaffected.

\section{Factor Analysis: Estimation and Inference }\label{sec:refined}
We now turn our attention to the estimation error  and asymptotic normality of the estimated factors and loadings.

\subsection{Error Bounds for Factors and Loadings}\label{subsec:factor-est-err}

To start with, let $\bH_{U,\bQ}:=\hat\bU^\top _\bQ\bU$, $\bH_{V,\bQ}:=\hat\bV^\top _\bQ\bV$ and  $\bR_{U,\bQ}:=\textsf{sign}\bbrac{\hat\bU^\top _\bQ\bU}$, $\bR_{V,\bQ}:=\textsf{sign}\bbrac{\hat\bV^\top _\bQ\bV}$. In addition, let $\bB\in\RR^{r\times r}$ be the invertible matrix obeying $\bV=T^{-1/2}\bF\bB$ and define $\bR_{L,\bQ}:=\brac{\bB^{-1}}^{\top }\bR_{V,\bQ}^\top $ and $\bR_{F,\bQ}:=\bB\bR^\top_{V,\bQ}$.  Then we get
\begin{theorem}\label{thm:factor-bound}
    Suppose that Assumption \ref{assump:iden} holds and {$\bE\sim \textsf{subG}_{K}(\bSigma_{\rm C},\bSigma_{\rm T})$.} There exist some universal constants $c_0,C_0>0$ such that if $\lambda_r^2\cdot \sigma_r\brac{\bF^\top\bQ\bF }\ge C_0 \sqbrac{\textsf{Err}_U(\bQ)+\inf_{\zeta\ge0}\op{\textsf{Tr}\brac{\bSigma_{\rm T}\bQ}\bSigma_{\rm C}-\zeta\bI_N}}$, then with probability at least $1-\eta_F-O\bbrac{e^{-c_0N}+e^{-c_0T}+ n ^{-9}}$, 
    \begin{align*}
       \frac{1}{\sqrt{N}}\op{\hat\bL_\bQ-\bL\bR_{L,\bQ}}
       &\lesssim\frac{\lambda_1+K\op{\bSigma_{\rm C}^{1/2}}\op{\bSigma_{\rm T}^{1/2}}}{\sqrt{N}}
       \brac{\op{\hat\bU_\bQ\hat\bU_\bQ^\top-\bU\bU^\top}
       +\op{\hat\bV_\bQ\hat\bV_\bQ^\top-\bV\bV^\top}^2}\\
       &\qquad+K \op{\bSigma_{\rm C}^{1/2}}\op{\bSigma_{\rm T}^{1/2}}\sqrt{\frac{r+\log n}{NT}}
      ,
    \end{align*}
    and
    $$
        \frac{1}{\sqrt{T}}\op{\hat\bF_\bQ-\bF\bR_{F,\bQ}}\lesssim\op{\hat\bU_\bQ\hat\bU_\bQ^\top -\bU\bU^\top }\brac{\kappa +\frac{K\op{\bSigma^{1/2}_{\rm T}}\op{\bSigma^{1/2}_{\rm C}}}{\lambda_r}\sqrt{\frac{n}{ T}}}+{\frac{K\op{\bSigma^{1/2}_{\rm T}}\op{\bSigma^{1/2}_{\rm C}}}{\lambda_r }}.
    $$
\end{theorem}

As an immediate consequence of  \Cref{thm:U-error}, \Cref{thm:V-error} and \Cref{thm:factor-bound}, we can obtain the estimation error for $\bL$ and $\bF$.  It is worth noting that $\lambda_r=\omega(1)$  is not necessary for consistency. Nonetheless, to simplify narratives, let us consider the regime $T\gtrsim N$,  $\op{\hat\bU_\bQ\hat\bU_\bQ^\top -\bU\bU^\top }=o(1)$, $\sigma_r\brac{\bF^\top\bQ\bF}\gtrsim T$, $\lambda_r=\omega(1)$, $\kappa, K, \op{\bSigma_{\rm T}},\op{\bSigma_{\rm C}}\asymp1$. Then under the assumptions of \Cref{thm:U-error} and  \Cref{thm:V-error} we have
	\begin{align}\label{eq:factor-simple-bound}
		&\frac{1}{\sqrt{N}}\op{\hat\bL_\bQ-\bL\bR_{L,\bQ}}=O_p\brac{\frac{1}{\sqrt{T}}\brac{1+\frac{\op{\bU^\top\bSigma_{\rm C}\bU_\perp}\ab{\textsf{Tr}\brac{\bSigma_{\rm T}\bQ}}}{\lambda_r\sqrt{NT}}}{+\frac{1}{\lambda_r\sqrt N}}},\notag\\
		&\frac{1}{\sqrt{T}}\op{\hat\bF_\bQ-\bF\bR_{F,\bQ}}=O_p\brac{\frac{1}{\lambda_r}\brac{1+\frac{\op{\bU^\top\bSigma_{\rm C}\bU_\perp}\ab{\textsf{Tr}\brac{\bSigma_{\rm T}\bQ}}}{\lambda_rT}}}.
	\end{align}
If either $\op{\bU^\top\bSigma_{\rm C}\bU_\perp}=0$ or $\bQ$ is chosen so that $\textsf{Tr}(\bSigma_{\rm T}\bQ)=0$,  then the bias term in \eqref{eq:factor-simple-bound} vanishes. Then, the factors converge at  rate $\lambda_r ^{-1}$, which matches the result in \cite{fan2024can}.  {For loadings, the convergence rate is $T^{-1/2}+\lambda_r^{-1}N^{-1/2}$. Note that the second term $\lambda_r^{-1}N^{-1/2}$ arises from recovering the loading scale in the second-stage SVD and is also present for the usual PC estimator \citep{fan2024can}. The gain from weighting appears through the reduction of the first-stage bias. More generally, beyond the bounded-covariance simplification above, relative to $\bQ=\bI_T$, this bias is reduced according to the ratio $\brac{\ab{\textsf{Tr}(\bSigma_{\rm T}\bQ)}/\sigma_r(\bF^\top\bQ\bF)}/\brac{\textsf{Tr}(\bSigma_{\rm T})/\sigma_r(\bF^\top\bF)}$. Thus a weight satisfying $\sigma_r(\bF^\top\bQ\bF)\asymp\sigma_r(\bF^\top\bF)$ and $\ab{\textsf{Tr}(\bSigma_{\rm T}\bQ)}=o\brac{\textsf{Tr}(\bSigma_{\rm T})}$ reduces this component by an asymptotic order. This reduction can lead to a smaller overall loading error when the first-stage terms are the leading terms.}

Similarly, we can also relate the perturbation bound in \Cref{subsec:two-to-inf-pertub} to derive the  estimation error in $\ell_2\to\ell_\infty$ norm for factors and  loadings.
	\begin{theorem}\label{thm:factor-bound-inf}
			Suppose Assumption \ref{assump:iden} holds and {$\bE\sim \textsf{subG}_{K}(\bSigma_{\rm C},\bSigma_{\rm T})$.} There exist some universal constants $c_0,C_0>0$ such that if $\lambda_r^2\cdot \sigma_r\brac{\bF^\top\bQ\bF }\ge C_0 \sqbrac{\textsf{Err}_U(\bQ)+\inf_{\zeta\ge0}\op{\textsf{Tr}\brac{\bSigma_{\rm T}\bQ}\bSigma_{\rm C}-\zeta\bI_N}}$, then with probability at least $1-\eta_F-O\bbrac{e^{-c_0N}+ n ^{-9}}$,
\begin{align*}
\infn{\hat\bL_\bQ-\bL\bR_{L,\bQ}}
&\lesssim\brac{\lambda_1+K\op{\bSigma_{\rm C}^{1/2}}\op{\bSigma_{\rm T}^{1/2}}}
\biggl[\infn{\hat\bU_\bQ\hat\bU_\bQ^\top-\bU\bU^\top}\\
&\hspace{13mm}+\infn{\bL\bLambda^{-1}}
\brac{\op{\hat\bU_\bQ\hat\bU_\bQ^\top-\bU\bU^\top}^2
+\op{\hat\bV_\bQ\hat\bV_\bQ^\top-\bV\bV^\top}^2}\biggr]\\
&\quad +K\infn{\bL\bLambda^{-1}}\sqrt{\frac{r+\log n}{T}}
\op{\bU^\top\bSigma_{\rm C}^{1/2}}\op{\bSigma_{\rm T}^{1/2}\bV},
\end{align*}
and
$$\infn{\hat\bF_\bQ-\bF\bR_{F,\bQ}}\lesssim \infn{\hat\bV_\bQ\hat\bV_\bQ^\top -\bV\bV^\top }\sqrt{T}.
$$
	\end{theorem}
	For simplification, consider the case when $T\gtrsim N$,  $K,\kappa,\bop{ \bSigma_{\rm C}}, \op{ \bSigma_{\rm T}}\asymp	1$, $\lambda_r=\omega(1)$ and $\infn{\hat\bU_\bQ\hat\bU_\bQ^\top -\bU\bU^\top }\lesssim\bop{\hat\bU_\bQ\hat\bU_\bQ^\top -\bU\bU^\top }\infn{\bL\bLambda^{-1}}$. Then $\lambda_1+K\op{\bSigma_{\rm C}^{1/2}}\op{\bSigma_{\rm T}^{1/2}}\lesssim\lambda_1$, and the bounds in  \Cref{thm:factor-bound-inf} reduce to  
		$$\infn{\hat\bL_\bQ-\bL\bR_{L,\bQ}}\lesssim \infn{\bL\bLambda^{-1}}\sqbrac{\lambda_1\op{\hat\bU_\bQ\hat\bU_\bQ^\top -\bU\bU^\top }{+\sqrt{\frac{r+\log n}{T}}+\frac{1}{\lambda_r}}}.$$
{After appropriate scaling, the first term in the loading bound is of the same order as the subspace estimation error, while the remaining two terms arise from recovering the loadings from the projected data.} More importantly, \Cref{thm:factor-bound-inf} provides a basis for us to develop an inferential theory for factors and loadings.

	\subsection{Asymptotic Normality}\label{subsec:inf}
	We are now in a position to establish the asymptotic normality of our factor and loading estimators. To this end,
	denote by $\bar\bU$ the leading singular vectors of $\bM\bQ\bM^\top $, and $\bar\bSigma$ the  diagonal matrix containing singular values of $\bM\bQ\bM^\top$. Note that there exists some  $\bar\bO\in\OO_{r}$ such that $\bar\bU\bar\bO=\bU$. The following theorem presents the asymptotic normality for  our estimate of the loadings.
	\begin{theorem}\label{thm:inf-L}
		Suppose that Assumption \ref{assump:iden} holds and {$\bE\sim \textsf{subG}_{K}(\bSigma_{\rm C},\bSigma_{\rm T})$.} In addition, assume that {$r$ is fixed,} $\infn{\bL\bLambda^{-1}}\lesssim\sqrt{r/N}$, $\sigma_r\brac{\bF^\top\bQ\bF }\gtrsim T$ and $\min_{i\in[N]}\sqbrac{\bSigma_{\rm C}}_{i,i}\sigma_r\brac{\bV^\top\bQ\bSigma_{\rm T}\bQ\bV}\gtrsim 1$, $N=\omega\big(K^2\kappa^4r\brac{r+\log n}\bop{\bSigma_{\rm C}^{1/2}}^2\allowbreak\bop{\bSigma_{\rm T }^{1/2}}^2\big)$, $\kappa\kappa_{\rm T}\infn{\bSigma_{\rm T}^{1/2}\bQ\bV}/\op{\bSigma_{\rm T}^{1/2}}=o(1)$ with $\kappa_{\rm T}:=\op{\bSigma_{\rm T}^{1/2}}\op{\bSigma_{\rm T}^{-1/2}}$,  and
		$$\begin{aligned}
		\lambda_r\gg&K\brac{\frac{n}{T}}^{1/2}\kappa^5\kappa_{\rm C, 1}^2\kappa_{\rm C, 2}^2r^{1/2}\log^2 n\op{\bSigma_{\rm C}^{1/2}}_1^2\op{\bSigma_{\rm T}^{1/2}}^2+K\frac{\ab{\textsf{Tr}\brac{\bSigma_{\rm T}\bQ}}}{N^{1/2}T^{1/2}}\bar\rho \kappa r^{1/2}\\
		&{+K^2\brac{\frac{T}{N}}^{1/2}\kappa^3r^{1/2}\op{\bSigma_{\rm C}^{1/2}}^2\op{\bSigma_{\rm T}^{1/2}}^2}.
		\end{aligned}$$
		Then for each $i\in[N]$, we  have
		\begin{align*}
		{\bSigma_{L,i}^{-1/2}\brac{\hat\bL_\bQ-\bL\bR_{L,\bQ}}_{i,\cdot}^\top \overset{d}{\rightarrow}\calN\brac{\mathbf0,\bI_r}},
		\end{align*}
		where $\bSigma_{L,i}:=T \sqbrac{\bSigma_{\rm C}}_{i,i}\bO_F^\top \bSigma\bV^\top \bQ\bSigma_{\rm T}\bQ\bV\bSigma\bO_F$ and $\bO_F:=\bar\bO^\top \bar\bSigma^{-1}\bar\bO\bSigma\bR_{V,\bQ}^\top.$
		{Here $\bSigma_{L,i}^{-1/2}$ denotes the compatible (not necessarily symmetric) whitening factor chosen so that $\bSigma_{L,i}^{-1/2}\bO_F^\top=\brac{T\sqbrac{\bSigma_{\rm C}}_{i,i}\bSigma\bV^\top\bQ\bSigma_{\rm T}\bQ\bV\bSigma}^{-1/2}$, where the inverse square root on the right is the symmetric one. In particular, $(\bSigma_{L,i}^{-1/2})^\top\bSigma_{L,i}^{-1/2}=\bSigma_{L,i}^{-1}$.}
	\end{theorem}
Note that $\lambda_r\gg 1$ is not required for the consistency of $\hat\bU_\bQ$ when $T\gg N$. But a diverging $\lambda_r$ is required by our asymptotic-normality result. Similar observations have been made earlier for PCA \citep{onatski2012asymptotics, choi2024high}.  {\Cref{thm:inf-L} separates the first-stage bias affected by $\bQ$ from the terms arising in the second-stage loading recovery. To fix ideas, consider the setting when $\kappa_{\rm C, 1},\kappa_{\rm C,2},\kappa,r,\op{\bSigma_{\rm C}^{1/2}}_1,\op{\bSigma_{\rm T}^{1/2}}\asymp 1$ and ignore logarithmic terms. The part of the SNR condition that is common to weighted and ordinary PCA is}
	\begin{align}\label{eq:inf-L-cond-simple}
		\lambda_r\gg \brac{\frac{n}{T}}^{1/2}+\brac{\frac{T}{N}}^{1/2}.
	\end{align}
{For ordinary PCA, $\bQ=\bI_T$, the trace term in \Cref{thm:inf-L} can contribute an additional first-stage bias. An appropriate weight can reduce this contribution while leaving the common baseline in \eqref{eq:inf-L-cond-simple} unchanged. More generally, provided that both choices satisfy the remaining signal and nondegeneracy requirements, the sufficient SNR condition is of smaller order when the PC trace term $K\kappa r^{1/2}\bar\rho\textsf{Tr}(\bSigma_{\rm T})/\sqrt{NT}$ dominates all other terms on the right-hand side of \Cref{thm:inf-L}, whereas $K\kappa r^{1/2}\bar\rho\ab{\textsf{Tr}(\bSigma_{\rm T}\bQ)}/\sqrt{NT}$ does not. A detailed rank-one Gaussian example in \Cref{pf-eq:inf-counterexmp} of the Supplementary Material (see \eqref{eq:inf-counterexmp}) shows that, under cross-sectional heteroskedasticity, the ordinary PC loading expansion can contain a first-stage bias of order $\sigma_{\rm C}^2\lambda^{-1}\sqrt{T/N}$. This illustrates that the trace-induced term in \Cref{thm:inf-L} can be active.}

\begin{figure}[!t]
	\centering
	\includegraphics[width=0.49\textwidth]{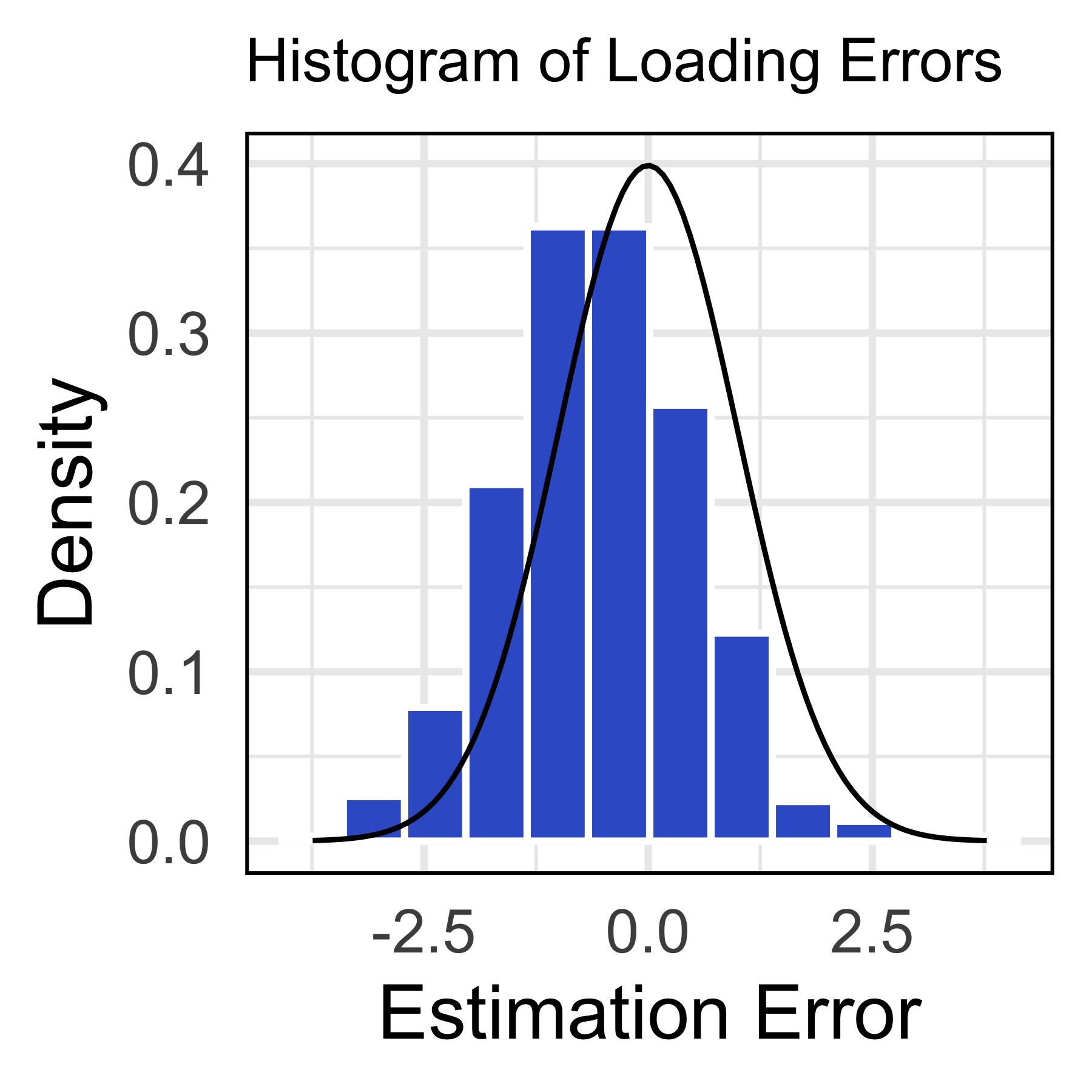}
	\includegraphics[width=0.49\textwidth]{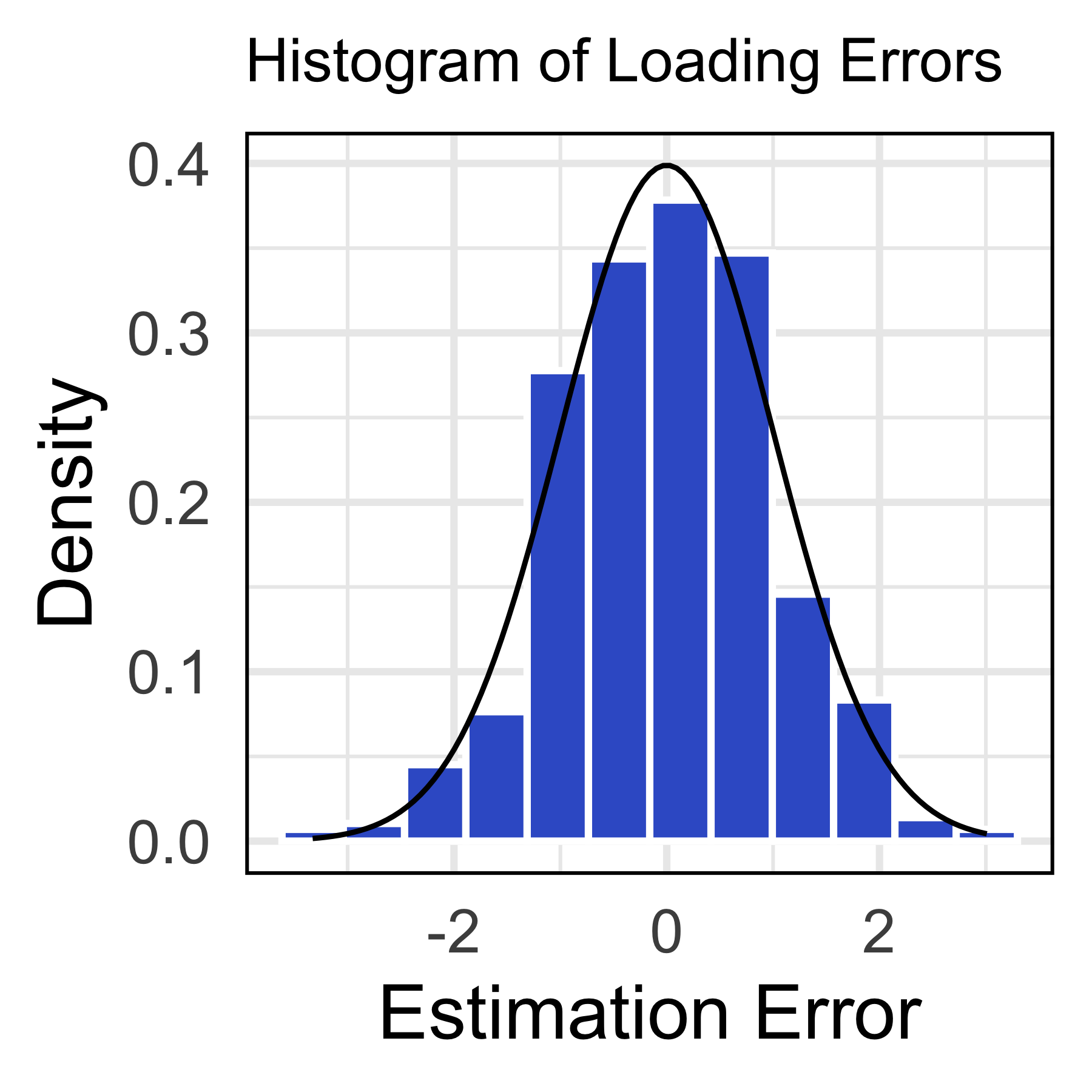}
		\caption{Distribution of a covariance-normalized coordinate of the PC estimator $\hat\bL_{\bI}-\bL\bR_{L,\bI}$ (Left) and the weighted PC estimator $\hat\bL_{\bQ}-\bL\bR_{L,\bQ}$ (Right) with $N=100$, $T=800$. See \Cref{sec:num} for details of the setting.}
	\label{fig:sim-inf-L-bias}
\end{figure}

To illustrate the same bias-reduction mechanism, we conducted a simulation study with $N=100$ and $T=800$. \Cref{fig:sim-inf-L-bias} shows the histogram of the estimation error for both estimates. The ordinary PC distribution is shifted, whereas the weighted PC distribution is centered closer to zero.

Similarly, the following theorem establishes the asymptotic normality of the estimated factors.
	\begin{theorem}\label{thm:inf-F}
		Suppose Assumption \ref{assump:iden} holds and $\bE\sim \textsf{subG}_{K}(\bSigma_{\rm C},\bSigma_{\rm T})$. In addition, assume {that $r$ is fixed,} $T= o\brac{e^{cN}}$ for a sufficiently small universal $c>0$, $\infn{\bL\bLambda^{-1}}\lesssim\sqrt{r/N}$, $\sigma_r\brac{\bF^\top\bQ\bF }\gtrsim T$, $\min_{t\in[T]}\sqbrac{\bSigma_{\rm T}}_{t,t}\sigma_r\brac{\bU^\top\bSigma_{\rm C}\bU}\gtrsim1$, $T=\omega\brac{K^2\brac{r+\log n}^2\bop{\bSigma_{\rm C}^{1/2}}^2\bop{\bSigma_{\rm T}^{1/2}}^2}$, $\kappa_{\rm C, 1}\infn{\bSigma_{\rm C}^{1/2}\bU}/\op{\bSigma_{\rm C}^{1/2}}=o(1)$ and
{\small
\begin{align*}
 \lambda_r\gg{}&K^2\frac nT\kappa_{\rm C,1}\kappa_{\rm C,2}\kappa_{\rm T,1}\kappa^5r\log^3n
  \Big(\op{\bSigma_{\rm C}^{1/2}}^2\op{\bSigma_{\rm T}^{1/2}}_1^2
       +\op{\bSigma_{\rm C}^{1/2}}_1^2\op{\bSigma_{\rm T}^{1/2}}^2+\op{\bSigma_{\rm C}^{1/2}}^2
       \op{\bSigma_{\rm T}^{1/2}\bQ\bSigma_{\rm T}^{1/2}}_1\Big).
\end{align*}
}
Then for each $t\in[T]$, we  have
		\begin{align*}
			\bSigma_{F,t}^{-1/2}\brac{\hat\bF_\bQ-\bF\bR_{F,\bQ}}_{t,\cdot}^\top \overset{d}{\rightarrow}\calN\brac{\mathbf0,\bI_r},
		\end{align*}
		where $\bSigma_{F,t}:=\sqbrac{\bSigma_{\rm T}}_{t,t}\bR_{V,\bQ}\bSigma^{-1}\bU^\top \bSigma_{\rm C}\bU\bSigma^{-1}\bR_{V,\bQ}^\top$.
		{Here $\bSigma_{F,t}^{-1/2}$ denotes the compatible (not necessarily symmetric) whitening factor chosen so that $\bSigma_{F,t}^{-1/2}\bR_{V,\bQ}\bSigma^{-1}=\brac{\sqbrac{\bSigma_{\rm T}}_{t,t}\bU^\top\bSigma_{\rm C}\bU}^{-1/2}$, where the inverse square root on the right is the symmetric one. In particular, $(\bSigma_{F,t}^{-1/2})^\top\bSigma_{F,t}^{-1/2}=\bSigma_{F,t}^{-1}$.}
	\end{theorem}
Note that the incoherence condition $\infn{\bL\bLambda^{-1}}$ is commonly adopted in the literature \citep[see, e.g.,][]{fan2024can, choi2024high}. 	In addition, \Cref{thm:inf-F} can be viewed as a generalization of the results from \cite{fan2024can} who considered the usual PCA estimate (i.e., $\bQ=\bI_T$) under the special case when $\bSigma_{\rm T}=\bI_T$.

\begin{remark}[How to interpret and use the CLTs]\label{rem:clt-interpretation}
The CLTs in Theorems~\ref{thm:inf-L}--\ref{thm:inf-F} are stated for the canonically aligned estimators \((\hat\bL_\bQ,\hat\bF_\bQ)\) (via \(\bR_{L,\bQ},\bR_{F,\bQ}\)) in order to make the limit law well-defined. This alignment is a statistical normalization and is not meant to impose an economic identification.

{A concrete rotation-invariant implication of the loading CLT concerns a fixed series $i$. Consider the row-zero null hypothesis $H_{0,i}:\l_i=\mathbf0$, which is equivalent to $\bM_{i,\cdot}=\mathbf0$. Under this null, the modeled low-rank common component for series $i$ is zero at every observed time point. Because the null sets all $r$ loading coordinates to zero, it is unchanged under every nonsingular factor reparameterization, which can be regarded as a  special case of a block-zero loading restriction. Such block-zero restrictions have been studied using PC-estimated loadings and applied to macroeconomic panels to examine multilevel loading structure \citep{han2019test}. Under $H_{0,i}$, $(\bL\bR_{L,\bQ})_{i,\cdot}=\mathbf0$, and \Cref{thm:inf-L} therefore gives
\[
\bSigma_{L,i}^{-1/2}(\hat\bL_\bQ)_{i,\cdot}^{\top}\overset{d}{\longrightarrow}\calN(\mathbf0,\bI_r).
\]
The compatible-whitening identity in that theorem and the continuous-mapping theorem yield
\[
(\hat\bL_\bQ)_{i,\cdot}\bSigma_{L,i}^{-1}(\hat\bL_\bQ)_{i,\cdot}^{\top}\overset{d}{\longrightarrow}\chi_r^2.
\]

For feasibility, as far as the noise covariances are concerned, $\bSigma_{L,i}$ involves $\bSigma_{\rm C}$ and $\bSigma_{\rm T}$ through the scalar $[\bSigma_{\rm C}]_{i,i}$ and the $r\times r$ projection $\bV^\top\bQ\bSigma_{\rm T}\bQ\bV$, rather than through the full covariance matrices. A feasible Wald test would follow by Slutsky's theorem if $\bSigma_{L,i}$ were estimated consistently in relative operator norm. Establishing such an estimator requires additional covariance structure; if $\bSigma_{\rm C}$ and $\bSigma_{\rm T}$ are estimated separately, a scale normalization is also needed. Possible settings include diagonal or sparse $\bSigma_{\rm C}$ and short-memory or parametrically structured $\bSigma_{\rm T}$, but sufficient conditions and a corresponding estimator are not developed here. We view this as an interesting direction for future work.}

\end{remark}

{Although the common component $M_{it}=\l_i^\top\f_t$ is rotation invariant, its limiting distribution is not an immediate consequence of the two marginal CLTs above. The loading and factor estimation errors are driven by the same noise matrix and are generally correlated, so one must derive their joint first-order representation, retain the loading--factor covariance term, and verify that the product of the two estimation errors is negligible. Under additional regularity conditions, \Cref{prop:common-component-clt} in the Supplementary Material provides this derivation and establishes a CLT for $\widehat M_{\bQ,it}=(\hat\bL_\bQ\hat\bF_\bQ^\top)_{it}$.}

\subsection{Examples}\label{sec:two-example}
To further illustrate the benefit of weighted PCA and the implication of our theoretical development, we now revisit the two motivating examples introduced in \Cref{sec:intro}. To fix ideas, we shall assume throughout this subsection that $\kappa,\kappa_{\rm C, 1},\kappa_{\rm C, 2},\op{\bSigma_{\rm C}}\asymp1$ for simplicity.

\paragraph{Multivariate Time Series Model.} 
Suppose the idiosyncratic noise $\bE$ is temporally uncorrelated and $\bF$ follows a multivariate autoregressive model that is independent of $\bE$:
\begin{align*}
    \f_t=\bA\f_{t-1}+\beps_{t},\qquad t=2,\cdots,T,
\end{align*}
where $\{\beps_t\}_{t=2}^T$ are independent, and are independent of $\f_1$, $\bI_r-\bA\bA^\top\succeq0$, and $\beps_t\sim \calN(0,\bI_r-\bA\bA^\top )$. In this scenario, a practical choice is to take $\bQ=\sum_t(\be_t\be_{t-1}^\top+\be_{t-1}\be_t^\top)$, which leverages the first-order auto-covariance information.

In light of the results above, we get \begin{corollary}\label{col:inf-L-exmp-ts}
		Suppose that Assumption \ref{assump:iden} holds, $\textsf{vec}\brac{\bE}\sim \calN \brac{0,\bSigma_{\rm C}\otimes \bI}$, $T=\omega\brac{r\brac{r+\log T}}$ and $\bar\bA:=(\bA+\bA^\top)/2$ is positive definite with $c\le\sigma_r(\bar\bA)\le\sigma_1(\bar\bA)\le 1-c$ for some constant $c>0$. 
\begin{itemize}
\item[(i)] If $\lambda_r=\omega\bbrac{\brac{{N}/{T}}^{1/2}\vee\brac{{N}/{T}}^{1/4}}$, then
	\begin{align}\label{eq:u-bound-ts}
		\op{\hat\bU_{\bQ}\hat\bU_{\bQ}^\top-\bU\bU^\top}=O_p\brac{\frac{1}{\lambda_r}\sqrt{\frac{N}{T}}\brac{1+{\lambda_r^{-1}}}}.
	\end{align}
\item[(ii)] If $\lambda_r=\omega\bbrac{1\vee\brac{{N}/{T}}^{1/2}}$, then 
			\begin{align*}
				\frac{1}{\sqrt{N}}\op{\hat\bL_\bQ-\bL\bR_{L,\bQ}}=O_p\brac{\frac{1}{\sqrt{T}}{+\frac{1}{\lambda_r\sqrt N}}}.
			\end{align*}
\item[(iii)] If \(r\) is fixed, $\infn{\bL\bLambda^{-1}}\lesssim\sqrt{r/N}$, $N\wedge T=\omega\brac{r\brac{r+\log n}}$ and $\lambda_r=\omega\bbrac{\brac{{n}/{T}}^{1/2}r^{1/2}\log^2 n\vee\allowbreak\brac{{T}/{N}}^{1/2}r^{1/2}}$, then for each $i\in[N]$,
		\begin{align*}
			\bSigma_{L,i}^{-1/2}\brac{\hat\bL_\bQ-\bL\bR_{L,\bQ}}_{i,\cdot}^\top \overset{d}{\rightarrow}\calN\brac{\mathbf0,\bI_r},
		\end{align*}
		where $\bSigma_{L,i}=\sqbrac{\bSigma_{\rm C}}_{i,i}\bO_F^\top \bU^\top \bL\sqbrac{2\bI_r+\brac{T-2}\sqbrac{2\bI_r+\bA^2+\brac{\bA^\top}^2}} \bL^\top\bU\bO_F$ and $\bO_F$ is as defined in \Cref{thm:inf-L}. Here $\bSigma_{L,i}^{-1/2}$ uses the compatible-whitening convention in \Cref{thm:inf-L}.
\end{itemize}
\end{corollary}

This setting connects our work with a fast growing literature in high dimensional time series analysis. See, e.g, \cite{lam2011estimation,gao2022modeling,qiao2025weight}.  In particular, \cite{lam2011estimation} and \cite{gao2022modeling} considered the auto-covariance-based method, whilst more recently \cite{qiao2025weight}  proposed a weight-calibrated estimation method. All their theoretical results  for estimating $\bU$ are restricted to the moderate-dimensional regime in that there exists some constant $\alpha\in(0,1]$ such that  $N^{1-\alpha}\ll T\lesssim N$ with the factor strength  $	\lambda_r^2\asymp N^{\alpha}$. They  established the rate of $O_p\bbrac{\sqrt{N^{1-\alpha}/{T}}}$, which is the same 
 rate entailed by our result under the same regime. However, our result can go beyond this regime, as we do not require the stringent dimension requirement $N^{1-\alpha}\ll T\lesssim N$. In fact, in terms of \eqref{eq:u-bound-ts} we only require $\lambda_r\gg \sqrt{{N}/{T}}\vee\brac{{N}/{T}}^{1/4}$ and hence $\lambda_r$ can go to $0$  if $T\gg N$. This captures a broader regime than previous work. Moreover, to our best knowledge, there is no asymptotic normality result similar to ours that exists in this line of work. We refer readers to \Cref{tab:wpca-compare} for a more detailed comparison with the existing literature. 
 
 \begin{remark}
    The fact that our estimation theory permits very weak, and in some regimes even vanishing, factor strengths should be interpreted as an identifiability and statistical possibility result, not as a blanket recommendation to retain arbitrarily weak factors in practice. When the incremental contribution of the last factor falls below the estimation noise level, including it may increase variance more than it reduces approximation bias. In such cases it is often preferable to work with an effective rank, selected by an eigenvalue-ratio or information criterion, and then apply WPCA on that choice. {Note that these rank-selection rules are heuristic here. All of our theoretical results condition on a correctly specified \(r\), and the asymptotic-normality results additionally treat \(r\) as fixed.} Our CLTs also require the stronger condition \(\lambda_r=\omega(1)\), in line with earlier inferential theory for weak factors.
 \end{remark}

\begin{table}[p]
\centering
\caption{Comparisons with other literature, stated in our notation.}
\label{tab:wpca-compare}
\renewcommand{\baselinestretch}{1.0}
\footnotesize
\setlength{\tabcolsep}{3.2pt}
\renewcommand{\arraystretch}{1.12}

\begin{tabular}{p{2.15cm} p{2.35cm} p{4.1cm} p{3.4cm} p{3.35cm}}
\hline
\textbf{Reference(s)} &
\textbf{Method} &
\textbf{(a) Signal strength and (b) noise structure} &
\textbf{Main comparable result} & \textbf{WPCA advantage} \\
\hline

\cite{onatski2012asymptotics} &
Usual PCA &
(a) $\lambda_r^2=\Omega(1)$;

(b) General $(\bSigma_{\rm C},\bSigma_{\rm T})$ &
PCA is inconsistent when the factor eigenvalues remain bounded &
WPCA can remain consistent under suitable weighting, even when $\lambda_r^2\to0$ \\

\cite{choi2024high} &
Usual PCA &
(a) $\lambda_r^2=\Theta\brac{N^\alpha}$, $\alpha\in(0,1)$;

(b) General weak dependence conditions on $\bE$ &
Asymptotic normality for usual PCA, requiring $\alpha>0$ for independent noise, and $\alpha>1/3$ for general dependence when $N\asymp T$&
Weaker asymptotic normality SNR as long as $\lambda_r^2=\tilde\Omega(1)$ when $N\asymp T$. \\

\cite{fan2024can} &
Usual PCA &
(a) $\lambda_r^2=\tilde\Omega\big(\op{\bSigma_{\rm C}^{\frac{1}{2}}}_1 \brac{\frac{n}{T}}^{\frac{1}{2}}\big)$;

(b) General $\bSigma_{\rm C}$ and $\bSigma_{\rm T}=\bI_T$ &
Non-asymptotic $\ell_2\!\to\!\ell_\infty$ bounds for subspace recovery&
Same SNR condition when $\bSigma_{\rm T}=\bI_T$; general $(\bSigma_{\rm C},\bSigma_{\rm T})$; weaker SNR for $\ell_2\!\to\!\ell_\infty$ bounds when $T\gg N$ \\

\cite{yan2021inference,agterberg2022entrywise} &
HeteroPCA &
(a) $\lambda_r^2=\Omega\big(\sigma_{\max}\brac{\frac{n}{T}}^{\frac{1}{4}} \brac{\frac{N}{T}}^{\frac{1}{4}}\big)$;

(b) Diagonal $(\bSigma_{\rm C},\bSigma_{\rm T})$ in \cite{yan2021inference}; diagonal $\bSigma_{\rm C}$ and general $\bSigma_{\rm T}$ in \cite{agterberg2022entrywise} &
Non-asymptotic $\ell_2\!\to\!\ell_\infty$ bounds for subspace recovery&
Same SNR condition when $\bSigma_{\rm T}=\bI_T$; general $(\bSigma_{\rm C},\bSigma_{\rm T})$ \\

\cite{bai2013statistical} &
Cross-sectional WLS &
(a) $\lambda_r^2=\Theta\brac{N}$;

(b) General weak dependence conditions on $\bE$; $ \bSigma_{\rm C}^{-1}$ must be estimable &
Efficient estimation/inference with feasible $\hat\bSigma_{\rm C}^{-1}$ weighting &
No need for estimation of $\bSigma_{\rm C}^{-1}$; weak factor allowed \\

\cite{lam2011estimation,gao2022modeling} &
Auto\-covariance-based PCA&
(a) $\lambda_r^2=\Theta\brac{N^\alpha}$, $\alpha\in(0,1)$;

(b) Block diagonal $(\bSigma_{\rm C},\bSigma_{\rm T})$ in \cite{lam2011estimation}; general $\bSigma_{\rm C}$ and $\bSigma_{\rm T}=\bI_{T}$ in \cite{gao2022modeling} &
Asymptotic  bound for subspace recovery of $O_p\!\brac{\sqrt{N^{1-\alpha}/T}}$ &
Same but non-asymptotic rate for subspace recovery; general $(\bSigma_{\rm C},\bSigma_{\rm T})$; $\ell_2\!\to\!\ell_\infty$ bounds; CLTs for $(\bL,\bF)$ \\

\cite{qiao2025weight} &
Weight-calibrated auto\-covariance-based PCA &
(a) $\lambda_r^2=\Theta\brac{N^\alpha}$, $\alpha\in(0,1)$

(b) General $(\bSigma_{\rm C},\bSigma_{\rm T})$  &
Asymptotic  bound for subspace recovery of $O_p\!\brac{\sqrt{N^{1-\alpha}/T}}$ provided that $T^{1-\alpha}\lesssim N\lesssim T$ &
Same but non-asymptotic rate for subspace recovery in a broader $(N,T)$ regime; $\ell_2\!\to\!\ell_\infty$ bounds; CLTs for $(\bL,\bF)$ \\
\hline
\end{tabular}
\vspace{0.35em}
\parbox{0.98\textwidth}{\footnotesize
\textit{Notes.}
Our advantage over usual PCA comes from the explicit bias term
$|\Tr(\bSigma_{\rm T}\bQ)|\,\|\bU^\top \bSigma_{\rm C}\bU_\perp\|$.
For the TS-factor line, the comparison is through the moderate-dimensional regime
$N^{1-\alpha}\ll T\lesssim N$ with $\lambda_r^2\asymp N^\alpha$, whereas our Cor.~1 only requires
$\lambda_r{\gg \sqrt{N/T}\vee(N/T)^{1/4}}$.
{For inference, Thm.~7 contains a common $(T/N)^{1/2}$ second-stage term for weighted and ordinary PCA. Choosing $\bQ$ so that $|\Tr(\bSigma_{\rm T}\bQ)|$ is small reduces the additional first-stage bias contribution.}}
\end{table}

\paragraph{Multivariate Functional Data.} Let $\bF$ be a smooth function so that $F_{t,k}=g_k(t/T)$, where $g_k\in C^1([0,1])$ for $k\in[r]$.  For identifiability, we assume that 
\begin{align*}
	\int_{0}^1g_k(u)g_l(u)=\delta_{kl},\qquad 1\le k,l\le r.
\end{align*}
We consider $\bQ\in\RR^{T\times T}$ such that $Q_{t,s}=\frac{1}{2B}\II\brac{1\le\ab{t-s}\le B}$, where $B$ is the number of sub-diagonals that are nonzero.
\begin{corollary}\label{col:inf-L-exmp-fd}
	Suppose the deterministic version of Assumption \ref{assump:iden} holds, $\textsf{vec}\brac{\bE}\sim \calN \brac{0,\bSigma_{\rm C}\otimes \bI}$ and $\lambda_r=\omega\bbrac{\brac{{N}/{T}}^{1/2}\vee\brac{{N}/(TB)}^{1/4}}$.
    \begin{itemize}
	    \item[(i)] If $T=\omega\brac{(\max_{k\in[r]}\ebrac{\op{g_k}_\infty\vee \op{g^\prime_k}_\infty})^2Br}$, then
	\begin{align}\label{eq:u-bound-fd}
		\op{\hat\bU_{\bQ}\hat\bU_{\bQ}^\top-\bU\bU^\top}=O_p\brac{\frac{1}{\lambda_r}\sqrt{\frac{N}{T}}+\frac{1}{\lambda_r^2}\sqrt{\frac{N}{TB}}}.
	\end{align}
	\item[(ii)] If $T\gtrsim N$ and $T=\omega\brac{(\max_{k\in[r]}\sqbrac{\op{g_k}_\infty\vee \op{g^\prime_k}_\infty})^2Br}$, then
		\begin{align}\label{eq:L-bound-fd}
				\frac{1}{\sqrt{N}}\op{\hat\bL_\bQ-\bL\bR_{L,\bQ}}=O_p\brac{\frac{1}{\sqrt{T}}\bbrac{1+\lambda_r^{-1}+\lambda_r^{-2}B^{-1/2}}+\frac{1}{\lambda_r\sqrt N}}.
		\end{align}
	\item[(iii)] If \(r\) is fixed, $\infn{\bL\bLambda^{-1}}\lesssim\sqrt{r/N}$, $T=\omega\brac{(\max_{k\in[r]}\sqbrac{\op{g_k}_\infty\vee \op{g^\prime_k}_\infty})^2Br}$, $N=\omega\brac{r\brac{r+\log n}}$ and $\lambda_r=\omega\brac{\brac{{n}/{T}}^{1/2}r^{1/2}\log^2 n\vee\brac{{T}/{N}}^{1/2}r^{1/2}}$, then for each $i\in[N]$,
		\begin{align*}
			\bSigma_{L,i}^{-1/2}\brac{\hat\bL_\bQ-\bL\bR_{L,\bQ}}_{i,\cdot}^\top \overset{d}{\rightarrow}\calN\brac{\mathbf0,\bI_r},
		\end{align*}
		where $\bSigma_{L,i}=T\sqbrac{\bSigma_{\rm C}}_{i,i}\bO_F^\top \bU^\top \bL \bL^\top\bU\bO_F$ and $\bO_F=T^{-1}\bar\bO^\top \bLambda^{-2}\bar\bO\bSigma\bR_{V,\bQ}^\top$.
		Here $\bSigma_{L,i}^{-1/2}$ uses the compatible-whitening convention in \Cref{thm:inf-L} with the displayed $\bO_F$.
        \end{itemize} 
\end{corollary}
As noted in \Cref{sec:intro}, requiring a divergent SNR for consistency is not intrinsic to factor models but a limitation of the PC estimator.   The error rate \eqref{eq:u-bound-fd} and \eqref{eq:L-bound-fd} imply that under this assumption of functional data for $\bF$, we shall choose $B\ge \lambda_r^{-2}$ to obtain a sharp rate for the subspace and the first-stage part of the loading estimation, when the factor signal is extremely small in that $\lambda_r\ll 1$. 

We can also derive properties of the estimated factors under this setting which relates to functional PCA \citep[see, e.g.,][]{hall2006propertiesa, hall2006propertiesb, shang2014survey, zhou2022theory}.
\begin{corollary}\label{col:inf-F-exmp-fd}
		Suppose the deterministic version of Assumption \ref{assump:iden} holds, $\textsf{vec}\brac{\bE}\sim \calN \brac{0,\bSigma_{\rm C}\otimes \bI}$ and $\lambda_r=\omega\bbrac{\brac{{N}/{T}}^{1/2}\allowbreak\vee \brac{{N}/(TB)}^{1/4}}$. Furthermore, assume that $T=\omega\brac{(\max_{k\in[r]}\ebrac{\op{g_k}_\infty\vee \op{g^\prime_k}_\infty})^2Br}$. Then
    \begin{itemize}
    \item[(i)] If $T\gtrsim N$, then
	\begin{align}\label{eq:v-F-bound-fd}
		\op{\hat\bV_{\bQ}\hat\bV_{\bQ}^\top-\bV\bV^\top}=O_p\brac{\frac{1}{\lambda_r}},\quad \frac{1}{\sqrt{T}}\op{\hat\bF_\bQ-\bF\bR_{F,\bQ}}=O_p\brac{\frac{1}{\lambda_r}}.
	\end{align}
    \item[(ii)] If \(r\) is fixed, $T= o\brac{e^{cN}}$,  $\infn{\bL\bLambda^{-1}}\lesssim\sqrt{r/N}$, $T=\omega\brac{\brac{r+\log n}^2}$ and $\lambda_r=\omega\brac{(n/T)r\log^3n}$, then for each $t\in[T]$,
		\begin{align*}
			\bSigma_{F,t}^{-1/2}\brac{\hat\bF_\bQ-\bF\bR_{F,\bQ}}_{t,\cdot}^\top \overset{d}{\rightarrow}\calN\brac{\mathbf0,\bI_r},
		\end{align*}
		where $\bSigma_{F,t}:=\bR_{V,\bQ}\bSigma^{-1}\bU^\top \bSigma_{\rm C}\bU\bSigma^{-1}\bR_{V,\bQ}^\top$.
		Here $\bSigma_{F,t}^{-1/2}$ uses the compatible-whitening convention in \Cref{thm:inf-F}.
\end{itemize}
\end{corollary}

We can compare these results with those for functional PCA in the literature. Consider, for example, the setting from \cite{zhou2022theory} where $r=O(1)$ and $T\gtrsim N$. In addition, it is assumed that $\lambda_k^2\asymp N$ for $k\in[r]$  and $\max_{k\in[r]}\ebrac{\bop{g_k}_{\infty}\vee\bop{g_k^\prime}_{\infty}\vee\bop{g_k^{\prime\prime}}_{\infty}}=O(1)$. By choosing the optimal tuning parameter and assuming suitable regularity conditions hold, \cite{zhou2022theory} (Corollary 2 therein) showed that the functional PCA satisfies
$$\op{\hat g_k^{\rm fPCA}-g_k}_{L_2}^2=\int_{0}^1\ab{\hat g_k(t)-g_k^{\rm fPCA}(t)}^2 dt=O_p\brac{\frac{1}{N}},\quad \forall k\in[r].$$  
To obtain the above bound, a strong eigen-gap condition was imposed to ensure that we can differentiate between the different PCs. Under the same condition, we can also show that the rotation matrix $\bR_{F,\bQ}$ in \Cref{col:inf-F-exmp-fd} obeys $\bR_{F,\bQ}=\bI_r+\bDelta_F$ with $\op{\bDelta_F}=O(1/T)$. This implies that $$\frac{1}{T}\fro{\hat\bF_{\bQ}-\bF}^2=\sum_{k=1}^r\frac{1}{T}\sum_{t=1}^T\sqbrac{\hat g_k\brac{\frac{t}{T}}-g\brac{\frac{t}{T}}}^2=O_p\brac{\frac{1}{N}},$$
for finite $r$. However, different from the functional PCA, this holds without the strong factor assumption and without the  boundedness condition on $\ebrac{g_sk^{\prime\prime}\brac{\cdot }}_{k=1}^r$.

\section{Adaptivity: CV}\label{sec:cv}
The choice of the weight matrix $\bQ$ is clearly essential to weighted PCA. In this section, we shall discuss how we can adaptively choose from a large class of weight matrices via CV. To this end, we shall assume in the rest of this section that $\ebrac{\f_t,t\in \ZZ}$ is \emph{weakly stationary} in the sense that $\EE \f_t=0$, $\EE \f_{t}\f^\top _{t^\prime}=\EE \f_{t-t^\prime}\f^\top _{0}$  and $\EE \op{\f_t}^2<\infty$ for $\forall t, t^\prime\in\ZZ$. Let  
\begin{align*}
	\bGamma_{\tau}:=\EE \f_{1}\f_{1+\tau }^\top \in\RR^{r\times r}
\end{align*}
 denote the auto-covariance at lag $\tau=1,\cdots,L$. Furthermore, we assume that  
	\begin{assumption}\label{assump:factor-ts}
		There exists some $1\le K_0\le T/2$ and $\bar\delta_F=o(1)$, $\bar\eta_F=o(1)$ such that $\bGamma_\tau\neq 0$ for $1\le \tau\le K_0$, and
		\begin{align*}
			\PP\brac{\bigcup_{\tau\in[K_0]}\ebrac{\aop{\frac{1}{T-\tau }\sum_{t=1}^{T-\tau}\f_{t}\f^\top _{t+\tau }-\bGamma_\tau }\ge \frac{\bar\delta_F}{K_0}}}\le \bar\eta_{F}.
		\end{align*}
	\end{assumption}

    Note that \Cref{assump:factor-ts} can be viewed as a relaxed version of  $\tau$-mixing condition with geometric decay. It holds, in particular, with $\bar\eta_F=o\brac{n^{-C}}$ for any constant $C>0$ under suitable moment conditions on $\bF$. See e.g., \cite{han2020moment}  (Assumptions (A1)-(A4) therein).

In light of \Cref{assump:factor-ts}, we shall consider the following class of weight matrices:
	\begin{align*}
		\scrQ_{T,K_0}:=\ebrac{\bQ_{\bgamma}:=\textsf{Toeplitz}(\gamma_0,\gamma_1,\ldots, \gamma_{T-1}): \bgamma\in\Delta_{T,K_0}},
	\end{align*}
where $\Delta_{T,K_0}:=\ebrac{\bgamma\in\RR^T: \gamma_i\ge 0,~\forall i\ge 0,~\gamma_j=0,~\forall j\ge K_0+1,~\sum_{i=0}^{K_0}\gamma_i=1}$. In particular, $\bgamma=\brac{1,0,\cdots,0}$ corresponds to classical PCA, and $\bgamma=\brac{0,\gamma_1,\cdots,\gamma_{K_0},0,\cdots 0}$ corresponds to using aggregated information from lag-$1$ to lag-$K_0$ auto-covariance. 
	
The signal strength under this setting can be characterized by the following quantity:
	\begin{align}\label{eq:mu-gamma-def}		\mu\brac{\bgamma}:=\lambda_{\min}\brac{\gamma_0\bI_r+\sum_{\tau=1}^{K_0}\gamma_\tau\brac{1-\frac{\tau}{T}}\brac{\bGamma_\tau+\bGamma_\tau^\top }}.
	\end{align}
		For candidates with $\mu(\bgamma)>0$, it is easy to see that $\sigma_r\brac{\EE\bM\bQ_\bgamma\bM^\top}=\sigma_r\brac{\bL\brac{\EE\bF^\top\bQ_\bgamma\bF}\bL^\top }\ge \lambda_r^2T \mu(\bgamma)$. {A quantitative lower bound for $\mu(\bgamma)$ requires corresponding assumptions on the lag auto-covariances $\ebrac{\bGamma_\tau}_{\tau=1}^{K_0}$.}
	
Now we shall choose $\bQ_\bgamma\in \scrQ_{T,K_0}$, or equivalently, $\bgamma\in \Delta_{T,K_0}$ via CV: a random subset of the entries of~$\bX$ is selected as the validation set and $\bgamma$ is set to the value that minimizes the validation error. More formally, independently of $\bX$, let $\bOmega$ be a $N$ by $T$ matrix with entries following i.i.d. \text{Ber}$(p_*)$ for a prespecified value of $p_\ast$ and $\Omega:=\ebrac{(i,t)\in [N]\times [T]:\Omega_{i,t}=1}$ be the index set. For each $\Omega$, define the projection operator $\calP_{\Omega}:\RR^{N\times T}\rightarrow \RR^{N\times T}$ as 
	\begin{align*}
[\calP_{\Omega}\brac{\bX}]_{i,t}=X_{i,t}\II\brac{\Omega_{i,t}=1},\quad \forall i\in[N], t\in[T].
	\end{align*}
	In addition, let $\bOmega_{\perp}:=\mathbf{1}_N\mathbf{1}_T^\top-\bOmega $ and similarly define $\Omega_{\perp}$.  Let $\hat\bL_\bgamma^{\Omega}=\sqbrac{\hat\l_{\bgamma,1}^{\Omega},\cdots,\hat\l_{\bgamma,N}^{\Omega}}^\top \in\RR^{N\times r}$ and $\hat\bF_\bgamma^{\Omega}=\sqbrac{\hat\f_{\bgamma,1}^{\Omega},\cdots,\hat\f_{\bgamma,T}^{\Omega}}^\top \in\RR^{T\times r}$ be our estimator using $p_*^{-1}\calP_{\Omega}\brac{\bX}$. Finally, we choose
	\begin{align*}
		\hat\bgamma:=\argmin_{\bgamma\in\calG_{K_0}}\textsf{CV}\brac{\bgamma},\quad \text{where}\quad \textsf{CV}\brac{\bgamma}:=\frac{1}{|\Omega_\perp|}\sum_{(i,t)\in\Omega_\perp}\bbrac{X_{i,t}-\hat\l_{\bgamma,i}^{\Omega\top }\hat\f^{\Omega}_{\bgamma,t}}^2.
	\end{align*}                                                            
	where $\calG_{K_0}=\ebrac{\bgamma_1,\cdots,\bgamma_M}\subset \Delta_{T,K_0}$ is a grid set of $\Delta_{T,K_0}$ with cardinality $M$. See \Cref{alg:AdaWPCA} for details. It is not difficult to check that an $\epsilon$-cover of $\Delta_{T,K_0}$ requires only $M=O\bbrac{\brac{C/\epsilon}^{K_0}}$ grid points.

    {Since the purpose of choosing $\bQ$ is to recover the loading space, an ideal population choice over the finite candidate grid would be
\[
    \bgamma^{\star}\in
    \argmin_{\bgamma\in\calG_{K_0}}
    \op{\bU_{\bQ_\bgamma}\bU_{\bQ_\bgamma}^{\top}
    -\bU\bU^\top},
\]
where $\bU_{\bQ_\bgamma}$ denotes the leading-$r$ population eigenspace of $\EE(\bX\bQ_\bgamma\bX^\top)$. This oracle choice is generally infeasible because it depends on the unknown loading space and population eigenspaces. AdaWPCA instead uses held-out reconstruction error to select a weight from the same finite grid. \Cref{thm:cv-consistency} below provides a partial theoretical justification for this data-driven criterion by showing that it separates two meaningful regimes of candidate weights.}

	\begin{algorithm}[!tbp]
		\footnotesize
		\caption{AdaWPCA}\label{alg:AdaWPCA}
		\KwData{ Mask parameter $p_*\in(0,1)$, candidate set $\bgamma=\ebrac{\bgamma_1,\cdots,\bgamma_M}$, rank $r$. }
		Let $\bOmega\in\ebrac{0,1}^{N\times T}$ consists of i.i.d. \text{Ber}$(p_*)$.
		
		\For{$\bgamma\in\calG_{K_0}$}{
			Let $\bbrac{\hat\bL_\bgamma^{\Omega},\hat\bF_\bgamma ^{\Omega}}=\text{Weighted-PCA}\brac{p_*^{-1}\calP_{\Omega}\brac{\bX},\bQ_\bgamma,r}$, as detailed in \Cref{alg:main}.
			\\ Calculate $\textsf{CV}\brac{\bgamma}={\frac{1}{|\Omega_\perp|}}\sum_{(i,t)\in\Omega_\perp}\brac{X_{i,t}-\hat\l_{\bgamma,i}^{\Omega\top }\hat\f^{\Omega}_{\bgamma,t}}^2$.
		}
		\KwResult{$\hat\bgamma=\argmin_{\bgamma\in \cal G_{K_0}}\textsf{CV}\brac{\bgamma}$, $\bbrac{\hat\bL,\hat\bF }=\text{Weighted-PCA}\brac{\bX,\bQ_{\hat\bgamma},r}$.}
	\end{algorithm}
	
		For the theoretical result below, we impose the following technical assumption. 
	\begin{assumption}\label{assump:digoanl-dominant}
		$\bSigma_{\rm C}$ and $\bSigma_{\rm T}$ satisfy that 
		\begin{itemize}
			\item[(1)] $\bSigma_{\rm C}=\textsf{diag}\bbrac{\sigma_{{\rm C},1}^2,\cdots, \sigma_{{\rm C},N}^2}$ where $\sigma_{{\rm C},i}>0$ for $i\in[N]$.
			\item[(2)]  $\sqbrac{\bSigma_{\rm T}}_{t,t}>(1+c)\sum_{t^\prime\in[T]\backslash\ebrac{t}}\ab{\sqbrac{\bSigma_{\rm T}}_{t,t^\prime}}$ for some universal constant $c>0$. 
            \item[(3)] Writing $\sigma_{{\rm T},t}^2:=[\bSigma_{\rm T}]_{t,t}$, there exists a constant $C>0$ such that, for all $i\in [N]$ and $t\in [T]$,
           $C^{-1}\sigma_{\rm C}\le \sigma_{{\rm C},i}\le C\sigma_{\rm C}$ and $C^{-1}\sigma_{\rm T}\le \sigma_{{\rm T},t}\le C\sigma_{\rm T}$ for some $\sigma_{\rm C},\sigma_{\rm T}>0$.
		\end{itemize}
	\end{assumption}
	\begin{theorem}\label{thm:cv-consistency}
			Suppose that $\bE\sim \textsf{subG}_{K}(\bSigma_{\rm C},\bSigma_{\rm T})$, where $\bSigma_{\rm T}=\textsf{diag}\bbrac{\sigma_{{\rm T},1}^2,\ldots,\sigma_{{\rm T},T}^2}$, and that Assumptions \ref{assump:iden}-\ref{assump:digoanl-dominant} hold, where $p_*\in[c,1-c]$, $T\ge N$, $T=o\brac{e^{c_*N}}$ for a sufficiently small universal constant $c_*>0$, and $K_0,\kappa=O(1)$. Assume that $\rho\gtrsim\sigma_{\rm C}^2$, $\lambda_1^2r/\brac{K \sigma_{\rm C}\sigma_{\rm T}}^2=o\brac{N}$, $\infn{\bL\bLambda^{-1}}\lesssim\sqrt{r/N}$, and
		\begin{align*}
			\frac{\lambda_r}{K\sigma_{\rm C}\sigma_{\rm T}}\gtrsim\kappa^2r\log^{3/2} n,\qquad \psi:=\frac{K\lambda_1}{\sigma_{\rm C}\sigma_{\rm T}}\sqrt{\frac{N}{T}}\brac{r+\log n}=o\brac{1}.
		\end{align*}
		Let $\bar\psi:=K\lambda_1(\sigma_{\rm C}\sigma_{\rm T})^{-1}\kappa\sqrt{r^2+r\log n}$. Fix $B_1,B_2\rightarrow\infty$ such that $\psi(B_1\vee B_2)=o(1)$, $\kappa^3\sqrt r{\psi B_1}/\brac{\bar\psi B_2}=o(1)$, and
		$\kappa^2\sqrt r\bar\psi B_2=o\brac{{\lambda_r^2}/\brac{\sigma_{\rm C}^2\sigma_{\rm T}^2}}$.
		For some constants $c_0,c_\mu,c_{\rm gap}\in(0,1)$, define
        \begin{align*}
            \calG_{K_0,1}&:=\ebrac{\bgamma\in \calG_{K_0}: \gamma_0\in[c_0\psi B_1,\psi B_1],\ \mu(\bgamma)\ge c_\mu},\\
            \calG_{K_0,2}&:=\ebrac{\bgamma\in \calG_{K_0}: \gamma_0\ge\psi B_2,\quad
        \bar\psi B_2\le\frac{\gamma_0}{\mu(\bgamma)}
        \le\frac{c_{\rm gap}\lambda_r^2}{\kappa^2\sqrt r\,\sigma_{\rm C}^2\sigma_{\rm T}^2}}.
        \end{align*}
        Suppose that both sets are nonempty, and assume for $\bgamma\in\calG_{K_0,1}\cup\calG_{K_0,2}$, the matrix inside the definition of $\mu(\bgamma)$ is positive definite with largest eigenvalue at most $C_\mu\mu(\bgamma)$, and that $\gamma_0\delta_F+\bar\delta_F=o(\mu(\bgamma))$.
        Then, for any $\bgamma_1\in\calG_{K_0,1}$ and $\bgamma_2\in\calG_{K_0,2}$,
		\begin{align*}
			\PP\brac{\textsf{CV}\brac{\bgamma_1}< \textsf{CV}\brac{\bgamma_2}}\ge 1-M\brac{\eta_F+\bar\eta_F}-O\brac{Mn^{-10}}.
		\end{align*}
	\end{theorem}
    {\Cref{thm:cv-consistency} shows that the CV criterion prefers any candidate in $\calG_{K_0,1}$ that preserves the factor-side signal to any candidate in $\calG_{K_0,2}$ having a positive but nearly cancelled signal and a much larger relative noise bias. Thus, the theorem establishes a pairwise separation guarantee over these two specified subsets of the finite candidate class. Note that it does not establish that $\hat\bgamma$ attains the population subspace oracle $\bgamma^{\star}$ over the entire grid. An explicit construction showing that the two comparison classes can be nonempty is given in \Cref{sec:cv-nonempty-construction}.}

	In practice, it is oftentimes not known  whether the heteroskedasticity of noise or the autocorrelation structure of  factors exists.  In general, a desirable goal is to consistently choose $\bgamma$s with  $\gamma_0=o(1)$ over those $\bgamma$s with  $\gamma_0=1$ when autocorrelation is strong,  and consistently choose $\bgamma$s with  $\gamma_0=1$ over $\bgamma$s with  $\gamma_0=o(1)$  when  autocorrelation is  weak. Nonetheless, it is unclear whether achieving this stronger consistency demands a new algorithm beyond the current CV scheme or simply more refined analysis,  and we leave it for future work.
	
	It is also  worth pointing out that similar CV approaches have been adopted by \cite{zeng2019double, wei2020determining, jin2021factor}. A nuanced difference between our goal and theirs is that they use CV to choose the number of factors $r$, and the inherent low-rankness can benefit analysis for delivering consistency of CV. In contrast, choosing the Toeplitz weights $\bgamma$ by CV brings out additional technical difficulty, which in turn hinges on a lower bound on singular subspace perturbation in the missing setting. For applications, we can apply multiple random draws of training set $\ebrac{\bOmega_{j}}_{j=1}^{K_{\sf cv}}$. For each $\bgamma\in\calG_{K_0}$, we can compute $\textsf{CV}_j\brac{\bgamma}$ and choose $\hat\bgamma=\argmin_{\bgamma\in\calG_{K_0}}\frac{1}{K_{\sf cv}}\sum_{j=1}^{K_{\sf cv}}\textsf{CV}_j\brac{\bgamma}$. 

    {Note that this repeated-CV version can be regarded as a heuristic stabilization.}

    		When $r$ is unknown, there has been a line of research on estimating it from the data \citep{bai2007determining,lam2012factor,wei2020determining, fan2022estimating}. In particular, {one heuristic is the ratio diagnostic} $\hat r:=\argmin_{j\in[R]}\sigma_{j+1}\brac{\bX\bX^\top}/\sigma_{j}\brac{\bX\bX^\top}$ in \cite{lam2012factor}, where $R$ is a reasonable upper bound of $r$ and can be taken as $R=N/2$ in practice.
		Alternatively, {another heuristic is to use CV over $\calG_{K_0}\times [R]$ to choose $\brac{\bgamma,r}$ jointly}.  

\section{Numerical Experiments}\label{sec:num}
    \subsection{Simulated-data Analysis}
	We first consider the factor model in \eqref{model:matrix-form} with \(N=100\) and $r=3$. The factors are assumed to be generated from the VAR model $\f_t=\bA\f_{t-1}+\beps_t$, where {$\bA=0.9\bI_r$}, and  $\beps_t\sim \calN(0,\bI_r-\bA\bA^\top )$. {We generate the idiosyncratic component by
$\bE=\bSigma_{\rm C}^{1/2} \bZ \bSigma_{\rm T}^{1/2}$, where \(\bZ=(Z_{it})\in\mathbb{R}^{N\times T}\) has i.i.d. entries and both \(\bSigma_{\rm C}\in\mathbb{R}^{N\times N}\) and \(\bSigma_{\rm T}\in\mathbb{R}^{T\times T}\) are symmetric positive definite. {The transition matrix has positive-definite symmetric part. The VAR recursion starts at zero and uses a burn-in of \(50\). Independently, draw \(\bG_L\in\RR^{N\times r}\) with i.i.d.\ \(\calN(0,1)\) entries and write its thin SVD as \(\bG_L=\bU_s\bD_L\bV_L^\top\), and we set \(\bL=1.2\bU_s\bD_L\). In an asymptotic interpretation with fixed \(r\), \(\bD_L/\sqrt N\to\bI_r\) in probability, so the loading singular values satisfy \(\lambda_k/\sqrt N\to1.2\) and have no \(T\)-dependence. We construct three orthonormal columns \(\bB_\perp\) by orthogonalizing an independent \(N\times3\) Gaussian matrix against \(\bU_s\), and set $\bSigma_{\rm C}=2.5\bU_s\bU_s^\top+220\bB_\perp\bB_\perp^\top+10(\bI_N-\bU_s\bU_s^\top-\bB_\perp\bB_\perp^\top)$. Its eigenvalues are \(2.5\) on the loading space, \(220\) on the three contaminating directions, and \(10\) on the remaining directions.}

	\paragraph{Estimation of Singular Vectors $\bU$ and $\bV$.}
    	To illustrate the fixed-$\bQ$ behavior described by the theory, the SPD spectral construction above places larger eigenvalues in directions orthogonal to the signal loading space, which induces strong contemporaneous contamination and yields a clear contrast between PCA and lag-1 WPCA. We consider two regimes for generating \(\bSigma_{\rm T}\) and the entries of \(\bZ\):

        \begin{itemize}
            \item    \(Z_{it}\sim\mathcal N(0,1)\) and \(\bSigma_{\rm T}=\bI_T\){, with \(T\in\{120,180,260,380,560\}\)}.
            \item  \(Z_{it}=\widetilde Z_{it}/\sqrt{5/3}\), where \(\widetilde Z_{it}\) follows a \(t_5\) distribution, and \(\bSigma_{\rm T}\) is AR(1)-Toeplitz with \((\bSigma_{\rm T})_{ts}=\rho_{\rm T}^{|t-s|}\) (\(\rho_{\rm T}=0.18\)){, with \(T\in\{150,220,320,480,720\}\).}
        \end{itemize}
{The prespecified lag-$1$ weight has \(Q_{t,t+1}=Q_{t+1,t}=\sqbrac{2(T-1)}^{-1}\) and all other entries zero. For this averaged lag-$1$ weight, \(T^{-1}\bF^\top\bF\to\bI_r\) and \(\bF^\top\bQ\bF\to0.9\bI_r\) in probability; equivalently, after the scalar normalization \(\|\bQ\|=1\) used in the theory, \(\sigma_r(\bF^\top\bQ\bF)\) is of order \(T\). We report \(\|\hat\bU\hat\bU^\top-\bU\bU^\top\|_{\rm F}\) and \(\|\hat\bV\hat\bV^\top-\bV\bV^\top\|_{\rm F}\). Each plotted mean uses \(300\) replications.} As shown by \Cref{fig:sim-line}, in both regimes, {the PCA errors remain large over the considered range, whereas the errors for the prespecified lag-$1$ weight are smaller and generally decline with \(T\)}.

    \begin{figure}[!t]
	\centering
	\includegraphics[width=\textwidth]{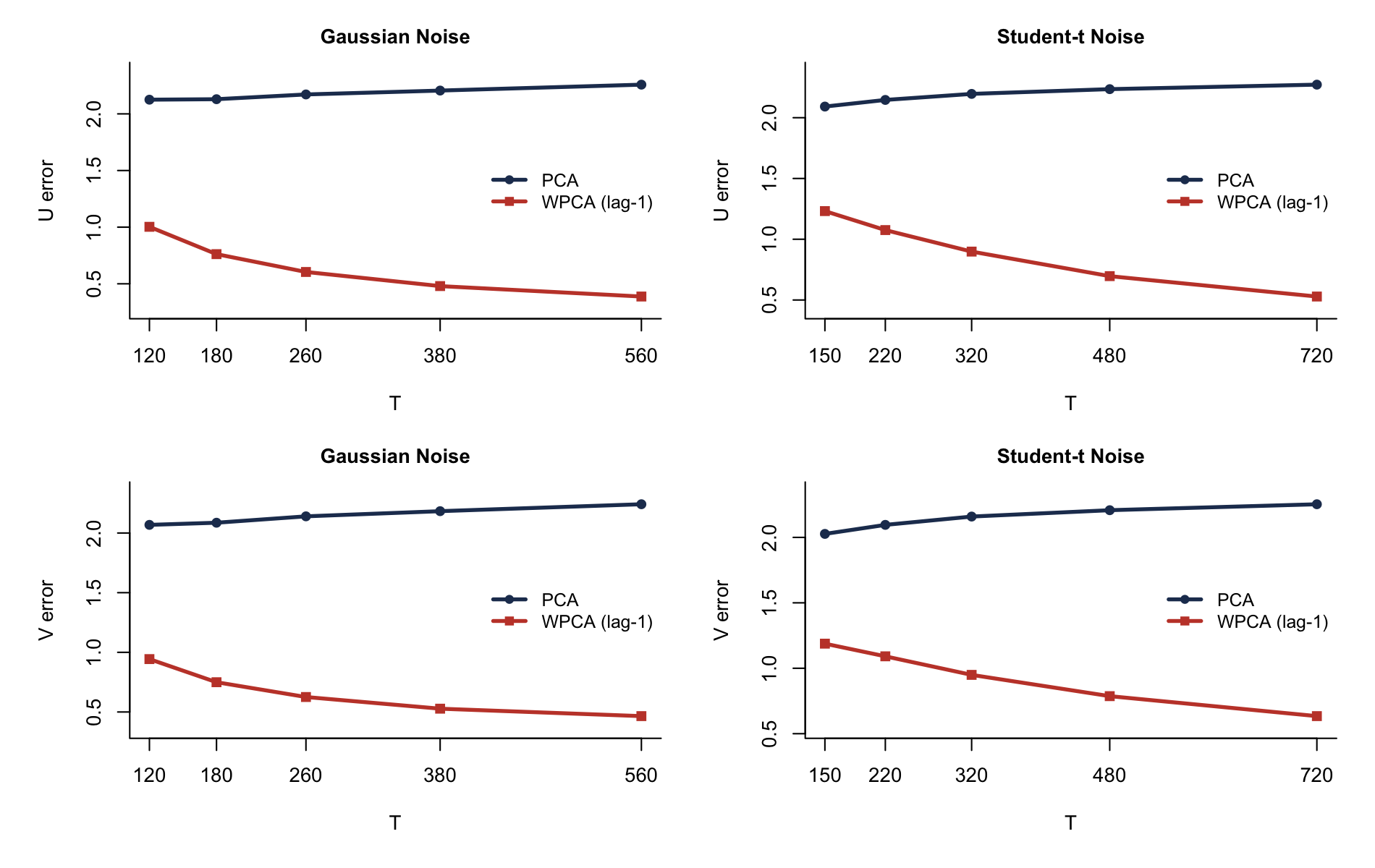}
	\caption{Estimation  of $\bU$ and $\bV$ in $\fro{\cdot}$ under Gaussian and Student's t noise, with $N=100$ and varying $T$. The results are averaged from ${300}$ simulation replicates.}
	\label{fig:sim-line}
	\end{figure}
	    }
    {Additional simulation results for adaptive estimation, inference, and CV selection are provided in \Cref{sec:supp-additional-simulations}.}
	
	\subsection{Real-data Analysis}
    
	We consider two datasets\footnote{The datasets are publicly available and updated in a timely manner at \url{https://www.stlouisfed.org/research/economists/mccracken/fred-databases.}} from the FRED (Federal Reserve Economic Data): FRED-MD \citep{mccracken2016fred} and FRED-QD \citep{mccracken2020fred}, which are monthly and quarterly macroeconomic databases consisting of U.S. indicators. FRED-MD dataset contains $126$ variables spanning from 1959-01-01 to 2025-01-01. FRED-QD dataset contains $245$ indicators spanning from 1959-Q1 to 2024-Q4. To obtain a balanced panel, we removed the variables that have more than $5\%$ missing values and kept the time periods without missing values. This leads to a complete data matrix $\bX_{\rm MD}\in\RR^{N\times T}$ with $N=121$ and $T=777$ for FRED-MD and $\bX_{\rm QD}\in\RR^{N\times T}$ with $N=207$ and $T=259$\footnote{The preprocessing procedure for the reconstruction-error comparison largely follows the example given by \url{https://github.com/cykbennie/fbi/blob/master/doc/factor_fred.html}; for that comparison, we do not apply the official FRED transformation codes.}. In the following reconstruction-error comparison, we use these balanced panels directly and write $\bX$ for either $\bX_{\rm MD}$ or $\bX_{\rm QD}$.

		Since we do not have access to true factors and loadings, we adopt a similar idea to CV to {compare the methods descriptively}. In particular, we randomly mask the entries of $\bX$ by $\bOmega^*\in\ebrac{0,1}^{N\times T}$ with $\Omega_{i,t}^*\overset{i.i.d.}{\sim}\text{Bern}\brac{q_{\sf tr}}$ with $q_{\sf tr}\in\ebrac{0.7,0.8,0.9}$. Then we treat  $\bX\circ\bOmega^*$ as our observed data and use $\bX\circ\brac{\mathbf{1}\mathbf{1}^\top-\bOmega^*}$ as the testing sample. {For AdaWPCA, both parts of every inner CV split are drawn only from $\{(i,t):\Omega^*_{i,t}=1\}$; the outer testing entries are excluded from inner fitting and validation, and the inner reconstruction uses the combined inverse-probability factor $(q_{\sf tr}p_*)^{-1}$.} 

        We first {use the ratio diagnostic on} the full data matrix $\bX$, which gives us $\hat r=1$ for both FRED-MD and FRED-QD datasets. {This rank choice is heuristic in the present weak-signal setting.} To incorporate {larger candidate ranks as a sensitivity analysis}, we apply three methods with rank $r\in \ebrac{1,2,3}$. The performance is  measured in terms of the relative reconstruction error defined as  $\sum_{(i,t)\in\Omega_\perp}\bbrac{X_{i,t}-\hat\l_{i}^{\Omega^*\top }\hat\f^{\Omega^*}_{t}}^2/\sum_{(i,t)\in\Omega_\perp}X_{i,t}^2$, where $(\hat\bL^{\Omega^* },\hat\bF^{\Omega^*})$ are corresponding estimators based on $\bX\circ \bOmega^*$. All reported results  shown in \Cref{tab:fred-md} are averaged over $100$ replicates in terms of the randomness of $\bOmega^*$.
	
		As shown in  \Cref{tab:fred-md}, {HeteroPCA has the smallest reported reconstruction error for $r=1$ at all three training proportions in both datasets, while AdaWPCA has the smallest reported error in all configurations with $r\in\{2,3\}$. The reported reconstruction errors generally decrease as $q_{\sf tr}$ increases, though not monotonically in every FRED-QD configuration. Because \(r\) is not selected by a validated weak-signal rank-selection theory here and the held-out exercise can penalize noisy higher-rank reconstructions, the rows for \(r=1,2,3\) should be read as fixed-rank sensitivity checks rather than as a monotone comparison across ranks.}

\begin{figure}[!htbp]
		\centering
		\includegraphics[width=\textwidth]{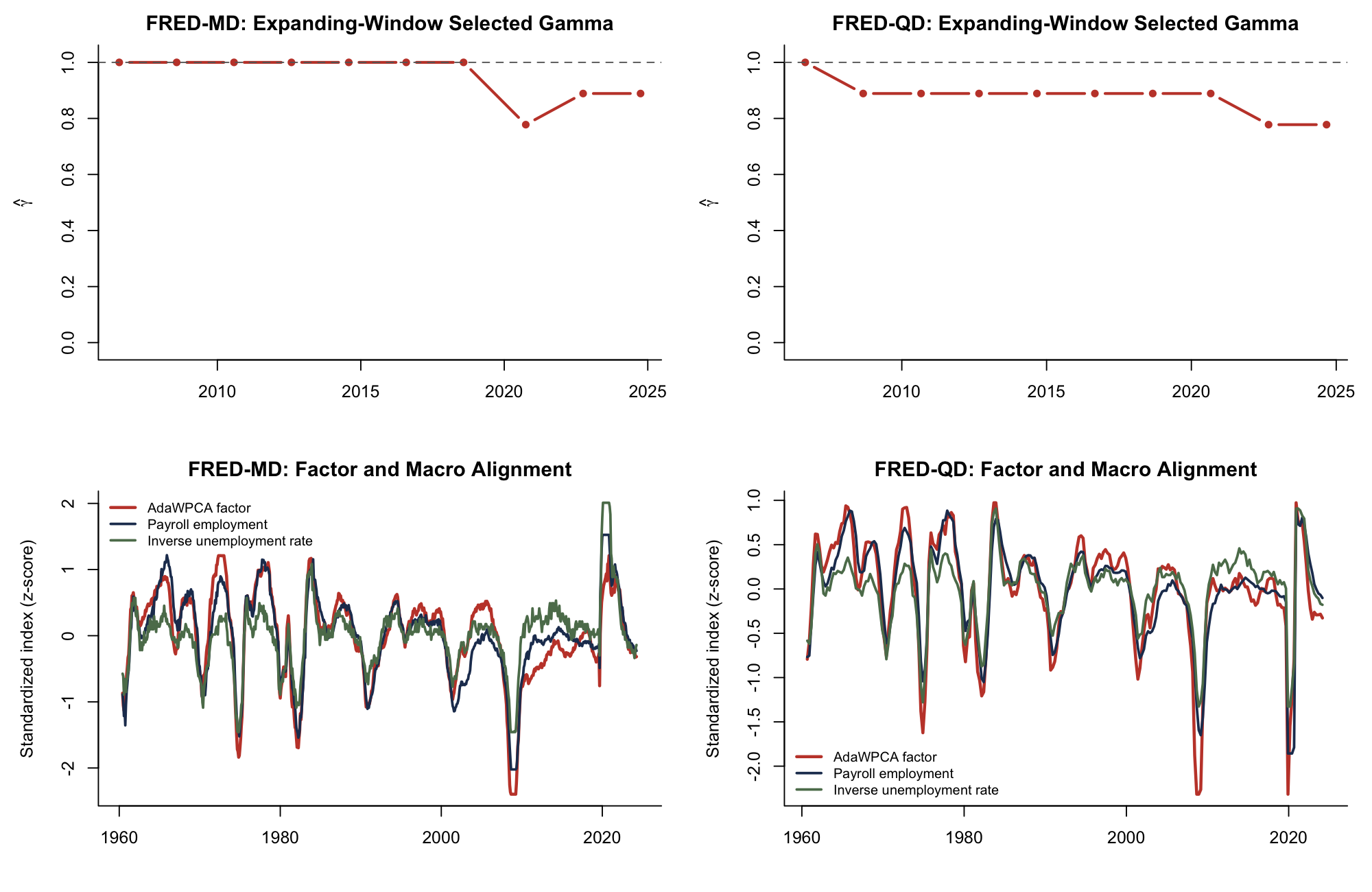}
		\caption{{In the factor-alignment panels, each series is first transformed to a standardized index (z-score), i.e., $(x_t-\bar x)/\mathrm{sd}(x)$, and then smoothed and clipped as described in the text. The factor sign is oriented to have nonnegative correlation with payroll employment.} The activity proxy is payroll employment (FRED code: PAYEMS), and labor-market slack is represented by the inverse unemployment rate (i.e., $-\,$UNRATE, where UNRATE is the civilian unemployment rate).}
		\label{fig:FRED}
	\end{figure}

    {
To complement the reconstruction-error comparison, we also provide a separate interpretive analysis of the estimated real-data factors. For this display, {we apply the official FRED transformation codes and then standardize the resulting FRED-MD and FRED-QD panels.} We compute the first AdaWPCA factor and examine its alignment with standard business-cycle indicators. We also report the CV-selected weight parameter $\hat\gamma$ {computed over a sequence of expanding estimation windows, each of which retains all earlier observations and extends the sample to a later endpoint,} in order to assess whether the data favor a nontrivial lag-based weighting over pure PCA. 
In \Cref{fig:FRED}, the resulting factor displays strong co-movement with payroll employment and it moves countercyclically with the unemployment rate. Moreover, the selected $\hat\gamma$ is often below $1$, especially for FRED-QD, indicating that the data frequently favor a substantial lag component rather than the PCA choice. {Hence, this provides a descriptive illustration that the weighted factor is aligned with conventional business-cycle indicators and that the selected weight differs from pure PCA in some of the  windows.}
{This descriptive alignment is consistent with a conventional business-cycle interpretation. It does not by itself identify a new structural factor or establish that the factor contains information unavailable to PCA.}
{For visualization, we orient the factor sign to have nonnegative correlation with payroll employment, apply a centered \(12\)-month moving average for FRED-MD and a centered \(4\)-quarter moving average for FRED-QD, and then clip each plotted series at its own first and ninety-ninth percentiles.}
}

	\begin{table}[!htbp]
		\centering
		\resizebox{0.95\linewidth}{!}{  
			\begin{tabular}{cccccccccc}
				rank & {Method}  &  $q_{\sf tr}=0.7$ & 
				$q_{\sf tr}=0.8$ & $q_{\sf tr}=0.9$ \\
				\hline
				& AdaWPCA &  0.221 & {0.193} & 
				{0.165} \\
				
				\multirow{1}{*}{$r=1$}  & PCA & 0.224  & {0.194} & {0.165}  \\
				& HeteroPCA & \textbf{0.220} & {\textbf{0.192}} & {\textbf{0.164}} \\\hline
				& AdaWPCA &  {\textbf{0.383}} & {\textbf{0.339}} & 
				{\textbf{0.311}} \\ 
				
				\multirow{1}{*}{$r=2$}  & PCA & {0.543}  & {0.514} & 0.442  \\
				& HeteroPCA & {0.544} & 0.528 & 0.508 \\\hline
				& AdaWPCA &  {\textbf{0.489}} & {\textbf{0.477}} & 
				\textbf{0.450} \\
				
				\multirow{1}{*}{$r=3$}  & PCA & {0.751}  &  {0.731} & {0.697}  \\
				& HeteroPCA & {0.760} & {0.748} & {0.738} 
			\end{tabular}
			\qquad \qquad 
			\begin{tabular}{cccccccccc}
				rank & {Method}  &  $q_{\sf tr}=0.7$ & 
				$q_{\sf tr}=0.8$ & $q_{\sf tr}=0.9$ \\
				\hline
				& AdaWPCA &  {0.461} & {0.429} & 
				{0.397} \\
				
				\multirow{1}{*}{$r=1$}  & PCA & {0.489}  & {0.448} & {0.406}  \\
				& HeteroPCA & \textbf{0.445} & {\textbf{0.406}} & {\textbf{0.368}}\\\hline
				& AdaWPCA &  {\textbf{0.701}} & {\textbf{0.658}} & 
				{\textbf{0.687}} \\ 
				
				\multirow{1}{*}{$r=2$}  & PCA & 0.906  & {0.903} & {0.898}  \\
				& HeteroPCA & 0.952 & 0.943 & {0.933} \\\hline
				& AdaWPCA &  {\textbf{0.714}} & {\textbf{0.707}} & 
				{\textbf{0.742}} \\
				
				\multirow{1}{*}{$r=3$}  & PCA & 0.972  &  {0.972} & {0.971}  \\
				& HeteroPCA & {0.981} & {0.981} & {0.979} 
			\end{tabular}
		}
		\caption{Reconstruction error of AdaWPCA, PCA and HeteroPCA on FRED-MD (left) and FRED-QD (right).}
		\label{tab:fred-md}
	\end{table}

\paragraph{Data and code availability.}
{Code and data for reproducing the numerical results are available at \url{https://github.com/ZhongyuanLyu/WPCA}. The repository README maps each numerical figure and table to its replication script, records the simulation seeds and data vintages used in this paper, and provides both full-replication and shorter test-run instructions.}

\clearpage
\bibliographystyle{plainnat}
\bibliography{ref}

\clearpage
\setcounter{page}{1}
\appendix
{\centering\section*{\LARGE Supplementary Material to ``Weak Factor Analysis via Temporally Weighted PCA''}}
This Supplementary Material  collects the detailed technical proofs and  derivations that support the main results in ``Weak Factor Analysis via Temporally Weighted PCA''.
\section{Summary}
A key ingredient  of our analysis is the representation formula of spectral projectors, i.e.,  \Cref{lem:eigen-expansion}, introduced in \cite{xia2021normal}. Although this tool has become standard in a variety of low-rank problems \citep[e.g.][]{agterberg2022entrywise,lyu2022optimal,xia2022inference,yang2025estimating,chen2025distributed}, our setting presents two additional layers of difficulty. First, we must accommodate both temporal and cross-sectional dependence in the noise matrix~$\bE$, whereas most existing analyses assume independence. Second, since we work with the weighted sample covariance $\bX\bQ\bX^\top$, we need concentration bounds for several new matrix-product structures. To overcome these challenges, we carry out a leave-one-out analysis on the higher-order terms in the expansion of spectral projectors for $\bU$ in  $\ell_2\rightarrow\ell_\infty$ norm (\Cref{lem:high-order-bound}), and we construct a parallel leave-one-out sequence for $\widetilde\bU_{\bQ}$ to control the corresponding $\ell_2\to\ell_\infty$ bound on $\bV$ (\Cref{lem:V-high-order-bound}). Finally, to analyze the pairwise CV separation result, we extend the theory to the missing-completely-at-random regime (\Cref{thm:U-two-infinty-mcar}) with a constant missing rate, accounting for the additional terms introduced by missingness.

The rest of the appendix is organized as follows.  \Cref{sec:supp-additional-simulations} provides additional simulation results for adaptive estimation, inference, and CV selection. \Cref{sec:pf-sec-pre} introduces a preparatory lemma and additional notation. \Cref{sec:pf-sec-1} presents the proofs for the subspace, factor, and loading estimators (\Cref{thm:U-error-gen} to \Cref{thm:factor-bound-inf}). \Cref{sec:pf-sec-2} develops the inferential theory for loadings and factors (\Cref{thm:inf-L} and \Cref{thm:inf-F}) and gives a detailed rank-one heteroskedastic example for the PC loading bias. \Cref{sec:pf-sec-3} analyzes the two motivating examples, establishing \Cref{col:inf-L-exmp-ts} and \Cref{col:inf-L-exmp-fd}. \Cref{sec:pf-sec-4} first gives an explicit nonempty-class construction and then proves the pairwise CV separation result in \Cref{thm:cv-consistency}. Finally, all auxiliary lemmas and their proofs appear in \Cref{sec:pf-tlem} to \Cref{sec:pf-pftlem}.

\section{Additional Simulations}\label{sec:supp-additional-simulations}
\subsection{Adaptive Estimation}
To illustrate the performance of our adaptive procedure, we then consider two regimes:
\begin{enumerate}
	\item[(1)] $\bSigma_{\rm T}=\bI_T$ and {$\bSigma_{\rm C}:=\textsf{diag}\bbrac{\ebrac{\omega_i}_{i=1}^N}+0.6\brac{\mathbf{1}_N\mathbf{1}_N^\top -\bI_N}$} with $\omega_i\overset{i.i.d.}{\sim} \textsf{Unif}[1,20]$.
	\item[(2)] $\bSigma_{\rm T}=\bI_T$ and {$\bSigma_{\rm C}=\textsf{diag}\bbrac{\ebrac{\omega_i}_{i=1}^N}$} with $\omega_i\overset{i.i.d.}{\sim} \textsf{Unif}[1,20]$.
\end{enumerate}
We set $N\in\ebrac{100,200}$ and $T\in\ebrac{250, 500, 750,1000}$. For ease of presentation, we consider $K_0=1$ and  $\bgamma=(\gamma,1-\gamma,0,\cdots,0)\in\calG_1$, for which we  use a single parameter  $\gamma\in \bar\calG:=\ebrac{0,1/9,2/9,\cdots,1}$ to represent the Toeplitz weights. Note that all results shown are based on  adaptive choice of the Toeplitz weights $\gamma$ by CV with $K_{\sf cv}=10$, {implemented as ten independent splits with $p_*=0.8$. Every split is scored by its mean squared error over its realized validation entries, and the ten split-level mean squared errors are averaged with equal weight. HeteroPCA uses $T_0=10$ diagonal-imputation iterations,} and all results are averaged from 100 simulation replicates.

Setting (1)  is a realistic and challenging setting where non-zero off-diagonal entries in $\bSigma_{\rm C}$ exist.  As shown in \Cref{tab:sim_est_off_diag}, AdaWPCA outperforms the other two in all different $(N,T)$ settings regarding the estimation of $\bU$ and $\bV$.

Setting (2)  is the preferable setting for HeteroPCA, which is specifically designed for diagonal-only heteroskedasticity. As shown in \Cref{tab:sim_est_diag}, AdaWPCA has almost comparable performance to PCA and HeteroPCA in terms of estimation of $\bU$, while it outperforms the other two regarding the estimation of $\bV$.

    \begin{center}
        \resizebox{\linewidth}{!}{  
            \begin{tabular}{cccccccccc}
                $N$ & $T$ &  AdaWPCA & 
                PCA & HeteroPCA \\
                \hline
                & $250$ &  {\textbf{0.539(0.188)}} & {0.660(0.246)} & {0.647(0.249)}\\
                
                \multirow{2.2}{*}{$100$} & $500$ & {\textbf{0.352(0.124)}} & {0.443(0.167)}  & {0.429(0.167)}  \\
                & $750$ & {\textbf{0.287(0.064)}} & {0.368(0.094)}  & {0.355(0.100)} &  \\
                & $1000$ & {\textbf{0.244(0.051)}} & {0.323(0.092)}  & {0.308(0.096)}  \\\hline
                & $250$ &  {\textbf{0.521(0.168)}} & {0.571(0.174)} & {0.568(0.179)}\\
                
                \multirow{2.2}{*}{$200$} & $500$ & {\textbf{0.347(0.103)}} & {0.391(0.117)}  & {0.388(0.120)}  \\
                & $750$ & {\textbf{0.281(0.061)}} & {0.322(0.075)}  & {0.319(0.082)}  \\
                & $1000$ & {\textbf{0.242(0.040)}} & {0.277(0.047)}  & {0.272(0.049)}  \\\hline
            \end{tabular}
            \quad
            \begin{tabular}{cccccccccc}
                $N$ & $T$ &  AdaWPCA & 
                PCA & HeteroPCA \\
                \hline
                & $250$ &  {\textbf{0.744(0.171)}} & {0.845(0.213)} & {0.835(0.214)}\\
                
                \multirow{2.2}{*}{$100$} & $500$ & {\textbf{0.688(0.105)}} & {0.752(0.134)}  & {0.741(0.133)}  \\
                & $750$ & {\textbf{0.678(0.066)}} & {0.728(0.081)}  & {0.718(0.082)} &  \\
                & $1000$ & {\textbf{0.664(0.059)}} & {0.709(0.077)}  & {0.699(0.078)}  \\\hline
                & $250$ &  {\textbf{0.562(0.159)}} & {0.612(0.164)} & {0.610(0.168)}\\
                
                \multirow{2.2}{*}{$200$} & $500$ & {\textbf{0.514(0.096)}} & {0.549(0.105)}  & {0.546(0.108)}  \\
                & $750$ & {\textbf{0.497(0.055)}} & {0.524(0.064)}  & {0.521(0.068)}  \\
                & $1000$ & {\textbf{0.490(0.042)}} & {0.511(0.046)}  & {0.507(0.047)}  \\\hline
            \end{tabular}
        }
		\captionof{table}{Estimation error in $\fro{\cdot }$ for singular vectors under setting (1). Left: error of $\hat\bU^{\hat\gamma}$; Right: error of $\hat\bV^{\hat\gamma}$. Each entry is averaged from $100$ simulation replicates, and the number in {parentheses} is the standard deviation. }
        \label{tab:sim_est_off_diag}
    \end{center}

    \begin{center}
        \resizebox{\linewidth}{!}{  
            \begin{tabular}{cccccccccc}
                $N$ & $T$ &  AdaWPCA & 
                PCA & HeteroPCA \\
                \hline
                & $250$ &  {0.480(0.053)} & {0.491(0.055)} & {\textbf{0.465(0.050)}}\\
                
                \multirow{2.2}{*}{$100$} & $500$ & {0.325(0.033)} & {0.337(0.036)}  & {\textbf{0.314(0.031)}}  \\
                & $750$ & {0.269(0.025)} & {0.281(0.027)}  & {\textbf{0.257(0.023)}} &  \\
                & $1000$ & {0.230(0.019)} & {0.247(0.023)}  &{\textbf{0.221(0.019)}}  \\\hline
                & $250$ &  {0.459(0.042)} & {0.457(0.042)} & {\textbf{0.449(0.040)}}\\
                
                \multirow{2.2}{*}{$200$} & $500$ & {0.319(0.025)} & {0.319(0.025)}  & {\textbf{0.312(0.023)}}  \\
                & $750$ & {0.260(0.020)} & {0.262(0.020)}  & {\textbf{0.253(0.018)}}  \\
                & $1000$ & {0.222(0.015)} & {0.223(0.015)}  & {\textbf{0.215(0.014)}}  \\\hline
            \end{tabular}
            \quad
            \begin{tabular}{cccccccccc}
                $N$ & $T$ &  AdaWPCA & 
                PCA & HeteroPCA \\
                \hline
                & $250$ &  {\textbf{0.686(0.064)}} & {0.722(0.072)} & {0.704(0.068)}\\
                
                \multirow{2.2}{*}{$100$} & $500$ & {\textbf{0.661(0.050)}} & {0.685(0.056)}  & {0.670(0.053)}  \\
                & $750$ & {\textbf{0.659(0.043)}} & {0.679(0.047)}  & {0.665(0.045)} &  \\
                & $1000$ & {\textbf{0.650(0.039)}} & {0.668(0.042)}  & {0.654(0.040)}  \\\hline
                & $250$ &  {\textbf{0.492(0.041)}} & {0.506(0.045)} & {0.500(0.043)}\\
                
                \multirow{2.2}{*}{$200$} & $500$ & {\textbf{0.480(0.035)}} & {0.489(0.036)}  & {0.483(0.035)}  \\
                & $750$ & {\textbf{0.477(0.030)}} & {0.485(0.031)}  & {0.479(0.030)}  \\
                & $1000$ & {\textbf{0.469(0.026)}} & {0.476(0.027)}  & {0.471(0.027)}  \\\hline
            \end{tabular}
        }
		\captionof{table}{ Estimation error in $\fro{\cdot }$ for singular vectors under setting (2).  Left: error of $\hat\bU^{\hat\gamma}$; Right: error of $\hat\bV^{\hat\gamma}$. Each entry is averaged from $100$ simulation replicates, and the number in {parentheses} is the standard deviation. }
        \label{tab:sim_est_diag}
    \end{center}

\subsection[Inference of Loadings and Factors]{Inference of Loadings $\bL$ and Factors $\bF$}
We consider the regime $\bSigma_{\rm T}=\bI_T$ and $\bSigma_{\rm C}=\textsf{diag}\bbrac{\ebrac{\omega_i}_{i=1}^N}+0.6\brac{\mathbf{1}_N\mathbf{1}_N^\top -\bI_N}$ with $\omega_i\overset{i.i.d.}{\sim} \textsf{Unif}[1,20]$, and set $\bQ=\sum_{t=1}^{T-1}\bbrac{\e_{t}\e^\top _{t+1}+\e_{t+1}\e^\top _{t}}$. In each simulation for loadings, we calculate $\bSigma_{L,N/2}^{-1/2}\bbrac{\hat\bL_{\bQ}-\bL\bR_{L,\bQ}}_{N/2,\cdot}^\top $ as in \Cref{thm:inf-L} and plot Q-Q plot and histogram of one coordinate. {For each displayed $(N,T)$ setting, the signal panel and $\bSigma_{\rm C}$, including the realized $\omega_i$'s, are drawn once and then held fixed. Each empirical distribution contains one normalized coordinate from each of $500$ independent noise replications, in which only $\bE$ is redrawn.} Similarly for factors, we calculate $\bSigma_{F,T/2}^{-1/2}\bbrac{\hat\bF_{\bQ}-\bF\bR_{F,\bQ}}_{T/2,\cdot}^\top $ as in \Cref{thm:inf-F} and plot Q-Q plot and histogram of one coordinate. The same setting applies for \Cref{fig:sim-inf-L-bias}. From \Cref{fig:sim-inf-L-1}-\ref{fig:sim-inf-F-2}, we can see that the standard normal density curve provides good approximations to the normalized histograms, which is consistent with the limiting approximations in \Cref{subsec:inf}.

\begin{center}
    \includegraphics[width=0.49\textwidth]{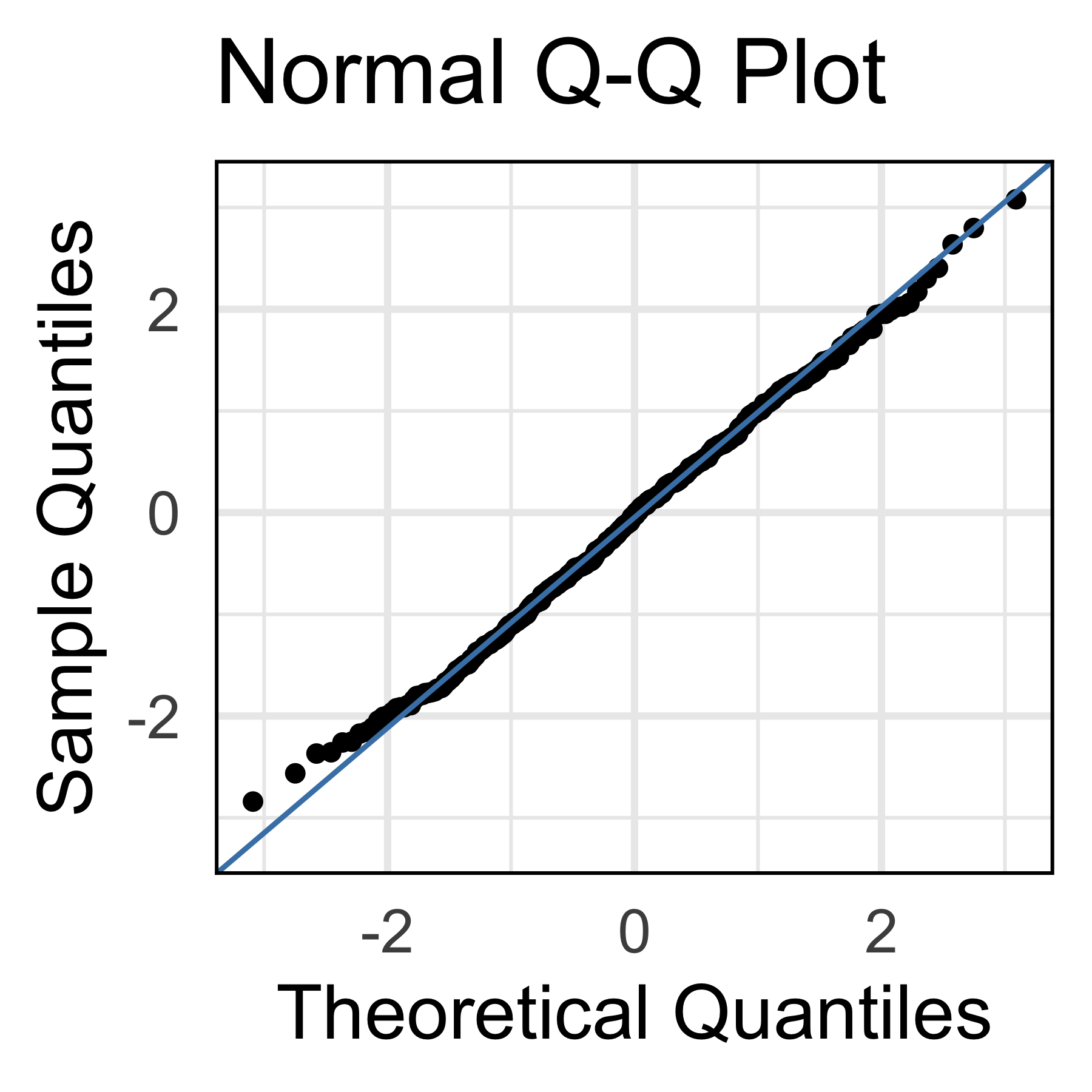}
    \includegraphics[width=0.49\textwidth]{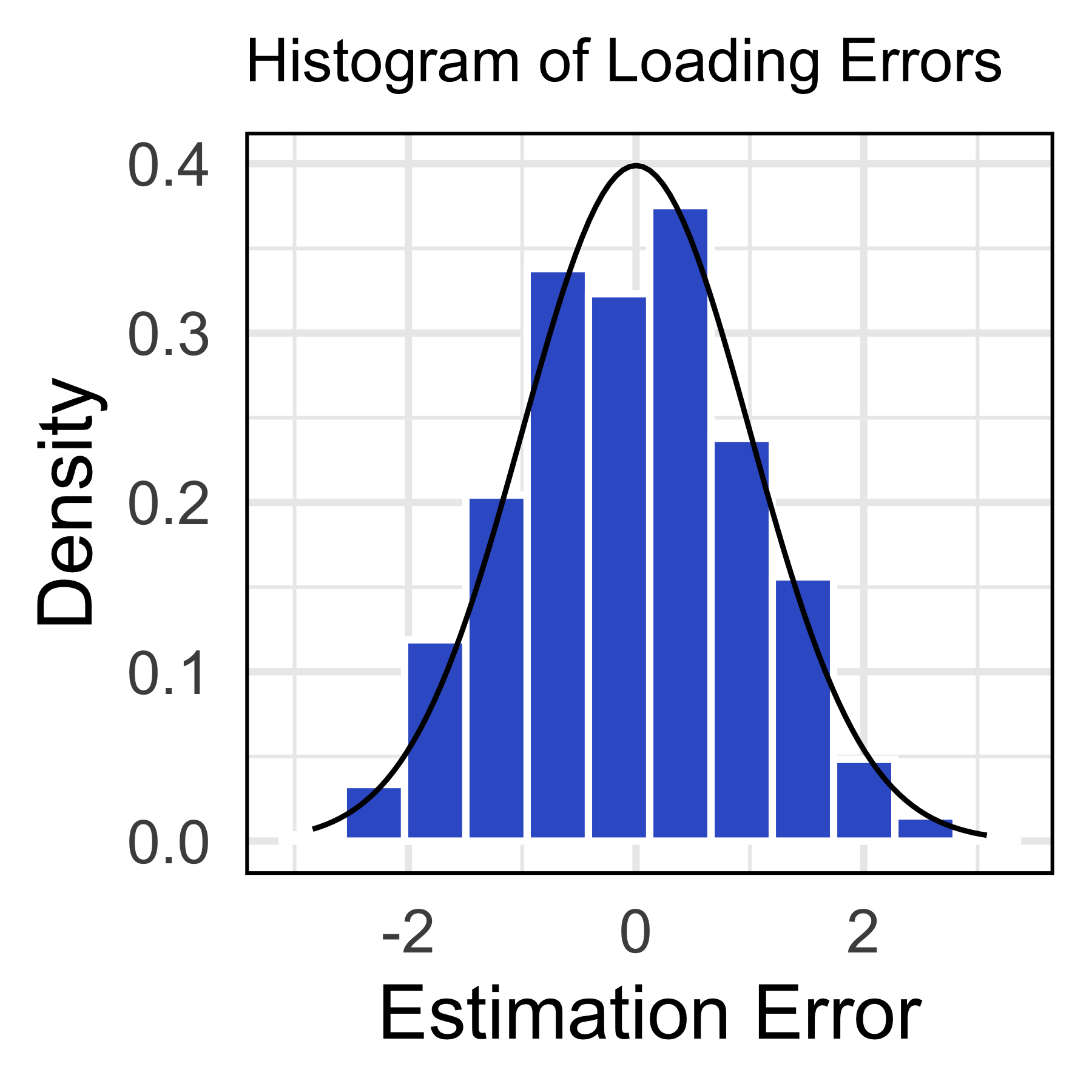}
    \captionof{figure}{Distribution  of a covariance-normalized coordinate of $\hat\bL_{\bQ}-\bL\bR_{L,\bQ}$ with $N=200$, $T=200$. Left: Q-Q plot. Right: Histogram with standard normal density overlaid.}
    \label{fig:sim-inf-L-1}
\end{center}

\begin{center}
    \includegraphics[width=0.49\textwidth]{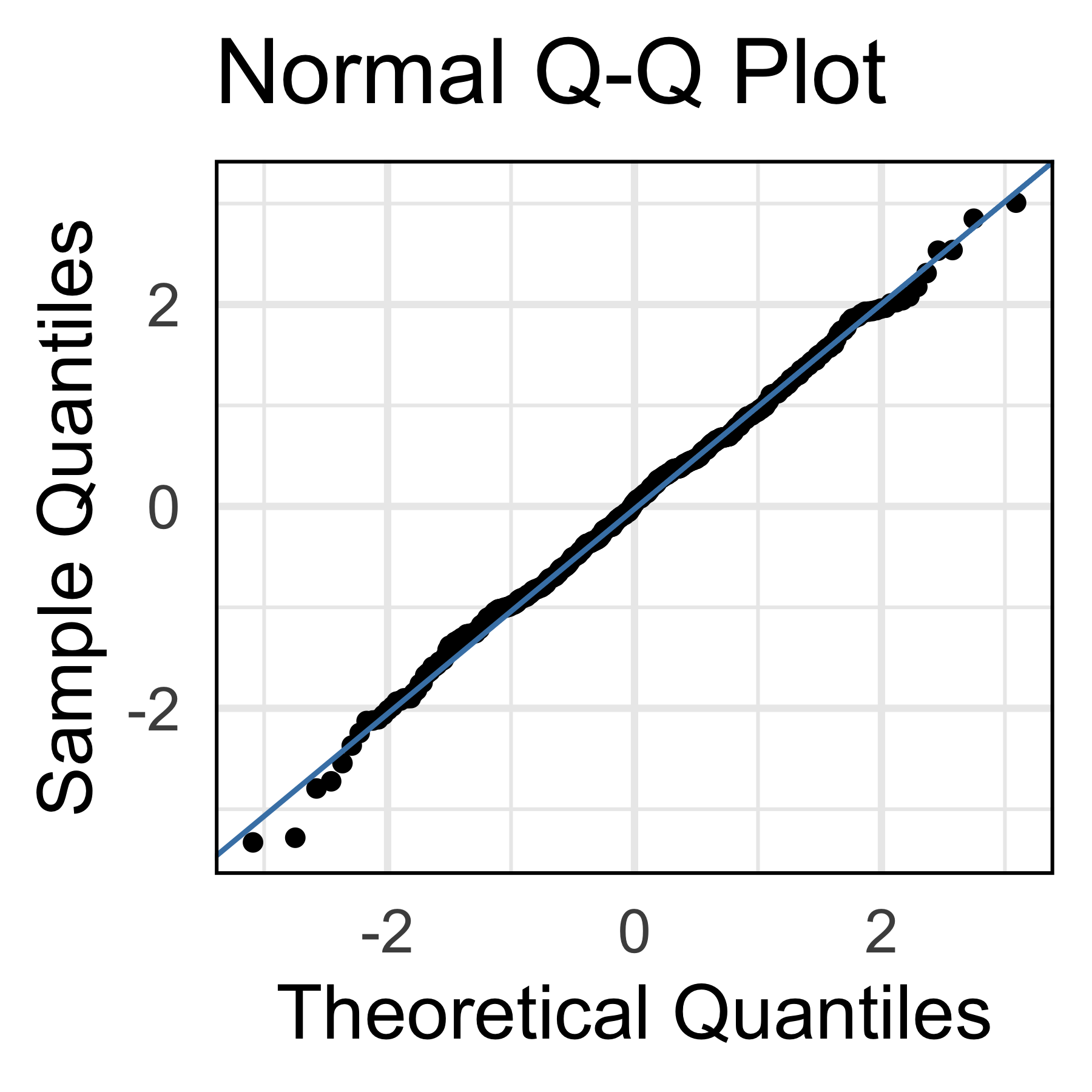}
    \includegraphics[width=0.49\textwidth]{figs/sim_inf_L_N100_T800_hist.png}
    \captionof{figure}{Distribution  of a covariance-normalized coordinate of $\hat\bL_{\bQ}-\bL\bR_{L,\bQ}$ with $N=100$, $T=800$. Left: Q-Q plot. Right: Histogram with standard normal density overlaid.}
    \label{fig:sim-inf-L-2}
\end{center}

\begin{center}
    \includegraphics[width=0.49\textwidth]{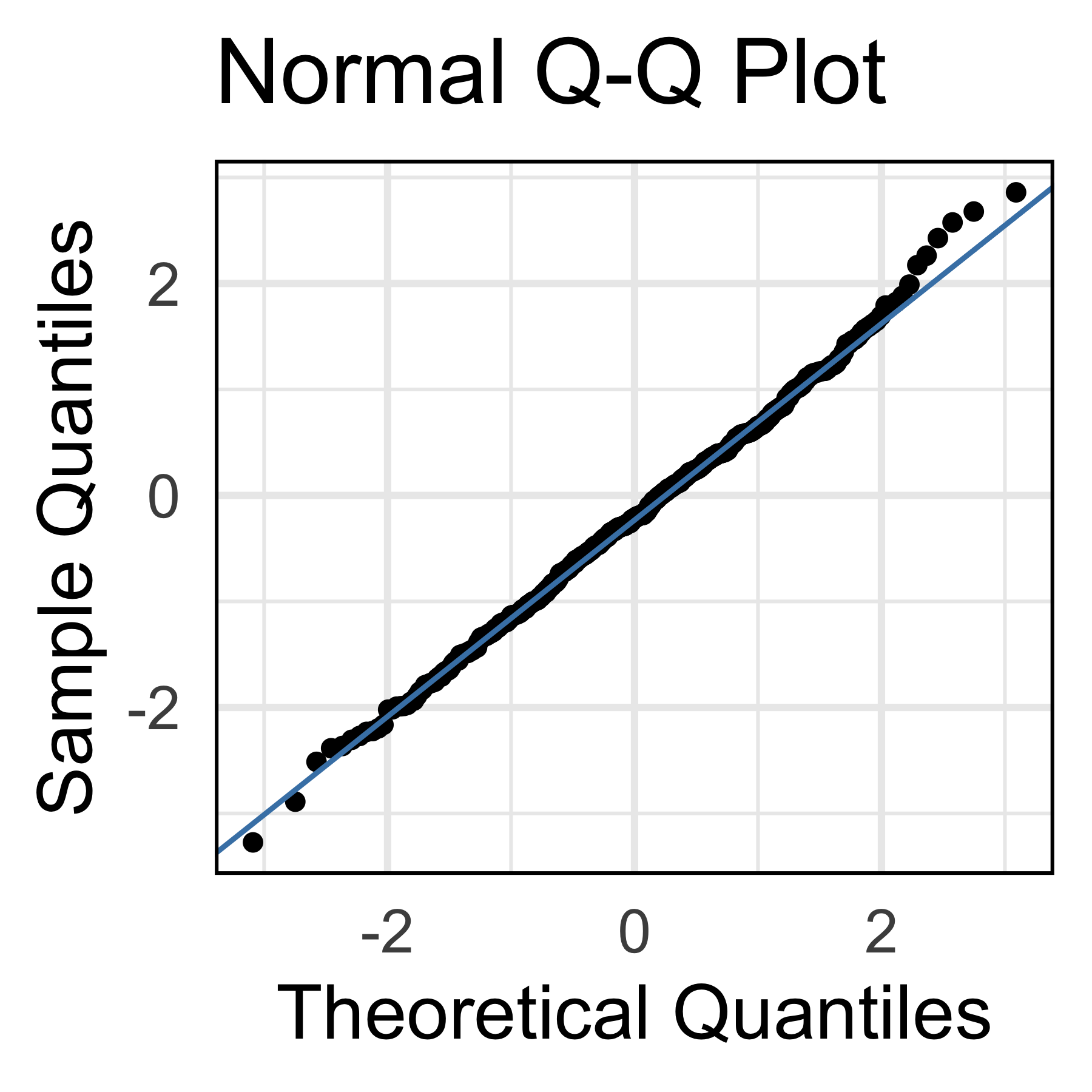}
    \includegraphics[width=0.49\textwidth]{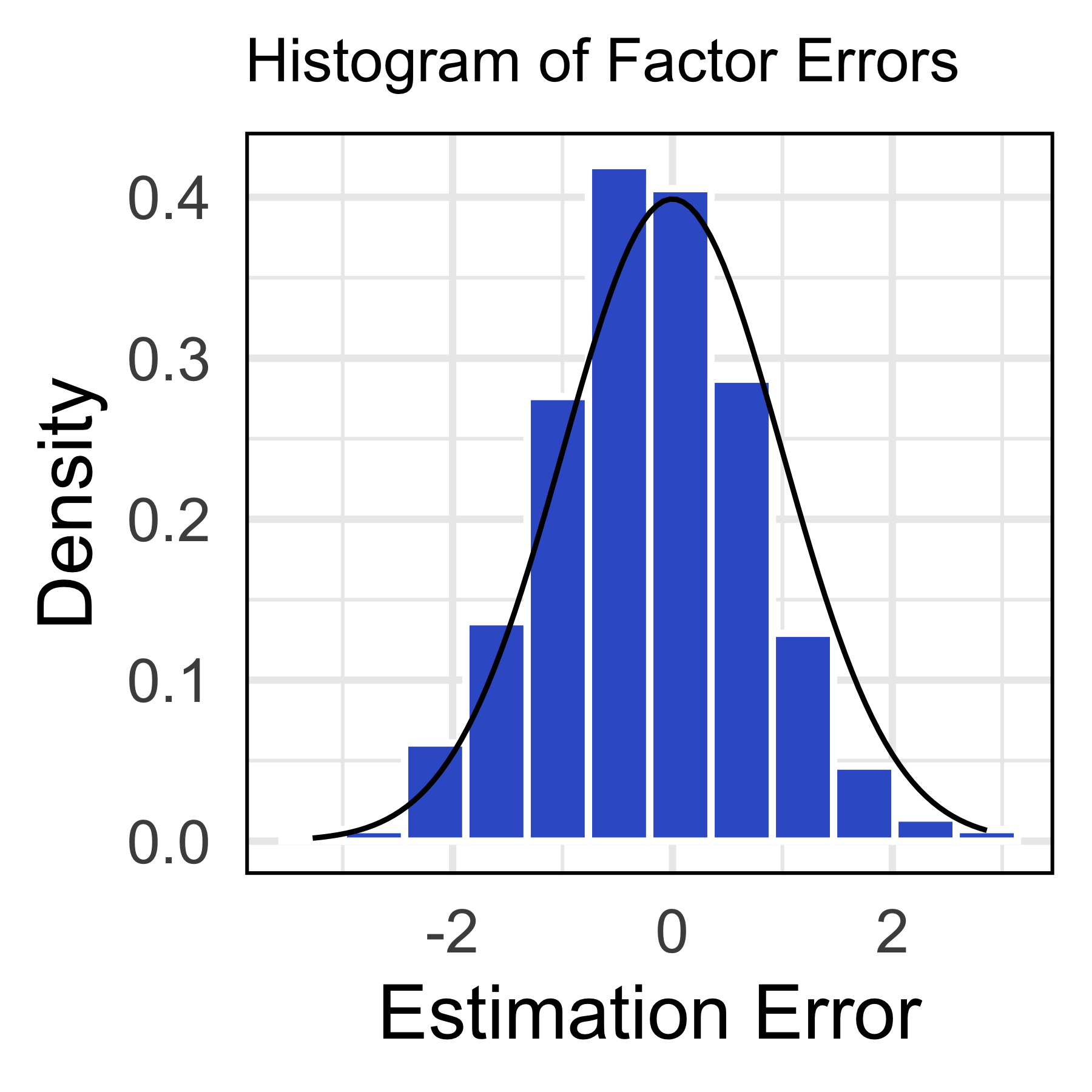}
    \captionof{figure}{Distribution  of a covariance-normalized coordinate of $\hat\bF_{\bQ}-\bF\bR_{F,\bQ}$ with $N=100$, $T=100$. Left: Q-Q plot. Right: Histogram with standard normal density overlaid.}
    \label{fig:sim-inf-F-1}
\end{center}

\begin{center}
    \includegraphics[width=0.49\textwidth]{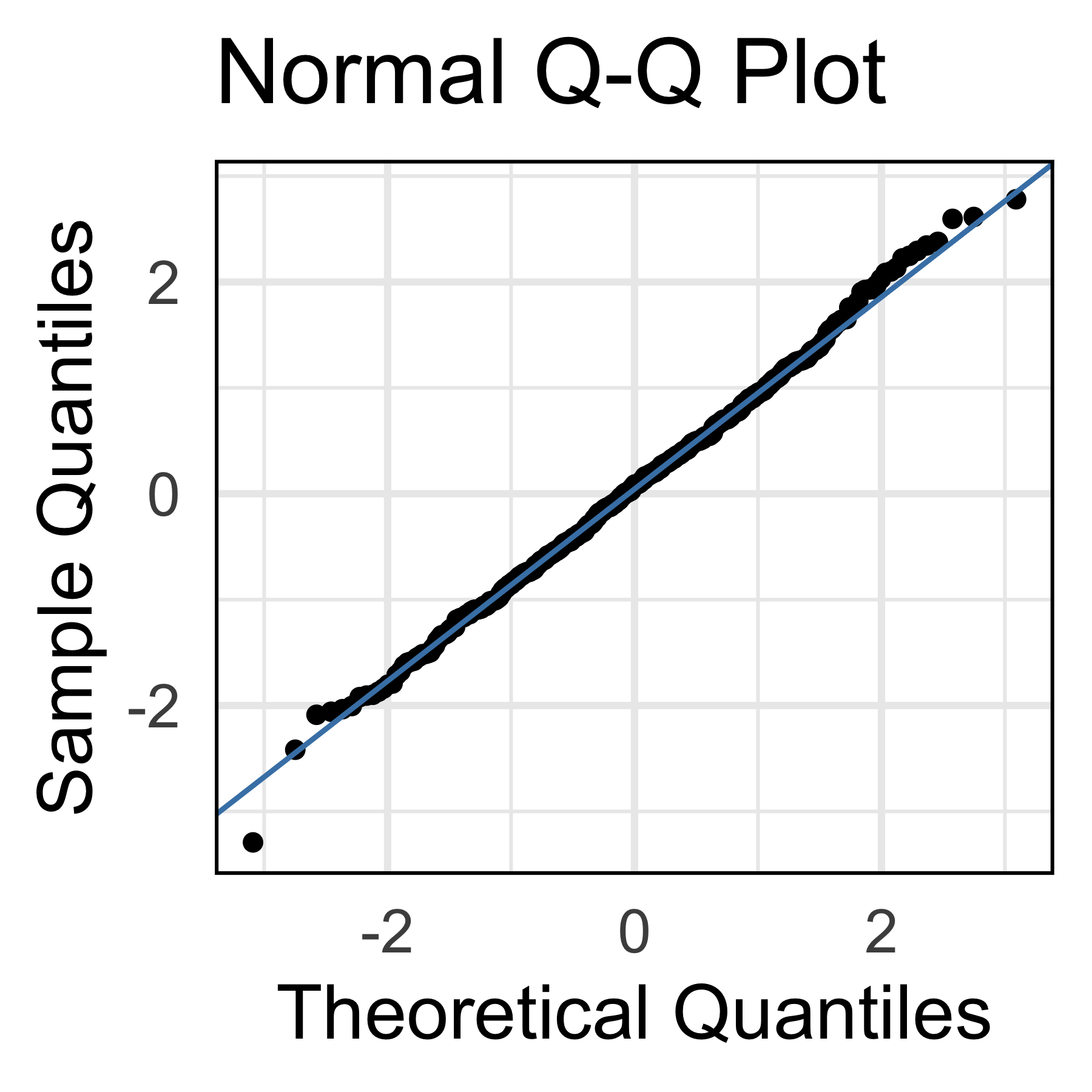}
    \includegraphics[width=0.49\textwidth]{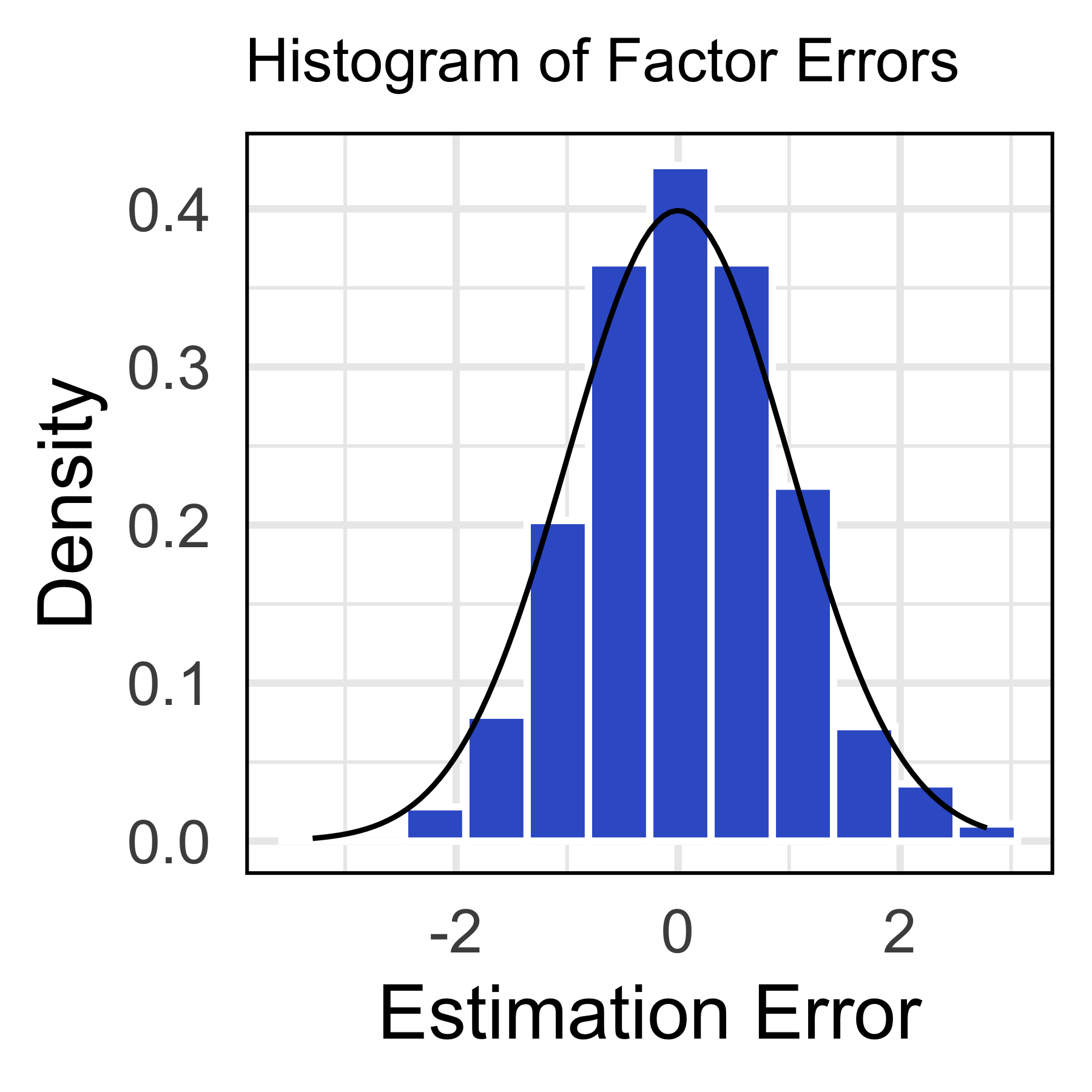}
    \captionof{figure}{Distribution  of a covariance-normalized coordinate of $\hat\bF_{\bQ}-\bF\bR_{F,\bQ}$ with $N=100$, $T=500$. Left: Q-Q plot. Right: Histogram with standard normal density overlaid.}
    \label{fig:sim-inf-F-2}
\end{center}

\subsection[{CV for Selecting a Toeplitz weight}]{{CV for Selecting a Toeplitz weight $\gamma$}}
 We consider two regimes, named explicitly here to avoid confusion with the ordering used in the preceding adaptive-estimation subsection:
\begin{enumerate}
    \item[(D)] $\bSigma_{\rm T}=\bI_T$ and $\bSigma_{\rm C}=\textsf{diag}\bbrac{\ebrac{\omega_i}_{i=1}^N}$ with $\omega_i\overset{i.i.d.}{\sim} \textsf{Unif}[1,20]$.
    \item[(O)] $\bSigma_{\rm T}=\bI_T$ and $\bSigma_{\rm C}=\textsf{diag}\bbrac{\ebrac{\omega_i}_{i=1}^N}+0.5\brac{\mathbf{1}_N\mathbf{1}_N^\top -\bI_N}$ with $\omega_i\overset{i.i.d.}{\sim} \textsf{Unif}[1,20]$.
\end{enumerate}
Similar to the estimation part, we consider {$K_0=1$} and $\gamma\in \bar\calG=\ebrac{0,1/9,2/9,\cdots,1}${, with the same $p_*=0.8$, $K_{\sf cv}=10$}. In particular, let  $R(\hat\gamma)$  denote  {the rank of $\hat\gamma$ according to the realized subspace estimation error $\fro{\hat\bU^\gamma\hat\bU^{\gamma\top}-\bU\bU^\top}$ for all $\gamma$s in  $\bar\calG$, with rank one corresponding to the smallest error}. We report the proportions of 100 simulations in which  CV chooses the top  three $\gamma$s in $\bar\calG$ (i.e., $R(\hat\gamma)\le 3$) and does not choose the bottom  three $\gamma$s in $\bar\calG$ (i.e., ${R(\hat\gamma)\le 7}$).  The results are summarized in \Cref{tab:sim_cv}. {Notably, CV does not generally choose one of the top three $\gamma$s under this infeasible ranking, but the Non-Bottom-$3$ proportion exceeds one half in 13 of the 16 settings. The three exceptions occur for the diagonal regime with $N=100$. This diagnostic is descriptive. }
\begin{center}
    \resizebox{0.95\linewidth}{!}{ 
    \begin{tabular}{ccccc}
            $N$ & $T$ & Top-$3$ & Non-Bottom-$3$  \\
            \hline
            & $250$ &  {0.03} & {0.46} \\
            
            \multirow{2.2}{*}{$100$} & $500$ & {0.03}    & {0.41}   \\
            & $750$ & {0.00} & {0.38} &  \\
            & $1000$ & {0.01} & {0.58}  \\\hline
            & $250$ &  {0.22}  & {1.00}\\
            
            \multirow{2.2}{*}{$200$} & $500$ & {0.24}   & {0.94} \\
            & $750$ & {0.16}   &  {0.87}  \\
            & $1000$ & {0.05}  & {0.73}  \\\hline
        \end{tabular}
        \qquad\qquad
    \begin{tabular}{ccccc}
            $N$ & $T$ & Top-$3$ & Non-Bottom-$3$  \\
            \hline
            & $250$ &  {0.30} & {0.80} \\
            
            \multirow{2.2}{*}{$100$} & $500$ & {0.28}    & {0.79}   \\
            & $750$ & {0.26} & {0.85} &  \\
            & $1000$ & {0.40} & {0.90}  \\\hline
            & $250$ &  {0.32}  & {0.83}\\
            
            \multirow{2.2}{*}{$200$} & $500$ & {0.44}   & {0.87} \\
            & $750$ & {0.38}   &  {0.86}  \\
            & $1000$ & {0.32}  & {0.85}  \\\hline
        \end{tabular}
    }
    \captionof{table}{Performance of CV under the diagonal regime (left) and the off-diagonal regime (right). ``Top-$3$" and ``Non-Bottom-$3$" represent the proportion of  
100 simulation replicates in which $\hat\gamma$ was among the top-3 and was not among the bottom-3 in {$\bar\calG$}, respectively, {under the infeasible ranking by the realized true-subspace error defined above}. }
    \label{tab:sim_cv}
\end{center}

\section{A Useful Lemma and Additional Notation}\label{sec:pf-sec-pre}
For convenience, we provide the following lemma  analogous to Lemma 2 in \cite{fan2024can}, establishing the connection between $\bL\bF^\top$ and $\bU\bSigma\bV^\top$.
\begin{lemma}\label{lem:good-event}
    Suppose that \Cref{assump:iden} holds, then there exists an event $\calE_0$ with $\PP\brac{\calE_0}\ge 1-\eta_F-O(n^{-10})$ such that the following properties hold when $\calE_0$ occurs:
    \begin{enumerate}
        \item[(1)] $\infn{\bV}\lesssim\sqrt{\dfrac{r+\log n}{T}}$.
        \item[(2)] $\sigma_i\brac{\bSigma
        }\asymp\lambda_i$ for $i\in[r]$.
    \end{enumerate}
    {Suppose in addition that \Cref{assump:factor-ts} holds. For the finite collection of candidates in \Cref{thm:cv-consistency}, there exists an event $\bar\calE_0$ with $\PP\brac{\bar\calE_0}\ge 1-M\brac{\eta_F+\bar\eta_F}-O(Mn^{-10})$ such that, under $\bar\calE_0$, (1) and (2) hold and, uniformly over these candidates,
    \begin{align*}
        \lambda_r\brac{\bM\bQ_{\bgamma}\bM^\top }\gtrsim \lambda_r^2 T\mu(\bgamma),\qquad
        \lambda_1\brac{\bM\bQ_{\bgamma}\bM^\top }\lesssim \lambda_1^2 T\mu(\bgamma),
    \end{align*}
    provided that $\gamma_0\delta_F+2\bar\delta_F=o\bbrac{\mu(\bgamma)}$ and the matrix inside the definition of $\mu(\bgamma)$ is positive definite with largest eigenvalue at most $C_\mu\mu(\bgamma)$.}
\end{lemma}
 The proof of the above lemma is deferred to \Cref{pf-lem:V-infn-bound}.
 
For ease of presentation, we define the de-mean operator $\frakD\brac{\bA}:=\bA-\EE\bA$ for any matrix $\bA$, and we denote the subspace estimation error of $\bU$ and $\bV$ in terms of spectral norm and $\ell_2\rightarrow\ell_\infty$ norm  as
\begin{align*}
\xi^U_{\bQ}&:=\op{\hat\bU_\bQ\hat\bU_\bQ^\top -\bU\bU^\top },\qquad\qquad ~\xi^V_{\bQ}:=\op{\hat\bV_\bQ\hat\bV_\bQ^\top -\bV\bV^\top },\\
\xi^U_{\bQ,\infty}&:=\infn{\hat\bU_\bQ\hat\bU_\bQ^\top -\bU\bU^\top },\qquad \xi^V_{\bQ,\infty}:=\infn{\hat\bV_\bQ\hat\bV_\bQ^\top -\bV\bV^\top }.
\end{align*}

\section{Proofs in \Cref{sec:main} and \Cref{subsec:factor-est-err}}\label{sec:pf-sec-1}
This section proves all results on subspace, factor, and loading estimation, namely Theorems \Cref{thm:U-error-gen}–\ref{thm:factor-bound-inf}.
The central ingredients are two entry-wise ($\ell_2\to\ell_\infty$) error bounds for the left and right singular subspaces, $\bU$ and $\bV$. To bound  $\xi^U_{\bQ,\infty}$, we adapt the strategy of \citet[Theorem 3]{agterberg2022entrywise}, but must now control additional terms that arise from the two-sided dependence encoded by $\bSigma_{\rm C}$, $\bSigma_{\rm T}$, and the weighting matrix $\bQ$. To bound $\xi^V_{\bQ,\infty}$, we construct a leave-one-out sequence $\big\{\wt\bU^{(-l)}_\bQ\big\}_{l\in[T]}$, obtained by a leave-one-out construction on the columns of $\bG$, to decouple the dependence, which allows us to replicate the row-wise analysis on the right singular vectors.

\subsection{Proof of \Cref{thm:U-error-gen}}\label{pf-thm:U-error-gen}
As explained in \Cref{sec:main}, we have 
\begin{align*}
\xi^U_\bQ\lesssim\frac{\inf_{\zeta\in\RR}\op{\bU^\top \brac{\bN-\zeta\bI}\bU_\perp}}{\lambda_r^2\sigma_r\brac{\bF^\top\bQ\bF } }\wedge 1\le\frac{\op{\bU^\top \bN\bU_\perp}}{\lambda_r^2\sigma_r\brac{\bF^\top\bQ\bF } }.
\end{align*}
It boils down to bound $\op{\bU^\top \bN\bU_\perp}\le 2\op{\bL\bF^\top\bQ\bE^\top}+\op{\frakD\bbrac{\bE\bQ\bE^\top}}+\op{\bU^\top\EE\bE\bQ\bE^\top\bU_\perp}$. The last term is nothing but $\op{\bU^\top\textsf{PTr}_{T}\ebrac{\bSigma_{\rm e}(\bI_N\otimes \bQ)}\bU_\perp}$, we next bound the first two terms separately.

{Recall that $\textsf{vec}\brac{\bE}=\bSigma_{\rm e}^{1/2}\g$, where the entries of $\g$ are independent, centered, unit-variance, and sub-Gaussian with $\|g_i\|_{\psi_2}\le K$.} 
We start with the bound for $\op{\bL\bF^\top\bQ\bE^\top}$.  Since $\bL\bF^\top\bQ$ is independent of $\bE$ and $\text{rank}(\bL\bF^\top\bQ)\le r$, it suffices to bound $\op{\bE\bV}$ with $\bV\in\OO_{T,r}$. To that end, we first notice that
for any fixed $\u\in \SS^{N-1}$ and $\v\in{\SS^{r-1}}$ and let $\bar\v:=\bV\v $, we have
\begin{align*}
\u^\top\bE\bV\v & =\textsf{Tr}\brac{\bar\v \u^\top\bE}=\textsf{vec}\brac{\u\bar \v ^\top}^\top\textsf{vec}\brac{\bE}=\inp{\bSigma_{\rm e}^{1/2}\brac{\u\otimes \bar\v }}{{\g} }.
\end{align*}
Thus we get  $\ab{\u^\top\bE\bV\v }\le {CK}\op{\bSigma_{\rm e}^{1/2}}\sqrt{N+r}$ with probability exceeding $1-O\brac{e^{-cN}}$ for any fixed $\u\in \SS^{N-1}$ and $\v\in{\SS^{r-1}}$. 
Using a standard $\epsilon$-net argument for $\u$ and $\v$, we obtain the same high probability bound for $\op{\bE\bV}$. Therefore, with probability at least $1-O\brac{e^{-cN}}$,
\begin{align*}
\op{\bL\bF^\top\bQ\bE^\top}\lesssim {K}\op{\bL\bF^\top\bQ}\cdot \sqrt{N}\op{\bSigma_{\rm e}^{1/2}}.
\end{align*}
Next, we are going to bound $\op{\frakD\bbrac{\bE\bQ\bE^\top}}$. For any fixed $\u\in \SS^{N-1}$, we have
\begin{align*}
\u^\top\bE\bQ\bE^\top\u
&={\g^\top  \bSigma_{\rm e}^{1/2}\brac{\u\u^\top\otimes\bQ}\bSigma_{\rm e}^{1/2}\g} .
\end{align*}
By Hanson-Wright inequality
and a standard $\epsilon$-net argument, we obtain that with probability exceeding $1-O\brac{e^{-cN}}$,
    \begin{align*}
        \op{\frakD\bbrac{\bE\bQ\bE^\top}}\lesssim {K^2}\op{\bSigma_{\rm e}^{1/2}}^2\brac{\fro{\bQ}\sqrt{N}+\op{\bQ}N}.
    \end{align*}
Combining these bounds with the population residual in the SNR condition and using $\op{\bL\bF^\top\bQ}\lesssim\lambda_1\sqrt{T}$ completes the proof.

\subsection{Proofs of \Cref{thm:U-error} and First Claim in \Cref{thm:UV-error-inf}}\label{pf-thm:U-error}
We will proceed our analysis on the event $\calE_0$  defined in \Cref{lem:good-event}.
\paragraph{Bound for $\hat\bU_\bQ$ in $\op{\cdot }$.}
Using Theorem 1 in \cite{xia2021normal}, we obtain that 
\begin{align*}
    \xi_{\bQ}^U\lesssim\frac{\inf_{\zeta\in\RR}\op{\bU^\top \brac{\bN-\zeta\bI_N}\bU_\perp}}{\sigma_{r}\brac{\bM\bQ\bM^\top }},
\end{align*}
where $\bN=\bM\bQ\bE^{\top}+\bE\bQ\bM^\top+\bE\bQ\bE^{\top}$.
Under Assumption \ref{assump:iden}, we arrive at with probability at least $1-O\brac{e^{-cN}}$,
\begin{align*}
    \inf_{\zeta\in\RR}\op{\bU^\top\brac{\bN-\zeta\bI_N}\bU_\perp}&\le \ab{\textsf{Tr}\brac{\bSigma_{\rm T}\bQ}}\inf_{\zeta\in\RR}\op{\bU^\top \brac{\bSigma_{\rm C}-\zeta\bI_N}\bU_\perp}\\
    &+C_0K^2\sqrt{N}\op{\bSigma_{\rm C}}\brac{\fro{\bSigma^{1/2}_{\rm T}\bQ\bSigma^{1/2}_{\rm T}}+\sqrt{N}\op{\bSigma^{1/2}_{\rm T}\bQ\bSigma^{1/2}_{\rm T}}}\\
    &+C_1K\lambda_1\sqrt{N T}\op{\bSigma_{\rm C}^{1/2}} \brac{\op{\bQ\bSigma^{1/2}_{\rm T}}+\op{\bSigma^{1/2}_{\rm T}\bQ}},
\end{align*}
where we've used \Cref{lem:GB-bound} and \Cref{lem:EQE-bound}.  Note that for any $\zeta\in\RR$, we have
\begin{align*}
    \op{\bU^\top \brac{\bSigma_{\rm C}-\zeta\bI_N}\bU_\perp}=\op{\bU^\top\bSigma_{\rm C}\bU_\perp}=\rho.
\end{align*}
In addition, we have $\fro{\bSigma^{1/2}_{\rm T}\bQ\bSigma^{1/2}_{\rm T}}\le \fro{\bQ}\op{\bSigma_{\rm T}}$ and $\op{\bSigma^{1/2}_{\rm T}\bQ\bSigma^{1/2}_{\rm T}}\le \op{\bQ}\op{\bSigma_{\rm T}}$.
The same bounds without the projections, together with $\inf_{\zeta\ge0}\op{\textsf{Tr}\brac{\bSigma_{\rm T}\bQ}\bSigma_{\rm C}-\zeta\bI_N}$, give the cluster separation required by Weyl's inequality.
The proof is completed by noticing that $\sigma_r\brac{\bM\bQ\bM^\top}\ge\lambda_r^2 \sigma_r\brac{\bF^\top\bQ\bF }$, $\op{\bQ}\lesssim 1$  and using a  standard union bound argument.
\paragraph{Bound for $\hat\bU_\bQ$  in $\infn{\cdot }$.}
The proof directly follows from the following theorem  whose proof can be found in \Cref{pf-thm:U-two-infinty}. 
\begin{theorem}\label{thm:U-two-infinty}
    Suppose Assumption \ref{assump:iden} holds. Fix  $\bQ\in\RR^{T\times T}$ with $\op{\bQ}=1$ and define 
    \begin{align}\label{eq:theta-def}
        \bar\vartheta :=\kappa_{\rm C, 1}\kappa_{\rm C, 2}&\left(
        K\sqrt{Tr}\log n\infn{\bU}^{-1}\op{\bSigma}\op{\bSigma_{\rm C}^{1/2}}_1\op{\bSigma_{\rm T}^{1/2}}\right.\notag\\
        &\left.+K^2\sqrt{nNr\log n}\op{\bSigma^{1/2}_{\rm T}}^2\op{\bSigma_{\rm C}^{1/2}}^2_{1} +\inf_{\zeta\in\RR}\op{\bDelta_{\zeta}}_1\right),
    \end{align}
    where $\bDelta_{\zeta}:=\textsf{Tr}\brac{\bSigma_{\rm T}\bQ}\bSigma_{\rm C}-\zeta\bI_N$.
    There exists some universal constants $c_0,C_0,C_1>0$ such that if $\lambda_r^2\sigma_r\brac{\bF^\top\bQ\bF}>C_0\bar\vartheta$, then we have
    \begin{align*}
        &\PP\brac{\xi^U_{\bQ,\infty}\le \frac{C_1\bar\vartheta \infn{\bU}}{\lambda_r^2\sigma_r\brac{\bF^\top\bQ\bF}}} \mid \bF\ge 1-O\brac{e^{-c_0N}+n^{-9}}.
    \end{align*}
\end{theorem}

\subsection{Proofs of \Cref{thm:V-error} and Second Claim in \Cref{thm:UV-error-inf}}\label{pf-thm:V-error}
In this proof, we will proceed on the event $\calE_0$  defined in \Cref{lem:good-event}.
\paragraph{Bound for $\hat\bV_\bQ$ in $\op{\cdot }$.} Observe that
\begin{align*}
    \wt\bU_\bQ\wt\bU_\bQ^\top\bX=\bM+\bU\bU^\top \bE+	\brac{\wt\bU_\bQ\wt\bU_\bQ^\top-\bU\bU^\top}\bM+\brac{\wt\bU_\bQ\wt\bU_\bQ^\top-\bU\bU^\top}\bE. 
\end{align*}
By the Davis-Kahan theorem and \Cref{lem:GB-bound}, with probability at least $1-O\brac{e^{-cT}}$,
\begin{align*}
    \op{\hat\bV_\bQ-\bV\bR^\top _{V,\bQ}}
    \lesssim\op{\hat\bU_\bQ\hat\bU_\bQ^\top -\bU\bU^\top }
    \brac{\frac{\lambda_1}{\lambda_r}+\frac{K\op{\bSigma^{1/2}_{\rm T}}\op{\bSigma^{1/2}_{\rm C}}}{\lambda_r}\sqrt{\frac{n}{ T}}}
    +\frac{K\op{\bSigma^{1/2}_{\rm T}}\op{\bSigma^{1/2}_{\rm C}}}{\lambda_r }.
\end{align*}
\paragraph{Bound for $\hat\bV_\bQ$ in $\infn{\cdot }$.}
By definition, $\hat\bV_\bQ$ is the leading $r$ left singular vectors of
{\small
\begin{align*}
    \bX^\top \wt\bU_\bQ\wt\bU_\bQ^\top=\sqrt{T}\bV\bSigma
    \bU^\top+\bE^\top\bU\bU^\top +	\sqrt{T}\bV\bSigma\bU^\top \bbrac{\wt\bU_\bQ\wt\bU_\bQ^\top-\bU\bU^\top}+\bE^\top \bbrac{\wt\bU_\bQ\wt\bU_\bQ^\top-\bU\bU^\top}. 
\end{align*}
}
Denote $\wt \bN:=\bE^\top\bU\bU^\top +	\sqrt{T}\bV\bSigma\bU^\top \bbrac{\wt\bU_\bQ\wt\bU_\bQ^\top-\bU\bU^\top}+\bE^\top \bbrac{\wt\bU_\bQ\wt\bU_\bQ^\top-\bU\bU^\top}$. By a standard dilation trick, we can obtain an asymmetric version of \Cref{lem:eigen-expansion} as 
\begin{align*}
    \begin{bmatrix}
        \hat\bV_\bQ\hat\bV_\bQ^\top -\bV\bV^\top & 0\\
        0 & \hat\bU_\bQ\hat\bU_\bQ^\top -\bU\bU^\top 
    \end{bmatrix} =\begin{bmatrix}
        \sum_{p\ge 1}\calS^V_{\bM,p}\bbrac{\wt \bN} & 0\\
        0 & \sum_{p\ge 1}\calS^U_{\bM,p}\bbrac{\wt \bN}
    \end{bmatrix}=\sum_{p\ge 1}\calS_{\bM,p}\bbrac{\wt \bN}
\end{align*}
provided  that $\sigma_{r}\brac{\bM}>4\bop{\wt \bN}$. Here,
$\s=\brac{s_1,\cdots,s_{p+1}}$, $\tau\brac{\s }:=\sum_{j=1}^{p+1}\II\brac{s_j>0}$, and 
\begin{align*}
    \calS_{\bM,p}\bbrac{\wt\bN}:=\sum_{\substack{\s :s_1+\cdots+s_{p+1}=p\\ s_i\ge 0,\forall i=1,\cdots,p+1}}\brac{-1}^{1+\tau\brac{\s}}\frakP^{-s_1}\wt\bN\frakP^{-s_2}\wt\bN\cdots \wt\bN\frakP^{-s_{p+1}}.
\end{align*}
In the block expansion, $\wt\bN$ denotes its symmetric dilation. We define $\wt\bSigma:=\sqrt{T}\bSigma$ and for $\forall s>0$, 
\begin{align*}
    \frakP^{-s}:=\begin{cases}
        \begin{bmatrix}
            0 & \bV\wt \bSigma^{-s}\bU^\top \\
            \bU\wt \bSigma^{-s}\bV^\top  & 0
        \end{bmatrix} & if~$s$~{is~odd}\\
        \begin{bmatrix}
            \bV\wt \bSigma^{-s}\bV^\top & 0 \\
            0  & \bU\wt \bSigma^{-s}\bU^\top
        \end{bmatrix} & if~$s$~{is~even}
    \end{cases},\qquad 
    \frakP^{0}:=\begin{bmatrix}
        \bV_\perp\bV_\perp^\top & 0\\
        0 &\bU_\perp\bU_\perp^\top 
    \end{bmatrix}.
\end{align*}
Note that for terms in $\calS_{\bM,p}\brac{\wt \bN}$ with $s_1\ge 1$ (which starts with $\bV\wt\bSigma^{-s}\bU^\top$), we can  bound them using $\infn{\bA\bB}\le \infn{\bA}\op{\bB}$. 
Denote $\bW_{2q-1}:=\bV_{\perp}\bV_{\perp}^\top\wt\bN$, $\bW_{2q}:=\bU_{\perp}\bU_{\perp}^\top\wt \bN^\top$, $\bB_{2q}:=\bU$, and $\bB_{2q+1}:=\bV$  for $q=1,2,\cdots$. It  suffices to bound $\infn{\bW_1\bW_2\cdots\bW_p\bB_{p+1}}$ for $p\ge 1$.

The signal condition of \Cref{thm:U-error} supplies the first-stage spectral separation. The second-stage condition also implies
\begin{align*}
 \frac{K^2N\op{\bSigma_{\rm C}^{1/2}}^2{\op{\bSigma_{\rm T}^{1/2}\bQ\bSigma_{\rm T}^{1/2}}_1}}
      {\lambda_r^2\sigma_r(\bF^\top\bQ\bF)}
 \le\frac{\overline{\textsf{Err}}_V(\bQ)}{\kappa{\kappa_{\rm T,1}}\lambda_r\sqrt T}
 <\frac1{C_1\kappa{\kappa_{\rm T,1}}}.
\end{align*}
Thus the  condition \eqref{eq:V-loo-separation}  follows from the stated SNR condition. We use the following lemma, proved in \Cref{pf-lem:V-high-order-bound}.
\begin{lemma}\label{lem:V-high-order-bound}
 Suppose that the signal condition of \Cref{thm:U-error} and \eqref{eq:V-loo-separation} hold, $T\gtrsim\log n$, and $T=o(e^{c_0N})$ for a sufficiently small universal constant $c_0>0$. On $\calE_0$, with conditional probability at least $1-O(e^{-cN}+n^{-10})$, simultaneously for all $p\ge1$,
 \begin{align*}
  \infn{\bW_1\bW_2\cdots\bW_p\bB_{p+1}}
  &\le C\infn{\bV}\overline{\textsf{Err}}_V(\bQ)\op{\wt\bN}^{p-1}
  \le C^p\infn{\bV}\bigl[\overline{\textsf{Err}}_V(\bQ)\bigr]^p,
 \end{align*}
 for some universal constants $c,C>0$.
\end{lemma}
For a term of degree $p$ in the projector expansion, a positive first exponent contributes at most $\infn{\bV}\op{\wt\bN}^p/\sigma_r(\bM)^p$. If the first exponent is zero, apply the first bound in \Cref{lem:V-high-order-bound} to the initial sequence of complementary projectors and use operator norms for the remaining factors. Since there are at most $4^p$ terms,
\begin{align*}
 \infn{\calS^V_{\bM,p}(\wt\bN)}
 \lesssim\frac{4^p\infn{\bV}\overline{\textsf{Err}}_V(\bQ)\op{\wt\bN}^{p-1}}{\sigma_r(\bM)^p}.
\end{align*}
The  bound $\op{\wt\bN}\lesssim\textsf{Err}_V(\bQ)$ in \Cref{pf-lem:V-high-order-bound} and the second-stage SNR condition ensure that $4\op{\wt\bN}/\sigma_r(\bM)$ is bounded away from one. Summing the geometric series gives
\begin{align}\label{eq:V-loo-projector-bound}
 \infn{\hat\bV_\bQ\hat\bV_\bQ^\top-\bV\bV^\top}
 &\lesssim\frac{\infn{\bV}\overline{\textsf{Err}}_V(\bQ)}{\sigma_r(\bM)},\\
 \infn{\sum_{p\ge2}\calS^V_{\bM,p}(\wt\bN)\bV}
 &\lesssim\frac{\infn{\bV}\overline{\textsf{Err}}_V(\bQ)\op{\wt\bN}}{\sigma_r(\bM)^2}.
 \label{eq:V-loo-projector-remainder}
\end{align}
Since $\sigma_r(\bM)\asymp\lambda_r\sqrt T$ on $\calE_0$, the first bound proves the second claim in \Cref{thm:UV-error-inf}. The second retains a spectral-norm factor for the higher-order remainder used in \Cref{pf-thm:inf-F}. A union bound with $\calE_0$ gives the stated probability, and $c_1$ may be chosen small enough that $e^{-cN}=O(n^{-10})$.

\subsection{Proofs of \Cref{thm:factor-bound} and \Cref{thm:factor-bound-inf}}\label{pf-thm:factor-bound}
We start with the following lemma, whose proof is deferred to \Cref{pf-lem:wtu-eq-hatu}.
\begin{lemma}\label{lem:wtu-eq-hatu}
    Assume that $\lambda_r^2\cdot \sigma_r\brac{\bF^\top\bQ\bF }\ge C_0 \sqbrac{\textsf{Err}_U(\bQ)+\inf_{\zeta\ge0}\op{\textsf{Tr}\brac{\bSigma_{\rm T}\bQ}\bSigma_{\rm C}-\zeta\bI_N}}$ for some sufficiently large universal constant $C_0>0$, we have $\wt\bU_\bQ \wt\bU_\bQ ^\top =\hat\bU_\bQ \hat\bU_\bQ ^\top$.
    \begin{align*}
        \PP\brac{\wt\bU_\bQ \wt\bU_\bQ ^\top =\hat\bU_\bQ \hat\bU_\bQ ^\top }\ge  1-O\brac{e^{-c_0N}},
    \end{align*}
    for some universal constant $c_0>0$.
\end{lemma}
\Cref{lem:wtu-eq-hatu} establishes the connection between $\wt\bU_\bQ$ and $\hat\bU_\bQ$. In other words,  $\hat\bU_\bQ$ spans the same subspace as $\wt\bU_\bQ$ with probability exceeding $1-O\brac{e^{-c_0N}}$. Next, we shall proceed on the event $\calE_0\bigcap\ebrac{\wt\bU_\bQ \wt\bU_\bQ ^\top =\hat\bU_\bQ \hat\bU_\bQ ^\top}$.
\paragraph{Bound for $\hat\bL_\bQ $ in $\op{\cdot}$.}
By definition, we get 
\begin{align}\label{eq:hatL-decomp}
    \hat\bL_\bQ -\bL\bR_{L,\bQ }&=\hat\bU_\bQ \hat\bSigma_\bQ -\bL\brac{\bB^{-1}}^{\top }\bR_{V,\bQ }^\top\notag\\
    &=\brac{\hat\bU_\bQ -\bU\bR^\top_{U,\bQ}}\hat\bSigma_\bQ +\bU\bR^\top_{U,\bQ}\brac{\hat\bSigma_\bQ -\bR_{U,\bQ }\bSigma\bR_{V,\bQ }^\top}.
\end{align}
We thus arrive at
\begin{align*}
    \op{\hat\bL_\bQ -\bL\bR_{L,\bQ }}\le \op{\hat\bU_\bQ -\bU\bR^\top_{U,\bQ}}\op{\hat\bSigma_\bQ }+\op{\hat\bSigma_\bQ -\bR_{U,\bQ }\bSigma\bR_{V,\bQ }^\top}.
\end{align*}
Notice that $\op{\hat\bSigma_\bQ -\bR_{U,\bQ }\bSigma\bR_{V,\bQ }^\top}\le \op{\bR_{U,\bQ }^\top \hat\bSigma_\bQ \bR_{V,\bQ }-\bH^\top _{U,\bQ}\hat\bSigma_\bQ \bH_{V,\bQ }}+\op{\bH^\top _{U,\bQ}\hat\bSigma_\bQ \bH_{V,\bQ }-\bSigma}$. We then observe that
\begin{align*}
    \op{\bH^\top _{U,\bQ}\hat\bSigma_\bQ \bH_{V,\bQ }-\bSigma}\le \op{\bH^\top _{U,\bQ}\hat\bSigma_\bQ \bH_{V,\bQ }-\bU^\top \brac{T^{-1/2}\bX}\bV}+\op{\bU^\top \brac{T^{-1/2}\bX}\bV-\bSigma}.
\end{align*}
For the first term, with probability at least $1-O\brac{e^{-c_0N}}$,
\begin{align*}
    &\op{\bH^\top _{U,\bQ}\hat\bSigma_\bQ \bH_{V,\bQ }-\bU^\top \brac{T^{-1/2}\bX}\bV}\\
    &=\op{\bU^\top \hat\bU_{\bQ }\hat\bSigma_{\bQ }\hat\bV_{\bQ }^\top \bV-\bU^\top \brac{T^{-1/2}\brac{\wt\bU_{\bQ }\wt\bU_{\bQ }^\top+\wt\bU_{\bQ ,\perp}\wt\bU_{\bQ ,\perp}^\top} \bX}\bV}\\
    &=\frac{1}{\sqrt{T}}\op{\bU^\top \wt\bU_{\bQ ,\perp}\wt\bU_{\bQ ,\perp}^\top \bX\bV}\\
    &\le \op{\bU^\top \wt\bU_{\bQ ,\perp}\wt\bU_{\bQ ,\perp}^\top \bU\bSigma}+\frac{1}{\sqrt{T}}\op{\bU^\top \wt\bU_{\bQ ,\perp}\wt\bU_{\bQ ,\perp}^\top\bSigma_{\rm C}^{1/2}\bG\bSigma_{\rm T}^{1/2}\bV}\\
    &\lesssim \brac{\xi^U_\bQ}^2\op{\bSigma}+K\sqrt{\frac{N}{T}}\xi^U_\bQ\op{\bSigma_{\rm C}^{1/2}}\op{\bSigma_{\rm T}^{1/2}\bV},
\end{align*}
where we have used $\op{\bU^\top\wt\bU_{\bQ,\perp}\wt\bU_{\bQ,\perp}^\top\bU}\le\brac{\xi^U_\bQ}^2$ and $\op{\bU^\top\wt\bU_{\bQ,\perp}\wt\bU_{\bQ,\perp}^\top}=\xi^U_\bQ$, and applied \Cref{lem:GB-bound} in the last inequality. For the second term, with probability at least $1-O\bbrac{ n ^{-10}}$,
\begin{align*}
    \op{\bU^\top \brac{T^{-1/2}\bX}\bV-\bSigma}=\frac{1}{\sqrt{T}}\op{\bU^\top \bSigma_{\rm C}^{1/2}\bG\bSigma_{\rm T}^{1/2}\bV}\lesssim K\sqrt{\frac{r+\log n }{T}}\op{\bU^\top \bSigma_{\rm C}^{1/2}}\op{\bSigma_{\rm T}^{1/2}\bV}.
\end{align*}
Hence we arrive at with probability at least $1-O\bbrac{e^{-c_0N}+n ^{-10}}$,
\begin{align*}
    \op{\bH^\top _{U,\bQ}\hat\bSigma_\bQ \bH_{V,\bQ }-\bSigma}
    &\lesssim \brac{\xi^U_\bQ }^2\op{\bSigma}
    +K \xi^U_\bQ \sqrt{\frac{N}{T}}\op{\bSigma_{\rm C}^{1/2}}\op{\bSigma_{\rm T}^{1/2}\bV}\\
    &\quad+K\sqrt{\frac{r+\log n }{T}}\op{\bU^\top \bSigma_{\rm C}^{1/2}}\op{\bSigma_{\rm T}^{1/2}\bV}.
\end{align*}
Note that if the singular values of $\bH_{U,\bQ}$ are $\cos\theta_j$, then $\op{\bR_{U,\bQ}-\bH_{U,\bQ}}=\max_j\brac{1-\cos\theta_j}\le\max_j\sin^2\theta_j$, and the same argument applies to $\bH_{V,\bQ}$. Therefore, we get $\op{\bR_{U,\bQ}-\bH_{U,\bQ}}\lesssim\brac{\xi^U_\bQ}^2$ and $\op{\bR_{V,\bQ}-\bH_{V,\bQ}}\lesssim\brac{\xi^V_\bQ}^2$. It follows that
\begin{align*}
    \op{\bR_{U,\bQ }^\top \hat\bSigma_\bQ \bR_{V,\bQ }-\bH^\top _{U,\bQ}\hat\bSigma_\bQ \bH_{V,\bQ }}\lesssim\op{\hat\bSigma_\bQ}\sqbrac{\brac{\xi^U_\bQ}^2+\brac{\xi^V_\bQ}^2}.
\end{align*}
Moreover, on $\calE_0$, we have $\sigma_r\brac{\bF^\top\bQ\bF}\le\op{\bF}^2\op{\bQ}\lesssim T$ and  $\brac{\fro{\bQ}+\sqrt N}\sqrt N\ge N$. The assumed SNR condition and the second term in $\textsf{Err}_U(\bQ)$ therefore imply
\begin{align*}
    K\sqrt{\frac NT}\op{\bSigma_{\rm C}^{1/2}}\op{\bSigma_{\rm T}^{1/2}}
    \lesssim\lambda_r\le\lambda_1.
\end{align*}
Moreover, by \Cref{lem:GB-bound}, with probability at least $1-O\bbrac{e^{-c_0(N+T)}}$,
\begin{align*}
    T^{-1/2}\op{\bE}
    \lesssim K\brac{1+\sqrt{\frac NT}}\op{\bSigma_{\rm C}^{1/2}}\op{\bSigma_{\rm T}^{1/2}}.
\end{align*}
Consequently, on the intersection of these events,
\begin{align*}
    \op{\hat\bSigma_\bQ}
    \le T^{-1/2}\op{\bX}
    \lesssim\lambda_1+K\op{\bSigma_{\rm C}^{1/2}}\op{\bSigma_{\rm T}^{1/2}}.
\end{align*}
Combining the preceding bounds  gives the second-order contribution. In addition, on $\calE_0$, $\op{\bSigma}\lesssim\lambda_1$ and
$K\xi^U_\bQ\sqrt{N/T}\op{\bSigma_{\rm C}^{1/2}}\op{\bSigma_{\rm T}^{1/2}\bV}
    \lesssim\lambda_1\xi^U_\bQ$. Therefore, combining all pieces gives
\begin{align}\label{eq:hatSigma-bound}
	\op{\hat\bSigma_\bQ -\bR_{U,\bQ }\bSigma\bR_{V,\bQ }^\top}
	&\lesssim\brac{\lambda_1+K\op{\bSigma_{\rm C}^{1/2}}\op{\bSigma_{\rm T}^{1/2}}}
	\sqbrac{\xi^U_\bQ+\brac{\xi^V_\bQ}^2}\notag\\
	&\quad+K\sqrt{\frac{r+\log n}{T}}\op{\bU^\top\bSigma_{\rm C}^{1/2}}\op{\bSigma_{\rm T}^{1/2}\bV}.
\end{align}
For $\op{\hat\bU_\bQ -\bU\bR^\top_{U,\bQ}}\op{\hat\bSigma_\bQ }$, we have {\small $\op{\hat\bU_\bQ -\bU\bR^\top_{U,\bQ}}\op{\hat\bSigma_\bQ }\lesssim \xi^U_\bQ  \brac{\op{\bSigma}+\op{\hat\bSigma_\bQ -\bR_{U,\bQ }\bSigma\bR_{V,\bQ }^\top}}$}. Hence we arrive at with probability $1-O\bbrac{ e^{-c_0N}+n ^{-10}}$,
\begin{align*}
    \frac{1}{\sqrt{N}}\op{\hat\bL_\bQ -\bL\bR_{L,\bQ }}
    &\lesssim\frac{\lambda_1+K\op{\bSigma_{\rm C}^{1/2}}\op{\bSigma_{\rm T}^{1/2}}}{\sqrt N}
    \brac{\xi^U_\bQ+\brac{\xi^V_\bQ}^2}\\
    &x\quad+K\sqrt{\frac{r+\log n}{NT}}\op{\bU^\top\bSigma_{\rm C}^{1/2}}\op{\bSigma_{\rm T}^{1/2}\bV}.
\end{align*}
\paragraph{Bound for $\hat\bL_\bQ $ in $\infn{\cdot}$.}
By \eqref{eq:hatL-decomp} and the orthogonality of $\bR_{U,\bQ}$ and $\bR_{V,\bQ}$, we get that
\begin{align*}
    \hat\bL_\bQ -\bL\bR_{L,\bQ }
    =\brac{\hat\bU_\bQ-\bU\bR_{U,\bQ}^\top}\hat\bSigma_\bQ
    +\bU\brac{\bR_{U,\bQ}^\top\hat\bSigma_\bQ\bR_{V,\bQ}-\bSigma}\bR_{V,\bQ}^\top.
\end{align*}
Notice that $\op{\bH_{U,\bQ }-\bR_{U,\bQ }}\lesssim\op{\hat \bU_\bQ-\bU\bR^\top_{U,\bQ}}^2\lesssim\brac{\xi^U_\bQ }^2$  by the proof of Lemma 1 in \cite{yan2021inference}, we have 
\begin{align*}
    \infn{\hat\bU_\bQ -\bU\bR^\top_{U,\bQ}}&\le \infn{\hat\bU_\bQ -\bU\bH^\top_{U,\bQ}}+\infn{\bU\brac{\bH_{U,\bQ }-\bR_{U,\bQ }}^\top}\\
    &\le \xi^U_{\bQ ,\infty}+\brac{\xi^U_\bQ }^2\infn{\bU}.
\end{align*}
Note that on the event $\wt\bU_\bQ\wt\bU_\bQ^\top=\hat\bU_\bQ\hat\bU_\bQ^\top$,
{\small
\begin{align*}
    &\bU\brac{\bH_{U,\bQ}^\top\hat\bSigma_\bQ\bH_{V,\bQ}-\bSigma}=-\bU\bU^\top\wt\bU_{\bQ,\perp}\wt\bU_{\bQ,\perp}^\top\bU\bSigma
    +\frac{1}{\sqrt T}\bU\bU^\top\bE\bV
    -\frac{1}{\sqrt T}\bU\bU^\top\wt\bU_{\bQ,\perp}\wt\bU_{\bQ,\perp}^\top\bE\bV.
\end{align*}
}
Here,
\begin{align*}
    \op{\bU^\top\wt\bU_{\bQ,\perp}\wt\bU_{\bQ,\perp}^\top\bU}
    \lesssim\brac{\xi^U_\bQ}^2,\qquad
    \infn{\bU\bU^\top\wt\bU_{\bQ,\perp}\wt\bU_{\bQ,\perp}^\top}
    \le\xi^U_{\bQ,\infty},
\end{align*}
where the second inequality follows from
$\bU\bU^\top\wt\bU_{\bQ,\perp}\wt\bU_{\bQ,\perp}^\top
=\brac{\bU\bU^\top-\wt\bU_\bQ\wt\bU_\bQ^\top}\wt\bU_{\bQ,\perp}\wt\bU_{\bQ,\perp}^\top$.
Moreover, by \Cref{lem:GB-bound},
\begin{align*}
    \frac{1}{\sqrt T}\op{\bE\bV}
    \lesssim K\sqrt{\frac{N+r}{T}}\op{\bSigma_{\rm C}^{1/2}}\op{\bSigma_{\rm T}^{1/2}\bV}
    \lesssim\lambda_1+K\op{\bSigma_{\rm C}^{1/2}}\op{\bSigma_{\rm T}^{1/2}}.
\end{align*}
It follows that
{\small
\begin{align*}
    &\infn{\bU\brac{\bH_{U,\bQ}^\top\hat\bSigma_\bQ\bH_{V,\bQ}-\bSigma}}\\
    &\quad\lesssim\brac{\lambda_1+K\op{\bSigma_{\rm C}^{1/2}}\op{\bSigma_{\rm T}^{1/2}}}
    \sqbrac{\xi^U_{\bQ,\infty}+\infn{\bU}\brac{\xi^U_\bQ}^2}
    \\ &\qquad+K\infn{\bU}\sqrt{\frac{r+\log n}{T}}
    \op{\bU^\top\bSigma_{\rm C}^{1/2}}\op{\bSigma_{\rm T}^{1/2}\bV}.
\end{align*}
}
Together with the factor bound above,
{\small
\begin{align*}
    \infn{\bU\brac{\bR_{U,\bQ}^\top\hat\bSigma_\bQ\bR_{V,\bQ}
    -\bH_{U,\bQ}^\top\hat\bSigma_\bQ\bH_{V,\bQ}}}\lesssim\infn{\bU}
    \brac{\lambda_1+K\op{\bSigma_{\rm C}^{1/2}}\op{\bSigma_{\rm T}^{1/2}}}
    \brac{\brac{\xi^U_\bQ}^2+\brac{\xi^V_\bQ}^2}.
\end{align*}
}
Combining the last two displays with the loading decomposition gives
\begin{align*}
    \infn{\hat\bL_\bQ-\bL\bR_{L,\bQ}}
    &\lesssim\brac{\lambda_1+K\op{\bSigma_{\rm C}^{1/2}}\op{\bSigma_{\rm T}^{1/2}}}
    \sqbrac{\xi^U_{\bQ,\infty}+\infn{\bU}
    \brac{\brac{\xi^U_\bQ}^2+\brac{\xi^V_\bQ}^2}}\\
    &\quad+K\infn{\bU}\sqrt{\frac{r+\log n}{T}}
    \op{\bU^\top\bSigma_{\rm C}^{1/2}}\op{\bSigma_{\rm T}^{1/2}\bV}.
\end{align*}
Finally, define $\bU_{L}:=\bL\bLambda^{-1}\in\RR^{N\times r}$. It is readily seen that $\bU_L^\top \bU_L=\bI_r$ and $\text{span}\brac{\bU_L}=\text{span}\brac{\bL}=\text{span}\brac{\bU}$. Hence there exists $\bO_L\in\OO_{r}$ such that $\bU=\bU_L\bO_L$. We thus get that
\begin{align}\label{eq:U-L-infty-relation}
    \infn{\bU}=\infn{\bU_L\bO_L}=\infn{\bL\bLambda^{-1}}.
\end{align}
Using \eqref{eq:U-L-infty-relation} gives the stated form in \Cref{thm:factor-bound-inf}.
\paragraph{Bound for $\hat\bF_\bQ $ in $\op{\cdot}$.}
By definition, we get 
\begin{align}\label{eq:hatF-decomp}
    \hat\bF_\bQ -\bF\bR_{F,\bQ }=\sqrt{T}\hat\bV_\bQ -\sqrt{T}\bV\bB^{-1}\bB\bR^\top_{V,\bQ}=\sqrt{T}\brac{\hat\bV_\bQ -\bV\bR_{V,\bQ }^\top}.
\end{align}
Using the same Wedin argument as in the proof of \Cref{thm:V-error}, we thus arrive at
\begin{align*}
    \frac{1}{\sqrt{T}}\op{\hat\bF_\bQ -\bF\bR_{F,\bQ }}
    &\lesssim \xi^U_\bQ \brac{\kappa +\frac{K\op{\bSigma^{1/2}_{\rm T}}\op{\bSigma^{1/2}_{\rm C}}}{\lambda_r}\sqrt{\frac{n}{ T}}}
    +\frac{K\op{\bSigma^{1/2}_{\rm T}}\op{\bSigma^{1/2}_{\rm C}}}{\lambda_r}.
\end{align*}
\paragraph{Bound for $\hat\bF_\bQ $ in $\infn{\cdot}$.}
From \eqref{eq:hatF-decomp}, we obtain 
\begin{align*}
    \infn{\hat\bF_\bQ -\bF\bR_{F,\bQ }}=\sqrt{T}\infn{\hat\bV_\bQ -\bV\bR_{V,\bQ }^\top}.
\end{align*}
Denote $\bD:=\hat\bV_\bQ\hat\bV_\bQ^\top-\bV\bV^\top$ and $\bC:=\bV^\top\hat\bV_\bQ\bR_{V,\bQ}\succeq0$. Then
\begin{align*}
    \hat\bV_\bQ\bR_{V,\bQ}-\bV=\bD\hat\bV_\bQ\bR_{V,\bQ}+\bV\brac{\bC-\bI_r},
    \qquad
    \bV\brac{\bI_r-\bC}=\bD^2\bV\brac{\bI_r+\bC}^{-1}.
\end{align*}
Since $\op{\bD}\le1$ and $\op{\brac{\bI_r+\bC}^{-1}}\le1$, it follows that
\begin{align*}
    \infn{\hat\bV_\bQ-\bV\bR_{V,\bQ}^\top}
    =\infn{\hat\bV_\bQ\bR_{V,\bQ}-\bV}\le2\infn{\bD},
\end{align*}
which gives the desired result by the second claim of \Cref{thm:UV-error-inf}.

\subsection{Proof of \Cref{thm:U-two-infinty}}\label{pf-thm:U-two-infinty}
The proof follows similar logic to that of Theorem 3 in \cite{agterberg2022entrywise}.
Let $\overline\bM:=\bM\bQ\bM^\top $, then by definition $\bar\bU\bar\bSigma \bar\bU^\top $ is the rank-r SVD of  $\overline\bM$, and $\bar\bSigma$ contains $\sigma_1\brac{\overline\bM}\ge \sigma_2\brac{\overline\bM}\ge \cdots \sigma_r\brac{\overline\bM}>0$. 
Note that $\wt\bU_\bQ$ can be viewed as the leading eigenvectors of the following  matrix:
\begin{align}\label{eq:main-decomp}
    &\underbrace{\bM\bQ\bM^\top}_{\overline{\bM}}+\underbrace{\bM\bQ\bE^{\top}+\bE\bQ\bM^\top}_{\bN_1}+\underbrace{\bE\bQ\bE^{\top}-\EE\bE\bQ\bE^{\top}}_{\bar\bN_2}+\underbrace{\EE\bE\bQ\bE^{\top}-\zeta\bI_N}_{\bDelta_{\zeta}},
\end{align}
 for any $\zeta\in\RR$. Denote $\bN:=\bN_1+\bN_2$ with $\bN_2:=\bar\bN_2+\bDelta_{\zeta}$. 
Adding a scalar multiple of the identity preserves the leading eigenspace. Taking the infimum over $\zeta\in\RR$ in the bounds below, the stated condition and $\op{\bDelta_\zeta}\le\op{\bDelta_\zeta}_1$ give $\sigma_r\brac{\overline\bM}>4\op{\bN}$.
The following lemma is  Theorem 1 in \cite{xia2021normal} adapted to our notation. 
\begin{lemma}[Theorem 1 in \cite{xia2021normal}]\label{lem:eigen-expansion}
    Suppose that $\sigma_{r}\brac{\overline\bM}>4\op{\bN}$, then we have
    \begin{align*}
        \wt\bU_\bQ \wt\bU_\bQ^\top -\bar\bU\bar\bU^\top =\sum_{p\ge 1}\calS_{\overline\bM,p}\brac{\bN}.
    \end{align*}
    Here, we define $\frakP^{-s}:=\bar\bU\bar\bSigma^{-s}\bar\bU^\top $ for $\forall s>0$, $\frakP^{0}:=\bar\bU_\perp\bar\bU_\perp^\top $,  $\s=\brac{s_1,\cdots,s_{p+1}}$, $\tau\brac{\s }:=\sum_{j=1}^{p+1}\II\brac{s_j>0}$, and 
    \begin{align*}
        \calS_{\overline\bM,p}\brac{\bN}:=\sum_{\substack{\s :s_1+\cdots+s_{p+1}=p\\ s_i\ge 0,\forall i=1,\cdots,p+1}}\brac{-1}^{1+\tau\brac{\s}}\frakP^{-s_1}\bN\frakP^{-s_2}\bN\cdots \bN\frakP^{-s_{p+1}}.
    \end{align*}
\end{lemma}
By \Cref{lem:eigen-expansion} and the condition $\sigma_{r}\brac{\overline\bM}>4\op{\bN}$, it suffices for us to bound the term $\infn{\frakP^{-s_1}\bN\frakP^{-s_2}\bN\cdots \bN\frakP^{-s_{p+1}}}$. For any $\s$ such that $s_1+\cdots+s_{p+1}=p$ and  $s_1\ge 1$, we have
\begin{align*}
    \infn{\frakP^{-s_1}\bN\frakP^{-s_2}\bN\cdots \bN\frakP^{-s_{p+1}}}\le \infn{\bar\bU}\brac{\frac{\op{\bN}}{\sigma_r\brac{\overline\bM} }}^p\le \infn{\bar\bU}\brac{\frac{\bar\vartheta}{\sigma_r\brac{\overline\bM} }}^p.
\end{align*}
With the aid of \Cref{lem:eigen-expansion}, we get the following lemma whose proof is deferred to \Cref{pf-lem:high-order-bound}.
\begin{lemma}\label{lem:high-order-bound} 
    Instate the conditions of \Cref{thm:U-two-infinty}. Consider the model defined in \eqref{eq:main-decomp}, there exists universal constants $C,C_0,C_1$ such that for any $p\ge 1$, we have with probability at least $1-O\bbrac{p  \brac{e^{-c_0N}+n^{-10}}}$ for any $1\le q\le p$ that 
    \begin{align*}
        \infn{\bar\bU_{\perp}\bar\bU_{\perp}^\top\brac{\bar\bU_{\perp}\bar\bU_{\perp}^\top\bN_2\bar\bU_{\perp}\bar\bU_{\perp}^\top }^{q-1}\bN\bar\bU}\le C_1\brac{C_0\bar\vartheta }^{q}\infn{\bar\bU},
    \end{align*}
    where $\bar\vartheta $ is defined in \eqref{eq:theta-def}.
\end{lemma}
For any $\s$ such that $s_1+\cdots+s_{p+1}=p$ and  $s_1=0$, by \Cref{lem:high-order-bound} and the fact that $\bar\bU_{\perp}\bar\bU_{\perp}^\top\bN_1\bar\bU_{\perp}\bar\bU_{\perp}^\top=0$, we have with probability at least $1-O\bbrac{p \brac{e^{-c_0N}+n ^{-10}}}$,
\begin{align*}
    \infn{\bar\bU_{\perp}\bar\bU_{\perp}^\top \bN\frakP^{-s_2}\bN\cdots \bN\frakP^{-s_{p+1}}}\le C_1\infn{\bar\bU}\brac{\frac{C_0\bar\vartheta}{\sigma_r\brac{\overline\bM} }}^{p}.
\end{align*}
With $p=2\lceil\log N\rceil$, we could arrive at with probability at least $1-O\bbrac{e^{-\wt c_0N}+n ^{-9}}$,
\begin{align*}
    \infn{\wt\bU_\bQ\wt\bU^\top_\bQ -\bar\bU\bar\bU^\top }&\le \sum_{p=1}^{2\lceil\log  N \rceil}\infn{\calS_{\overline\bM,p}\brac{\bN}}+\sum_{p\ge 2\log  N }\infn{\calS_{\overline\bM,p}\brac{\bN}}\\
    &\le C_1\infn{\bar\bU}\sum_{p=1}^{2\lceil\log  N \rceil}\brac{\frac{C_0\bar\vartheta}{\sigma_r\brac{\overline\bM} }}^{p}+\sum_{p\ge 2\log  N }\brac{\frac{4\op{\bN}}{\sigma_r\brac{\overline\bM} }}^p\\
    &\lesssim \frac{\bar\vartheta}{\sigma_r\brac{\overline\bM} }\brac{\infn{\bU}+ N ^{-2}}\lesssim \frac{\bar\vartheta}{\sigma_r\brac{\overline\bM} }\infn{\bU}.
\end{align*} 
where the last inequality holds due to $\infn{\bU}\gtrsim \sqrt{r/N}$. 

\section{Proofs in \Cref{subsec:inf}}\label{sec:pf-sec-2}
This section develops the inference results in Theorems \Cref{thm:inf-L} and \Cref{thm:inf-F}, as well as a rank-one heteroskedastic example for the PC loading bias.  Our arguments blend the spectral-projector expansion of  \Cref{lem:eigen-expansion} with the $\ell_2\rightarrow\ell_\infty$ error bounds for the singular subspaces $\bU$ and $\bV$.   In all these proofs, we will conduct our analysis on event $\calE_0$ defined in \Cref{lem:good-event}.

\subsection{Proof of \Cref{thm:inf-L}}\label{pf-thm:inf-L}
 From \eqref{eq:U-L-infty-relation} we know that $\infn{\bU}=\infn{\bL\bLambda^{-1}}\lesssim\sqrt{r/N}$. By definition, we have $\bar\rho\le \op{\bSigma_{\rm C}}_1\le \op{\bSigma_{\rm C}^{1/2}}^2_1$ and $\ab{\textsf{Tr}\brac{\bSigma_{\rm T}\bQ}}\le T\op{\bSigma_{\rm T}}$. We first state a weaker condition on $\lambda_r$ that is implied by the condition of \Cref{thm:inf-L} via the nondegeneracy condition, $\op{\bQ}=1$, and the inequality $ab\le a^p/p+b^{q}/q$:
\begin{align*}
    \frac{\lambda_r}{K} &\gg \kappa^3r^{1/2}\op{\bSigma_{\rm C}^{1/2}}^2{\op{\bSigma_{\rm T}^{1/2}}^2}+\brac{\frac{n}{T}}^{1/2}r\log^{1/2}n\op{\bSigma_{\rm C}^{1/2}}^2\op{\bSigma_{\rm T }^{1/2}}^2\\
    &+\brac{\frac{n}{T}}^{1/3}\brac{\frac{N}{T}}^{1/6} \kappa\kappa_{\rm C, 1}^{2/3}\kappa_{\rm C, 2}^{2/3}r^{1/2}\log n\op{\bSigma_{\rm C}^{1/2}}_1^{4/3}\op{\bSigma_{\rm T}^{1/2}}^{4/3}\\
    &+\brac{\frac{N}{T}}^{1/2}\kappa^5\kappa_{\rm C, 1}^2\kappa_{\rm C, 2}^2r^{1/2}\log^2 n\op{\bSigma_{\rm C}^{1/2}}_1^2\op{\bSigma_{\rm T}^{1/2}}^2\\
    &+\bar\rho r^{1/2}\frac{\ab{\textsf{Tr}\brac{\bSigma_{\rm T}\bQ}}}{\brac{NT}^{1/2}}+\bar\rho^{2/3}\kappa r^{1/6}\frac{\ab{\textsf{Tr}\brac{\bSigma_{\rm T}\bQ }}^{2/3}}{\brac{N}^{1/6}T^{1/2}}\\
    &+\rho^{1/2}\kappa r^{1/4}\frac{\ab{\textsf{Tr}\brac{\bSigma_{\rm T}\bQ}}^{1/2}}{\brac{NT}^{1/4}}\op{\bSigma_{\rm C}^{1/2}}^{1/2}\op{\bSigma_{\rm T}^{1/2}}^{1/2}\\
    &{+K\brac{\frac{T}{N}}^{1/2}\kappa^3r^{1/2}\op{\bSigma_{\rm C}^{1/2}}^2\op{\bSigma_{\rm T}^{1/2}}^2+\brac{\frac{n}{T}}^{1/2}\kappa\op{\bSigma_{\rm C}^{1/2}}\op{\bSigma_{\rm T}^{1/2}}}.
\end{align*}
Note that the above condition also implies the condition to apply  \Cref{lem:eigen-expansion} and \Cref{thm:U-error}, which will be used in the proof.

Recall that $\bH_{U,\bQ}=\hat\bU_\bQ^\top\bU$, by \Cref{lem:eigen-expansion}  we have
\begin{align*}
    \hat\bU_\bQ\bH_{U,\bQ} -\bU=\brac{\hat\bU_\bQ\hat\bU_\bQ^\top -\bU\bU^\top}\bU =\bU_{\perp}\bU_{\perp}^\top \bN\bU\bar\bO^\top \bar\bSigma^{-1}\bar\bO+\sum_{p\ge 2}\calS_{\overline\bM,p}\brac{\bN}\bU. 
\end{align*}
We can get the following decomposition:
\begin{align*}
    \hat\bL_\bQ-\bL\bR_{L,\bQ}&=\hat\bU_\bQ\hat\bSigma_\bQ-\bL\brac{\bB^{-1}}^{\top }\bR_{V,\bQ}^\top=\hat\bU_\bQ\hat\bSigma_\bQ-\bU\bSigma\bR_{V,\bQ}^\top\notag\\
    &=\brac{\hat\bU_\bQ\bH_{U,\bQ}-\bU}\bSigma\bR_{V,\bQ}^\top\\
    &+\underbrace{\hat\bU_\bQ\brac{\bR_{U,\bQ}-\bH_{U,\bQ}}\bSigma\bR_{V,\bQ}^\top+\hat\bU_\bQ\bR_{U,\bQ}\brac{\bR_{U,\bQ}^\top \hat\bSigma_\bQ-\bSigma\bR_{V,\bQ}^\top}}_{=:\mPhi_{L}}.
\end{align*}
\paragraph{Characterizing the first order term in the leading term.} 
Let $\bar\zeta=\argmin_{\zeta\in\RR}\op{\bDelta_\zeta}_1$ and $\bO_F:=\bar\bO^\top \bar\bSigma^{-1}\bar\bO\bSigma\bR_{V,\bQ}^\top$. We first characterize the distribution of each row of $\bU_{\perp}\bU_{\perp}^\top \bN\bU$ instead of $\bU_{\perp}\bU_{\perp}^\top \bN\bU\bO_F$. For $i\in[N]$, by \Cref{lem:eigen-expansion} we have
\begin{align}\label{eq:inf-L-first-order-form}
    \sqbrac{\bU_{\perp}\bU_{\perp}^\top \bN\bU}_{i,\cdot}&=\e_i^\top \bE\bQ\bM^\top\bU+\e_i^\top\frakD\brac{ \bE\bQ\bE^{\top}}\bU+\e_i^\top \bU_{\perp}\bU_{\perp}^\top\bDelta_{\bar\zeta}\bU\notag\\
    &-\e_i^\top \bU\bU^\top \sqbrac{\bE\bQ\bM^\top\bU+\frakD\brac{\bE\bQ\bE^{\top}}\bU+\bDelta_{\bar\zeta}\bU}.
\end{align}
We treat each term in \eqref{eq:inf-L-first-order-form} separately.  
Denote $\bZ=\bE\bQ\bM^\top\bU$ and $$\bY_{i,jt}:=G_{j,t}\sqbrac{\bSigma_{\rm C}^{1/2}}_{i,j} \sqbrac{\bSigma_{\rm T}^{1/2}\bQ\bM^\top\bU}_{t,\cdot}^\top. $$ For the first term in \eqref{eq:inf-L-first-order-form}, we have
\begin{align*}
    \bZ_{i,\cdot}^\top =\sum_{j\in[N]}\sum_{t\in[T]}\bY_{i,jt}=\sum_{j\in[N]}\sum_{t\in[T]}G_{j,t}\underbrace{\sqbrac{\bSigma_{\rm C}^{1/2}}_{i,j} \sqbrac{\bSigma_{\rm T}^{1/2}\bQ\bM^\top\bU}_{t,\cdot}^\top}_{\w_{j,t}^{(i)}}.
\end{align*}
Notice that 
\begin{align*}
\sum_{j\in[N]}\sum_{t\in[T]}\textsf{Cov}\brac{\bY_{i,jt}}&=\sum_{j\in[N]}\sum_{t\in[T]}\sqbrac{\bSigma_{\rm C}^{1/2}}_{i,j}^2\bU^\top\bM\bQ\sqbrac{\bSigma_{\rm T}^{1/2}}_{\cdot,t}\sqbrac{\bSigma_{\rm T}^{1/2}}_{\cdot,t}^\top\bQ\bM^\top\bU\\
    &= \underbrace{T\cdot \sqbrac{\bSigma_{\rm C}}_{i,i}\bSigma\bV^\top \bQ\bSigma_{\rm T}\bQ\bV\bSigma}_{=:\bar\bSigma_{L,i}}.
\end{align*}
By the nondegeneracy condition in \Cref{thm:inf-L},  we have $\textsf{Cov}\brac{\bar\bSigma_{L,i}^{-1/2}\bZ_{i,\cdot}^\top}=\bI_r$. We then show that  $\bar\bSigma_{L,i}^{-1/2}\bZ_{i,\cdot }^\top \overset{d}{\rightarrow} \calN\brac{\mathbf0,\bI_r}$ by Lemma~\ref{lem:subg_clt_weighted}. To that end, it remains to verify 
\begin{align*}
\delta_{L,i}:=\max_{j,t}\big\|\bar\bSigma_{L,i}^{-1/2} \w_{j,t}^{(i)}\big\|_2 \to 0.
\end{align*}
{By definition, $\op{\bar\bSigma_{L,i}^{-1/2}}\lesssim \bbrac{\lambda_r\sqrt{T}\sqbrac{\bSigma_{\rm C}}_{i,i}^{1/2}\sigma_r\brac{\bSigma_{\rm T}^{1/2}\bQ\bV}}^{-1}$. Moreover, on $\calE_0$, $\sigma_r\brac{\bF^\top\bQ\bF}\gtrsim T$ and $\op{T^{-1}\bF^\top\bF-\bI_r}\le\delta_F=o(1)$ imply $\sigma_r\brac{\bV^\top\bQ\bV}\gtrsim1$. Hence
\[
\sigma_r\brac{\bSigma_{\rm T}^{1/2}\bQ\bV}\ge \sigma_r\brac{\bSigma_{\rm T}^{1/2}}\sigma_r\brac{\bQ\bV}\ge \frac{\op{\bSigma_{\rm T}^{1/2}}}{\kappa_{\rm T}}\sigma_r\brac{\bV^\top\bQ\bV}\gtrsim\frac{\op{\bSigma_{\rm T}^{1/2}}}{\kappa_{\rm T}}.
\]}
Combining with $\|(\bSigma_{\rm T}^{1/2}\bQ \bM^\top \bU)_{t,\cdot}\|_2\le \sqrt{T}\lambda_1\infn{\bSigma_{\rm T}^{1/2}\bQ\bV}$, we obtain
\begin{align*}
\delta_{L,i}\le \|\bar\bSigma_{L,i}^{-1/2}\|\max_{j,t}\|\w_{j,t}^{(i)}\|_2
\lesssim {\kappa\frac{\infn{\bSigma_{\rm T}^{1/2}\bQ\bV}}{\sigma_r\brac{\bSigma_{\rm T}^{1/2}\bQ\bV}}\lesssim\kappa\kappa_{\rm T}\frac{\infn{\bSigma_{\rm T}^{1/2}\bQ\bV}}{\op{\bSigma_{\rm T}^{1/2}}}}=o(1),
\end{align*}
by the condition in \Cref{thm:inf-L}. Therefore Lemma~\ref{lem:subg_clt_weighted} yields {$\bar\bSigma_{L,i}^{-1/2}\bZ_{i,\cdot}^\top \overset{d}{\rightarrow} \mathcal N(\mathbf0,\bI_r)$ conditionally on $\bF$}. {By the compatible-whitening convention in \Cref{thm:inf-L},
\begin{align*}
    \bSigma_{L,i}^{-1/2}\bO_F^\top\bZ_{i,\cdot}^\top=\bar\bSigma_{L,i}^{-1/2}\bZ_{i,\cdot}^\top\overset{d}{\rightarrow}\calN\brac{\mathbf0,\bI_r}.
\end{align*}}
It remains to show that the remaining terms in  \eqref{eq:inf-L-first-order-form} are negligible. Using {$\op{\bar\bSigma_{L,i}^{-1/2}}\lesssim \bbrac{\lambda_r \sqrt{T}}^{-1}$ by the nondegeneracy condition} and \Cref{lem:GBGU-two-inf-bound}, we obtain that with probability at least $1-O\brac{n^{-10}}$,
\begin{align*}
    \op{\bar\bSigma_{L,i}^{-1/2}\sqbrac{\e_i^\top \frakD\brac{\bE\bQ\bE^{\top}}\bU}^\top}\lesssim  \frac{K^2}{\lambda_r}\sqrt{\frac{nNr\log n}{T}}\infn{\bU}\op{\bSigma_{\rm C}^{1/2}}_1^2\op{\bSigma_{\rm T}^{1/2}}^2=o\brac{1},
\end{align*}
and
\begin{align*}
    \op{\bar\bSigma_{L,i}^{-1/2}\sqbrac{\e_i^\top \bDelta_{\bar\zeta}\bU}^\top}\lesssim \frac{\op{\bDelta_{\bar\zeta}}_1\infn{\bU}}{\lambda_r\sqrt{T}}=\frac{\bar\rho \ab{\textsf{Tr}\brac{\bSigma_{\rm T}\bQ}}\infn{\bU}}{\lambda_r\sqrt{T}}=o\brac{1},
\end{align*}
provided that 
\begin{align}\label{eq:inf-L-cond-0}
    \lambda_r\gg K^2\sqrt{\frac{n}{T}}r\log^{1/2}n\op{\bSigma_{\rm C}^{1/2}}^2\op{\bSigma_{\rm T }^{1/2}}^2+\frac{\ab{\textsf{Tr}\brac{\bSigma_{\rm T}\bQ}}}{\brac{NT}^{1/2}}\bar\rho r^{1/2}.
\end{align}
In addition, we have with probability at least $1-O\brac{n^{-10}}$,
\begin{align*}
    \op{\bar\bSigma_{L,i}^{-1/2}\sqbrac{\e_i^\top \bU\bU^\top \brac{\bE\bQ\bM^\top\bU}}^\top }\lesssim K\infn{\bU}\sqrt{r+\log n}\op{\bSigma_{\rm C}^{1/2}}\op{\bSigma_{\rm T }^{1/2}}=o\brac{1},
\end{align*}
and
\begin{align*}
    \op{\bar\bSigma_{L,i}^{-1/2}\sqbrac{\e_i^\top\bU\bU^\top  \frakD\brac{\bE\bQ\bE^{\top}}\bU}^\top}\lesssim  \frac{K^2N}{\lambda_r\sqrt{T}}\infn{\bU}\op{\bSigma_{\rm C}^{1/2}}^2\op{\bSigma^{1/2}_{\rm T}}^2=o\brac{1},
\end{align*}
provided that $N\gg Kr\brac{r+\log n}\op{\bSigma_{\rm C}^{1/2}}^2\op{\bSigma_{\rm T }^{1/2}}^2$ and
\begin{align*}
    \lambda_r\gg K^2\sqrt{\frac{N}{T}}r^{1/2}\op{\bSigma_{\rm C}^{1/2}}\op{\bSigma_{\rm T }^{1/2}},
\end{align*}
which is implied by \eqref{eq:inf-L-cond-0}.
Moreover, we have $\op{\bar\bSigma_{L,i}^{-1/2}\sqbrac{\e_i^\top\bU\bU^\top  \bDelta_{\bar\zeta}\bU}^\top}\lesssim \frac{\op{\bDelta_{\bar\zeta}}\infn{\bU}}{\lambda_r\sqrt{T}}=o\brac{1}$.
{Combining these bounds with the compatible-whitening identity above yields
\begin{align*}
    \bSigma_{L,i}^{-1/2}\sqbrac{\bU_{\perp}\bU_{\perp}^\top \bN\bU\bO_F}_{i,\cdot}^\top \overset{d}{\rightarrow} \calN\brac{\mathbf0,\bI_r}.
\end{align*}}

\paragraph{Bounding the higher order terms in the leading term.} Notice that $\op{\bO_F^{-1}}\le \op{\bSigma^{-1}}\op{ \bar\bSigma}\lesssim{\lambda_1^2T}/{\lambda_r}$. {By the nondegeneracy condition, $\op{\bar\bSigma_{L,i}^{-1/2}}\lesssim\brac{\lambda_r\sqrt T}^{-1}$. Hence
\begin{align*}
    \op{\bSigma_{L,i}^{-1/2}}\lesssim  \frac{\kappa\lambda_1T}{\lambda_r\sqrt T}\lesssim \kappa^2\sqrt{T}.
\end{align*}}
For the higher order terms in $\brac{\hat\bU_\bQ\bH_{U,\bQ}-\bU}\bSigma\bR_{V,\bQ}^\top$, by \Cref{thm:U-two-infinty}, we get   with probability at least $1-\eta_F-O\brac{e^{-c_0N}+n^{-9}}$,
{\small
\begin{align*}
    \aop{\bSigma_{L,i}^{-1/2}\sqbrac{\e_i^\top \brac{\hat\bU_\bQ\bH_{U,\bQ}-\bU-\bU_{\perp}\bU_{\perp}^\top \bN\bU\bar\bO^\top \bar\bSigma^{-1}\bar\bO}\bSigma\bR_{V,\bQ}^\top}^\top }\lesssim\frac{\kappa^2\sqrt{T}\op{\bSigma}\infn{\bU}\sqbrac{\overline{\textsf{Err}}_U(\bQ)}^2}{\brac{\lambda_r^2\sigma_r\brac{\bF^\top\bQ\bF }}^2}.
\end{align*}
}
which is $o(1)$ on the same event
provided that  $\sigma_r\brac{\bF^\top\bQ\bF }\gtrsim T$,
\begin{align}\label{eq:inf-L-cond-1}
    \lambda_r\gg K\sqrt{\frac{N}{T}}\kappa^5\kappa_{\rm C, 1}^2\kappa_{\rm C, 2}^2r^{1/2}\log^2 n\op{\bSigma_{\rm C}^{1/2}}_1^2\op{\bSigma_{\rm T}^{1/2}}^2,
\end{align}
and
\begin{align}\label{eq:inf-L-cond-2}
    \lambda_r^3 \gg K^3\frac{n}{T}\sqrt{\frac{N}{T}} \kappa^3\kappa_{\rm C, 1}^2\kappa_{\rm C, 2}^2r^{3/2}\log n\op{\bSigma_{\rm C}^{1/2}}_1^4\op{\bSigma_{\rm T}^{1/2}}^4 +\frac{\ab{\textsf{Tr}\brac{\bSigma_{\rm T}\bQ}}^2}{T\sqrt{NT}}\bar\rho^2\kappa^3r^{1/2}.
\end{align}
\paragraph{Bounding the remaining terms.} 
It remains to bound $\op{\bSigma_{L,i}^{-1/2}\brac{\e_i^\top\mPhi_{L}}^\top}$. Following the argument in Section C.3.1 in \cite{chen2021spectral} we can get that $\op{\bR_{U,\bQ}-\bH_{U,\bQ}}\lesssim\brac{\xi_\bQ^U}^2$. By the bounds used to derive \eqref{eq:hatSigma-bound}, we have
{\footnotesize
\begin{align*}
    \op{\bR_{U,\bQ}^\top \hat\bSigma_\bQ-\bSigma\bR_{V,\bQ}^\top}=\op{\hat\bSigma_\bQ-\bR_{U,\bQ}\bSigma\bR_{V,\bQ}^\top}&\lesssim K\xi^U_\bQ\brac{1+{\sqrt{\frac{N}{T}}}}\op{ \bSigma_{\rm C}^{1/2}}\op{ \bSigma_{\rm T}^{1/2}}\\
    &{+\brac{\xi^U_\bQ}^2\lambda_1+K\frac{\brac{\xi^U_\bQ}^2}{\lambda_r}\brac{1+\sqrt{\frac{N}{T}}+\xi^U_\bQ\frac{N}{T}}\op{\bSigma_{\rm C}}\op{\bSigma_{\rm T}}}\\
    &{+K\sqrt{\frac{r+\log n}{T}}\op{\bU^\top\bSigma_{\rm C}^{1/2}}\op{\bSigma_{\rm T}^{1/2}\bV}}\\
    &{+\sqbrac{\lambda_1+K\brac{1+\sqrt{N/T}}\op{\bSigma_{\rm C}^{1/2}}\op{\bSigma_{\rm T}^{1/2}}}\brac{\xi^V_\bQ}^2}.
\end{align*}
}
{The signal-to-noise condition together with \Cref{thm:U-error} gives $\xi^U_\bQ=o(1)$ on the corresponding high-probability event. Since $1+\sqrt{N/T}+\xi^U_\bQ N/T\le (2+\xi^U_\bQ)n/T$ and $\lambda_1=\kappa\lambda_r$, the last term in the preceding weaker condition implies $\frac{K}{\lambda_r\lambda_1}\brac{1+\sqrt{N/T}+\xi^U_\bQ N/T}\op{\bSigma_{\rm C}}\op{\bSigma_{\rm T}}=o(1)$ on the same event. Hence the covariance term is absorbed by $\brac{\xi^U_\bQ}^2\lambda_1$.} Using the fact that $\xi_{\bQ,\infty}^U=o\brac{\infn{\bU}}$, we obtain that 
\begin{align*}
    \infn{\mPhi_{L}}&\lesssim\xi_\bQ^U\infn{\bU}\sqbrac{\xi_\bQ^U\lambda_1+K\brac{1+{\sqrt{\frac{N}{T}}}}\op{ \bSigma_{\rm C}^{1/2}}\op{ \bSigma_{\rm T}^{1/2}}}\\
    &{+K\infn{\bU}\sqrt{\frac{r+\log n}{T}}\op{\bU^\top\bSigma_{\rm C}^{1/2}}\op{\bSigma_{\rm T}^{1/2}\bV}}\\
    &{+\infn{\bU}\sqbrac{\lambda_1+K\brac{1+\sqrt{N/T}}\op{\bSigma_{\rm C}^{1/2}}\op{\bSigma_{\rm T}^{1/2}}}\brac{\xi^V_\bQ}^2}.
\end{align*}
Hence we have
\begin{align*}
\op{\bSigma_{L,i}^{-1/2}\brac{\e_i^\top\mPhi_{L}}^\top}
&\lesssim  \calR_1+\calR_2+\calR_3+\calR_4+\calR_5,
\end{align*}
where we define
\begin{align*}
    &\calR_1:=\kappa^2\sqrt{T}\brac{\xi_\bQ^U}^2\lambda_1\infn{\bU},\quad \calR_2:=K\kappa^2\sqrt{T}\xi_\bQ^U\infn{\bU}\op{ \bSigma_{\rm C}^{1/2}}\op{ \bSigma_{\rm T}^{1/2}},\\
	    &\calR_3:=K\kappa^2\sqrt{T}{\xi_\bQ^U}\infn{\bU}\sqrt{\frac{N}{T}}\op{ \bSigma_{\rm C}^{1/2}}\op{ \bSigma_{\rm T}^{1/2}},\\
	    &\calR_4:=K\kappa^2\sqrt{T}\infn{\bU}\sqrt{\frac{r+\log n}{T}}\op{\bU^\top\bSigma_{\rm C}^{1/2}}\op{\bSigma_{\rm T}^{1/2}\bV},\\
	    &\calR_5:=\kappa^2\sqrt{T}\infn{\bU}\sqbrac{\lambda_1+K\brac{1+\sqrt{N/T}}\op{\bSigma_{\rm C}^{1/2}}\op{\bSigma_{\rm T}^{1/2}}}\brac{\xi^V_\bQ}^2.
\end{align*}
It boils down to show {that $\calR_k=o\brac{1}$ for $k\in[5]$ on an event whose probability tends to one}. 
By \Cref{thm:U-error}, we get   with probability at least $1-\eta_F-O\brac{e^{-c_0N}}$,
\begin{align*}
    \calR_1&\lesssim\frac{\kappa^3r^{1/2}}{\lambda_r}\sqrt{\frac{N}{T}}\op{\bSigma_{\rm C}^{1/2}}^2\op{\bSigma_{\rm T}^{1/2}}^2+\frac{\kappa r^{1/2}}{\lambda_r^3}\brac{\frac{N}{T}}^{3/2}\op{\bSigma_{\rm C}^{1/2}}^4\op{\bSigma_{\rm T}^{1/2}}^4+\frac{\rho^2\kappa r^{1/2}\ab{\textsf{Tr}\brac{\bSigma_{\rm T}\bQ}}^2}{\lambda_r^3 T\sqrt{N T}},
\end{align*}
and
\begin{align*}
    \calR_2&\lesssim\frac{K\kappa^3 r^{1/2}}{\lambda_r}\op{\bSigma_{\rm C}^{1/2}}^2\op{\bSigma_{\rm T}^{1/2}}^2+\frac{K\kappa^2 r^{1/2}}{\lambda_r^2}\sqrt{\frac{N}{T}}\op{\bSigma_{\rm C}^{1/2}}^3\op{\bSigma_{\rm T}^{1/2}}^3\\
    &+\frac{K\rho\kappa^2r^{1/2}\ab{\textsf{Tr}\brac{\bSigma_{\rm T}\bQ}}}{\lambda_r^2}\sqrt{\frac{1}{NT}}\op{\bSigma_{\rm C}^{1/2}}\op{\bSigma_{\rm T}^{1/2}},
\end{align*}
{For $\calR_3$, if $N\le T$, then $\calR_3\lesssim\calR_2=o\brac{1}$. If $N>T$, \Cref{thm:U-error} and $\sigma_r\brac{\bF^\top\bQ\bF}\gtrsim T$ give directly
\begin{align*}
    \calR_3\lesssim{}&\frac{K^2\kappa^3r^{1/2}}{\lambda_r}\sqrt{\frac NT}\op{\bSigma_{\rm C}^{1/2}}^2\op{\bSigma_{\rm T}^{1/2}}^2
    +\frac{K^3\kappa^2r^{1/2}}{\lambda_r^2}\frac NT\op{\bSigma_{\rm C}^{1/2}}^3\op{\bSigma_{\rm T}^{1/2}}^3\\
    &+\frac{K\kappa^2r^{1/2}\rho\ab{\textsf{Tr}\brac{\bSigma_{\rm T}\bQ}}}{\lambda_r^2T}
    \op{\bSigma_{\rm C}^{1/2}}\op{\bSigma_{\rm T}^{1/2}}=o\brac{1}.
\end{align*}
Here the first two terms are controlled by the first term in the stated SNR condition, and the last by its first and trace terms together, using $\rho\le\bar\rho$ and the nondegeneracy condition.}
We can conclude that $\calR_1=o\brac{1}$ provided that \eqref{eq:inf-L-cond-1} and \eqref{eq:inf-L-cond-2} hold, $\calR_2=o\brac{1}$ provided that 
\begin{align}\label{eq:inf-L-cond-3}
    \begin{gathered}
        \lambda_r\gg K\kappa^3r^{1/2}\op{\bSigma_{\rm C}^{1/2}}^2\op{\bSigma_{\rm T}^{1/2}}^2,\\
        \lambda_r\gg K\kappa r^{1/4}\brac{\frac{N}{T}}^{1/4}\op{\bSigma_{\rm C}^{1/2}}^{3/2}\op{\bSigma_{\rm T}^{1/2}}^{3/2}+K\rho^{1/2}\kappa r^{1/4}\frac{\ab{\textsf{Tr}\brac{\bSigma_{\rm T}\bQ}}^{1/2}}{\brac{NT}^{1/4}}\op{\bSigma_{\rm C}^{1/2}}^{1/2}\op{\bSigma_{\rm T}^{1/2}}^{1/2},
    \end{gathered}
\end{align}
{Finally,}
\begin{align*}
{\calR_4}&{\lesssim K\kappa^2\sqrt{\frac{r\brac{r+\log n}}{N}}\op{\bSigma_{\rm C}^{1/2}}\op{\bSigma_{\rm T}^{1/2}}=o\brac{1}},
\end{align*}
{and, by \Cref{thm:V-error},}
\begin{align*}
{\calR_5}&{\lesssim \frac{K^2\kappa^3}{\lambda_r}\sqrt{\frac{rT}{N}}\op{\bSigma_{\rm C}^{1/2}}^2\op{\bSigma_{\rm T}^{1/2}}^2+o\brac{1}=o\brac{1}}.
\end{align*}
Combining \eqref{eq:inf-L-cond-0}, \eqref{eq:inf-L-cond-1}, \eqref{eq:inf-L-cond-2} and  \eqref{eq:inf-L-cond-3}, we complete the proof.

\subsection{Proof of \Cref{thm:inf-F}}\label{pf-thm:inf-F}

We condition on $\bF$ and work on $\calE_0$ from \Cref{lem:good-event}. The nondegeneracy condition implies $\op{\bSigma_{\rm C}^{1/2}}\op{\bSigma_{\rm T}^{1/2}}\gtrsim1$. Since $\op{\bQ}=1$, $\fro{\bQ}\le\sqrt T$ and $|\textsf{Tr}(\bSigma_{\rm T}\bQ)|\le T\op{\bSigma_{\rm T}}$, the first SNR summand in \Cref{thm:inf-F} implies the first-stage separation required by \Cref{thm:U-error}. That theorem gives, with conditional probability at least $1-O(e^{-cN}+n^{-10})$,
\begin{align}\label{eq:factor-loo-U-rate}
 \op{\wt\bU_\bQ\wt\bU_\bQ^\top-\bU\bU^\top}&\lesssim\frac{K\kappa}{\lambda_r}\sqrt{\frac nT}
       \op{\bSigma_{\rm C}^{1/2}}\op{\bSigma_{\rm T}^{1/2}}_1
       +\frac{K^2}{\lambda_r^2}\frac nT
       \op{\bSigma_{\rm C}^{1/2}}^2\op{\bSigma_{\rm T}^{1/2}}_1^2\notag\\
 &\quad\lesssim\frac{K\kappa}{\lambda_r}\sqrt{\frac nT}
       \op{\bSigma_{\rm C}^{1/2}}\op{\bSigma_{\rm T}^{1/2}}_1=o(1).
\end{align}
The last  summand in the SNR condition of \Cref{thm:inf-F} also gives
\begin{align*}
 \frac{K^2N\op{\bSigma_{\rm C}^{1/2}}^2
                 \op{\bSigma_{\rm T}^{1/2}\bQ\bSigma_{\rm T}^{1/2}}_1}
      {\sigma_r(\bM\bQ\bM^\top)}
 \lesssim\frac{K^2N\op{\bSigma_{\rm C}^{1/2}}^2
        \op{\bSigma_{\rm T}^{1/2}\bQ\bSigma_{\rm T}^{1/2}}_1}
        {T\lambda_r^2}=o(1).
\end{align*}
Thus \eqref{eq:V-loo-separation} holds. Also $T\gg\log^2n$, so the spectral bound $\op{\wt\bN}\lesssim\textsf{Err}_V(\bQ)$ used in \Cref{pf-lem:V-high-order-bound} gives
\begin{align}\label{eq:factor-loo-N-rate}
 \op{\wt\bN}
 \lesssim K\kappa^2\sqrt n\,
       \op{\bSigma_{\rm C}^{1/2}}\op{\bSigma_{\rm T}^{1/2}}_1,
 \qquad
 \frac{\op{\wt\bN}}{\lambda_r\sqrt T}=o(1).
\end{align}
Consequently the projector expansion in \Cref{pf-thm:V-error} is valid. Recall that $\bH_{V,\bQ}=\hat\bV_\bQ^\top\bV$. In addition, from \eqref{eq:U-L-infty-relation} we know that $\infn{\bU}=\infn{\bL\bLambda^{-1}}\lesssim\sqrt{r/N}$. From the proof of \Cref{thm:factor-bound}, we have
\begin{align*}
    \hat\bF_\bQ-\bF\bR_{F,\bQ}&=\sqrt{T}\brac{\hat\bV_\bQ-\bV\bR_{V,\bQ}^\top}=\sqrt{T}\brac{\hat\bV_\bQ\bR_{V,\bQ} -\bV}\bR_{V,\bQ}^\top\\
    &=\sqrt{T}\brac{\hat\bV_\bQ\bH_{V,\bQ} -\bV}\bR_{V,\bQ}^\top+\underbrace{\sqrt{T}\hat\bV_\bQ\brac{\bR_{V,\bQ}-\bH_{V,\bQ}}\bR_{V,\bQ}^\top}_{=:\mPhi_{F}}.
\end{align*}
By the proof of \Cref{thm:V-error}, we get that 
\begin{align*}
    \hat\bV_\bQ\bH_{V,\bQ} -\bV=\bV_\perp\bV_\perp^\top \wt\bN\bU\wt \bSigma^{-1}+\sum_{p\ge 2}\calS^V_{\bM,p}\brac{\wt \bN}\bV.
\end{align*}
\paragraph{Characterizing the first order term in the leading term.}
\begin{align*}
    \wt \bN=\bE^\top\bU\bU^\top +	\sqrt{T}\bV\bSigma\bU^\top \brac{\wt\bU_\bQ\wt\bU_\bQ^\top-\bU\bU^\top}+\bE^\top \brac{\wt\bU_\bQ\wt\bU_\bQ^\top-\bU\bU^\top}
\end{align*}
Similar to the proof of \Cref{thm:inf-L}, it suffices to characterize the distribution of each row of $\bV_\perp\bV_\perp^\top \wt\bN\bU$. Fix $t\in[T]$, we have 
\begin{align}\label{eq:inf-F-first-order-form}
    \sqbrac{\bV_\perp\bV_\perp^\top \wt\bN\bU }_{t,\cdot}&=\e_t^\top \bE^\top\bU+\e_t^\top \bE^\top \brac{\wt\bU_\bQ\wt\bU_\bQ^\top-\bU\bU^\top}\bU\notag\\
    &-\e_t^\top\bV\bV^\top \sqbrac{\bE^\top\bU\bU^\top +	\bE^\top \brac{\wt\bU_\bQ\wt\bU_\bQ^\top-\bU\bU^\top}}\bU.
\end{align}
Denote $\wt\bZ=\bE^\top \bU$ and $\wt\bY_{t,il}:=G_{i,l}\sqbrac{\bSigma_{\rm T}^{1/2}}_{t,l} \sqbrac{\bSigma_{\rm C}^{1/2}\bU}_{i,\cdot}^\top$. For the first term in \eqref{eq:inf-F-first-order-form}, we have 
\begin{align*}
    \wt\bZ_{t,\cdot}^\top =\sum_{i\in[N]}\sum_{l\in[T]}\wt\bY_{t,il}=\sum_{i\in[N]}\sum_{l\in[T]}G_{i,l}\underbrace{\sqbrac{\bSigma_{\rm T}^{1/2}}_{t,l} \sqbrac{\bSigma_{\rm C}^{1/2}\bU}_{i,\cdot}^\top}_{\wt\w_{i,l}^{(t)}} 
\end{align*}
We thus arrive at
\begin{align*}
    \sum_{i\in[N]}\sum_{l\in[T]}\textsf{Cov}\brac{\wt\bY_{t,il}}&=\sum_{i\in[N]}\sum_{l\in[T]}\sqbrac{\bSigma_{\rm T}^{1/2}}_{t,l}^2\bU^\top\sqbrac{\bSigma_{\rm C}^{1/2}}_{\cdot,i}\sqbrac{\bSigma_{\rm C}^{1/2}}_{\cdot,i}^\top\bU=\underbrace{\sqbrac{\bSigma_{\rm T}}_{t,t}\bU^\top \bSigma_{\rm C}\bU}_{=:\bar\bSigma _{F,t}}.
\end{align*}
By the nondegeneracy condition, $\op{\bar\bSigma_{F,t}^{-1/2}}\lesssim1$. In particular, $\bar\bSigma_{F,t}$ is positive definite and $\textsf{Cov}\bbrac{\bar\bSigma_{F,t}^{-1/2}\wt\bZ_{t,\cdot}^\top}=\bI_r$. 
We then show that  $\bar\bSigma_{F,t}^{-1/2}\wt \bZ_{t,\cdot }^\top \overset{d}{\rightarrow} \calN\brac{\mathbf0,\bI_r}$ by Lemma~\ref{lem:subg_clt_weighted}. To that end, it remains to verify  $\delta_{F,t}:=\max_{i,l}\big\|\bar\bSigma_{F,t}^{-1/2} \wt\w_{i,l}^{(t)}\big\|_2 \to 0$.
Since $\op{\bar\bSigma_{F,t}^{-1/2}}\le \op{\bSigma_{\rm C}^{-1/2}}\sqbrac{\bSigma_{\rm T}}_{t,t}^{-1/2}$,we obtain
\begin{align*}
\delta_{F,t}\le \|\bar\bSigma_{F,t}^{-1/2}\|\max_{i,l}\|\wt\w_{i,l}^{(t)}\|
\lesssim \kappa_{\rm C, 1}\frac{\infn{\bSigma_{\rm C}^{1/2}\bU}}{\op{\bSigma_{\rm C}^{1/2}}}=o(1),
\end{align*}
by the condition in \Cref{thm:inf-F}. Therefore Lemma~\ref{lem:subg_clt_weighted} yields $\bar\bSigma_{F,t}^{-1/2}\wt\bZ_{t,\cdot}^\top \overset{d}{\rightarrow} \mathcal N(\mathbf0,\bI_r)$ conditionally on $\bF$. It remains to bound the other terms in \eqref{eq:inf-F-first-order-form}. The first summand in the SNR condition implies $K\op{\bSigma_{\rm C}^{1/2}}\op{\bSigma_{\rm T}^{1/2}}_1=o(\lambda_r)$. By \Cref{lem:V-first-order-bound}, \eqref{eq:factor-loo-U-rate}, $\infn{\bV}\lesssim\sqrt{(r+\log n)/T}$ and $\sigma_r(\bF^\top\bQ\bF)\gtrsim T$,
\begin{align}\label{eq:factor-loo-product-rate}
 &\infn{\bE^\top(\wt\bU_\bQ\wt\bU_\bQ^\top-\bU\bU^\top)}\notag\\
 &\quad\lesssim\frac{K^2\kappa\sqrt{r+\log n}}{\lambda_r}
       \op{\bSigma_{\rm C}^{1/2}}^2\left(\sqrt{\frac nT}\op{\bSigma_{\rm T}^{1/2}}_1^2
       +\frac NT\kappa_{\rm T,1}
          \op{\bSigma_{\rm T}^{1/2}\bQ\bSigma_{\rm T}^{1/2}}_1\right)=o(1).
\end{align}
The two terms are controlled by the first and last  summands in the SNR condition of \Cref{thm:inf-F}, respectively. Hence, because $\op{\bar\bSigma_{F,t}^{-1/2}}\lesssim1$,
\begin{align*}
 \op{\bar\bSigma_{F,t}^{-1/2}
   [\e_t^\top\bE^\top(\wt\bU_\bQ\wt\bU_\bQ^\top-\bU\bU^\top)\bU]^\top}
 =o(1).
\end{align*}
The two terms preceded by $\bV\bV^\top$ satisfy
\begin{align*}
 &\op{\bar\bSigma_{F,t}^{-1/2}
          [\e_t^\top\bV\bV^\top\bE^\top\bU]^\top}\lesssim\frac{K(r+\log n)}{\sqrt T}
       \op{\bSigma_{\rm C}^{1/2}}\op{\bSigma_{\rm T}^{1/2}}=o(1),\\
 &\op{\bar\bSigma_{F,t}^{-1/2}
    [\e_t^\top\bV\bV^\top\bE^\top
       (\wt\bU_\bQ\wt\bU_\bQ^\top-\bU\bU^\top)\bU]^\top}\\
 &\quad\lesssim K\sqrt{\frac{N(r+\log n)}{T}}
       \op{\bSigma_{\rm C}^{1/2}}\op{\bSigma_{\rm T}^{1/2}}
       \op{\wt\bU_\bQ\wt\bU_\bQ^\top-\bU\bU^\top}=o(1).
\end{align*}
Here the first bound uses concentration of the $r\times r$ matrix $\bV^\top\bE^\top\bU$ and the sample-size condition. The second uses $\op{\bE\bV}\lesssim K\op{\bSigma_{\rm C}^{1/2}}\op{\bSigma_{\rm T}^{1/2}}\sqrt N$ and \eqref{eq:factor-loo-U-rate}. By the compatible-whitening convention in \Cref{thm:inf-F},
$\bSigma_{F,t}^{-1/2}\bR_{V,\bQ}\bSigma^{-1}=\bar\bSigma_{F,t}^{-1/2}$.
Since $\wt\bSigma=\sqrt T\bSigma$, the leading weighted-sum CLT and the preceding bounds imply
\begin{align*}
 \bSigma_{F,t}^{-1/2}
  [\sqrt T\bV_\perp\bV_\perp^\top\wt\bN\bU\wt\bSigma^{-1}\bR_{V,\bQ}^\top]_{t,\cdot}^\top
 \overset d\longrightarrow\calN(\mathbf0,\bI_r).
\end{align*}

\paragraph{Bounding the higher order terms.}
By \eqref{eq:factor-loo-U-rate}, $\textsf{Err}_V(\bQ)\lesssim K\kappa^2\sqrt n\op{\bSigma_{\rm C}^{1/2}}\op{\bSigma_{\rm T}^{1/2}}_1$. The last term in $\infn{\bV}\overline{\textsf{Err}}_V(\bQ)$ is $o(1)$ by the last  summand in the SNR condition. Therefore,
\begin{align}\label{eq:factor-loo-row-rate}
 \infn{\bV}\overline{\textsf{Err}}_V(\bQ)
 \lesssim K\kappa^2\sqrt{\frac nT(r+\log n)}\,
       \op{\bSigma_{\rm C}^{1/2}}\op{\bSigma_{\rm T}^{1/2}}_1.
\end{align}
In particular, $\overline{\textsf{Err}}_V(\bQ)/(\lambda_r\sqrt T)=o(1)$, since $\infn{\bV}\ge\sqrt{r/T}$. Thus the second claim of \Cref{thm:UV-error-inf} applies. The sharper remainder bound \eqref{eq:V-loo-projector-remainder}, together with $\op{\bSigma_{F,t}^{-1/2}}\lesssim\lambda_1$, gives
\begin{align}\label{eq:factor-loo-higher-remainder}
 \op{\bSigma_{F,t}^{-1/2}
       \big[\e_t^\top\sqrt T
          \sum_{p\ge2}\calS^V_{\bM,p}(\wt\bN)\bV\bR_{V,\bQ}^\top\big]^\top}&\lesssim\lambda_1\sqrt T\,
       \frac{\infn{\bV}\overline{\textsf{Err}}_V(\bQ)\op{\wt\bN}}
            {\lambda_r^2T}\notag\\
 &\lesssim\frac{K^2\kappa^5}{\lambda_r}\frac nT\sqrt{r+\log n}\,
       \op{\bSigma_{\rm C}^{1/2}}^2\op{\bSigma_{\rm T}^{1/2}}_1^2=o(1).
\end{align}
The last step follows from \eqref{eq:factor-loo-N-rate}, \eqref{eq:factor-loo-row-rate}, and the first summand in the SNR condition.

\paragraph{Bounding the alignment correction.}
The projector bound implies $\infn{\hat\bV_\bQ}\lesssim\sqrt{(r+\log n)/T}$: indeed,
$\infn{\hat\bV_\bQ}\le\infn{\bV}+\infn{\hat\bV_\bQ\hat\bV_\bQ^\top-\bV\bV^\top}$.
Also, it follows that
\begin{align*}
 \op{\bR_{V,\bQ}-\bH_{V,\bQ}}
 \lesssim\op{\hat\bV_\bQ\hat\bV_\bQ^\top-\bV\bV^\top}^2
 \lesssim\frac{\op{\wt\bN}^2}{\lambda_r^2T}.
\end{align*}
Consequently,
\begin{align*}
 &\op{\bSigma_{F,t}^{-1/2}
     [\e_t^\top\sqrt T\hat\bV_\bQ(\bR_{V,\bQ}-\bH_{V,\bQ})\bR_{V,\bQ}^\top]^\top}\\
 &\quad\lesssim\lambda_1\sqrt{r+\log n}\,
                     \frac{\op{\wt\bN}^2}{\lambda_r^2T}\lesssim\frac{K^2\kappa^5}{\lambda_r}\frac nT\sqrt{r+\log n}\,
       \op{\bSigma_{\rm C}^{1/2}}^2\op{\bSigma_{\rm T}^{1/2}}_1^2=o(1).
\end{align*}
Combining the preceding bounds establishes the rowwise linearization
\begin{align}\label{eq:factor-loo-linearization}
 \op{\bSigma_{F,t}^{-1/2}\left[
       (\hat\bF_\bQ-\bF\bR_{F,\bQ})_{t,\cdot}^\top
       -\bR_{V,\bQ}\bSigma^{-1}\bU^\top\bE\e_t\right]}=o_p(1).
\end{align}
Lemma~\ref{lem:subg_clt_weighted} applied to the leading weighted sum, together with Slutsky's theorem, proves the conclusion.

\subsection{Inference for the common component}

Fix $(i,t)\in[N]\times[T]$.  Note that the rotation-invariant common
component and its weighted-PCA estimator by $M_{it}:=\l_i^\top\f_t=\e_i^\top\bM\e_t$ and $\widehat M_{\bQ,it}:=\e_i^\top\hat\bL_{\bQ}\hat\bF_{\bQ}^\top\e_t$.
Let  $\bA_{\bQ}:=\bF^\top\bQ\bF$, $\bP:=\bU\bU^\top$, $\h_{\bQ,t}:=\bQ\bF\bA_{\bQ}^{-1}\f_t$,
where $\h_{\bQ,t}$ is defined on the event that $\bA_{\bQ}$ is
invertible.  Set
\begin{align*}
 v_{L,it}(\bQ)
 &:= [\bSigma_{\rm C}]_{i,i}\,
       \h_{\bQ,t}^\top\bSigma_{\rm T}\h_{\bQ,t},\\
 v_{F,it}(\bQ)
 &:= [\bSigma_{\rm T}]_{t,t}\,
       \e_i^\top\bP\bSigma_{\rm C}\bP\e_i,\\
 c_{LF,it}(\bQ)
 &:= \bigl(\e_i^\top\bSigma_{\rm C}\bP\e_i\bigr)
       \bigl(\h_{\bQ,t}^\top\bSigma_{\rm T}\e_t\bigr),\\
 v_{M,it}(\bQ)
 &:=v_{L,it}(\bQ)+v_{F,it}(\bQ)+2c_{LF,it}(\bQ).
\end{align*}

\begin{proposition}[Common-component CLT]
\label{prop:common-component-clt}
Assume the following conditions.
\begin{enumerate}[label=\textnormal{(\roman*)},leftmargin=2.4em]
 \item Assumption~\ref{assump:iden} and all conditions of
 Theorems~\ref{thm:inf-L} and~\ref{thm:inf-F} hold simultaneously.
 \item The weight matrix $\bQ$ is symmetric with
 $\op{\bQ}=1$ and $r,K=O(1)$.
 \item The factor row satisfies
 $\f_t\sim\calN(\mathbf0,\bI_r)$ for $t\in[T]$.
 \item There exist constants $0<c_\Sigma\le C_\Sigma<\infty$ such that
 \begin{align}\label{eq:cc-uniform-spectrum}
  c_\Sigma\bI_N\preceq\bSigma_{\rm C}\preceq C_\Sigma\bI_N,
  \qquad
  c_\Sigma\bI_T\preceq\bSigma_{\rm T}\preceq C_\Sigma\bI_T.
 \end{align}
\end{enumerate}
Then $\PP(\bA_{\bQ}\text{ is invertible})\to1$ and
$\PP(v_{M,it}(\bQ)>0)\to1$. Moreover,
\begin{align}\label{eq:cc-linear-representation}
 \widehat M_{\bQ,it}-M_{it}
 =\e_i^\top\bE\h_{\bQ,t}
  +\e_i^\top\bP\bE\e_t
  +o_p\left(v_{M,it}(\bQ)^{1/2}\right),
\end{align}
and
\begin{align}\label{eq:cc-clt}
 \frac{\widehat M_{\bQ,it}-M_{it}}
      {v_{M,it}(\bQ)^{1/2}}
 \overset{d}{\longrightarrow}\calN(0,1).
\end{align}
\end{proposition}
\subsection{Proof of \Cref{prop:common-component-clt}}
For notational convenience, we define
\begin{align*}
 \a_{\bQ,t}&:=(\bF\bR_{F,\bQ})_{t,\cdot}^\top,
 &\b_{\bQ,i}&:=(\bL\bR_{L,\bQ})_{i,\cdot}^\top,\\
 \bDelta_{L,i}&:=(\hat\bL_{\bQ}-\bL\bR_{L,\bQ})_{i,\cdot}^\top,
 &\bDelta_{F,t}&:=(\hat\bF_{\bQ}-\bF\bR_{F,\bQ})_{t,\cdot}^\top.
\end{align*}
Since $\bR_{L,\bQ}\bR_{F,\bQ}^\top=\bI_r$,
$M_{it}=\b_{\bQ,i}^\top\a_{\bQ,t}$ and
\begin{align}\label{eq:cc-exact-product-expansion}
 \widehat M_{\bQ,it}-M_{it}
 =\a_{\bQ,t}^\top\bDelta_{L,i}
  +\b_{\bQ,i}^\top\bDelta_{F,t}
  +\bDelta_{L,i}^\top\bDelta_{F,t}.
\end{align}
The rowwise expansions established in the proofs of
Theorems~\ref{thm:inf-L} and~\ref{thm:inf-F} give
\begin{align}\label{eq:cc-rowwise-linearizations}
 \bDelta_{L,i}
 &=\boldsymbol\zeta_{L,i}+\br_{L,i},
 &\op{\bSigma_{L,i}^{-1/2}\br_{L,i}}&=o_p(1),\notag\\
 \bDelta_{F,t}
 &=\boldsymbol\zeta_{F,t}+\br_{F,t},
 &\op{\bSigma_{F,t}^{-1/2}\br_{F,t}}&=o_p(1),
\end{align}
where  $\boldsymbol\zeta_{L,i}
:=\bO_F^\top\bU^\top\bM\bQ\bE^\top\e_i$ and $\boldsymbol\zeta_{F,t}:=\bR_{V,\bQ}\bSigma^{-1}\bU^\top\bE\e_t$. The first relation follows from the estimates after
\eqref{eq:inf-L-first-order-form}, and the second from those after
\eqref{eq:factor-loo-linearization}. Note that the alignment identities imply
\begin{align}\label{eq:cc-alignment-cancellation}
 \a_{\bQ,t}^\top\boldsymbol\zeta_{L,i}
 =\e_i^\top\bE\h_{\bQ,t},\qquad \b_{\bQ,i}^\top\boldsymbol\zeta_{F,t}=\e_i^\top\bP\bE\e_t.
\end{align}
Conditionally on $\bF$, separability gives, for
$\u_1,\u_2\in\RR^N$ and $\v_1,\v_2\in\RR^T$,
\begin{align}\label{eq:cc-separable-covariance}
 \textsf{Cov}(\u_1^\top\bE\v_1,\u_2^\top\bE\v_2\mid\bF)
 =(\u_1^\top\bSigma_{\rm C}\u_2)
  (\v_1^\top\bSigma_{\rm T}\v_2).
\end{align}
Hence the conditional variance of the sum in
\eqref{eq:cc-alignment-cancellation} is $v_{M,it}(\bQ)$. Let
$p_i:=\e_i^\top\bP\e_i$ and define
\begin{align*}
 \rho_{{\rm C},i}
 :=\frac{\e_i^\top\bSigma_{\rm C}\bP\e_i}
 {[[\bSigma_{\rm C}]_{i,i}
    \e_i^\top\bP\bSigma_{\rm C}\bP\e_i]^{1/2}},
\end{align*}
with $\rho_{{\rm C},i}:=0$ when $p_i=0$. Condition
\eqref{eq:cc-uniform-spectrum} implies, for $p_i>0$,
\begin{align*}
 1-\rho_{{\rm C},i}^2
 &=\frac{\det\!\left([\e_i,\bP\e_i]^\top
       \bSigma_{\rm C}[\e_i,\bP\e_i]\right)}
 {(\e_i^\top\bSigma_{\rm C}\e_i)
  (\e_i^\top\bP\bSigma_{\rm C}\bP\e_i)}
 \ge\frac{c_\Sigma^2}{C_\Sigma^2}(1-p_i).
\end{align*}
The loading incoherence condition gives
$p_i\le\infn{\bU}^2\lesssim r/N=o(1)$, so there exists
$q_\Sigma\in(0,1)$ such that $|\rho_{{\rm C},i}|\le q_\Sigma$ for all
sufficiently large $(N,T)$. Cauchy--Schwarz in the temporal covariance
metric then gives $|c_{LF,it}(\bQ)|
 \le q_\Sigma\sqbrac{v_{L,it}(\bQ)v_{F,it}(\bQ)}^{1/2}$. Therefore,
\begin{align}\label{eq:cc-variance-lower}
 v_{M,it}(\bQ)
 &\ge v_{L,it}(\bQ)+v_{F,it}(\bQ)
      -2q_\Sigma\sqbrac{v_{L,it}(\bQ)v_{F,it}(\bQ)}^{1/2}\notag\\
 &\ge\varepsilon_\Sigma
      \sqbrac{v_{L,it}(\bQ)+v_{F,it}(\bQ)},
 \qquad \varepsilon_\Sigma:=1-q_\Sigma>0.
\end{align}
The common relevance condition in Theorems~\ref{thm:inf-L} and
\ref{thm:inf-F} gives $\sigma_r(\bA_{\bQ})\gtrsim T$, so
$\PP(\bA_{\bQ}\text{ is invertible})\to1$. On this event,
$\f_t\ne\mathbf0$ almost surely and hence $\h_{\bQ,t}\ne\mathbf0$ almost surely.
Thus $v_{L,it}(\bQ)>0$ almost surely, and
\eqref{eq:cc-variance-lower} yields $\PP(v_{M,it}(\bQ)>0)\to1$. We thus have
\begin{align}\label{eq:cc-weighted-sum}
 \e_i^\top\bE\h_{\bQ,t}+\e_i^\top\bP\bE\e_t
 =\sum_{j=1}^N\sum_{s=1}^T G_{js}\omega_{js}^{(it)},
\end{align}
where $\omega_{js}^{(it)}:=[\bSigma_{\rm C}^{1/2}]_{i,j}
    [\bSigma_{\rm T}^{1/2}\h_{\bQ,t}]_s
   +[\bSigma_{\rm C}^{1/2}\bP\e_i]_j
    [\bSigma_{\rm T}^{1/2}\e_t]_s$. Equation~\eqref{eq:cc-separable-covariance} gives
$\sum_{j,s}(\omega_{js}^{(it)})^2=v_{M,it}(\bQ)$. The maximal-weight
bounds in the proofs of Theorems~\ref{thm:inf-L} and~\ref{thm:inf-F},
applied respectively to the scalar directions $\bO_F\a_{\bQ,t}$ and
$\bU^\top\e_i$, give
\begin{align*}
 &\max_{j,s}
 \frac{|[\bSigma_{\rm C}^{1/2}]_{i,j}
 [\bSigma_{\rm T}^{1/2}\h_{\bQ,t}]_s|}
 {v_{L,it}(\bQ)^{1/2}}=o_p(1),\qquad \max_{j,s}
 \frac{|[\bSigma_{\rm C}^{1/2}\bP\e_i]_j
 [\bSigma_{\rm T}^{1/2}\e_t]_s|}
 {v_{F,it}(\bQ)^{1/2}}=o_p(1).
\end{align*}
Hence,
by \eqref{eq:cc-variance-lower}, $\max_{j,s}|\omega_{js}^{(it)}|/{v_{M,it}(\bQ)^{1/2}}
 =o_p(1)$. Applying Lemma~\ref{lem:subg_clt_weighted} yields convergence of the conditional characteristic functions in
probability. Since these characteristic functions are bounded by one,
taking expectations gives
\begin{align}\label{eq:cc-leading-clt}
 \frac{\e_i^\top\bE\h_{\bQ,t}+\e_i^\top\bP\bE\e_t}
 {v_{M,it}(\bQ)^{1/2}}
 \overset{d}{\longrightarrow}\calN(0,1).
\end{align}
It suffices to bound the remainder terms. Condition~\eqref{eq:cc-uniform-spectrum}, $\op{\bQ}=1$,
$T^{-1}\bF^\top\bF\to_p\bI_r$, and
$\sigma_r(\bA_{\bQ})\gtrsim T$ imply $c_\Sigma\bA_{\bQ}(\bF^\top\bF)^{-1}\bA_{\bQ}
 \preceq\bF^\top\bQ\bSigma_{\rm T}\bQ\bF
 \preceq C_\Sigma\bF^\top\bF$, and hence $\bF^\top\bQ\bSigma_{\rm T}\bQ\bF\asymp T\bI_r$. Using the
covariance formulas in Theorems~\ref{thm:inf-L} and~\ref{thm:inf-F},
together with $\op{\bR_{F,\bQ}}+\op{\bR_{F,\bQ}^{-1}}=O_p(1)$, gives
\begin{align}\label{eq:cc-variance-rates}
 v_{L,it}(\bQ)&\asymp T^{-1}\op{\f_t}^2,
 \quad \op{\bSigma_{L,i}}\lesssim T^{-1},
 \quad \op{\bSigma_{F,t}}\lesssim\lambda_r^{-2}.
\end{align}
The scalar remainders satisfy
$|\a_{\bQ,t}^\top\br_{L,i}|=o_p\brac{v_{L,it}(\bQ)^{1/2}}$  and $|\b_{\bQ,i}^\top\br_{F,t}|=o_p\brac{v_{F,it}(\bQ)^{1/2}}$ by \eqref{eq:cc-rowwise-linearizations} and Cauchy--Schwarz in the
corresponding covariance metrics. Equation~\eqref{eq:cc-variance-lower}
therefore makes both terms $o_p\brac{v_{M,it}(\bQ)^{1/2}}$. Since $r$ is
fixed, \eqref{eq:cc-rowwise-linearizations} and
\eqref{eq:cc-variance-rates} also give
\begin{align*}
 \op{\bDelta_{L,i}}=O_p(T^{-1/2}),
 \qquad
 \op{\bDelta_{F,t}}=O_p(\lambda_r^{-1}).
\end{align*}
The signal condition in Theorem~\ref{thm:inf-F} implies
$\lambda_r\to\infty$. Since $\f_t\sim\calN(\mathbf0,\bI_r)$ and $r$ is
fixed, $(\lambda_r\op{\f_t})^{-1}=o_p(1)$. Thus,
\begin{align*}
 \frac{|\bDelta_{L,i}^\top\bDelta_{F,t}|}
 {v_{M,it}(\bQ)^{1/2}}
 &\lesssim_p
 \frac{T^{-1/2}\lambda_r^{-1}}
      {T^{-1/2}\op{\f_t}}
 =o_p(1).
\end{align*}
Substituting \eqref{eq:cc-rowwise-linearizations} and
\eqref{eq:cc-alignment-cancellation} into
\eqref{eq:cc-exact-product-expansion} proves
\eqref{eq:cc-linear-representation}. Combining
\eqref{eq:cc-linear-representation} with
\eqref{eq:cc-leading-clt} and Slutsky's theorem proves
\eqref{eq:cc-clt}.

\subsection{A rank-one heteroskedastic example for the PC loading bias}\label{pf-eq:inf-counterexmp}
Consider the rank-one model $\bX=\l\f^\top+\bE$ with $\textsf{vec}\brac{\bE}\sim\calN\brac{0,\bSigma_{\rm C}\otimes\bSigma_{\rm T}}$, where $\bSigma_{\rm C}=\textsf{diag}\bbrac{\sigma_{{\rm C},1}^2,\ldots,\sigma_{{\rm C},N}^2}$ and $\bSigma_{\rm T}=\bI_T$. Let $\lambda:=\op{\l}$, $\sigma_{\rm C}:=\max_{i\in[N]}\sigma_{{\rm C},i}$, and assume $\sigma_{\rm C}\lesssim\min_{i\in[N]}\sigma_{{\rm C},i}$. Write $\u:=\l/\lambda$ and let $\bU_\perp$ be an orthonormal basis for the orthogonal complement of $\u$. Suppose that Assumption \ref{assump:iden} holds, {$\infn{\u}\lesssim N^{-1/2}$,} $\op{\bU_\perp^\top\bSigma_{\rm C}\u}\gtrsim\sigma_{\rm C}^2$, and $\lambda=\omega\bbrac{\sigma_{\rm C}\sqrt{n/T}\log n}$.

Let $\hat\l^{\rm PC}$ denote the ordinary PC loading estimator. Then, for some possibly data-dependent scalar $c\in\RR$ representing the rank-one loading alignment,
\nopagebreak[4]
\begin{align}
    \sqrt{T}\bbrac{\hat\l^{\rm PC}-c\l}=\g_L+\b_L+\boldsymbol{\delta}_L,\label{eq:inf-counterexmp}
\end{align}
where $\g_L\sim N\bbrac{0,\bSigma_{\rm C}}$ {conditional on $\bF$. Moreover,} with probability at least $1-\eta_F-O\brac{e^{-c_0N}+n^{-9}}$,
\begin{align*}
    \op{\b_L}_\infty\asymp\frac{\sigma_{\rm C}^2}{\lambda}\sqrt{\frac{T}{N}},\qquad
    \op{\boldsymbol{\delta}_L}_\infty\lesssim {\sigma_{\rm C}}\sqrt{\frac{\log n}{N}}{+\frac{\sigma_{\rm C}^2\sqrt{\log n}}{\lambda}}+\frac{\sigma_{\rm C}^2\log^2n}{\lambda}\sqrt{\frac{N}{T}}+\frac{\sigma_{\rm C}^4}{\lambda^3}\sqrt{\frac{T}{N}}.
\end{align*}
Thus the first-stage PC bias need not vanish unless $\lambda\gg\sigma_{\rm C}^2\sqrt{T/N}$. {In the separable model, $\EE\sqbrac{\bE\bQ\bE^\top}=\textsf{Tr}\brac{\bSigma_{\rm T}\bQ}\bSigma_{\rm C}$. For ordinary PCA this equals $T\bSigma_{\rm C}$. Hence the deterministic term generating $\b_L$ is the ordinary-PC instance of the trace term in \Cref{thm:inf-L}. If an admissible $\bQ$ preserves the factor signal, satisfies the remaining nondegeneracy conditions, and reduces $\ab{\textsf{Tr}\brac{\bSigma_{\rm T}\bQ}}$, then the bound in \Cref{thm:inf-L} correspondingly reduces this particular first-stage contribution, while the common second-stage requirement remains.}

\paragraph{Proof of \eqref{eq:inf-counterexmp}.}
Notice that the condition $\kappa\kappa_{\rm T}\infn{\bSigma_{\rm T}^{1/2}\bQ\bV}/\op{\bSigma_{\rm T}^{1/2}}=o(1)$ in \Cref{thm:inf-L} is  not needed since $\bE$ is Gaussian.
{Write $T^{-1/2}\l\f^\top=\sigma\u\v^\top$, and let $(\hat\sigma,\hat\u,\hat\v)$ be the leading singular triplet of $T^{-1/2}\bX$, so that $T^{-1/2}\bX\hat\v=\hat\sigma\hat\u$. Without loss of generality, assume $h_U:=\hat\u^\top\u>0$. Let $d_F^{-1}$ be the scalar such that $\v=d_F^{-1}T^{-1/2}\f$ and set $c:=d_F\hat\sigma h_U/\sigma$. Since $d_F\l=\sigma\u$, we have the exact decomposition
\begin{align}\label{eq:hatl-l-decomp}
    \hat\l^{\rm PC}-c\l
    =\hat\sigma\brac{\hat\u-h_U\u}
    =\frac{\hat\sigma}{h_U}\bU_\perp\bU_\perp^\top\brac{\hat\u h_U-\u}.
\end{align}}
Fix $\zeta\in\RR$. Note that $\bX\bX^\top-\zeta\bI_N=T\sigma^2\u\u^\top +\bN_1+\bN_2  $, where $\bN_1:=T^{1/2}\sigma\u \v^\top\bE^\top +T^{1/2}\sigma\bE\v \u^\top $, $\bN_2:=\bar\bN_2+\bDelta_\zeta$, $\bar\bN_2:=\frakD\bbrac{\bE\bE^\top}$ and $\bDelta_\zeta:=T \bSigma_{\rm C}-\zeta\bI_N$. In addition, denote $\bN:=\bN_1+\bN_2$. 
We can  apply \Cref{lem:eigen-expansion} and get that 
\begin{align*}
    \hat\u h_U-\u=\sum_{p\ge 1}\calS_{T\sigma^2\u\u^\top,p}\bbrac{\bN}\u ,
\end{align*}
provided that $\sigma^2 T\ge 4\op{\bN}$. Notice that $\calS_{T\sigma^2\u\u^\top,1}\bbrac{\bN}\u=\bbrac{\sigma ^2T}^{-1}\bU_{\perp}\bU_{\perp}^\top \bN\u$.
We start with considering the following first-order term: 
\begin{align*}
    \sigma\calS_{T\sigma^2\u\u^\top,1}\bbrac{\bN}\u={\brac{\sigma T}^{-1}\bU_{\perp}\bU_{\perp}^\top \bN\u}=\brac{\sigma T}^{-1}\bU_{\perp}\bU_{\perp}^\top \bbrac{T^{1/2}\sigma\bE\v\u^\top+\frakD\bbrac{\bE\bE^\top}+\bDelta_\zeta}\u.
\end{align*}
For the first term in the first-order term, we have
\begin{align*}
    \bU_{\perp}\bU_{\perp}^\top \bE\v=\bbrac{\bI_N-\u\u^\top} \bSigma_{\rm C}^{1/2}\bG\v=\bSigma_{\rm C}^{1/2}\bG\v-\u\u^\top\bSigma_{\rm C}^{1/2}\bG\v.
\end{align*}
Note that $\bSigma_{\rm C}^{1/2}\bG\v$ is the leading term that characterizes the distribution as $T^{-1/2}\sqbrac{\bSigma_{\rm C}^{1/2}\bG\v}_{i} \sim N\bbrac{0,T^{-1}\sqbrac{\bSigma_{\rm C}}_{i,i}}$. Moreover, we have 
\begin{align*}
    \op{T^{-1/2}\u\u^\top \bE\v}_{\infty}=\op{T^{-1/2}\u\u^\top \bSigma_{\rm C}^{1/2}\bG\v}_{\infty}\lesssim {\sigma_{\rm C}}\sqrt{\frac{\log n}{NT}}. 
\end{align*}
For the second term in the first-order term, by \Cref{lem:GBGU-two-inf-bound} we have
\begin{align*}
    &\op{\brac{\sigma T}^{-1}\bU_{\perp}\bU_{\perp}^\top \frakD\bbrac{\bE\bE^\top}\u}_{\infty}\\
    &\le \brac{\sigma T}^{-1}\brac{\op{\u\u^\top \bSigma_{\rm C}^{1/2}\frakD\bbrac{\bG\bG^\top}\bSigma_{\rm C}^{1/2}\u}_{\infty}+\op{\bSigma_{\rm C}^{1/2}\frakD\bbrac{\bG\bG^\top}\bSigma_{\rm C}^{1/2}\u}_{\infty}}\lesssim\frac{\sigma_{\rm C}^2}{\sigma }\sqrt{\frac{\log n}{T}}.
\end{align*}
For the last term in the first-order term, notice that 
\begin{align*}
    \op{\sigma ^{-1}\bU_{\perp}\bU_{\perp}^\top \bbrac{\bSigma_{\rm C}-\zeta T^{-1}\bI}\u}_{\infty }=\op{\sigma ^{-1}\bU_{\perp}\bU_{\perp}^\top\bSigma_{\rm C}\u}_{\infty }
    &\gtrsim  \frac{\sigma_{\rm C}^2}{\sqrt{N}\sigma}.
\end{align*}
The second inequality holds as $\rho=\op{\bU_{\perp}\bU_{\perp}^\top\bSigma_{\rm C}\u}\gtrsim  \sigma_{\rm C}^2$ by assumption. 
{Conversely, since $\bSigma_{\rm C}$ is diagonal and $\infn{\u}\lesssim N^{-1/2}$,
\begin{align*}
    \op{\bU_{\perp}\bU_{\perp}^\top\bSigma_{\rm C}\u}_{\infty}
    \le 2\sigma_{\rm C}^2\infn{\u}\lesssim \frac{\sigma_{\rm C}^2}{\sqrt N}.
\end{align*}}
For high-order terms, from the proof of \Cref{thm:inf-L} we obtain that 
\begin{align*}
    \op{\sigma{\bU_\perp\bU_\perp^\top}\sum_{p\ge 2}\calS_{T\sigma^2\u\u^\top,p}\bbrac{\bN}\u}_{\infty}\lesssim\frac{\sigma_{\rm C}^2\log^2n}{\sigma }\frac{\sqrt{N}}{T}+\frac{\sigma_{\rm C}^4	}{\sigma^3}\sqrt{\frac{1}{N}}.
\end{align*}
where we've used  $\bar\rho\le \sigma_{\rm C}^2$.
{It remains to account for the factor $\hat\sigma/(\sigma h_U)$ in \eqref{eq:hatl-l-decomp}. Set $a:={\hat\sigma}/\brac{\sigma h_U}$. On the event $\op{T^{-1/2}\bE}\le c_1\sigma$ for a sufficiently small constant $c_1>0$, the standard perturbation bounds give
\begin{align*}
    \ab{\hat\sigma-\sigma}
    &\lesssim \ab{\u^\top T^{-1/2}\bE\v}
    +\frac{\op{T^{-1/2}\bE}^2}{\sigma},
    &1-h_U&\lesssim\frac{\op{T^{-1/2}\bE}^2}{\sigma^2}.
\end{align*}
Gaussian scalar and spectral-norm concentration therefore give, with probability at least $1-O(e^{-c_0N})$,
\begin{align*}
    \ab{a-1}
    &\lesssim \frac{\sigma_{\rm C}}{\sigma}\sqrt{\frac{N}{T}}
    +\frac{\sigma_{\rm C}^2}{\sigma^2}\brac{1+\frac{N}{T}}
    =o(1).
\end{align*}
The assumed lower bound on $\lambda$, together with $\sigma\asymp\lambda$ on the event in \Cref{lem:good-event}, ensures the small-perturbation event above. Taking $\zeta=0$, we also have $\op{\bN}\lesssim \sigma\sigma_{\rm C}\sqrt{NT}+\sigma_{\rm C}^2(N+T)$; hence the condition $\sigma^2T\ge4\op{\bN}$ used above holds. Now define
\begin{align*}
    \g_L&:=\bE\v,\qquad
    \b_L:=a\frac{\sqrt T}{\sigma}\bU_\perp\bU_\perp^\top\bSigma_{\rm C}\u,\\
    \boldsymbol{\delta}_L
    &:=\brac{a\bU_\perp\bU_\perp^\top-\bI_N}\bE\v
    +\frac{a}{\sigma\sqrt T}\bU_\perp\bU_\perp^\top\frakD\bbrac{\bE\bE^\top}\u\\
    &\quad+\frac{\hat\sigma\sqrt T}{h_U}\bU_\perp\bU_\perp^\top
    \sum_{p\ge2}\calS_{T\sigma^2\u\u^\top,p}\bbrac{\bN}\u .
\end{align*}
Then \eqref{eq:hatl-l-decomp} and the first-order expansion above give
\[
    \sqrt T\brac{\hat\l^{\rm PC}-c\l}=\g_L+\b_L+\boldsymbol{\delta}_L
\]
exactly. Conditional on $\bF$, $\g_L=\bE\v\sim N(0,\bSigma_{\rm C})$. Moreover, the preceding lower bound, the upper bound following it, and $a=1+o(1)$ yield
\[
    \op{\b_L}_\infty\asymp\frac{\sigma_{\rm C}^2}{\sigma}\sqrt{\frac{T}{N}}.
\]
Finally, Gaussian concentration gives $\op{\bE\v}_\infty\lesssim\sigma_{\rm C}\sqrt{\log n}$ and $\ab{\u^\top\bE\v}\lesssim\sigma_{\rm C}\sqrt{\log n}$. In particular,
\[
    \op{\brac{a\bU_\perp\bU_\perp^\top-\bI_N}\bE\v}_\infty
    \le\ab{a-1}\op{\bE\v}_\infty+a\infn{\u}\ab{\u^\top\bE\v}.
\]
The first term on the right is absorbed by the next two terms in the bound below under the assumed lower bound on $\lambda$, and the second is of order $\sigma_{\rm C}\sqrt{\log n/N}$. Combining these bounds with the bounds for the centered quadratic and higher-order terms above gives
\[
    \op{\boldsymbol{\delta}_L}_\infty
    \lesssim \sigma_{\rm C}\sqrt{\frac{\log n}{N}}
    +\frac{\sigma_{\rm C}^2\sqrt{\log n}}{\sigma}
    +\frac{\sigma_{\rm C}^2\log^2n}{\sigma}\sqrt{\frac{N}{T}}
    +\frac{\sigma_{\rm C}^4}{\sigma^3}\sqrt{\frac{T}{N}}.
\]
The result follows from $\sigma\asymp\lambda$ on the event in \Cref{lem:good-event}.}

\section{Proofs in \Cref{sec:two-example}}\label{sec:pf-sec-3}
In this section, we provide the proofs of  \Cref{col:inf-L-exmp-ts} and \Cref{col:inf-L-exmp-fd}.  Notice that the conditions $\kappa\kappa_{\rm T}\infn{\bSigma_{\rm T}^{1/2}\bQ\bV}/\op{\bSigma_{\rm T}^{1/2}}=o(1)$ in \Cref{thm:inf-L} and $\kappa_{\rm C, 1}\infn{\bSigma_{\rm C}^{1/2}\bU}/\op{\bSigma_{\rm C}^{1/2}}=o(1)$ in \Cref{thm:inf-F} are not needed if $\bE$ is Gaussian. 

\subsection{Proof of \Cref{col:inf-L-exmp-ts}}\label{pf-col:inf-L-exmp-ts}
{Assumption \ref{assump:iden} is imposed in the corollary. We first establish two factor-side Gram-matrix bounds used in all three parts. Let $\calF_1:=\sigma(\f_1)$ and $\calF_t:=\sigma(\f_1,\beps_2,\ldots,\beps_t)$ for $t\ge2$.}
{Write $\bR_t:=\f_t\f_t^\top-\bI_r$. By Assumption \ref{assump:iden}, with probability at least $1-\eta_F$, $\op{\sum_{t=1}^T\bR_t}\le T\delta_F=o(T)$.}
{The uniform sub-Gaussian bound on $\f_t$ and a union bound imply that there is an event $\calE_F$ with $\PP\brac{\calE_F}\ge1-\eta_F-O\brac{T^{-20}}$ on which
\begin{align*}
\op{\sum_{t=1}^T\bR_t}\le T\delta_F=o(T),\qquad
\max_{t\in[T]}\op{\f_t}\lesssim\sqrt{r+\log T},\qquad
\op{\bF^\top\bF}\lesssim T.
\end{align*}
Each partial sum below differs from $\sum_{t=1}^T\bR_t$ by at most two endpoint terms. On $\calE_F$, each such term has operator norm at most $O\brac{r+\log T}=o(T)$. It follows that
\begin{align*}
\op{\sum_{t=1}^{T-1}\bR_t}\vee
\op{\sum_{t=1}^{T-2}\bR_t}\vee
\op{\sum_{t=2}^{T-1}\bR_t}=o(T).
\end{align*}
}

\paragraph{A lag-two Gram-matrix bound.}
We first show that $\sigma_r\brac{\bF^\top\bQ^2\bF}\gtrsim T$. By definition, we have
\begin{align*}
[\bQ^2]_{i,j}=\begin{cases}
    2, & i=j\in\ebrac{2,\cdots,T-1}\\
    1, & i=j\in\ebrac{1,T}\text{~or~}\ab{i-j}=2\\
    0, & \text{o.w.}
\end{cases}
\end{align*}
In particular, we have
\begin{align*}
\EE\f_{t+2}\f_t^\top=\EE\brac{\bA\f_{t+1}+\beps_{t+2}}\f_t^\top=\EE\brac{\bA^2\f_{t}+\bA\beps_{t+1}+\beps_{t+2}}\f_t^\top=\bA^2.
\end{align*}
{Moreover, $\bI_r-\bA\bA^\top\succeq0$ implies $\op{\bA}\le1$, and
\[
2\bI_r+\bA^2+\brac{\bA^\top}^2=(\bA+\bA^\top)^2+\brac{\bI_r-\bA\bA^\top}+\brac{\bI_r-\bA^\top\bA}\succeq4c^2\bI_r.
\]}
It follows that
\begin{align*}
\EE\bF^\top  \bQ^2\bF&=\EE\sqbrac{\f_1\f_1^\top+\f_T\f_T^\top+2\sum_{t=2}^{T-1}\f_t\f_t^\top+\sum_{t=1}^{T-2}\brac{\f_t\f_{t+2}^\top+\f_{t+2}\f_t^\top}}\\
&=2\bI_r+\brac{T-2}\sqbrac{2\bI_r+\bA^2+\brac{\bA^\top}^2}{\succeq\sqbrac{2+4c^2\brac{T-2}}\bI_r}.
\end{align*}
{For later use, write $\bG_{2,T}:=\bF^\top\bQ^2\bF$ and $\bG_{2,0}:=\EE\bG_{2,T}$. We now give the martingale bounds used to control the deviations from the population Gram matrices. For $k\in\ebrac{1,2}$, define
\begin{align*}
\bM_{k,m}:=\sum_{t=1}^{m}\beps_{t+k}\f_t^\top,\qquad
\bZ_{k,t}:=\begin{bmatrix}
\mathbf0&\beps_{t+k}\f_t^\top\\
\f_t\beps_{t+k}^\top&\mathbf0
\end{bmatrix},
\end{align*}
where $m\le T-1$ for $k=1$ and $m\le T-2$ for $k=2$. For $k=1$, let $\calH_{1,t}:=\calF_{t+1}$, and for $k=2$, let $\calH_{2,t}:=\calF_{t+2}$. {For each fixed $k$, we apply Theorem 2.1 of \cite{kroshnin2024bernstein} under a regular conditional law given $\calH_{k,0}$, identifying the theorem's $X_t$ and $\calF_t$ with $\bZ_{k,t}$ and $\calH_{k,t}$, respectively, and taking $n=T-k$, $d=2r$, and $\alpha=1$. In the notation of that theorem, set
\begin{align*}
\bSigma_t:=\EE\sqbrac{\bZ_{k,t}^2\mid\calH_{k,t-1}},\qquad
U_t:=\op{\lambda_{\max}\brac{\bZ_{k,t}}_+\mid\calH_{k,t-1}}_{\psi_1},
\end{align*}
It remains to verify, for some deterministic $\sigma>0$ and $U\ge K>0$, the three bounds defining the event
\begin{align*}
\calE:=\ebrac{
\lambda_{\max}\brac{\sum_{t=1}^{T-k}\bSigma_t}\le\sigma^2,\qquad
\sum_{t=1}^{T-k}U_t^2\le U^2,\qquad
\max_{t\le T-k}U_t\le K}.
\end{align*}
}
The matrix $\bZ_{k,t}$ is self-adjoint, $\calH_{k,t}$-measurable, and integrable. Moreover, $\f_t$ is $\calH_{k,t-1}$-measurable, and $\beps_{t+k}$ is independent of $\calH_{k,t-1}$ with mean zero. Hence $\EE\sqbrac{\bZ_{k,t}\mid\calH_{k,t-1}}=\mathbf0$, verifying the supermartingale-difference requirement. Set $\bSigma_\epsilon:=\bI_r-\bA\bA^\top$. For both $k=1$ and $k=2$, we have
\begin{align*}
U_t
=\op{\op{\bZ_{k,t}}\mid\calH_{k,t-1}}_{\psi_1}
\lesssim\sqrt r\,\op{\f_t},\qquad \bSigma_t=\begin{bmatrix}
\op{\f_t}^2\bSigma_\epsilon&\mathbf0\\
\mathbf0&\textsf{Tr}\brac{\bSigma_\epsilon}\f_t\f_t^\top
\end{bmatrix}.
\end{align*}
{On $\calE_F$, the preceding bound for $U_t$ gives
\begin{align*}
\max_{t\le T-k}U_t&\lesssim\sqrt{r\brac{r+\log T}},\\
\sum_{t=1}^{T-k}U_t^2
&\lesssim r\sum_{t=1}^{T-k}\op{\f_t}^2
\le r\textsf{Tr}\brac{\bF^\top\bF}\lesssim Tr^2.
\end{align*}
The formula for $\bSigma_t$, together with $\op{\bF^\top\bF}\lesssim T$, also gives
\begin{align*}
\lambda_{\max}\brac{\sum_{t=1}^{T-k}\bSigma_t}
&\le\op{\bSigma_\epsilon}\sum_{t=1}^{T-k}\op{\f_t}^2
\vee\textsf{Tr}\brac{\bSigma_\epsilon}
\op{\sum_{t=1}^{T-k}\f_t\f_t^\top}
\lesssim Tr.
\end{align*}
Therefore, with a sufficiently large constant $C$, we may take the deterministic parameters
\begin{align*}
K:=C\sqrt{r\brac{r+\log T}},\qquad
U:=Cr\sqrt T,\qquad
\sigma:=C\sqrt{Tr}.
\end{align*}
For all sufficiently large $T$, the condition $r\brac{r+\log T}=o(T)$ implies $U\ge K$, and $\calE_F\subseteq\calE$. In the notation of Theorem 2.1, $z=z\brac{U,\sigma;1}=4\log\brac{eU/\sigma}\lesssim\log(er)$. where $\log x:=\max\brac{\ln x,1}$, as in that paper. For $\alpha=1$, its Bernstein bound (Ber) is $\sigma\sqrt{2x}+3Kzx/4$. The self-adjoint dilation identity gives $\lambda_{\max}\brac{\sum_{t=1}^m\bZ_{k,t}}=\op{\bM_{k,m}}$. Applying the event-localized inequality (26) and the subsequent optimization in its proof, with $x=C_0\log T$, gives
\begin{align*}
\max_{m\le T-k}\op{\bM_{k,m}}
&\lesssim \sqrt{Tr\log T}
+\sqrt{r\brac{r+\log T}}\log(er)\log T,\qquad k\in\ebrac{1,2}.
\end{align*}
Let $\calB_k$ denotes the event on which the above bound fails. Almost surely, $\PP\brac{\calE_F\cap\calB_k\mid\calH_{k,0}}\le2rT^{-C_0}$.} 
Hence $\max_{m\le T-k}\op{\bM_{k,m}}=o(T)$ outside $\calB_k$. Integrating the probability bound and taking a union bound over $k\in\ebrac{1,2}$ gives $\PP\bbrac{\calE_F\cap\brac{\calB_1\cup\calB_2}}\le4rT^{-C_0}$. Thus, after redefining $\calE_F$ as the intersection with the complements of $\calB_1$ and $\calB_2$, its probability remains at least $1-\eta_F-O(T^{-20})$, after increasing $C_0$.} {The lag-two expansion is now exact:
\begin{align*}
\bG_{2,T}-\bG_{2,0}
&=\sum_{t=1}^T\bR_t+\sum_{t=2}^{T-1}\bR_t
+\bA^2\sum_{t=1}^{T-2}\bR_t
+\brac{\sum_{t=1}^{T-2}\bR_t}\brac{\bA^\top}^2\\
&\quad+\bA\bM_{1,T-2}+\bM_{1,T-2}^\top\bA^\top
+\bM_{2,T-2}+\bM_{2,T-2}^\top.
\end{align*}
The preceding partial-sum and martingale bounds, together with $\op{\bA}\le1$, therefore give
\begin{align*}
\op{\bG_{2,T}-\bG_{2,0}}=o(T),\qquad
c_1T\bI_r\preceq\bG_{2,T}\preceq C_1T\bI_r
\end{align*}
on $\calE_F$, for some constants $c_1,C_1>0$.

\paragraph{Verification of $\sigma_r\brac{\bF^\top\bQ\bF}\gtrsim T$.}
The corresponding exact expansion for the signal matrix is
\begin{align*}
\bF^\top\bQ\bF-\brac{T-1}\brac{\bA+\bA^\top}
&=\bA\sum_{t=1}^{T-1}\bR_t
+\brac{\sum_{t=1}^{T-1}\bR_t}\bA^\top
+\bM_{1,T-1}+\bM_{1,T-1}^\top.
\end{align*}
It follows from the preceding partial-sum and martingale bounds that
\begin{align*}
\op{\bF^\top\bQ\bF-2\brac{T-1}\bar\bA}=o(T).
\end{align*}
Since $c\bI_r\preceq\bar\bA\preceq(1-c)\bI_r$, Weyl's inequality yields
\begin{align*}
c_2T\bI_r\preceq\bF^\top\bQ\bF\preceq C_2T\bI_r
\end{align*}
on $\calE_F$, for some constants $c_2,C_2>0$.}

{\paragraph{Proof of part (i).}
For the lag-one weight, $\textsf{Tr}\brac{\bQ}=0$, $\fro{\bQ}=\sqrt{2(T-1)}$, and $\op{\bQ}\le2$. Replacing $\bQ$ by $\bQ/\op{\bQ}$ if necessary changes only universal constants and does not change the WPCA estimator. Under the bounded quantities assumed throughout this subsection, the factor-signal bound above gives $\lambda_r^2\sigma_r\brac{\bF^\top\bQ\bF}\gtrsim\lambda_r^2T$, whereas $\textsf{Err}_U\brac{\bQ}\lesssim\lambda_r\sqrt{NT}+\sqrt{NT}+N$. Both the bias term in $\textsf{Err}_U\brac{\bQ}$ and the additional scalar-shift term in the SNR condition of \Cref{thm:U-error} vanish because $\textsf{Tr}\brac{\bQ}=0$ and one may take $\zeta=0$. Therefore, $\lambda_r\gg(N/T)^{1/2}\vee(N/T)^{1/4}$ verifies the SNR condition of \Cref{thm:U-error}. Applying that theorem gives
\begin{align*}
\op{\hat\bU_\bQ\hat\bU_\bQ^\top-\bU\bU^\top}
&=O_p\brac{
\frac{1}{\lambda_r}\sqrt{\frac NT}
+\frac{1}{\lambda_r^2}\sqrt{\frac NT}
+\frac{N}{\lambda_r^2T}}.
\end{align*}
When $N\le T$, the last term is bounded by $\lambda_r^{-2}\sqrt{N/T}$. When $N>T$, it is $o\brac{\lambda_r^{-1}\sqrt{N/T}}$ because $\lambda_r\gg\sqrt{N/T}$. Consequently,
\begin{align*}
\op{\hat\bU_\bQ\hat\bU_\bQ^\top-\bU\bU^\top}
=O_p\brac{\frac{1}{\lambda_r}\sqrt{\frac NT}
\brac{1+\frac{1}{\lambda_r}}}.
\end{align*}
This proves part (i).}

{\paragraph{Proof of part (ii).}
Write compactly $\xi^U_\bQ:=\op{\hat\bU_\bQ\hat\bU_\bQ^\top-\bU\bU^\top}$ and 
$\xi^V_\bQ:=\op{\hat\bV_\bQ\hat\bV_\bQ^\top-\bV\bV^\top}$. The condition in part (ii) implies the SNR condition used in part (i), and the preceding calculation gives
\begin{align*}
\xi^U_\bQ=O_p\brac{\frac{1}{\lambda_r}\sqrt{\frac NT}}=o_p(1).
\end{align*}
By \Cref{lem:wtu-eq-hatu}, the equality $\wt\bU_\bQ\wt\bU_\bQ^\top=\hat\bU_\bQ\hat\bU_\bQ^\top$ holds on an event whose intersection with $\calE_0$ has probability tending to one. We work on this intersection.
For the right singular space, we invoke the operator-norm perturbation argument in the proof of \Cref{thm:V-error}. The condition $T=o(e^{cN})$ in that theorem is used only in the subsequent $\ell_2\to\ell_\infty$ argument, but not in its operator-norm proof. Since $\sqrt{n/T}=1\vee\sqrt{N/T}$, the preceding bound for $\xi^U_\bQ$ and the condition $\lambda_r\gg1\vee\sqrt{N/T}$ imply that every term in the perturbation-to-gap ratio tends to zero. Thus, this ratio is
\begin{align*}
O_p\brac{\frac{1}{\lambda_r}+\kappa\xi^U_\bQ
+\frac{1\vee\sqrt{N/T}}{\lambda_r}\xi^U_\bQ}=o_p(1).
\end{align*}
This establishes the required separation. The same argument bounds $\op{\hat\bV_\bQ-\bV\bR_{V,\bQ}^\top}$ by the expression below. Since $\xi^V_\bQ\le2\op{\hat\bV_\bQ-\bV\bR_{V,\bQ}^\top}$, it follows that
\begin{align*}
\xi^V_\bQ
&=O_p\brac{\frac{1}{\lambda_r}+\kappa\xi^U_\bQ
+\frac{1\vee\sqrt{N/T}}{\lambda_r}\xi^U_\bQ}
=O_p\brac{\frac{1\vee\sqrt{N/T}}{\lambda_r}}.
\end{align*}

For the loading error, we invoke the proof of \Cref{thm:factor-bound}, in particular \eqref{eq:hatL-decomp} and the polar-factor bounds used to derive \eqref{eq:hatSigma-bound}. The only refinement needed here concerns the final projected-noise term in \eqref{eq:hatSigma-bound}. Conditional on $\bF$, $\textsf{Cov}\bbrac{\textsf{vec}\brac{\bU^\top\bE\bV}\mid\bF} =\brac{\bU^\top\bSigma_{\rm C}\bU}\otimes\bI_r$, and $\op{\bU^\top\bSigma_{\rm C}\bU}=O(1)$. Hence
\begin{align*}
\frac{1}{\sqrt T}\op{\bU^\top\bE\bV}=O_p\brac{\sqrt{\frac rT}}.
\end{align*}
Thus, in the proof of \Cref{thm:factor-bound}, we may replace the $1-O\brac{n^{-10}}$ high-probability bound for this term by the preceding Gaussian $O_p$ bound. Together with $\lambda_1\asymp\lambda_r$, $\xi^U_\bQ=o_p(1)$, and $\sqrt{N/T}=o(\lambda_r)$, this yields
\begin{align*}
\frac{1}{\sqrt N}\op{\hat\bL_\bQ-\bL\bR_{L,\bQ}}
=O_p\brac{\frac{\lambda_r}{\sqrt N}\brac{\xi^U_\bQ+(\xi^V_\bQ)^2}
+\sqrt{\frac{r}{NT}}}.
\end{align*}
Substituting the preceding two subspace bounds into this display gives
\begin{align*}
\frac{1}{\sqrt N}\op{\hat\bL_\bQ-\bL\bR_{L,\bQ}}
=O_p\brac{\frac{1}{\sqrt T}+\frac{1}{\lambda_r\sqrt N}
+\frac{1}{\lambda_r\sqrt T}+\frac{\sqrt N}{\lambda_rT}
+\sqrt{\frac{r}{NT}}}.
\end{align*}
Here $\lambda_r\gg1$ absorbs the third term into $T^{-1/2}$, $\lambda_r\gg\sqrt{N/T}$ implies $\sqrt N/(\lambda_rT)=o(T^{-1/2})$, and $r\le N$ implies $\sqrt{r/(NT)}\le T^{-1/2}$. This proves part (ii).}

{\paragraph{Proof of part (iii).}
We now verify each condition of \Cref{thm:inf-L}. The factor-signal bound above gives $\sigma_r\brac{\bF^\top\bQ\bF}\gtrsim T$. Recall that $\bV=T^{-1/2}\bF\bB$.  Assumption \ref{assump:iden} implies $\sigma_r\brac{\bB}\asymp\sigma_1\brac{\bB}\asymp1$. It follows from the lag-two Gram-matrix bound that
\begin{align*}
\sigma_r\brac{\bV^\top\bQ^2\bV}
=\sigma_r\brac{T^{-1}\bB^\top\bG_{2,T}\bB}\gtrsim1.
\end{align*}
The assumptions $\op{\bSigma_{\rm C}}\asymp1$ and $\kappa_{\rm C,1}\asymp1$ imply that the eigenvalues of $\bSigma_{\rm C}$ are bounded above and away from zero. They also imply $\op{\bSigma_{\rm C}^{1/2}}_1+\op{\bSigma_{\rm C}^{-1/2}}_1=O(1)$. Hence $[\bSigma_{\rm C}]_{i,i}\asymp1$ uniformly in $i$, and the preceding display verifies the covariance nondegeneracy condition in \Cref{thm:inf-L}. The condition $N\wedge T=\omega\brac{r\brac{r+\log n}}$ then implies the required growth condition on $N$. Finally, $\textsf{Tr}\brac{\bQ}=0$, so the trace term in the SNR condition of \Cref{thm:inf-L} vanishes. Under the bounded quantities assumed throughout this subsection, the remaining terms are bounded by
\begin{align*}
\brac{\frac nT}^{1/2}r^{1/2}\log^2n
+\brac{\frac TN}^{1/2}r^{1/2}.
\end{align*}
{The signal-strength condition in part (iii) therefore implies the SNR condition of \Cref{thm:inf-L}, which gives the CLT with the covariance specified in that theorem.}

It remains to show that this covariance may be replaced by the one displayed in the corollary. Let $\bC_L:=\bU^\top\bL$. The SVD identity $T^{-1/2}\bL\bF^\top=\bU\bSigma\bV^\top$ gives the exact equality
\begin{align*}
T\bSigma\bV^\top\bQ^2\bV\bSigma
=\bC_L\bG_{2,T}\bC_L^\top.
\end{align*}
Thus, denoting the exact covariance in \Cref{thm:inf-L} by $\widetilde\bSigma_{L,i}$, the preceding identity gives
\begin{align*}
\widetilde\bSigma_{L,i}
=[\bSigma_{\rm C}]_{i,i}\bO_F^\top\bC_L\bG_{2,T}\bC_L^\top\bO_F.
\end{align*}
Set $\bS_T:=[\bSigma_{\rm C}]_{i,i}\bC_L\bG_{2,T}\bC_L^\top$ and 
$\bS_0:=[\bSigma_{\rm C}]_{i,i}\bC_L\bG_{2,0}\bC_L^\top$.
Since the singular values of $\bC_L$ are $\lambda_1,\ldots,\lambda_r$, the lag-two Gram-matrix bound, $\bG_{2,0}\succeq cT\bI_r$, and $\kappa=O(1)$ imply
\begin{align*}
\op{\bS_T-\bS_0}=o\brac{\sigma_r\brac{\bS_0}},\qquad
\frac{\sigma_1\brac{\bS_0}}{\sigma_r\brac{\bS_0}}=O(1)
\end{align*}
on $\calE_F$. Under the compatible-whitening convention in \Cref{thm:inf-L}, whitening factors for $\widetilde\bSigma_{L,i}=\bO_F^\top\bS_T\bO_F$ and $\bSigma_{L,i}=\bO_F^\top\bS_0\bO_F$ are, respectively,
$\widetilde\bW_i:=\bS_T^{-1/2}\brac{\bO_F^\top}^{-1}$ and  $\bW_i:=\bS_0^{-1/2}\brac{\bO_F^\top}^{-1}$.
The positive-definite square-root perturbation inequality, e.g., \cite{carlsson2018perturbation}, gives
\begin{align*}
\op{\bS_T^{1/2}-\bS_0^{1/2}}
\le\frac{\op{\bS_T-\bS_0}}
{\sigma_r\brac{\bS_T}^{1/2}+\sigma_r\brac{\bS_0}^{1/2}}.
\end{align*}
Together with the preceding relative bound, this yields $\bW_i\widetilde\bW_i^{-1}
=\bS_0^{-1/2}\bS_T^{1/2}=\bI_r+o(1)$ on $\calE_F$. Therefore,
\begin{align*}
\bW_i\brac{\hat\bL_\bQ-\bL\bR_{L,\bQ}}_{i,\cdot}^\top
=\brac{\bI_r+o(1)}\widetilde\bW_i
\brac{\hat\bL_\bQ-\bL\bR_{L,\bQ}}_{i,\cdot}^\top,
\end{align*}
and Slutsky's theorem proves the CLT with the covariance in the corollary. This proves part (iii).}

\subsection{Proof of \Cref{col:inf-L-exmp-fd}}\label{pf-col:inf-L-exmp-fd}
We start with the following lemma, whose proof can be found in \Cref{pf-lem:function-signal}.
\begin{lemma}\label{lem:function-signal}
Assume that $\max_{k\in[r]}\ebrac{\op{g_k}_\infty\vee \op{g^\prime_k}_\infty}\le C_g$ for some $C_g>0$, then we have  
\begin{enumerate}
    \item[(1)] $\op{\bF^\top\bF-T\bI_r}\lesssim C_g^2r$ provided that $T=\omega\brac{C_g^2r}$.
    \item[(2)] $\op{\bF^\top\bQ\bF-T\bI_r}\lesssim C_g^2Br$ provided that $T=\omega\brac{C_g^2Br}$.
    \item[(3)] $\op{\bF^\top\bQ^2\bF-T\bI_r}\lesssim C_g^2Br$ provided that $T=\omega\brac{C_g^2Br}$.
\end{enumerate}
\end{lemma}
{\Cref{lem:function-signal} verifies the Gram condition in the deterministic version of \Cref{assump:iden}, with $\delta_F\lesssim C_g^2r/T$ and $\eta_F=0$, and also gives $\op{\bF}\lesssim \sqrt{T}$ provided that $T\gg C_g^2r$. The row bound is included in the corollary's deterministic Assumption~\ref{assump:iden}.} Therefore, by \Cref{thm:U-error} we obtain that with probability at least $1-O(e^{-cN})$,
\begin{align*}
        \textsf{Err}_U\brac{\bQ}&\lesssim \lambda_1\sqrt{NT}+\brac{\sqrt{T/B}+\sqrt{N}}\sqrt{N},
\end{align*}
where we have used $\fro{\bQ}\le \sqrt{T/B}$ by construction. 
From \Cref{lem:function-signal} we also obtain that $\sigma_r\brac{\bF^\top\bQ\bF}\gtrsim T$, and hence we get \eqref{eq:u-bound-fd}.

\paragraph{Proof of part (ii).}
By part (i), we have
\[
\xi^U_\bQ
=O_p\!\left(
\frac1{\lambda_r}\sqrt{\frac NT}
+\frac1{\lambda_r^2}\sqrt{\frac N{TB}}
\right)=o_p(1).
\]
Since \(T\gtrsim N\) and the signal condition implies
\(\lambda_r\gg[N/(TB)]^{1/4}\ge[N/(TB)]^{1/2}\), because \(N/(TB)\le1\), the same bound also gives
\(\xi^U_\bQ=O_p(\lambda_r^{-1})\). The operator-norm perturbation argument in the proof of \Cref{thm:V-error}, combined with the deterministic bound \(\xi^V_\bQ\le1\), yields
\[
\xi^V_\bQ
=O_p\left(
1\wedge\left[
\frac1{\lambda_r}+\xi^U_\bQ+\frac{\xi^U_\bQ}{\lambda_r}
\right]
\right)
=O_p\left(1\wedge\frac1{\lambda_r}\right).
\]
Because the noise is Gaussian and \(\bF\) is deterministic, we have 
\[
\frac1{\sqrt T}\op{\bU^\top\bE\bV}
=O_p\!\left(\sqrt{\frac rT}\right).
\]
Using the exact loading decomposition in the proof of \Cref{thm:factor-bound}, with this Gaussian bound in place of the uniform \(1-O(n^{-10})\) bound, and using \(\lambda_1\asymp\lambda_r\), gives
\[
\frac1{\sqrt N}\op{\hat\bL_\bQ-\bL\bR_{L,\bQ}}
=O_p\left(
\frac{1+\lambda_r}{\sqrt N}
\left[\xi^U_\bQ+(\xi^V_\bQ)^2\right]
+\sqrt{\frac r{NT}}
\right).
\]
For the first subspace term, we have 
\[
\frac{1+\lambda_r}{\sqrt N}\xi^U_\bQ
=O_p\left(
\frac1{\sqrt T}
\left(
1+\lambda_r^{-1}
+\lambda_r^{-2}B^{-1/2}
\right)
\right).
\]
For the right-subspace term, if \(\lambda_r\ge1\), then
\((1+\lambda_r)(\xi^V_\bQ)^2=O_p(\lambda_r^{-1})\); if \(\lambda_r<1\), then \(\xi^V_\bQ\le1\) and \(1+\lambda_r\lesssim\lambda_r^{-1}\). Hence
\[
\frac{1+\lambda_r}{\sqrt N}(\xi^V_\bQ)^2
=O_p\!\left(\frac1{\lambda_r\sqrt N}\right).
\]
Finally, \(\sqrt{r/(NT)}\le T^{-1/2}\) because \(r\le N\). Combining the bounds proves \eqref{eq:L-bound-fd}.

\paragraph{Verification for part (iii).}
By \Cref{lem:function-signal},
\(\sigma_r(\bF^\top\bQ\bF)\gtrsim T\) and
\(\sigma_r(\bF^\top\bQ^2\bF)\gtrsim T\). Since
\(\bV=T^{-1/2}\bF\bB\), with
\(\sigma_r(\bB)\asymp\sigma_1(\bB)\asymp1\), the latter bound gives
\(\sigma_r(\bV^\top\bQ^2\bV)\gtrsim1\). Under the bounded covariance quantities assumed throughout this subsection, this verifies the covariance nondegeneracy condition in \Cref{thm:inf-L}; the condition on \(N\) in part (iii) gives the required growth condition. It remains to verify temporal incoherence. Note that
\[
\infn{\bQ\bV}
\le \frac{\op{\bB}}{\sqrt T}\infn{\bQ\bF}
\le \frac{\op{\bB}}{\sqrt T}\infn{\bF}
\lesssim C_g\sqrt{\frac rT}
=o(1),
\]
where the last step follows from \(T=\omega(C_g^2Br)\) and \(B\ge1\). Because \(\bSigma_{\rm T}=\bI_T\), \(\kappa=O(1)\), and \(\kappa_{\rm T}=1\), this is precisely the temporal-incoherence condition in \Cref{thm:inf-L}.

Finally, \(\textsf{Tr}\brac{\bQ}=0\), so the trace term in the SNR condition of \Cref{thm:inf-L} vanishes. With the other quantities bounded, the remaining requirement reduces to
\[
\lambda_r\gg
\left(\frac nT\right)^{1/2}r^{1/2}\log^2n
+\left(\frac TN\right)^{1/2}r^{1/2},
\]
which is implied by the condition in part (iii). Thus \Cref{thm:inf-L} applies. 

Finally, it suffices to note that we can replace $\bar\bSigma$  with the singular values of $T\bL\bL^\top$, which is equivalent to $T\bLambda^2$, provided that $T\gg C_g^2Br$. {Indeed, \Cref{lem:function-signal} gives relative $o(1)$ perturbations of both $\bF^\top\bQ\bF$ and $\bF^\top\bQ^2\bF$ from their positive-definite targets. Since $\kappa=O(1)$ throughout this subsection, the perturbations remain relative $o(1)$ after congruence by $\bU^\top\bL$. Weyl's inequality and Slutsky's theorem then justify the replacement of $\bar\bSigma$ and of the covariance in the standardized statistic respectively.} This completes the proof.

\subsection{Proof of \Cref{col:inf-F-exmp-fd}}\label{pf-col:inf-F-exmp-fd}
Under \(T\gtrsim N\), the argument in the proof of part (ii) of \Cref{col:inf-L-exmp-fd} gives
\[
\op{\hat\bV_\bQ\hat\bV_\bQ^\top-\bV\bV^\top}
=O_p\!\left(1\wedge\frac1{\lambda_r}\right)
=O_p\!\left(\frac1{\lambda_r}\right).
\]
The standard aligned-singular-vector inequality and the identity
\[
\frac1{\sqrt T}\op{\hat\bF_\bQ-\bF\bR_{F,\bQ}}
=
\op{\hat\bV_\bQ-\bV\bR_{V,\bQ}^\top}
\]
therefore prove part (i).

For part (ii), \Cref{lem:function-signal} gives
\(\sigma_r(\bF^\top\bQ\bF)\gtrsim T\) and verifies the deterministic version of Assumption~\ref{assump:iden}. Under the bounded quantities assumed throughout this subsection, the eigenvalues of \(\bSigma_{\rm C}\) are bounded above and away from zero. Hence
\[
\min_{t\in[T]}[\bSigma_{\rm T}]_{t,t}
\,\sigma_r(\bU^\top\bSigma_{\rm C}\bU)
=\sigma_r(\bU^\top\bSigma_{\rm C}\bU)
\gtrsim1.
\]
Moreover, using symmetry of \(\bSigma_{\rm C}^{1/2}\),
\[
\kappa_{\rm C,1}
\frac{\infn{\bSigma_{\rm C}^{1/2}\bU}}
{\op{\bSigma_{\rm C}^{1/2}}}
\le C\infn{\bU}
=C\infn{\bL\bLambda^{-1}}
\lesssim\sqrt{\frac rN}
=o(1),
\]
where \(r\) is fixed. With \(\bSigma_{\rm T}=\bI_T\), fixed \(r\), and the bounded condition numbers and covariance norms assumed here, the growth and SNR conditions of \Cref{thm:inf-F} reduce to
\[
T=\omega\!\left((r+\log n)^2\right),
\qquad
\lambda_r=\omega\!\left(\frac nT\,r\log^3n\right),
\]
which are exactly the conditions in part (ii). The band weight satisfies $\op{\bSigma_{\rm T}^{1/2}\bQ\bSigma_{\rm T}^{1/2}}_1=O(1)$ after operator-norm normalization, with $\bSigma_{\rm T}=\bI_T$ and $\kappa_{\rm T,1}=1$. Applying \Cref{thm:inf-F} gives the stated result.

\subsection{Proof of \Cref{lem:function-signal}}\label{pf-lem:function-signal}
Without loss of generality we rescale $\bQ$ such that $Q_{t,s}=\II\ebrac{1\le \ab{t-s}\le B}$. 
It  suffices to show that $\op{\bF^\top\bF-T\bI_r}\lesssim C_g^2r$, $\op{\bF^\top\bQ\bF-2BT\bI_r}\lesssim C_g^2B^2r$ and $\op{\bF^\top\bQ^2\bF-4B^2T\bI_r}\lesssim C_g^2B^3r$.
\\[2ex] For $0\le h\le B$, define
\begin{align*}
S_h(k,l):=\sum_{t=1}^{T-h}g_k\brac{\frac{t+h}{T}}g_l\brac{\frac{t}{T}},\qquad 1\le k,l\le r.
\end{align*}
\paragraph{Proof of claim (1)}	Notice that $\sqbrac{\bF^\top\bF}_{k,l}= S_0(k,l)$ for $k,l\in[r]$, for which  have
\begin{align*}
\ab{T^{-1}S_0(k,l)-\delta_{kl}}=\ab{\frac{1}{T}\sum_{t=1}^{T}g_k\brac{\frac{t}{T}}g_l\brac{\frac{t}{T}}-\int_0^1g_k(u)g_l(u)du}\le \frac{C_g^2}{2T}.
\end{align*}
This implies that $\op{\bF^\top\bF-{T\bI_r}}\le r\ab{[\bF^\top\bF-{T\bI_r}]_{k,l}}\le C_g^2r$.

\paragraph{Proof of claim (2)}	
Next, note that $\sqbrac{\bF^\top\bQ\bF}_{k,l}=\sum_{h=1}^B S_h(k,l)+\sum_{h=1}^B S_h(l,k)$ for $k,l\in[r]$. Notice that for any $h=o(T)$ and $1\le t\le T-h$, we have
\begin{align*}
g_k\brac{\frac{t+h}{T}}=g_k\brac{\frac{t}{T}}+\frac{h}{T}g_k^\prime \brac{\xi_{t,h}},
\end{align*}
where $\xi_{t,h}\in[t/T,(t+h)/T]$. Therefore, we have
\begin{align*}
\sum_{h=1}^BS_h(k,l)&=\sum_{h=1}^B\sum_{t=1}^{T-h}g_k\brac{\frac{t+h}{T}}g_l\brac{\frac{t}{T}}\\
&=\sum_{h=1}^B\sum_{t=1}^{T-h}g_k\brac{\frac{t}{T}}g_l\brac{\frac{t}{T}}+\sum_{h=1}^B\frac{h}{T}\sum_{t=1}^{T-h}g_k^\prime\brac{\xi_{t,h}}g_l\brac{\frac{t}{T}}.
\end{align*}

For any $u\in[0,1]$, let $G_{k,l}(u):=g_k(u)g_l(u)$, then $G_{k,l}^\prime(u)=g_k^\prime(u) g_l(u)+g_k(u) g_l^\prime(u)$. Therefore, we have $\op{G^\prime}_{\infty}\le C_g^2$.

For any $h\in[T]$, we have 
\begin{align*}
\ab{\sum_{t=1}^{T-h}g_k\brac{\frac{t}{T}}g_l\brac{\frac{t}{T}}-\sum_{t=1}^{T}g_k\brac{\frac{t}{T}}g_l\brac{\frac{t}{T}}}\le {\sum_{t=T-h+1}^{T}\ab{g_k\brac{\frac{t}{T}}g_l\brac{\frac{t}{T}}}}\le C_g^2h.
\end{align*}
Hence we arrive at 
\begin{align*}
\ab{\frac{1}{T}\sum_{t=1}^{T-h}g_k\brac{\frac{t}{T}}g_l\brac{\frac{t}{T}}-\delta_{kl}}\le \frac{C_g^2\brac{h+1/2}}{T}.
\end{align*}
We can bound the leading term as
\begin{align*}
\ab{\sum_{h=1}^B\sum_{t=1}^{T-h}g_k\brac{\frac{t}{T}}g_l\brac{\frac{t}{T}}-BT\delta_{kl}}\le C_g^2B\brac{B+\frac{1}{2}}.
\end{align*}
It remains to bound
\begin{align*}
{\ab{\sum_{h=1}^B\frac{h}{T}\sum_{t=1}^{T-h}g_k^\prime\brac{\xi_{t,h}}g_l\brac{\frac{t}{T}}}}\le \frac{1}{2}C_g^2B\brac{B+1}.
\end{align*}
Collecting these two bounds we arrive at 
\begin{align*}
\ab{\sum_{h=1}^BS_h(k,l)-BT\delta_{kl}}&\lesssim C_g^2B^2.
\end{align*}
Similarly we can obtain the same bound for $\ab{\sum_{h=1}^BS_h(l,k)-BT\delta_{kl}}$. We arrive at
\begin{align*}
\ab{\sum_{h=1}^BS_h(k,l)+\sum_{h=1}^BS_h(l,k)-2BT\delta_{kl}}&\lesssim C_g^2B^2.
\end{align*}
Let $\bR:=\bF^\top\bQ\bF-2BT\bI$, then we have
\begin{align*}
\op{\bR}\le r\max_{k,l}\ab{\sum_{h=1}^BS_h(k,l)+\sum_{h=1}^BS_h(l,k)-2BT\delta_{kl}}\lesssim C_g^2B^2r.
\end{align*}
We complete the proof by using Weyl's inequality and that $T\gg C^2_gBr$.
\paragraph{Proof of claim (3)} {For $t\in[T]$, let $\calN_t:=\ebrac{s\in[T]:1\le\ab{t-s}\le B}$ and $d_t:=|\calN_t|$. By definition,
\begin{align*}
[\bQ\bF]_{t,k}=\sum_{s\in\calN_t}g_k\brac{\frac{s}{T}}.
\end{align*}
Set $\bR:=\bQ\bF-2B\bF$. If $B+1\le t\le T-B$, then $d_t=2B$ and the Lipschitz property gives
\begin{align*}
\ab{R_{t,k}}
\le\sum_{h=1}^B\ab{g_k\brac{\frac{t+h}{T}}-g_k\brac{\frac{t}{T}}}
+\sum_{h=1}^B\ab{g_k\brac{\frac{t-h}{T}}-g_k\brac{\frac{t}{T}}}
\lesssim \frac{C_gB^2}{T}.
\end{align*}
For the at most $2B$ boundary rows, we instead use
\begin{align*}
\ab{R_{t,k}}
&\le\sum_{s\in\calN_t}\ab{g_k\brac{\frac{s}{T}}-g_k\brac{\frac{t}{T}}}
+(2B-d_t)\ab{g_k\brac{\frac{t}{T}}}
\lesssim C_gB.
\end{align*}
Consequently, for every $k,l\in[r]$,
\begin{align*}
\ab{[\bF^\top\bR]_{k,l}}\lesssim C_g^2B^2,\qquad \ab{[\bR^\top\bR]_{k,l}}\lesssim C_g^2\brac{\frac{B^4}{T}+B^3}\lesssim C_g^2B^3.
\end{align*}
Since $\bQ\bF=2B\bF+\bR$,
\begin{align*}
\bF^\top\bQ^2\bF-4B^2\bF^\top\bF
=2B\brac{\bF^\top\bR+\bR^\top\bF}+\bR^\top\bR.
\end{align*}
Using $\op{\bA}\le r\max_{k,l}|A_{k,l}|$ for an $r\times r$ matrix, we obtain
\begin{align*}
\op{\bF^\top\bQ^2\bF-4B^2\bF^\top\bF}\lesssim C_g^2B^3r.
\end{align*}
Combined with claim (1), this proves the desired result after restoring the normalization factor $(2B)^{-2}$.}

\section{Proofs in \Cref{sec:cv}}\label{sec:pf-sec-4}
In this section, we {first give a nonempty-class construction and then} provide the proof of  \Cref{thm:cv-consistency}.

\subsection{{A nonempty-class construction}}\label{sec:cv-nonempty-construction}
{We verify that all conditions of \Cref{thm:cv-consistency}, including the nonemptiness of its two comparison classes, can hold jointly. Consider an asymptotic sequence with even $N$, $r=1$, $K_0=2$, fixed $p_*\in(0,1)$, $T\asymp N\log^6T$, and
\[
\lambda_1=C_\lambda\log^{3/2}T,\qquad \bU=N^{-1/2}\mathbf1_N,\qquad \bL=\lambda_1\bU,\qquad \bSigma_{\rm T}=\bI_T,
\]
where $C_\lambda$ is a sufficiently large constant. Let $\bSigma_{\rm C}$ be diagonal with its first $N/2$ diagonal entries equal to $1$ and its remaining entries equal to $4$, and take the entries of $\bG$ to be standard Gaussian. Then $K,\kappa,\sigma_{\rm C},\sigma_{\rm T}\asymp1$, $\infn{\bL\bLambda^{-1}}=N^{-1/2}$, and
\[
\rho=\op{\bU^\top\bSigma_{\rm C}\bU_\perp}
=\sqrt{\bU^\top\bSigma_{\rm C}^2\bU-(\bU^\top\bSigma_{\rm C}\bU)^2}=\frac32\asymp\sigma_{\rm C}^2.
\]
Thus the covariance and incoherence conditions hold. Moreover, $\lambda_1^2=o(N)$, $T=o(e^{cN})$, and the lower SNR condition holds upon choosing $C_\lambda$ sufficiently large.

Let $f_t=\sqrt2\cos(\pi t/3+\Phi)$, where $\Phi$ is uniform on $[0,2\pi]$ and independent of the noise. This bounded process is stationary, $\EE f_t^2=1$, and $\Gamma_\tau=\cos(\pi\tau/3)$. In particular, $\Gamma_1=1/2$ and $\Gamma_2=-1/2$. The geometric-series formula shows, uniformly in $\Phi$, that the sample variance and the sample lag-$1$ and lag-$2$ covariances differ from their expectations by $O(T^{-1})$. Hence Assumptions~\ref{assump:iden} and \ref{assump:factor-ts} hold with $\delta_F,\bar\delta_F=O(T^{-1})$ and $\eta_F=\bar\eta_F=0$. Writing $\ell:=\log T$, we have
\[
\psi\asymp\ell^{-1/2},\qquad \bar\psi\asymp\ell^2.
\]
Take $B_1=\ell^{1/16}$ and $B_2=\ell^{1/8}$. Then
\[
\psi(B_1\vee B_2)\asymp\ell^{-3/8}=o(1),\qquad
\frac{\psi B_1}{\bar\psi B_2}\asymp\ell^{-41/16}=o(1),\qquad
\bar\psi B_2\asymp\ell^{17/8}=o(\ell^3)=o(\lambda_1^2).
\]
Let the finite grid $\calG_2$ contain the two candidates
\[
\bgamma^{(1)}=(g_1,1-g_1,0,\ldots),\qquad
\bgamma^{(2)}=(g_2,0,1-g_2,0,\ldots),
\]
where $g_1:=\psi B_1$, $a_T:=1-2/T$, $m_T:=(\bar\psi B_2^2)^{-1}$, and $g_2:=(a_T+m_T)/(1+a_T)$. Direct calculation gives
\[
\mu\brac{\bgamma^{(1)}}=g_1+(1-g_1)\brac{1-\frac1T}\longrightarrow1,\qquad
\mu\brac{\bgamma^{(2)}}=g_2-a_T(1-g_2)=m_T\asymp\ell^{-9/4},
\]
with $g_2\to1/2$. Therefore, for fixed constants $c_0,c_\mu,c_{\rm gap}\in(0,1)$ and all sufficiently large $T$, $\bgamma^{(1)}\in\calG_{2,1}$, while $g_2\ge\psi B_2$. Moreover,
\[
\frac{g_2}{\mu\brac{\bgamma^{(2)}}}\asymp\ell^{9/4},\qquad
\bar\psi B_2\asymp\ell^{17/8}\ll\ell^{9/4}\ll\ell^3\asymp\lambda_1^2,
\]
so $\bgamma^{(2)}\in\calG_{2,2}$. Since $r=1$, the matrix positivity and largest-eigenvalue requirements reduce to positivity of the displayed scalars. Finally, uniformly over these two candidates,
\[
\frac{\gamma_0\delta_F+2\bar\delta_F}{\mu(\bgamma)}=O\brac{T^{-1}\ell^{9/4}}=o(1).
\]
This verifies the remaining conditions and the nonemptiness of both comparison classes.}

We shall introduce additional notation for the proofs in \Cref{sec:cv}.
{Write $p:=p_*$, let $\bX^\natural:=p^{-1}\calP_{\Omega}\brac{\bX}$ and }$\bE^\natural:={\bX^\natural-\bM=}p^{-1}\calP_{\Omega}\brac{\bE}+\bDelta_M$, where $\bDelta_M:=\brac{p^{-1}\bOmega-\mathbf{1}_N\mathbf{1}^\top _T}\circ \bM$, and let 
 $\wt\bU_\bgamma^\Omega$ {contain the eigenvectors associated with the $r$ largest algebraic eigenvalues of} $\bX^\natural\bQ_{\bgamma}\bX^{\natural\top}$. In addition, let $\bar\kappa_{\bgamma}:={\lambda_1\brac{\bM\bQ_\bgamma\bM^\top}/\lambda_{r}\brac{\bM\bQ_\bgamma\bM^\top}}$ {for any candidate considered below.}

\subsection{Proof of \Cref{thm:cv-consistency}}\label{pf-thm:cv-consistency}
We  conduct our analysis on $\calE_0\bigcap \bar\calE_0$ defined in \Cref{lem:good-event}. {The condition $T=o(e^{c_*N})$, with $c_*$ sufficiently small, allows the $e^{-cN}$ terms in the probability bounds below to be absorbed into $n^{-10}$.} To start with, we state the following theorem characterizing the accuracy of $\wt\bU_\bgamma^\Omega$ in $\ell_2\rightarrow\ell_\infty$ norm, which is the key tool for proving {the pairwise CV comparison in \Cref{thm:cv-consistency}}. This result is analogous to \Cref{thm:U-two-infinty} under the  missing completely at random regime with a constant missing rate. We defer its proof to  \Cref{pf-thm:U-two-infinty-mcar}.
\begin{theorem}\label{thm:U-two-infinty-mcar}
    Suppose $\bE\sim \textsf{subG}_{K}(\bSigma_{\rm C},\bSigma_{\rm T})$, where {$\bSigma_{\rm T}=\textsf{diag}\bbrac{\sigma_{{\rm T},1}^2,\ldots,\sigma_{{\rm T},T}^2}$}, and that Assumptions \ref{assump:iden}-\ref{assump:digoanl-dominant} hold{, with $T=o(e^{c_*N})$ as in \Cref{thm:cv-consistency}}. Let $C>0$ be some universal constant and {$\bgamma\in\calG_{K_0,1}\cup\calG_{K_0,2}$}. Define 
    \begin{align}\label{eq:theta-def-mcar}
        \vartheta_\bgamma:=C&\left[K^2\sqrt{nNr}\log n\brac{\sigma_{\rm C}\sigma_{\rm T}\vee \op{\bM}_{\sf max}}^2+\sqrt{NT}\op{\bSigma}\brac{\sigma_{\rm C}\sigma_{\rm T}\log n+\op{\bM}_{\sf max}}\right.\notag\\
        &\left.+T\op{\bSigma}^2\infn{\bV}\sqrt{r\log n}+\inf_{\zeta\in\RR}\op{\bDelta_{\bgamma,\zeta}}_1\right].
    \end{align}
    There exist some universal constants $C_1,C_2>0$ such that if {$\lambda_{r}\brac{\bM\bQ_\bgamma\bM^\top}>C_1\bar\kappa_\bgamma\sqrt r\,\vartheta_\bgamma$}, then
    \begin{align*}&\PP\brac{\infn{\wt\bU^\Omega_\bgamma\wt\bU_\bgamma^{\Omega\top} -\bU\bU^\top }\le \frac{C_2\vartheta_\bgamma\infn{\bU}}{{\lambda_{r}}\brac{\bM\bQ_\bgamma\bM^\top}} \mid \bF}\ge 1-O\brac{n^{-10}}.
    \end{align*}
\end{theorem}

\paragraph{Upper bounds for $\op{\bM}_{\sf max}$ and $\infn{\bU}$.}
Recall \eqref{eq:U-L-infty-relation} in the proof of \Cref{thm:factor-bound},  we have $\bU_{L}=\bL\bLambda^{-1}\in\RR^{N\times r}$  and there exists $\bO_L\in\OO_{r}$ such that $\bU=\bU_L\bO_L$. We thus get that
\begin{align*}
    \infn{\bU}=\infn{\bL\bLambda^{-1}}\lesssim\sqrt{\frac{r}{N}}.
\end{align*}
Combined with $\infn{\bV}\lesssim\sqrt{\frac{r+\log n}{T}}$, we obtain
\begin{align*}
    \op{\bM}_{\sf max}\le \sqrt{T}\infn{\bU}\op{\bSigma}\infn{\bV}\lesssim\lambda_1\sqrt{\frac{r\brac{r+\log n}}{N}}=o\brac{K\sigma_{\rm C}\sigma_{\rm T}\sqrt{r+\log n}}.
\end{align*}
where we've used $\lambda_1^2 r/\brac{K\sigma_{\rm C}\sigma_{\rm T}}^2=o\brac{N}$.
\paragraph{Lower bound for $\sigma_r\brac{\bM\bQ_\bgamma\bM^\top}$.}

{For $\bgamma\in\calG_{K_0,1}\cup\calG_{K_0,2}$, Assumptions \ref{assump:iden} and \ref{assump:factor-ts} give
\begin{align*}
\op{\frac{1}{T}\bF^\top\bQ_\bgamma\bF-\brac{\gamma_0\bI_r+\sum_{\tau=1}^{K_0}\gamma_\tau\brac{1-\frac{\tau}{T}}\brac{\bGamma_\tau+\bGamma_\tau^\top}}}\le\gamma_0\delta_F+2\bar\delta_F=o\brac{\mu(\bgamma)}.
\end{align*}
Consequently, on $\bar\calE_0$,
\[
\lambda_r\brac{\bM\bQ_\bgamma\bM^\top}\gtrsim\lambda_r^2T\mu(\bgamma),\qquad
\lambda_1\brac{\bM\bQ_\bgamma\bM^\top}\lesssim\lambda_1^2T\mu(\bgamma),
\]
and hence $\bar\kappa_\bgamma=O(\kappa^2)$.} In addition, the quantity defined in \eqref{eq:theta-def-mcar} in \Cref{thm:U-two-infinty-mcar} admits 
\begin{align*}
    \vartheta_\bgamma\lesssim &\sqrt{NTr}\log n\brac{K\sigma_{\rm C}\sigma_{\rm T}\vee \op{\bM}_{\sf max}}^2+\sqrt{NT}\lambda_1\brac{K\sigma_{\rm C}\sigma_{\rm T}\log n+\op{\bM}_{\sf max}}\notag\\
    &+\lambda_1^2T\infn{\bV}\sqrt{r\log n}+\gamma_0 \sigma_{\rm C}^2\sigma_{\rm T}^2T.
\end{align*}
Using the condition that $\lambda_r/\brac{K\sigma_{\rm C}\sigma_{\rm T}}\gtrsim\kappa^2r\log^{3/2} n$ and 
\begin{align*}
    \psi={\frac{K\lambda_1}{\sigma_{\rm C}\sigma_{\rm T}}}\sqrt{\frac{N}{T}}\brac{r+\log n}=o\brac{\gamma_0},\quad {\bgamma\in\calG_{K_0,1}\cup\calG_{K_0,2}},
\end{align*}
we obtain that   $\vartheta_\bgamma\lesssim \gamma_0 \sigma_{\rm C}^2\sigma_{\rm T}^2T$.
{For $\calG_{K_0,1}$, the required signal gap follows from $\mu(\bgamma)\ge c_\mu$, $\gamma_0=o(1)$, and the stated SNR condition. For $\calG_{K_0,2}$, it follows from the upper bound on $\gamma_0/\mu(\bgamma)$ by choosing $c_{\rm gap}$ sufficiently small. Hence}
\begin{align*}
    {\lambda_r\brac{\bM\bQ_\bgamma\bM^\top}\ge C_0\bar\kappa_\bgamma r^{1/2}\vartheta_\bgamma,\quad \forall\bgamma\in\calG_{K_0,1}\cup\calG_{K_0,2}.}
\end{align*}
In the following, we thus proceed our analysis conditional on $\bF$ with the condition {$\lambda_r\brac{\bM\bQ_\bgamma\bM^\top}\ge C_0\bar\kappa_{\bgamma}r^{1/2}\vartheta_\bgamma$ for $\forall\bgamma\in\calG_{K_0,1}\cup\calG_{K_0,2}$.}

\paragraph{Analysis for CV.}
Recall that $\bar\bU\bar\bSigma\bar\bU^\top$ is the {rank-$r$ eigendecomposition of the positive-semidefinite matrix} $\bM\bQ_\bgamma\bM^\top$, and we have $\bU=\bar\bU\bar\bO$ for some $\bar\bO\in\OO_{r}$. {Because $\wt\bU_\bgamma^\Omega\wt\bU_\bgamma^{\Omega\top}\bX^\natural$ has rank at most $r$, the second-stage rank-$r$ SVD in \Cref{alg:main} gives
\[
\hat\bL_\bgamma^\Omega\hat\bF_\bgamma^{\Omega\top}
=\wt\bU_\bgamma^\Omega\wt\bU_\bgamma^{\Omega\top}\bX^\natural.
\]
Hence} $\textsf{CV}\brac{\bgamma}$ can be written as 
\begin{align*}
	\textsf{CV}\brac{\bgamma}&={\frac{1}{|\Omega_\perp|}}\fro{\calP_{\Omega_\perp}\brac{\bU\bU^\top p_*^{-1}\calP_{\Omega}\brac{\bX} -\wt\bU^\Omega_\bgamma\wt\bU_\bgamma^{\Omega\top} p_*^{-1}\calP_{\Omega}\brac{\bX}+\bE-\bU\bU^\top \bE^\natural}}^2,
\end{align*}
where $\bE^\natural=p_*^{-1}\calP_{\Omega}\brac{\bX}-\bM$. For a realized split, $|\Omega_\perp|$ is common to all candidates. It therefore cancels from every pairwise CV comparison, so the remaining argument can work with the unnormalized squared norms below.
Denote $\mXi_\bgamma:=\bbrac{\wt\bU^\Omega_\bgamma\wt\bU^{\Omega\top}_\bgamma-\bU\bU^\top }{\bX^\natural}$. It suffices to show that for any $\bgamma_1\in\calG_{K_0,1}$, $\bgamma_2\in\calG_{K_0,2}$ and $\bgamma\in\calG_{K_0,1}\cup\calG_{K_0,2}$,
\begin{align}
    \ab{\inp{\calP_{\Omega_\perp}\brac{\bE-\bU\bU^\top \bE^\natural}}{\calP_{\Omega_\perp}\brac{\mXi_\bgamma}}}&\ll \fro{\calP_{\Omega_\perp}\brac{\mXi_{\bgamma_2}}}^2\label{eq:cv-cond-2}\\
    \fro{\calP_{\Omega_\perp}\brac{\mXi_{\bgamma_1}}}^2&\ll \fro{\calP_{\Omega_\perp}\brac{\mXi_{\bgamma_2}}}^2.\label{eq:cv-cond-1}
\end{align}
To see \eqref{eq:cv-cond-2}, we first consider the term $\ab{\inp{\calP_{\Omega_\perp}\brac{\bE}}{\calP_{\Omega_\perp}\brac{\mXi_{\bgamma}}}}$. Note that 
\begin{align*}
    \ab{\inp{\calP_{\Omega_\perp}\brac{\bE}}{\calP_{\Omega_\perp}\brac{\mXi_{\bgamma}}}}
    &=\ab{\inp{\calP_{\Omega_\perp}\brac{\bE}}{\mXi_\bgamma}}
    \le \op{\calP_{\Omega_\perp}\brac{\bE}}\op{\mXi_\bgamma}_*
    \le \sqrt{2r}\op{\calP_{\Omega_\perp}\brac{\bE}}\fro{\mXi_\bgamma}\\
    &\lesssim K\sigma_{\rm C}\sigma_{\rm T}\sqrt{nr \log n}\fro{\calP_{\Omega_\perp}\brac{\mXi_\bgamma}},
\end{align*}
{where $\operatorname{rank}(\mXi_\bgamma)\le2r$, and the last inequality follows from \Cref{lem:GB-bound-mcar,lem:cv-Pdeltagam-lb} with probability at least $1-O\brac{n^{-10}}$.} For $\ab{\inp{\calP_{\Omega_\perp}\brac{\bU\bU^\top \bE^\natural}}{\calP_{\Omega_\perp}\brac{\mXi_{\bgamma}}}}$, we have 
\begin{align*}
    &\ab{\inp{\calP_{\Omega_\perp}\brac{\bU\bU^\top \bE^\natural}}{\calP_{\Omega_\perp}\brac{\mXi_{\bgamma}}}}\le \fro{\calP_{\Omega_\perp}\brac{\bU\bU^\top \brac{p_*^{-1}\calP_{\Omega}\brac{\bE}+\bDelta_M}}}\fro{\calP_{\Omega_\perp}\brac{\mXi_\bgamma}}\\
    &\le \fro{\bU\bU^\top \brac{p_*^{-1}\calP_{\Omega}\brac{\bE}+\bDelta_M}}\fro{\calP_{\Omega_\perp}\brac{\mXi_\bgamma}}\\
    &\le \sqrt{r}\op{p_*^{-1}\calP_{\Omega}\brac{\bE}+\bDelta_M}\fro{\calP_{\Omega_\perp}\brac{\mXi_\bgamma}}\\
    &\lesssim\brac{K\sigma_{\rm C}\sigma_{\rm T}\sqrt{\log n}+\op{\bM}_{\sf max}}\sqrt{nr}\fro{\calP_{\Omega_\perp}\brac{\mXi_\bgamma}}.
\end{align*}
where the last inequality holds with probability at least $1-O\brac{n^{-10}}$ due to \Cref{lem:GB-bound-mcar}. 
Combined with the upper bound for $\op{\bM}_{\sf max}$, we can conclude that \eqref{eq:cv-cond-2} holds as long as \eqref{eq:cv-cond-1} and the following condition hold:  
\begin{align}
    \fro{\calP_{\Omega_\perp}\brac{\mXi_{\bgamma_2}}}\gg \sqrt{nr}\brac{K\sigma_{\rm C}\sigma_{\rm T}\sqrt{r+\log n}}.\label{eq:cv-cond-2-prime}
\end{align}
It remains to show \eqref{eq:cv-cond-1} and \eqref{eq:cv-cond-2-prime}, for which the following two lemmas are needed and their proofs are deferred to \Cref{pf-lem:Deltagam-fro-lb}  and \Cref{pf-lem:Deltagam-fro-ub}.
\begin{lemma}\label{lem:Deltagam-fro-lb}
    Under the conditions of \Cref{thm:cv-consistency}, for any {$\bgamma\in\calG_{K_0,1}\cup\calG_{K_0,2}$}, with probability at least $1-O\bbrac{n^{-10}}$ we have
    \begin{align*}
        \fro{\mXi_\bgamma}\ge  c\cdot \frac{\gamma_0 \lambda_r \sigma_{\rm C}^2\sigma_{\rm T}^2 T^{3/2}}{{\lambda_1} {\brac{\bM\bQ_\bgamma\bM^\top}}}
        {\gtrsim\frac{\gamma_0}{\mu(\bgamma)}
        \frac{\sigma_{\rm C}^2\sigma_{\rm T}^2\sqrt T}{\kappa\lambda_1}},
    \end{align*}
    for some universal constant $c>0$.
\end{lemma}
\begin{lemma}\label{lem:Deltagam-fro-ub}
    Under the conditions of \Cref{thm:cv-consistency}, for any {$\bgamma\in\calG_{K_0,1}\cup\calG_{K_0,2}$}, with probability at least $1-O\bbrac{n^{-10}}$ we have
    \begin{align*}
        \fro{\mXi_\bgamma}\le  C\cdot \frac{\gamma_0 \lambda_1\sigma_{\rm C}^2\sigma_{\rm T}^2T^{3/2}r^{1/2}}{{\lambda_r}\brac{\bM\bQ_\bgamma\bM^\top}}
        {\lesssim\frac{\gamma_0}{\mu(\bgamma)}
        \frac{\kappa^2\sqrt r\,\sigma_{\rm C}^2\sigma_{\rm T}^2\sqrt T}{\lambda_1}},
    \end{align*}
    for some universal constant $C>0$.
\end{lemma}
In addition, we also need the following lemma, which is a result of combining \Cref{thm:U-two-infinty-mcar} and  Theorem B.4 in \cite{jin2021factor}, whose proof can be found in \Cref{pf-lem:cv-Pdeltagam-lb}.
\begin{lemma} \label{lem:cv-Pdeltagam-lb}
    Under the conditions of \Cref{thm:cv-consistency}, for any {$\bgamma\in\calG_{K_0,1}\cup\calG_{K_0,2}$}, there are universal constants $c_0,c_1,c_2,{c_3}>0$ such that
    \begin{align*}
        {\frac{1}{\sqrt{1-p_*}}\fro{\calP_{\Omega_\perp}\brac{\mXi_\bgamma}}
        \ge\frac18\fro{\mXi_\bgamma}
        \brac{1-\frac{c_3\alpha_{\sf sp}\brac{\mXi_\bgamma}}{\sqrt{NT}}},}
    \end{align*}
    with probability at least  ${1-O\brac{n^{-10}}-c_1\exp\brac{-c_2 n^{-1}NT\log n}}$, where
    \begin{align*}
        \alpha_{\sf sp}\brac{\mXi_\bgamma}:=\frac{\sqrt{NT}\op{\mXi_\bgamma}_{\sf max}}{\fro{\mXi_\bgamma}}.	
    \end{align*}
\end{lemma}
Combining \Cref{lem:Deltagam-fro-lb}, \Cref{lem:Deltagam-fro-ub} and \Cref{lem:cv-Pdeltagam-lb}, {for $\bgamma_1\in\calG_{K_0,1}$ and $\bgamma_2\in\calG_{K_0,2}$ we obtain}
\begin{align*}
    {\frac{\fro{\mXi_{\bgamma_1}}}{\fro{\mXi_{\bgamma_2}}}
    \lesssim\kappa^3\sqrt r\,
    \frac{\gamma_{1,0}/\mu(\bgamma_1)}{\gamma_{2,0}/\mu(\bgamma_2)}
    \lesssim\kappa^3\sqrt r\,\frac{\psi B_1}{\bar\psi B_2}=o(1),}
\end{align*}
{and hence \eqref{eq:cv-cond-1} holds with probability at least $1-O(n^{-10})$. In addition,}
\begin{align*}
    {\fro{\mXi_{\bgamma_2}}\gtrsim
    \frac{\gamma_{2,0}}{\mu(\bgamma_2)}
    \frac{\sigma_{\rm C}^2\sigma_{\rm T}^2\sqrt T}{\kappa\lambda_1}
    \ge B_2K\sigma_{\rm C}\sigma_{\rm T}\sqrt{n\brac{r^2+r\log n}},}
\end{align*}
{where $n=T$ because $T\ge N$. Thus \eqref{eq:cv-cond-2-prime} holds. The proof is completed by taking a union bound over $\calG_{K_0,1}\cup\calG_{K_0,2}$.}

\subsection{Proof of \Cref{thm:U-two-infinty-mcar}}\label{pf-thm:U-two-infinty-mcar}
With slight abuse of notation, we adopt the same notations as in the proof of \Cref{thm:U-two-infinty} and we  mainly follow  the arguments therein, but with more additional terms to treat. 

Let $\bar\bU\bar\bSigma \bar\bU^\top$ be the {rank-$r$ eigendecomposition of the positive-semidefinite matrix} $\overline\bM:=\bM\bQ_\bgamma\bM^\top$, {where $\bar\bSigma$ contains its positive eigenvalues in decreasing order.} Then there exists some  $\bar\bO\in\OO_{r}$ such that $\bar\bU\bar\bO=\bU$.  Notice that we can write $\bE^\natural=p^{-1}\calP_{\Omega}\brac{\bE}+\bDelta_M$ with $\bDelta_M:=p^{-1}\bM\circ \bOmega-\bM=p^{-1}\frakD\brac{\bOmega}\circ \bM$, and {$\wt\bU^\Omega_\bgamma$ can be viewed as the eigenvectors associated with the $r$ largest algebraic eigenvalues of} the following matrix for any $\zeta\in\RR$:
\begin{align}\label{eq:main-decomp-mcar}
    &\underbrace{\bM\bQ_\bgamma\bM^\top}_{\bar\bU\bar\bSigma \bar\bU^\top}+\underbrace{\bM\bQ_\bgamma\bE^{\natural\top}+\bE^\natural\bQ_\bgamma\bM^\top}_{\bN_1}+\underbrace{\bE^\natural\bQ_\bgamma\bE^{\natural\top}-\EE\bE^\natural\bQ_\bgamma\bE^{\natural\top}}_{\bar\bN_2}+\underbrace{\EE\bE^\natural\bQ_\bgamma\bE^{\natural\top}-\zeta\bI_N}_{\bDelta_{\bgamma,\zeta}}. 
\end{align}
Denote $\bN:=\bN_1+\bN_2$ with $\bN_2:=\bar\bN_2+\bDelta_{\bgamma,\zeta}$. Define
\begin{align*}
    \bN_{2,1}&:=p^{-1}\calP_{\Omega}\bbrac{\bE}\bQ_\bgamma\sqbrac{p^{-1}\calP_{\Omega}\bbrac{\bE}}^\top, \quad \bN_{2,2}:=p^{-1}\calP_{\Omega}\bbrac{\bE}\bQ_\bgamma\bDelta_M^{\top},\\
    \bN_{2,3}&:=\bDelta_M\bQ_\bgamma\sqbrac{p^{-1}\calP_{\Omega}\bbrac{\bE}}^\top,\quad \bN_{2,4}:=\bDelta_M\bQ_\bgamma\bDelta_M^{\top}.
\end{align*}
By definition we have $\bar\bN_2=\frakD\brac{\sum_{k=1}^4\bN_{2,k}}$.
Note that for any $i,j\in[N]$ and $\bQ\in\RR^{T\times T}$,
\begin{align*}
    &\EE\sqbrac{p^{-2}\calP_{\Omega}\brac{\bE}\bQ\sqbrac{\calP_{\Omega}\brac{\bE}}^\top }_{i,j}
=p^{-2}\sigma_{{\rm C},i}\sigma_{{\rm C},j}\sum_{t_1,t_2\in[T]}\EE\Omega_{i,t_1}\Omega_{j,t_2}Q_{t_1,t_2}\sqbrac{\bSigma_{\rm T}}_{t_1,t_2}.
\end{align*}
We thus have $\EE\sqbrac{p^{-2}\calP_{\Omega}\brac{\bE}\bQ\sqbrac{\calP_{\Omega}\brac{\bE}}^\top }_{i,j}=0$ for $i\ne j$. Therefore, $\EE\sqbrac{p^{-2}\calP_{\Omega}\brac{\bE}\bQ\sqbrac{\calP_{\Omega}\brac{\bE}}^\top }$ is a diagonal matrix with the $(i,i)$-th entry being
\begin{align*}
    \EE\sqbrac{p^{-2}\calP_{\Omega}\brac{\bE}\bQ\sqbrac{\calP_{\Omega}\brac{\bE}}^\top }_{i,i}
    =p^{-1}\sigma_{{\rm C},i}^2\textsf{Tr}\brac{\bQ\circ \bSigma_{\rm T}}+\sigma_{{\rm C},i}^2\sum_{t_1\ne t_2\in[T]}Q_{t_1,t_2}\sqbrac{\bSigma_{\rm T}}_{t_1,t_2}.
\end{align*}
Moreover, we have
\begin{align*}
    \EE\sqbrac{\bDelta_M\bQ\bDelta_M^\top}_{i,j}
    =p^{-2}\EE\sum_{t\in[T]}\frakD\brac{\Omega_{i,t}}\frakD\brac{\Omega_{j,t}}M_{i,t}M_{j,t}Q_{t,t}.
\end{align*}
Hence we have $\EE\sqbrac{\bDelta_M\bQ\bDelta_M^\top}_{i,j}=0$ for $i\ne j$ and  $\EE\sqbrac{\bDelta_M\bQ\bDelta_M^\top}$ is a diagonal matrix with the $(i,i)$-th entry being
\begin{align*}
    \EE\sqbrac{\bDelta_M\bQ\bDelta_M^\top}_{i,i}&=p^{-1}\brac{1-p}\sum_{t\in[T]}M_{i,t}^2Q_{t,t}.
\end{align*}
We have the following lemma, whose proof can be found in \Cref{pf-lem:high-order-bound-mcar}.
\begin{lemma}\label{lem:high-order-bound-mcar} 
    Instate the conditions of \Cref{thm:U-two-infinty-mcar}. Consider the model defined in \eqref{eq:main-decomp-mcar}, there exists universal constants $C,C_0,C_1$ such that for any ${q_0\ge 1}$, we have with probability at least $1-O\bbrac{{q_0} n^{-10}}$ for any $1\le q\le {q_0}$ that 
    \begin{align*}
        \infn{\bar\bU_{\perp}\bar\bU_{\perp}^\top\brac{\bar\bU_{\perp}\bar\bU_{\perp}^\top\bN_2\bar\bU_{\perp}\bar\bU_{\perp}^\top }^{q-1}\bN\bar\bU}\le C_1\brac{C_0\vartheta_\bgamma }^{q}\infn{\bar\bU}.
    \end{align*}
\end{lemma}
The remaining proof is exactly the same as that of \Cref{thm:U-two-infinty} by replacing  \Cref{lem:high-order-bound} therein with \Cref{lem:high-order-bound-mcar}. We omit the details for brevity.

\section{Technical Lemmas}\label{sec:pf-tlem}
In this section, we collect all necessary technical lemmas for the proofs of results in \Cref{sec:main}-\ref{sec:cv}. All these lemmas are concentration inequalities for various forms of matrix product in either spectral norm or $\ell_2\rightarrow\ell_\infty$ norm.    
\subsection{Lemmas for \Cref{sec:main} and \Cref{sec:refined}}
{\begin{lemma}\label{lem:GB-bound}
    Let $\bG\in \RR^{N\times T}$ be a random matrix with independent   entries such that $\EE G_{ij}=0$, $\EE G^2_{ij}=1$ and $\bop{G_{ij}}_{\psi_2}\le K$, and $\bB$ be a fixed $T\times T_0$ matrix. There exist some universal constants $c_0,C_0,C_1>0$ such that
    \begin{align*}
        \PP\brac{\op{\bG\bB}\ge C_0K\sqrt{N+T\wedge T_0}\op{\bB}}\le C_1e^{-c_0\brac{N+T\wedge T_0}}.
    \end{align*}
\end{lemma}}
\textit{Proof: }See \Cref{pf-lem:GB-bound}.
  {\begin{lemma}\label{lem:EQE-bound}
     Let $\bG\in \RR^{N\times T}$ be a random matrix with independent entries such that $\EE G_{ij}=0$, $\EE G^2_{ij}=1$ and $\bop{G_{ij}}_{\psi_2}\le K$, and $\bE:=\bSigma_{\rm C}^{1/2}\bG\bSigma_{\rm T}^{1/2}$. There exist some universal constants $c_0,C_0,C_1>0$ such that for any $\bQ\in\RR^{T\times T}$, we have
    {\small
\begin{align*}
        \PP\brac{\op{\bE\bQ\bE^\top-\EE\bE\bQ\bE^\top}\ge C_0K^2\sqrt{N}\op{\bSigma_{\rm C}^{1/2}}^2\brac{\fro{\bSigma^{1/2}_{\rm T}\bQ\bSigma^{1/2}_{\rm T}}+\sqrt{N}\op{\bSigma^{1/2}_{\rm T}\bQ\bSigma^{1/2}_{\rm T}}}}\le C_1e^{-c_0N}. 
    \end{align*}
}
\end{lemma}}
\textit{Proof: }See \Cref{pf-lem:EQE-bound}.

{\begin{lemma}\label{lem:GBGU-two-inf-bound}
    Let $\bG\in \RR^{N\times T}$ be a random matrix with independent entries such that $\EE G_{ij}=0$, $\EE G^2_{ij}=1$ and $\bop{G_{ij}}_{\psi_2}\le K$, and $1\le r\le N\wedge T$. There exist some universal constants $C_0,C_1>0$ such that the following conclusions hold:
    \begin{itemize}
        \item [(1)] $	\PP\brac{\infn{ \bB \bG\bV}\ge C_0K\infn{\bV^\top }\infn{ \bB}\sqrt{r\log n} }\le C_1 n ^{-10}$ for any fixed $\bB\in\RR^{N\times N}$, $\bV\in\RR^{T\times r}$. 
        \item [(2)] {\small $\PP\brac{\op{\e_i^\top \bG\bB \bG^\top\bU^{(-i)}-\EE \e_i^\top \bG\bB \bG^\top\bU^{(-i)}}\ge C_0K^2\sqrt{nNr\log n }\op{\bB}\infn{\bU^{(-i)}}\mid \bU^{(-i)}}\le \allowbreak \\C_1 (e^{-c_0 n}+n ^{-11})$}
for any $i\in[N]$, fixed $\bB\in\RR^{T\times T}$ and (possibly) random $\bU^{(-i)}\in\RR^{N\times r}$ independent of $\bG_{i,\cdot}$.
    \end{itemize}
\end{lemma}}
\textit{Proof: }See \Cref{pf-lem:GBGU-two-inf-bound}.

{\begin{lemma}[Multivariate CLT for sub-Gaussian weighted sums]\label{lem:subg_clt_weighted}
Fix $r\in\mathbb N$.  Let $\{g_k\}_{k=1}^{m}$ be independent,
$\mathbb E[g_k]=0$, $\mathbb E[g_k^2]=1$, and $\|g_k\|_{\psi_2}\le K$.
Let $w_{k}\in\mathbb R^{r}$ be deterministic vectors (possibly depending on $m$), and define $\bZ:=\sum_{k=1}^{m} g_k \w_k$, $ 
\bar\bSigma:=\sum_{k=1}^{m} \w_k \w_k^\top$. Assume $\bar\bSigma$ is positive definite and $\delta_m:=\max_{1\le k\le m}\big\|\bar\bSigma^{-1/2}\w_k\big\| \rightarrow 0$ as $m\to\infty$, then
\begin{align*}
\bar\bSigma^{-1/2}\bZ \overset{d}{\rightarrow} \mathcal N(\mathbf0,\bI_r).
\end{align*}
\end{lemma}
\textit{Proof: }See \Cref{pf-lem:subg_clt_weighted}.}

\subsection{Lemmas for \Cref{sec:cv}}
\begin{lemma}\label{lem:GB-bound-mcar}
	     {Let $\bG\in \RR^{N\times T}$ be a random matrix with independent entries such that $\EE G_{ij}=0$, $\EE G^2_{ij}=1$ and $\bop{G_{ij}}_{\psi_2}\le K$,  $\bE:=\bSigma_{\rm C}^{1/2}\bG\bSigma_{\rm T}^{1/2}$,} and $\bOmega\in \RR^{N\times T}$ be independent of $\bE$ with $[\bOmega]_{i,j}\overset{i.i.d.}{\sim }\text{Ber}\brac{p}$ with $p\in[c,1-c]$ for some universal constant $c\in(0,1/2)$, where $\bSigma_{\rm C}$ and $\bSigma_{\rm T}$ satisfy \Cref{assump:digoanl-dominant}. Let $\bM\in\RR^{N\times T}$ be fixed and denote $\bDelta_M:=p^{-1}\frakD\brac{\bOmega}\circ \bM$. There exist some universal constants $c_0,C_0,C_1>0$ such that
\begin{itemize}
            {\item[(1)] $\PP\brac{\op{\calP_{\Omega}\brac{\bE}\bB}\ge C_0\sigma_{\rm C}\sigma_{\rm T}K\op{\bB}\sqrt{\brac{N+T\wedge T_0}\log n}}\le C_1\brac{e^{-c_0\brac{N+T\wedge T_0}}+n^{-10}}$, for any fixed $\bB\in\RR^{T\times T_0}$; }
        \item[(2)] $\PP\brac{\op{\bDelta_M\bB}\ge C_0\op{\bM}_{\sf max}\op{\bB}\sqrt{\brac{N+T\wedge T_0}}}\le C_1e^{-c_0\brac{N+T\wedge T_0}}$, for any fixed $\bB\in\RR^{T\times T_0}$.
    \end{itemize}
    \end{lemma}
    \textit{Proof: }See \Cref{pf-lem:GB-bound-mcar}.
    \begin{lemma}\label{lem:quad-ineq-mcar}
        Instate the conditions of \Cref{lem:GB-bound-mcar} and let $\bQ_\bgamma$ defined as in \Cref{sec:cv}. There exist some universal constants $c_0,C_0,C_1>0$ such that
        \begin{itemize}
            {\item[(1)] {\small $\PP\brac{\op{\calP_{\Omega}\brac{\bE}\bQ_\bgamma\brac{\calP_{\Omega}\brac{\bE}}^\top-\EE\sqbrac{\calP_{\Omega}\brac{\bE}\bQ_\bgamma\brac{\calP_{\Omega}\brac{\bE}}^\top}}\ge   C_0\sigma_{\rm C}^2\sigma_{\rm T}^2K^2\brac{\sqrt{NT\log n}+N}}\allowbreak\allowbreak\le C_1\brac{n^{-10}+e^{-c_0N}}$};
            \item[(2)]  $\PP\brac{\op{\sqbrac{\calP_\Omega\brac{\bE}}\bQ_{\bgamma}\bDelta_M^\top }\ge C_0\sigma_{\rm C}\sigma_{\rm T}K\infn{\bM}\sqrt{N\log n}}\le C_1\brac{n^{-10}+e^{-c_0N}}$;}
            \item[(3)]  $\PP\brac{\op{\bDelta_M\bQ_{\bgamma}\bDelta_M^\top -\EE\bDelta_M\bQ_{\bgamma}\bDelta_M^\top }\ge C_0\op{\bM}_{\sf max}^2\bbrac{\sqrt{NT}+N}}\le C_1e^{-c_0N}$.
            
        \end{itemize}
    \end{lemma}
    \textit{Proof: }See \Cref{pf-lem:quad-ineq-mcar}.
    \begin{lemma}\label{lem:first-order-two-inf-bound-mcar}
        Instate the conditions of \Cref{lem:GB-bound-mcar}. For any $\bV\in\RR^{T\times r}$ with $r\le N\wedge T$,  there exist some universal constants $C_0,C_1>0$ such that
        \begin{itemize}
             {\item[(1)] $\PP\brac{\infn{\calP_{\Omega}\brac{\bE}\bV}\ge C_0\sigma_{\rm C}\sigma_{\rm T}K\infn{\bV^\top }r^{1/2}{\log n} }\le C_1 n ^{-15}$;}
            \item[(2)] $\PP\brac{\infn{\bDelta_M\bV}\ge C_0\infn{\bV}\infn{\bM }(r\log n)^{1/2} }\le C_1 n ^{-15}$.
        \end{itemize}
    \end{lemma}
    \textit{Proof: }See \Cref{pf-lem:first-order-two-inf-bound-mcar}.
    \begin{lemma}\label{lem:GBGU-two-inf-bound-mcar}
        Instate the conditions of \Cref{lem:quad-ineq-mcar}. Let $1\le r\le N\wedge T$.  For any $i\in[N]$ and (possibly) random $\bU^{(-i)}\in\RR^{N\times r}$ independent of $\bE_{i,\cdot}$ and $\bOmega_{i,\cdot }$, there exist some universal constants $C_0,C_1>0$ such that
        \begin{itemize}
            \item[(1)]
            {$\PP\brac{\op{\e_i^\top\frakD\brac{ \calP_{\Omega}\brac{\bE}\bQ_\bgamma \sqbrac{\calP_{\Omega}\brac{\bE}}^\top }\bU^{(-i)}}\ge C_0\sigma_{{\rm C}}^2\sigma_{{\rm T}}^2K^2\sqrt{nNr}\infn{\bU^{(-i)}}\log n \mid \bU^{(-i)}} \allowbreak~~~~~~~\allowbreak\le  C_1 \brac{e^{-c_0n}+n^{-15}}$;
            \item[(2)] $\PP\brac{\op{\e_i^\top \calP_{\Omega}\brac{\bE}\bQ_\bgamma \bDelta_M^\top \bU^{(-i)}}\ge C_0\sigma_{\rm C}\sigma_{\rm T}K\op{\bM}_{\sf max}\infn{\bU^{(-i)}}\sqrt{nNr\log n}\mid \bU^{(-i)} }\allowbreak\le  C_1 \brac{e^{-c_0n}+n^{-15}}$;
            \item[(3)] $\PP\brac{\op{\e_i^\top\bDelta_M\bQ_\bgamma \sqbrac{\calP_{\Omega}\brac{\bE}}^\top \bU^{(-i)}}\ge C_0\sigma_{\rm C}\sigma_{\rm T}K\op{\bM}_{\sf max}\infn{\bU^{(-i)}}\sqrt{nNr}\log n \mid \bU^{(-i)}}\allowbreak\allowbreak\le C_1 \brac{e^{-c_0n}+n^{-15}}$;}
            \item[(4)] $\PP\brac{\op{\e_i^\top\frakD\brac{ \bDelta_M\bQ_\bgamma \bDelta_M^\top }\bU^{(-i)}}\ge C_0\op{\bM}_{\sf max}^2\infn{\bU^{(-i)}}\sqrt{nNr}\log n \mid \bU^{(-i)}}\allowbreak\le C_1 \brac{e^{-c_0n}+n^{-15}}$.
        \end{itemize}
    \end{lemma}
    \textit{Proof: }See \Cref{pf-lem:GBGU-two-inf-bound-mcar}.

    \section{Proofs of Lemmas}\label{sec:pf-pflem}
    
    \subsection{Proof of \Cref{lem:good-event} }\label{pf-lem:V-infn-bound}
    Let $\bU^\sharp:=\bL\bLambda^{-1}$. Notice that the columns of $\bU$ are the left singular vectors of $\bL\bF^\top$ and the columns of $\bU^\sharp$ are the left singular vectors of $\bL$. Since $\bL\bF^\top $ is of rank $r$, we conclude that there exists some $\bO^\sharp\in\OO_{r}$ such that $\bU=\bU^\sharp\bO^\sharp$. 
    Also, we have $\bV=\brac{T^{-1/2}\bL\bF^\top }^\top\bU\bSigma^{-1}=T^{-1/2}\bF\bB$ where $\bB:=\bL^\top \bU\bSigma^{-1}=\bLambda\bU^{\sharp\top}\bU\bSigma^{-1}=\bLambda\bO^\sharp\bSigma^{-1}$. Hence we have $T^{-1/2}\bF=\bV\bB^{-1}$ and $T^{-1}\bF^{\top }\bF=\brac{\bB^{-1}}^\top\bB^{-1} $. We thereby arrive at 
    \begin{align*}
        \op{\brac{\bB^{-1}}^\top\bB^{-1}-\bI_r}=\op{T^{-1}\bF^{\top }\bF-\bI_r}\lesssim \delta_F=o(1),
    \end{align*} 
    with probability at least $1-\eta_F$ by \Cref{assump:iden}. We thus conclude that with probability at least $1-\eta_F$, $\sigma_i\brac{\bB}\asymp 1$ for $i\in[r]$ and hence
    \begin{align*}
        \infn{\bV}=\frac{1}{\sqrt{T}}\infn{\bF\bB}\lesssim \frac{1}{\sqrt{T}}\infn{\bF}.
    \end{align*}
    Notice that $\infn{\bF}\lesssim\sqrt{r+\log n}$ with probability at least $1-O(n^{-10})$ by using the sub-gaussianity of $\f_t$. We can conclude the proof for (1). To see (2), it suffices to apply Theorem A.2 in \cite{braun2006accurate} to have 
    \begin{align*}
        \ab{\lambda_i-\sigma_i\brac{\bSigma}}=\ab{\lambda_i\brac{T^{-1}\bF\bL^\top\bL\bF^\top }-\lambda_i\brac{\bL^\top\bL}}\le \sigma_i^2\brac{\bSigma}\op{T^{-1}\bF^\top \bF-\bI_r}=o\brac{\sigma_i^2\brac{\bSigma}},
    \end{align*}
    with probability at least $1-\eta_F$. {For the additional claim, write
    \[
    \bA_\bgamma:=\gamma_0\bI_r+\sum_{\tau=1}^{K_0}\gamma_\tau\brac{1-\frac{\tau}{T}}\brac{\bGamma_\tau+\bGamma_\tau^\top}.
    \]
    Assumptions \ref{assump:iden} and \ref{assump:factor-ts} imply $\op{T^{-1}\bF^\top\bQ_\bgamma\bF-\bA_\bgamma}
    \le\gamma_0\delta_F+2\bar\delta_F=o\brac{\mu(\bgamma)}$. Weyl's inequality, the positivity and conditioning assumptions on $\bA_\bgamma$, and the singular values of $\bL$ therefore give
    \[
    \lambda_r\brac{\bM\bQ_\bgamma\bM^\top}\gtrsim\lambda_r^2T\mu(\bgamma),\qquad
    \lambda_1\brac{\bM\bQ_\bgamma\bM^\top}\lesssim\lambda_1^2T\mu(\bgamma).
    \]
    A union bound over the finite candidate set gives the stated probability.}

\subsection{Proof of \Cref{lem:V-high-order-bound}}\label{pf-lem:V-high-order-bound}
We use the following column leave-one-out bound, proved in \Cref{pf-lem:V-first-order-bound}.
\begin{lemma}\label{lem:V-first-order-bound}
 Suppose that the signal condition of \Cref{thm:U-error} holds, $T=o(e^{c_0N})$ for a sufficiently small universal constant $c_0>0$, and
 \begin{align}\label{eq:V-loo-separation}
  \lambda_r^2\sigma_r(\bF^\top\bQ\bF)
  \ge C_0K^2N\op{\bSigma_{\rm C}^{1/2}}^2\op{\bSigma_{\rm T}^{1/2}\bQ\bSigma_{\rm T}^{1/2}}_1
 \end{align}
 for a sufficiently large universal constant $C_0>0$. On $\calE_0$, with conditional probability at least $1-O(e^{-cN}+n^{-10})$,
 \begin{align}\label{eq:loo-noise-projector}
 &\infn{\bE^\top(\wt\bU_\bQ\wt\bU_\bQ^\top-\bU\bU^\top)}\lesssim K\op{\bSigma_{\rm C}^{1/2}}\op{\bSigma_{\rm T}^{1/2}}_1\sqrt{r+\log n}\,\xi^U_\bQ\notag\\
 &\qquad+\frac{K^2N\op{\bSigma_{\rm C}^{1/2}}^2\op{\bSigma_{\rm T}^{1/2}\bQ\bSigma_{\rm T}^{1/2}}_1}
{\lambda_r^2\sigma_r(\bF^\top\bQ\bF)}\left(\kappa_{\rm T,1}\lambda_1\sqrt T\infn{\bV}
+K\op{\bSigma_{\rm C}^{1/2}}\op{\bSigma_{\rm T}^{1/2}}_1\sqrt{r+\log n}\right).
 \end{align}
\end{lemma}
Sub-Gaussian concentration with standard epsilon-net argument give, on an event with the same probability,
\begin{align*}
 \infn{\bE^\top\bU}
 \lesssim K\op{\bSigma_{\rm C}^{1/2}}\op{\bSigma_{\rm T}^{1/2}}_1\sqrt{r+\log n},\quad 
 \max\bigl\{\op{\wt\bN},\op{\bE^\top\wt\bU_\bQ\wt\bU_\bQ^\top}\bigr\}
 \lesssim\textsf{Err}_V(\bQ).
\end{align*}
For the second bound, $T\gtrsim\log n$ gives $\op{\bE^\top\bU}\lesssim K\op{\bSigma_{\rm C}^{1/2}}\op{\bSigma_{\rm T}^{1/2}}\sqrt T$. Combine this with $\op{\bE}\lesssim K\op{\bSigma_{\rm C}^{1/2}}\op{\bSigma_{\rm T}^{1/2}}\sqrt n$ and the definition of $\wt\bN$. By \Cref{lem:wtu-eq-hatu}, the two left projectors agree on this event. Since $\xi^U_\bQ\le1$ and the fraction in \eqref{eq:loo-noise-projector} is at most $C_0^{-1}$ by \eqref{eq:V-loo-separation}, its first summand and quadratic-noise term are absorbed by $K\op{\bSigma_{\rm C}^{1/2}}\op{\bSigma_{\rm T}^{1/2}}_1\sqrt{r+\log n}$. Hence
\begin{align*}
 \infn{\bE^\top\wt\bU_\bQ\wt\bU_\bQ^\top}
 &\lesssim K\op{\bSigma_{\rm C}^{1/2}}\op{\bSigma_{\rm T}^{1/2}}_1\sqrt{r+\log n}\\
 &\quad+\frac{K^2\kappa\kappa_{\rm T,1} N\sqrt T
              \op{\bSigma_{\rm C}^{1/2}}^2\op{\bSigma_{\rm T}^{1/2}\bQ\bSigma_{\rm T}^{1/2}}_1}
                   {\lambda_r\sigma_r(\bF^\top\bQ\bF)}\infn{\bV}.
\end{align*}
Also $\bW_1=\bV_\perp\bV_\perp^\top\bE^\top\wt\bU_\bQ\wt\bU_\bQ^\top$, since $\bV_\perp^\top\bM^\top=\mathbf0$. Therefore,
\begin{align*}
 \infn{\bW_1}
 &\le\infn{\bE^\top\wt\bU_\bQ\wt\bU_\bQ^\top}
      +\infn{\bV}\op{\bE^\top\wt\bU_\bQ\wt\bU_\bQ^\top}
 \lesssim\infn{\bV}\overline{\textsf{Err}}_V(\bQ).
\end{align*}
For every $p\ge1$, the remaining factors satisfy $\op{\bW_j}\le\op{\wt\bN}$ and $\op{\bB_{p+1}}=1$. Thus, on the same event,
\begin{align*}
 \infn{\bW_1\bW_2\cdots\bW_p\bB_{p+1}}
 \le\infn{\bW_1}\op{\wt\bN}^{p-1}
 \lesssim\infn{\bV}\overline{\textsf{Err}}_V(\bQ)\op{\wt\bN}^{p-1}.
\end{align*}
This proves both assertions.

    \subsection{Proof of \Cref{lem:wtu-eq-hatu}}\label{pf-lem:wtu-eq-hatu}
	Note that the bounds used in the proof of \Cref{thm:U-error} and Weyl's inequality yield $\lambda_r\brac{\bX\bQ\bX^\top}\ge \lambda_r^2\sigma_r\brac{\bF^\top\bQ\bF}-C\sqbrac{\textsf{Err}_U(\bQ)+\inf_{\zeta\ge0}\op{\textsf{Tr}\brac{\bSigma_{\rm T}\bQ}\bSigma_{\rm C}-\zeta\bI_N}}>0$ with probability at least $1-O\brac{e^{-c_0N}}$. Hence the $r$ eigenvectors selected in \Cref{alg:main} belong to $\text{Col}\bbrac{\bX\bQ\bX^\top}\subseteq\text{Col}\bbrac{\bX}$. Thus,
\begin{align*}
    \text{Col}\bbrac{\wt\bU_\bQ}\subseteq \text{Col}\bbrac{\bX\bQ\bX^\top} \subseteq \text{Col}\bbrac{\bX}
\end{align*}
Moreover, we claim that $\text{rank}\bbrac{\wt\bU_\bQ^\top\bX}=r$. To see this, suppose there exists $\a\in\RR^r\backslash\ebrac{\mathbf{0}}$ such that  $ \a^\top \wt\bU_\bQ^\top\bX=\mathbf{0}^\top$, equivalently, $\bX^\top \b=\mathbf{0}$ where $\b:=\wt\bU_\bQ\a$. This implies that $\b\in\text{Null}\bbrac{\bX^\top}=\text{Col}\bbrac{\bX}^\perp$. Since $\wt\bU_\bQ$ has orthonormal columns, $\a\ne \mathbf{0}$ implies $\b\ne \mathbf{0}$. Also, $\b \in  \text{Col}\bbrac{\wt\bU_\bQ}\subseteq \text{Col}\bbrac{\bX}$. Thus we have $\b\in \text{Col}\bbrac{\bX}\cap \text{Col}\bbrac{\bX}^\perp=\{\mathbf{0}\}$, which is a contradiction.

Now, $\wt\bU_\bQ\wt\bU_\bQ^\top\bX$ has rank $r$ and clearly $\text{Col}\bbrac{\wt\bU_\bQ\wt\bU_\bQ^\top\bX}\subseteq \text{Col}\bbrac{\wt\bU_\bQ}$. Since $\text{dim}\text{Col}\bbrac{\wt\bU_\bQ}=r$, we must have $\text{Col}\bbrac{\wt\bU_\bQ\wt\bU_\bQ^\top\bX}=\text{Col}\bbrac{\wt\bU_\bQ}$, and hence $\wt\bU_\bQ \wt\bU_\bQ ^\top =\hat\bU_\bQ \hat\bU_\bQ ^\top$.

    \subsection{Proof of \Cref{lem:high-order-bound}}\label{pf-lem:high-order-bound}
    We will prove by induction. For notational simplicity, define
    \begin{align}\label{eq:thetas-def}
        \vartheta_{1}&:=K\sqrt{NT} \op{\bSigma}\op{\bSigma_{\rm C}^{1/2}}\op{\bSigma^{1/2}_{\rm T}},\quad \vartheta_{2}:=K^2\sqrt{nN}\op{\bSigma^{1/2}_{\rm T}}^2\op{\bSigma_{\rm C}^{1/2}}^2+\op{\bDelta_{\gamma,\zeta}},\notag\\
        \vartheta&:=C\left(\vartheta_{1}+\vartheta_{2}+K^2\sqrt{nNr\log n}\op{\bSigma^{1/2}_{\rm T}}^2\op{\bSigma_{\rm C}^{1/2}}_{1}^2\right.\notag\\
        &\left.+K\infn{\bU}^{-1}\sqrt{T}\op{\bSigma}\op{\bSigma_{\rm C}^{1/2}}_{1}\op{\bSigma^{1/2}_{\rm T}}r^{1/2}\log n+\op{\bDelta_{\gamma,\zeta}}_1\right).
    \end{align}
    By definition we have $\bar\vartheta =\kappa_{\rm C, 1}\kappa_{\rm C, 2}\vartheta$. Define the events
    \begin{align}\label{eq:hp-event}
        \calE_1:=&\ebrac{\max\ebrac{\op{\bN_1},\op{\bar\bN_2}}\le C_1\brac{\vartheta_{1}+\vartheta_{2}}},\notag\\
        \calE_2:=&\ebrac{\max\ebrac{\op{\bN_1^{(-i)}},\op{\bar\bN^{(-i)}_2}}\le C_2\brac{\vartheta_{1}+\vartheta_{2}},\quad \forall i\in[N]},\notag\\
        \calE_3:=&\ebrac{\op{\frakD\brac{\bG\bSigma_{\rm T}^{1/2}\bQ\bSigma_{\rm T}^{1/2}\bG^{\top}}}\le C_3\vartheta_{2}^* },\notag\\
        \calE_4:=&\ebrac{\infn {\bG\bSigma^{1/2}_{\rm T}\bQ\bM^\top \bar\bU}\le C_4K\sqrt{T}\op{\bSigma}\op{\bSigma^{1/2}_{\rm T}}r^{1/2}\log n },
    \end{align}
    for some sufficiently large universal constants $C_1,C_2,C_3,C_4>0$, where $\vartheta_{2}^*:=K^2\sqrt{nN}\op{\bSigma^{1/2}_{\rm T}\bQ}\op {\bSigma^{1/2}_{\rm T}}$. In addition, for any   $p\ge 2$ let 
    \begin{align*}
        \calE_{0,p}:=&\ebrac{\infn{\bar\bU_{\perp}\bar\bU_{\perp}^\top\brac{\bar\bU_{\perp}\bar\bU_{\perp}^\top\bN_2\bar\bU_{\perp}\bar\bU_{\perp}^\top }^{q-1}\bN\bar\bU}\le C\brac{C_0\bar\vartheta}^{q}\infn{\bar\bU}\text{~for~} 1\le q\le p-1}.
    \end{align*}
    for some sufficiently universal constants $C,C_0>0$. We introduce the following lemmas, whose proofs are deferred to  \Cref{pf-lem:hp-event} and \Cref{pf-lem:first-order-bound}. 
    \begin{lemma}\label{lem:hp-event}
        For $\ebrac{\calE_j}_{j\in[4]}$ defined in \eqref{eq:hp-event},  we have $\PP\brac{\bigcap_{j\in[4]}\calE_j}\ge 1-O\brac{e^{-c_0N}+n^{-20}}$.
    \end{lemma}
    \begin{lemma}\label{lem:first-order-bound} 
        We have with probability at least $1-O\bbrac{e^{-c_0N}+n ^{-10}}$ that 
        \begin{align*}
            \infn{\bar\bU_{\perp}\bar\bU_{\perp}^\top\bN\bar\bU}\le CC_0\vartheta\infn{\bar\bU},
        \end{align*}
        where $\vartheta$ is defined in \eqref{eq:thetas-def} and $C,C_0>0$ are the same constants defined in $\calE_{0,p}$.
    \end{lemma}
    By \Cref{lem:hp-event} and \Cref{lem:first-order-bound}, we have $\PP\brac{\calE_{0,p}}\ge 1-O\bbrac{p \brac{e^{-c_0N}++n ^{-10}}}$ with $p=2$. Suppose that $\PP\brac{\calE_{0,p}}\ge 1-O\bbrac{p \brac{e^{-c_0N}++n ^{-10}}}$ for $p\ge 2$. Notice that
    \begin{align}\label{eq:induction-bound}
        &\infn{\brac{\bar\bU_{\perp}\bar\bU_{\perp}^\top\bN_2\bar\bU_{\perp}\bar\bU_{\perp}^\top }^{p-1}\bN\bar\bU}= \infn{\bar\bU_{\perp}\bar\bU_{\perp}^\top\bN_2\bar\bU_{\perp}\bar\bU_{\perp}^\top\brac{\bar\bU_{\perp}\bar\bU_{\perp}^\top\bN_2\bar\bU_{\perp}\bar\bU_{\perp}^\top }^{p-2}\bN\bar\bU}\notag\\
        &\le \vartheta^{p}\infn{\bar\bU}+\op{\bSigma_{\rm C}^{1/2}}_{1}\infn{\bSigma^{-1/2}_{\rm C}\bN_2\bar\bU_{\perp}\bar\bU_{\perp}^\top\brac{\bar\bU_{\perp}\bar\bU_{\perp}^\top\bN_2\bar\bU_{\perp}\bar\bU_{\perp}^\top }^{p-2} \bN\bar\bU}.
    \end{align}
    Fix $i\in[N]$, we are going to bound the $\ell_2$ norm of  the the $i$-th row of $$\bSigma^{-1/2}_{\rm C}\bN_2 \brac{\bar\bU_{\perp}\bar\bU_{\perp}^\top\bN_2\bar\bU_{\perp}\bar\bU_{\perp}^\top }^{p-2}\bN\bar\bU.$$ Let $\bG^{(-i)}$ be the same as $\bG$ except zeroing the $i$-th row, i.e, $\bG^{(-i)}=\brac{\bI_N-\e_i\e_i^\top }\bG$. Notice that $\EE\bE\bQ\bE^\top=\textsf{Tr}\brac{\bSigma_{\rm T}\bQ}\bSigma_{\rm C}$ and $\EE\bG^{(-i)}\bSigma_{\rm T}^{1/2}\bQ\bSigma_{\rm T}^{1/2}\bG^{(-i)\top}=\textsf{Tr}\brac{\bSigma_{\rm T}\bQ}\brac{\bI_N-\e_i\e_i^\top }$.  Define
    \begin{align*}
        \bN^{(-i)}_1&=\bM\bQ\bSigma_{\rm T}^{1/2}\bG^{(-i)\top}\bSigma_{\rm C}^{1/2}+\bSigma_{\rm C}^{1/2}\bG^{(-i)}\bSigma^{1/2}_{\rm T}\bQ\bM^\top,\\
        \bar\bN^{(-i)}_2&=\bSigma_{\rm C}^{1/2}\bG^{(-i)}\bSigma_{\rm T}^{1/2}\bQ\bSigma_{\rm T}^{1/2}\bG^{(-i)\top}\bSigma_{\rm C}^{1/2}-\textsf{Tr}\brac{\bSigma_{\rm T}\bQ}\bSigma_{\rm C}^{1/2}\brac{\bI_N-\e_i\e_i^\top }\bSigma_{\rm C}^{1/2}.
    \end{align*}
    
    Define $\bN^{(-i)}:=\bN^{(-i)}_1+\bN^{(-i)}_2$ with $\bN^{(-i)}_2:=\bar\bN^{(-i)}_2+\bDelta_{\zeta}$ and $\mPsi^{(-i)}:=\bN-\bN^{(-i)}$. Denote
    \begin{align*}
        \bC:&=\bar\bU_{\perp}\bar\bU_{\perp}^\top\brac{\bar\bU_{\perp}\bar\bU_{\perp}^\top\bN_2\bar\bU_{\perp}\bar\bU_{\perp}^\top }^{p-2}\bN\bar\bU,\\
        \bC^{(-i)}:&=\bar\bU_{\perp}\bar\bU_{\perp}^\top\brac{\bar\bU_{\perp}\bar\bU_{\perp}^\top\bN^{(-i)}_2\bar\bU_{\perp}\bar\bU_{\perp}^\top }^{p-2}\bN^{(-i)}\bar\bU.
    \end{align*}
    Notice that $\bSigma^{-1/2}_{\rm C}\bN_2 \bar\bU_{\perp}\bar\bU_{\perp}^\top\brac{\bar\bU_{\perp}\bar\bU_{\perp}^\top\bN_2\bar\bU_{\perp}\bar\bU_{\perp}^\top }^{p-2}\bN\bar\bU=\bSigma^{-1/2}_{\rm C}\bN_2\bC=\bSigma^{-1/2}_{\rm C}\bar\bN_2\bC+\bSigma^{-1/2}_{\rm C}\bDelta_{\zeta}\bC$. We have	 
    \begin{align*}
        \sqbrac{\bSigma^{-1/2}_{\rm C}\bar\bN_2\bC}_{i,:}&=\e_i^\top\frakD\brac{ \bG\bSigma_{\rm T}^{1/2}\bQ\bSigma_{\rm T}^{1/2}\bG^{\top}}\bSigma_{\rm C}^{1/2}\bC^{(-i)}+\e_i^\top \bSigma^{-1/2}_{\rm C}\bar\bN_2\brac{\bC-\bC^{(-i)}}.
    \end{align*}
    
    \paragraph{Bound for $\op{\e_i^\top \bSigma^{-1/2}_{\rm C}\bar\bN_2\brac{\bC-\bC^{(-i)}}}$.} 
    First we  note that for $p\ge 4$, we have
    \begin{align*}
        &\brac{\bar\bU_{\perp}\bar\bU_{\perp}^\top\bN_2\bar\bU_{\perp}\bar\bU_{\perp}^\top }^{p-2}-\brac{\bar\bU_{\perp}\bar\bU_{\perp}^\top\bN^{(-i)}_2\bar\bU_{\perp}\bar\bU_{\perp}^\top }^{p-2}\\
        &=\sqbrac{\brac{\bar\bU_{\perp}\bar\bU_{\perp}^\top\bN_2\bar\bU_{\perp}\bar\bU_{\perp}^\top }^{p-4}-\brac{\bar\bU_{\perp}\bar\bU_{\perp}^\top\bN^{(-i)}_2\bar\bU_{\perp}\bar\bU_{\perp}^\top }^{p-4}}\brac{\bar\bU_{\perp}\bar\bU_{\perp}^\top\bN_2\bar\bU_{\perp}\bar\bU_{\perp}^\top }^2\\
        &+\brac{\bar\bU_{\perp}\bar\bU_{\perp}^\top\bN^{(-i)}_2\bar\bU_{\perp}\bar\bU_{\perp}^\top }^{p-4}\brac{\bar\bU_{\perp}\bar\bU_{\perp}^\top\mPsi^{(-i)}\bar\bU_{\perp}\bar\bU_{\perp}^\top }\brac{\bar\bU_{\perp}\bar\bU_{\perp}^\top\bN_2\bar\bU_{\perp}\bar\bU_{\perp}^\top }\\
        &+\brac{\bar\bU_{\perp}\bar\bU_{\perp}^\top\bN^{(-i)}_2\bar\bU_{\perp}\bar\bU_{\perp}^\top }^{p-3}\brac{\bar\bU_{\perp}\bar\bU_{\perp}^\top\mPsi^{(-i)}\bar\bU_{\perp}\bar\bU_{\perp}^\top }
    \end{align*}
    where we've used $\bar\bU_{\perp}\bar\bU_{\perp}^\top\mPsi^{(-i)}\bar\bU_{\perp}\bar\bU_{\perp}^\top =\bar\bU_{\perp}\bar\bU_{\perp}^\top\brac{\bN_2-\bN_2^{(-i)}}\bar\bU_{\perp}\bar\bU_{\perp}^\top$.
    Hence we can arrive at 
    \begin{align}\label{eq:A-A(-i)-decomp}
        &\bC-\bC^{(-i)}
        =\sum_{m=1}^{p-2}\bar\bU_{\perp}\bar\bU_{\perp}^\top\brac{\bS^{(-i)} }^{m-1}\bS_{\Psi}^{(-i)}\bS^{p-m-2}\bN\bar\bU+\bar\bU_{\perp}\bar\bU_{\perp}^\top\brac{\bS^{(-i)} }^{p-2}\mPsi^{(-i)}\bar\bU,
    \end{align}
    where $\bS:=\bar\bU_{\perp}\bar\bU_{\perp}^\top\bN_2\bar\bU_{\perp}\bar\bU_{\perp}^\top$, $\bS^{(-i)}:=\bar\bU_{\perp}\bar\bU_{\perp}^\top\bN^{(-i)}_2\bar\bU_{\perp}\bar\bU_{\perp}^\top$, $\bS^{(-i)}_{\Psi}:=\bar\bU_{\perp}\bar\bU_{\perp}^\top\mPsi^{(-i)}\bar\bU_{\perp}\bar\bU_{\perp}^\top$, and the summation is empty if $p=2$. It is readily seen that the decomposition \eqref{eq:A-A(-i)-decomp} still holds for $p=3$. We then get   
    \begin{align}\label{eq:decomp-S-Psi}
        \bS^{(-i)}_{\Psi}&=\bar\bU_{\perp}\bar\bU_{\perp}^\top\mPsi^{(-i)}\bar\bU_{\perp}\bar\bU_{\perp}^\top=\bar\bU_{\perp}\bar\bU_{\perp}^\top\brac{\bar\bN_2-\bar\bN_2^{(-i)}}\bar\bU_{\perp}\bar\bU_{\perp}^\top\notag\\
        &=\bar\bU_{\perp}\bar\bU_{\perp}^\top\bSigma_{\rm C}^{1/2}\e_i\e_i^\top \frakD\brac{\bG\bSigma_{\rm T}^{1/2}\bQ\bSigma_{\rm T}^{1/2}\bG^{\top}}\bSigma_{\rm C}^{1/2}\bar\bU_{\perp}\bar\bU_{\perp}^\top\notag\\
        &+\bar\bU_{\perp}\bar\bU_{\perp}^\top\bSigma_{\rm C}^{1/2}\frakD\brac{\bG\bSigma_{\rm T}^{1/2}\bQ\bSigma_{\rm T}^{1/2}\bG^\top} \e_i\e_i^\top\bSigma_{\rm C}^{1/2}\bar\bU_{\perp}\bar\bU_{\perp}^\top\notag\\
        &-\bar\bU_{\perp}\bar\bU_{\perp}^\top\bSigma_{\rm C}^{1/2}\e_i\e_i^\top\frakD\brac{\bG\bSigma_{\rm T}^{1/2}\bQ\bSigma_{\rm T}^{1/2}\bG^\top} \e_i\e_i^\top\bSigma_{\rm C}^{1/2}\bar\bU_{\perp}\bar\bU_{\perp}^\top.
    \end{align}
    where the last equality holds since $\EE\brac{\bar\bN_2-\bar\bN_2^{(-i)}}=0$.

    We are going to bound $\op{\bar\bU_{\perp}\bar\bU_{\perp}^\top\brac{\bS^{(-i)} }^{m-1}\bS_{\Psi}^{(-i)}\bS^{p-m-2}\bN\bar\bU}$ for $p\ge 3$, $m\ge 1$. For the term related to the  first term in  the decomposition \eqref{eq:decomp-S-Psi}, we have  
    {\footnotesize
\begin{align*}
        &\op{\bar\bU_{\perp}\bar\bU_{\perp}^\top\brac{\bS^{(-i)} }^{m-1}\bar\bU_{\perp}\bar\bU_{\perp}^\top\bSigma_{\rm C}^{1/2}\e_i\e_i^\top \frakD\brac{\bG\bSigma_{\rm T}^{1/2}\bQ\bSigma_{\rm T}^{1/2}\bG^{\top}}\bSigma_{\rm C}^{1/2}\bar\bU_{\perp}\bar\bU_{\perp}^\top\bS^{p-m-2}\bN\bar\bU}\\
        &\le \op{ \brac{\bS^{(-i)} }^{m-1}\bar\bU_{\perp}\bar\bU_{\perp}^\top\bSigma_{\rm C}^{1/2}\e_i}\op {\e_i^\top \bSigma_{\rm C}^{-1/2}\bar\bU\bar\bU^\top  \bSigma_{\rm C}^{1/2}\frakD\brac{\bG\bSigma_{\rm T}^{1/2}\bQ\bSigma_{\rm T}^{1/2}\bG^{\top}}\bSigma_{\rm C}^{1/2}\bar\bU_{\perp}\bar\bU_{\perp}^\top\bS^{p-m-2}\bN\bar\bU}\\
        &+\op{ \brac{\bS^{(-i)} }^{m-1}\bar\bU_{\perp}\bar\bU_{\perp}^\top\bSigma_{\rm C}^{1/2}\e_i}\op {\e_i^\top\bSigma_{\rm C}^{-1/2}\bar\bU_{\perp}\bar\bU_{\perp}^\top \bSigma_{\rm C}^{1/2}\frakD\brac{\bG\bSigma_{\rm T}^{1/2}\bQ\bSigma_{\rm T}^{1/2}\bG^{\top}}\bSigma_{\rm C}^{1/2}\bar\bU_{\perp}\bar\bU_{\perp}^\top\bS^{p-m-2}\bN\bar\bU}\\
        &\le \op{ \brac{\bS^{(-i)} }^{m-1}}\op{\bSigma_{\rm C}^{1/2}\e_i}\op{\e_i^\top\bSigma_{\rm C}^{-1/2}}_{\infty}\left(\infn{\bar\bU}\op{\bSigma_{\rm C}^{1/2}\frakD\brac{\bG\bSigma_{\rm T}^{1/2}\bQ\bSigma_{\rm T}^{1/2}\bG^{\top}}\bSigma_{\rm C}^{1/2}\bar\bU_{\perp}\bar\bU_{\perp}^\top\bS^{p-m-2}\bN\bar\bU}\right.\\
        &+\left.\infn {\bar\bU_{\perp}\bar\bU_{\perp}^\top\bSigma_{\rm C}^{1/2} \frakD\brac{\bG\bSigma_{\rm T}^{1/2}\bQ\bSigma_{\rm T}^{1/2}\bG^{\top}}\bSigma_{\rm C}^{1/2}\bar\bU_{\perp}\bar\bU_{\perp}^\top\bS^{p-m-2}\bN\bar\bU}\right).
    \end{align*}
}
    Notice that $\op{\bSigma_{\rm C}^{1/2}\frakD\brac{\bG\bSigma_{\rm T}^{1/2}\bQ\bSigma_{\rm T}^{1/2}\bG^{\top}}\bSigma_{\rm C}^{1/2}\bar\bU_{\perp}\bar\bU_{\perp}^\top\bS^{p-m-2}\bN\bar\bU}\lesssim \op{\bSigma_{\rm C}^{1/2}}^2\vartheta_{2}^* \vartheta^{p-m-1}\lesssim \vartheta^{p-m}$.
    Moreover, by $\bar\bN_2=\bN_2-\bDelta_{\zeta}$ and induction we have 
    \begin{align*}
        &\infn {\bar\bU_{\perp}\bar\bU_{\perp}^\top\bSigma_{\rm C}^{1/2} \frakD\brac{\bG\bSigma_{\rm T}^{1/2}\bQ\bSigma_{\rm T}^{1/2}\bG^{\top}}\bSigma_{\rm C}^{1/2}\bar\bU_{\perp}\bar\bU_{\perp}^\top\bS^{p-m-2}\bN\bar\bU}\\
        &\le \infn {\bar\bU_{\perp}\bar\bU_{\perp}^\top\bN_2\bar\bU_{\perp}\bar\bU_{\perp}^\top\bS^{p-m-2}\bN\bar\bU}+\infn {\bar\bU_{\perp}\bar\bU_{\perp}^\top\bDelta_{\zeta}\bar\bU_{\perp}\bar\bU_{\perp}^\top\bS^{p-m-2}\bN\bar\bU}\\
        &\le \infn {\bS^{p-m-1}\bN\bar\bU}+\infn {\bDelta_{\zeta}\bS^{p-m-2}\bN\bar\bU}+\infn {\bU\bU^\top \bDelta_{\zeta}\bS^{p-m-2}\bN\bar\bU}\\
        &\le C\brac{C_0\bar\vartheta }^{p-m}\infn{\bar\bU}+\op{\bDelta_{\zeta}}_{1}\infn {\bar\bU_\perp\bar\bU_\perp^\top\bS^{p-m-2}\bN\bar\bU}+\vartheta^{p-m}\infn{\bar\bU}\\
        &\lesssim \brac{C_0\bar\vartheta }^{p-m}\infn{\bar\bU},
    \end{align*}
    where we've used ${\bar\vartheta }^{-1}\op{\bDelta_{\zeta}}_{1}\le 1$ by definition of $\bar\vartheta $ in \eqref{eq:theta-def}. Hence we conclude that
    \begin{align*}
        &\op{ \bar\bU_{\perp}\bar\bU_{\perp}^\top\brac{\bS^{(-i)} }^{m-1}\bar\bU_{\perp}\bar\bU_{\perp}^\top\bSigma_{\rm C}^{1/2}\e_i\e_i^\top \frakD\brac{\bG\bSigma_{\rm T}^{1/2}\bQ\bSigma_{\rm T}^{1/2}\bG^{\top}}\bSigma_{\rm C}^{1/2}\bar\bU_{\perp}\bar\bU_{\perp}^\top\bS^{p-m-2}\bN\bar\bU}\\
        &\lesssim\op{ \brac{\bS^{(-i)} }^{m-1}}\op{\bSigma_{\rm C}^{1/2}\e_i}\op{\e_i^\top \bSigma_{\rm C}^{-1/2}}_{\infty}\brac{C_0\bar\vartheta}^{p-m}\infn{\bar\bU}.
    \end{align*}
    For the term related to the second term in  the decomposition \eqref{eq:decomp-S-Psi}, we have that 
    \begin{align*}
        &\op{\bar\bU_{\perp}\bar\bU_{\perp}^\top\brac{\bS^{(-i)} }^{m-1}\bar\bU_{\perp}\bar\bU_{\perp}^\top\bSigma_{\rm C}^{1/2} \frakD\brac{\bG\bSigma_{\rm T}^{1/2}\bQ\bSigma_{\rm T}^{1/2}\bG^{\top}}\e_i\e_i^\top\bSigma_{\rm C}^{1/2}\bar\bU_{\perp}\bar\bU_{\perp}^\top\bS^{p-m-2}\bN\bar\bU}\\
        &\le  \op{ \brac{\bS^{(-i)} }^{m-1}}\op {\bSigma_{\rm C}^{1/2} \frakD\brac{\bG\bSigma_{\rm T}^{1/2}\bQ\bSigma_{\rm T}^{1/2}\bG^{\top}}}\op {\e_i^\top\bSigma_{\rm C}^{1/2}}_{\infty}\infn{\bar\bU_{\perp}\bar\bU_{\perp}^\top\bS^{p-m-2}\bN\bar\bU}\\
        &\lesssim  \op{ \brac{\bS^{(-i)} }^{m-1}}\op {\e_i^\top\bSigma_{\rm C}^{1/2}}_{\infty}\brac{\op{\bSigma_{\rm C}^{1/2}}\vartheta_{2}^* }\brac{C_0\bar\vartheta }^{p-m-1}\infn{\bar\bU}.
    \end{align*}
    For the term related to the last term in  the decomposition \eqref{eq:decomp-S-Psi}, we have that 
    \begin{align*}
        &\op{\bar\bU_{\perp}\bar\bU_{\perp}^\top \brac{\bS^{(-i)} }^{m-1}\bar\bU_{\perp}\bar\bU_{\perp}^\top\bSigma_{\rm C}^{1/2}\e_i\e_i^\top\frakD\brac{\bG\bSigma_{\rm T}^{1/2}\bQ\bSigma_{\rm T}^{1/2}\bG^\top} \e_i\e_i^\top\bSigma_{\rm C}^{1/2}\bar\bU_{\perp}\bar\bU_{\perp}^\top\bS^{p-m-2}\bN\bar\bU}\\
        &\le \op{ \brac{\bS^{(-i)} }^{m-1}\bar\bU_{\perp}\bar\bU_{\perp}^\top\bSigma_{\rm C}^{1/2}\e_i}\op {\e_i^\top\frakD\brac{\bG\bSigma_{\rm T}^{1/2}\bQ\bSigma_{\rm T}^{1/2}\bG^\top} \e_i}\op {\e_i^\top\bSigma_{\rm C}^{1/2}\bar\bU_{\perp}\bar\bU_{\perp}^\top\bS^{p-m-2}\bN\bar\bU}\\
        &\lesssim  \op{ \brac{\bS^{(-i)} }^{m-1}}\op {\e_i^\top\bSigma_{\rm C}^{1/2}}_{\infty}\brac{\op{\bSigma_{\rm C}^{1/2}\e_i}\vartheta_{2}^* }\brac{C_0\bar\vartheta }^{p-m-1}\infn{\bar\bU}.
    \end{align*}
    Collecting all pieces, we arrive at the following bound for {\small $\op{\bar\bU_{\perp}\bar\bU_{\perp}^\top \brac{\bS^{(-i)} }^{m-1}\bS_{\Psi}^{(-i)}\bS^{p-m-2}\bN\bar\bU}$}:
    \begin{align}\label{eq:sum-term-bound}
        &\op{\bar\bU_{\perp}\bar\bU_{\perp}^\top \brac{\bS^{(-i)} }^{m-1}\bS_{\Psi}^{(-i)}\bS^{p-m-2}\bN\bar\bU}\notag\\
        &\lesssim\brac{C_0\bar\vartheta }^{p-m}\infn{\bar\bU}\op{ \brac{\bS^{(-i)} }^{m-1}}\brac{\op{\e_i^\top \bSigma_{\rm C}^{-1/2}}_{\infty}\op{\bSigma_{\rm C}^{1/2}\e_i}+\bar\vartheta^{-1}\vartheta_{2}^* \op {\e_i^\top\bSigma_{\rm C}^{1/2}}_{\infty}\op{\bSigma_{\rm C}^{1/2}}}.
    \end{align}
    Next, we are going to  bound the term  $\op{\bar\bU_{\perp}\bar\bU_{\perp}^\top\brac{\bS^{(-i)} }^{p-2}\mPsi^{(-i)}\bar\bU}$.  Notice that
    \begin{align}\label{eq:boundary-term-decomp}
        &\op{\bar\bU_{\perp}\bar\bU_{\perp}^\top\brac{\bS^{(-i)} }^{p-2}\mPsi^{(-i)}\bar\bU}\notag\\
        &\le \op{\bar\bU_{\perp}\bar\bU_{\perp}^\top\brac{\bS^{(-i)} }^{p-2}\brac{\bN_1-\bN_1^{(-i)}}\bar\bU}+\op{\bar\bU_{\perp}\bar\bU_{\perp}^\top\brac{\bS^{(-i)} }^{p-2}\brac{\bar\bN_2-\bar\bN_2^{(-i)}}\bar\bU}.
    \end{align}
    The second term of \eqref{eq:boundary-term-decomp} can be bounded in the same way as 
\[
\op{ \bar\bU_{\perp}\bar\bU_{\perp}^\top\brac{\bS^{(-i)} }^{m-1}\bS_{\Psi}^{(-i)}\bS^{p-m-2}\bN\bar\bU}
\]
 in view of \eqref{eq:decomp-S-Psi}, and a same bound as in \eqref{eq:sum-term-bound}  can be obtained. We only bound the first term of \eqref{eq:boundary-term-decomp}. Notice that 
    \begin{align}\label{eq:boundary-term-bound}
        &\op{\bar\bU_{\perp}\bar\bU_{\perp}^\top\brac{\bS^{(-i)} }^{p-2}\brac{\bN_1-\bN_1^{(-i)}}\bar\bU}\le \op{\brac{\bS^{(-i)} }^{p-2}\bSigma_{\rm C}^{1/2}\e_i\e_i^\top \bG\bSigma^{1/2}_{\rm T}\bQ\bM^\top\bar\bU}\notag\\
        &\le \op{\brac{\bS^{(-i)} }^{p-2}}\op {\bSigma_{\rm C}^{1/2}\e_i}\infn {\bG\bSigma^{1/2}_{\rm T}\bQ\bM^\top \bar\bU}\notag\\
        &\lesssim\vartheta^{p-2}K\sqrt{T}\op{\bSigma}\op {\bSigma_{\rm C}^{1/2}\e_i}\op{\bSigma^{1/2}_{\rm T}\bQ}r^{1/2}\log n .
    \end{align}
    where we've used the bound in $\calE_4$ in the last inequality. We then  bound $\op{\e_i^\top \bSigma^{-1/2}_{\rm C}\bar\bN_2\brac{\bC-\bC^{(-i)}}}$ on $\calE_{0,p}\bigcap\ebrac{\bigcap_{j=1}^4\calE_j}$: 
    {\footnotesize
\begin{align}\label{eq:loo-remain-bound}
        &\op{\e_i^\top \bSigma^{-1/2}_{\rm C}\bar\bN_2\brac{\bC-\bC^{(-i)}}}\notag\\
        &\overset{\eqref{eq:A-A(-i)-decomp}}{\le} \sum_{m=1}^{p-2}\op{\e_i^\top \bSigma^{-1/2}_{\rm C}\bar\bN_2\bar\bU_{\perp}\bar\bU_{\perp}^\top\brac{\bS^{(-i)} }^{m-1}\bS_{\Psi}^{(-i)}\bS^{p-m-2}\bN\bar\bU}+\op{\e_i^\top \bSigma^{-1/2}_{\rm C}\bar\bN_2\bar\bU_{\perp}\bar\bU_{\perp}^\top\brac{\bS^{(-i)} }^{p-2}\mPsi^{(-i)}\bar\bU}\notag\\
        &\overset{\eqref{eq:sum-term-bound},\eqref{eq:boundary-term-bound}}{\lesssim} \sum_{m=1}^p\brac{C_0\bar\vartheta }^{p-m}\vartheta^{m}\infn{\bar\bU}\op{\bSigma^{-1/2}_{\rm C}\e_i}\brac{\op{\e_i^\top \bSigma_{\rm C}^{-1/2}}_{\infty}\op{\bSigma_{\rm C}^{1/2}\e_i}+\bar \vartheta^{-1}\vartheta_{2}^* \op {\e_i^\top\bSigma_{\rm C}^{1/2}}_{\infty}\op{\bSigma_{\rm C}^{1/2}}}\notag\\
        &+\vartheta^{p-1}K\op{\bSigma^{-1/2}_{\rm C}\e_i}\op {\bSigma_{\rm C}^{1/2}\e_i}\op{\bSigma^{1/2}_{\rm T}\bQ}\sqrt{T}\op{\bSigma}r^{1/2}\log n \notag\\
        &\le  \brac{C_0\bar\vartheta }^{p}\bar\vartheta ^{-1}\vartheta\infn{\bar\bU}\op{\bSigma^{-1/2}_{\rm C}\e_i}\Bigg[\op{\e_i^\top \bSigma_{\rm C}^{-1/2}}_{\infty}\op{\bSigma_{\rm C}^{1/2}\e_i}+\bar\vartheta ^{-1}\vartheta_{2}^* \op {\e_i^\top\bSigma_{\rm C}^{1/2}}_{\infty}\op{\bSigma_{\rm C}^{1/2}}\Bigg]\notag\\
        &+\vartheta^{p-1}K\op{\bSigma^{-1/2}_{\rm C}\e_i}\op {\bSigma_{\rm C}^{1/2}\e_i}\op{\bSigma^{1/2}_{\rm T}\bQ}\sqrt{T}\op{\bSigma}r^{1/2}\log n.
    \end{align}
}
    where the last inequality holds due to $\bar\vartheta ^{-1}\vartheta=\brac{\kappa_{\rm C, 1}\kappa_{\rm C, 2}}^{-1}<1$ and $C_0>0$ sufficiently large.
    \paragraph{Bound for $\op{\e_i^\top\frakD\brac{ \bG\bSigma_{\rm T}^{1/2}\bQ\bSigma_{\rm T}^{1/2}\bG^{\top}}\bSigma_{\rm C}^{1/2}\bC^{(-i)}}$.} 
    We first upper bound $\infn{\bSigma_{\rm C}^{1/2}\bC^{(-i)}}$. Using the expansion of $\bC-\bC^{(-i)}$ in \eqref{eq:A-A(-i)-decomp} and bound for each term obtained before, we have
    \begin{align}
        &\op{\bC-\bC^{(-i)}}\notag\\
        &\overset{\eqref{eq:A-A(-i)-decomp}}{\le} \sum_{m=1}^{p-2}\op{\bar\bU_{\perp}\bar\bU_{\perp}^\top\brac{\bS^{(-i)} }^{m-1}\bS_{\Psi}^{(-i)}\bS^{p-m-2}\bN\bar\bU}+\op{\bar\bU_{\perp}\bar\bU_{\perp}^\top\brac{\bS^{(-i)} }^{p-2}\mPsi^{(-i)}\bar\bU}\notag\\
        &\overset{\eqref{eq:sum-term-bound},\eqref{eq:boundary-term-bound}}{\lesssim} \sum_{m=1}^p\brac{C_0\bar\vartheta }^{p-m}\vartheta^{m-1}\infn{\bar\bU}\brac{\op{\e_i^\top \bSigma_{\rm C}^{-1/2}}_{\infty}\op{\bSigma_{\rm C}^{1/2}\e_i}+\bar\vartheta^{-1}\vartheta_{2}^* \op {\e_i^\top\bSigma_{\rm C}^{1/2}}_{\infty}\op{\bSigma_{\rm C}^{1/2}}}\notag\\
        &+\vartheta^{p-2}K\op {\bSigma_{\rm C}^{1/2}\e_i}\op{\bSigma^{1/2}_{\rm T}\bQ}\sqrt{T}\op{\bSigma}r^{1/2}\log n \notag\\
        &\le  \brac{C_0\bar\vartheta }^{p-1}\bar\vartheta ^{-1}\vartheta\infn{\bar\bU}\Bigg[\op{\e_i^\top \bSigma_{\rm C}^{-1/2}}_{\infty}\op{\bSigma_{\rm C}^{1/2}\e_i}+\bar\vartheta ^{-1}\vartheta_{2}^* \op {\e_i^\top\bSigma_{\rm C}^{1/2}}_{\infty}\op{\bSigma_{\rm C}^{1/2}}\Bigg]\notag\\
        &+\vartheta^{p-2}K\op {\bSigma_{\rm C}^{1/2}\e_i}\op{\bSigma^{1/2}_{\rm T}\bQ}\sqrt{T}\op{\bSigma}r^{1/2}\log n\notag\\
        &\le  \brac{C_0\bar\vartheta }^{p-1}\infn{\bar\bU}+\vartheta^{p-2}K\op {\bSigma_{\rm C}^{1/2}\e_i}\op{\bSigma^{1/2}_{\rm T}\bQ}\sqrt{T}\op{\bSigma}r^{1/2}\log n.
    \end{align}
    where we've used \eqref{eq:kappa-def}, \eqref{eq:thetas-def} and the fact that $\op{\bSigma_{\rm C}^{1/2}}_1/\op{\bSigma_{\rm C}^{1/2}}\le
    \kappa_{\rm C, 1}$.  Notice that  the definition of $\vartheta$ implies  
    \begin{align*}
        K\op {\bSigma_{\rm C}^{1/2}\e_i}\op{\bSigma^{1/2}_{\rm T}\bQ}\sqrt{T}\op{\bSigma}r^{1/2}\log n\lesssim \vartheta\infn{\bar\bU},
    \end{align*}
    We conclude that $\op{\bC-\bC^{(-i)}}\lesssim \brac{C_0\bar\vartheta }^{p-1}\infn{\bar\bU}$. Hence on $\calE_{0,p}\bigcap\ebrac{\bigcap_{j=1}^4\calE_j}$, we get  
    \begin{align*}
        \infn{\bSigma_{\rm C}^{1/2}\bC^{(-i)}}\le\infn{\bSigma_{\rm C}^{1/2}\bC}+\infn{\bSigma_{\rm C}^{1/2}\bbrac{\bC-\bC^{(-i)}}}\lesssim \op{\bSigma_{\rm C}^{1/2}}_1\brac{C_0\bar\vartheta }^{p-1}\infn{\bar\bU}.
    \end{align*}
    By claim (2) in \Cref{lem:GBGU-two-inf-bound}, we  obtain that with probability at least $1-O\brac{e^{-c_0n}+n^{-11}}$, 
    {\small
\begin{align}\label{eq:loo-term-bound}
        \op{\e_i^\top \frakD\brac{ \bG\bSigma_{\rm T}^{1/2}\bQ\bSigma_{\rm T}^{1/2}\bG^{\top}}\bSigma_{\rm C}^{1/2}\bC^{(-i)}}&\lesssim K^2\infn{\bSigma_{\rm C}^{1/2}\bC^{(-i)}}\sqrt{nNr\log n}\op{\bSigma_{\rm T}^{1/2}\bQ\bSigma_{\rm T}^{1/2}}\notag \\
        &\lesssim\brac{C_0\bar\vartheta }^{p-1}K^2\infn{\bar\bU}\sqrt{nNr\log n}\op{\bSigma_{\rm C}^{1/2}}_1\op{\bSigma_{\rm T}^{1/2}}^2\op{\bQ}.
    \end{align}
}
    \paragraph{Bound for $\infn{\brac{\bar\bU_{\perp}\bar\bU_{\perp}^\top\bN_2\bar\bU_{\perp}\bar\bU_{\perp}^\top }^{p-1}\bN\bar\bU}$.}
    By \eqref{eq:induction-bound}, \eqref{eq:loo-remain-bound} and \eqref{eq:loo-term-bound}, on $\calE_{0,p}\bigcap\ebrac{\bigcap_{j=1}^4\calE_j}$ and a union bound over $i\in[N]$,  we can obtain that with probability at least $1-O\brac{p n ^{-10}}$,
    \begin{align*}
        &\infn{\brac{\bar\bU_{\perp}\bar\bU_{\perp}^\top\bN_2\bar\bU_{\perp}\bar\bU_{\perp}^\top }^{p-1}\bN\bar\bU}\\
        &\le  \vartheta^{p}\infn{\bar\bU}+\op{\bSigma_{\rm C}^{1/2}}_{1}\max_{i\in[N]}\op{\e_i^\top \bSigma^{-1/2}_{\rm C}\bN_2\brac{\bar\bU_{\perp}\bar\bU_{\perp}^\top\bN_2\bar\bU_{\perp}\bar\bU_{\perp}^\top }^{p-2} \bN\bar\bU}\\
        &\lesssim  \brac{C_0\bar\vartheta }^{p-1}K^2\infn{\bar\bU}\op{\bSigma_{\rm C}^{1/2}}_{1}^2\sqrt{nNr\log n}\op{\bSigma^{1/2}_{\rm T}}^2\op{\bQ}+\brac{C_0\bar\vartheta }^{p}\infn{\bar\bU}\\
        &+\kappa_{\rm C, 2}\vartheta^{p-1}K\sqrt{T}\op{\bSigma}\op{\bSigma_{\rm C}^{1/2}}_{1}\op{\bSigma^{1/2}_{\rm T}\bQ}r^{1/2}\log n +\kappa_{\rm C, 1}\brac{C_0\bar \vartheta}^{p-1}\infn{\bar\bU}\op{\bDelta_{\zeta}}_1.
    \end{align*}
    where we've used \eqref{eq:kappa-def}, \eqref{eq:thetas-def} and the fact that $\op{\bSigma_{\rm C}^{1/2}}_1/\op{\bSigma_{\rm C}^{1/2}}\le\kappa_{\rm C, 1}$. Using the definitions of $\vartheta$ and $\bar\vartheta$, we obtain  
    \begin{align*}
        &K^2\sqrt{nNr\log n}\op{\bSigma_{\rm C}^{1/2}}_{1}^2\op{\bSigma^{1/2}_{\rm T}}^2 \lesssim \vartheta,\quad \kappa_{\rm C, 2}K\sqrt{T}\op{\bSigma}\op{\bSigma_{\rm C}^{1/2}}_{1}\op{\bSigma^{1/2}_{\rm T}\bQ}r^{1/2}\log n\lesssim\bar\vartheta\infn{\bar\bU}.
    \end{align*}
    We can thus conclude that  there exists some universal constant $C^\prime>0$ such that
    \begin{align*}
        \PP\brac{\infn{\brac{\bar\bU_{\perp}\bar\bU_{\perp}^\top\bN_2\bar\bU_{\perp}\bar\bU_{\perp}^\top }^{p-1}\bN\bar\bU}\ge C^\prime\brac{C_0\bar\vartheta }^{p}\infn{\bar\bU}}\lesssim\brac{p+1} \brac{e^{-c_0N}+n ^{-10}}.
    \end{align*}
    Therefore,  $\PP\brac{\calE_{0,p+1}}\ge 1-O\brac{\brac{p+1} \brac{e^{-c_0N}+n ^{-10}}}$. The proof is completed by choosing sufficiently large $C_0$ such that $C^\prime/C$ is absorbed into $C_0$.

\subsection{Proof of \Cref{lem:Deltagam-fro-lb}}\label{pf-lem:Deltagam-fro-lb}
Recall the decomposition \eqref{eq:main-decomp-mcar}:
\begin{align*}
    &\underbrace{\bM\bQ_\bgamma\bM^\top}_{\bar\bU\bar\bSigma \bar\bU^\top}+\underbrace{\bM\bQ_\bgamma\bE^{\natural\top}+\bE^\natural\bQ_\bgamma\bM^\top}_{\bN_1}+\underbrace{\bE^\natural\bQ_\bgamma\bE^{\natural\top}-\EE\bE^\natural\bQ_\bgamma\bE^{\natural\top}}_{\bar\bN_2}+\underbrace{\EE\bE^\natural\bQ_\bgamma\bE^{\natural\top}-\zeta\bI_N}_{\bDelta_{\bgamma,\zeta}}. 
\end{align*}
{where $\bE^\natural=p^{-1}\calP_{\Omega}\brac{\bE}+p^{-1}\frakD\brac{\bOmega}\circ \bM$. Let $\bD_M:=\operatorname{diag}\brac{\op{\bM_{1,\cdot}}^2,\ldots,\op{\bM_{N,\cdot}}^2}$. Since $\bSigma_{\rm T}$ is diagonal and every diagonal entry of $\bQ_\bgamma$ equals $\gamma_0$, direct calculation gives
\begin{align*}
    \EE\bE^\natural\bQ_\bgamma\bE^{\natural\top}
    =\frac{\gamma_0}{p}\bbrac{\textsf{Tr}\brac{\bSigma_{\rm T}}\bSigma_{\rm C}+(1-p)\bD_M}.
\end{align*}
Choose $\zeta$ as the midpoint of the largest and smallest diagonal entries of this matrix. Because scalar shifts vanish between $\bU$ and $\bU_\perp$,
\begin{align*}
    \fro{\bU^\top\bDelta_{\bgamma,\zeta}\bU_\perp}
    \ge\op{\bU^\top\bDelta_{\bgamma,\zeta}\bU_\perp}\ge\frac{\gamma_0}{p}\bbrac{\textsf{Tr}\brac{\bSigma_{\rm T}}\rho-(1-p)\op{\bD_M}}
    \gtrsim\gamma_0T\sigma_{\rm C}^2\sigma_{\rm T}^2.
\end{align*}
Here $\op{\bD_M}\le T\lambda_1^2\infn{\bU}^2\lesssim T\lambda_1^2r/N=o\brac{T\sigma_{\rm C}^2\sigma_{\rm T}^2}$. The same choice also gives $\inf_{\zeta\in\RR}\op{\bDelta_{\bgamma,\zeta}}_1\lesssim\gamma_0T\sigma_{\rm C}^2\sigma_{\rm T}^2$.}
{Moreover,
\begin{align*}
\op{\bE^\natural}&\lesssim
\brac{\sigma_{\rm C}\sigma_{\rm T}+\op{\bM}_{\sf max}}\sqrt{T\log n},\qquad \sigma_r\brac{\bM}&\gtrsim\lambda_r\sqrt T
\gtrsim\brac{\sigma_{\rm C}\sigma_{\rm T}+\op{\bM}_{\sf max}}\sqrt{T\log n}.
\end{align*}}
Therefore, we have
\begin{align}\label{eq:Xi-gamma-lb-decomp}
    \fro{\mXi_\bgamma }
    &\ge \fro{\brac{\wt\bU^\Omega_\bgamma\wt\bU^{\Omega\top}_\bgamma-\bU\bU^\top}\bM}
    -\fro{\brac{\wt\bU^\Omega_\bgamma\wt\bU^{\Omega\top}_\bgamma-\bU\bU^\top}\bE^\natural}\notag\\
    &\ge\brac{\frac{\sigma_r\brac{\bM}}{\sqrt2}-\op{\bE^\natural}}
    \fro{\wt\bU^\Omega_\bgamma\wt\bU^{\Omega\top}_\bgamma-\bU\bU^\top}
    \gtrsim\sigma_r\brac{\bM}
    \fro{\wt\bU^\Omega_\bgamma\wt\bU^{\Omega\top}_\bgamma-\bU\bU^\top}.
\end{align} 
{Here we used $\fro{\brac{\wt\bU^\Omega_\bgamma\wt\bU^{\Omega\top}_\bgamma-\bU\bU^\top}\bU}=2^{-1/2}\fro{\wt\bU^\Omega_\bgamma\wt\bU^{\Omega\top}_\bgamma-\bU\bU^\top}$ and $\sigma_r(\bM)\gtrsim\lambda_r\sqrt T\gg\op{\bE^\natural}$.} Moreover, we have
\begin{align*}
    \fro{\calS_{\overline\bM,1}\brac{\bN}}\ge   \fro{\calS_{\overline\bM,1}\brac{\bDelta_{\bgamma,\zeta}}}-\fro{\calS_{\overline\bM,1}\brac{\bN_1}}-\fro{\calS_{\overline\bM,1}\brac{\bar\bN_2}}.
\end{align*}
Notice that $\calS_{\overline\bM,1}\brac{\bDelta_{\bgamma,\zeta}}=\bar\bU\bar\bSigma^{-1}\bar\bU^\top\bDelta_{\bgamma,\zeta}\bar\bU_{\perp}\bar\bU_{\perp}^\top+\bar\bU_{\perp}\bar\bU_{\perp}^\top \bDelta_{\bgamma,\zeta}\bar\bU\bar\bSigma^{-1}\bar\bU^\top$, and we have 
\begin{align*}
    \fro{\calS_{\overline\bM,1}\brac{\bDelta_{\bgamma,\zeta}}}^2&=\fro{\bar\bU\bar\bSigma^{-1}\bar\bU^\top\bDelta_{\bgamma,\zeta}\bar\bU_{\perp}\bar\bU_{\perp}^\top}^2+\fro{\bar\bU_{\perp}\bar\bU_{\perp}^\top \bDelta_{\bgamma,\zeta}\bar\bU\bar\bSigma^{-1}\bar\bU^\top}^2\\
    &+2\inp{\bar\bU\bar\bSigma^{-1}\bar\bU^\top\bDelta_{\bgamma,\zeta}\bar\bU_{\perp}\bar\bU_{\perp}^\top}{\bar\bU_{\perp}\bar\bU_{\perp}^\top \bDelta_{\bgamma,\zeta}\bar\bU\bar\bSigma^{-1}\bar\bU^\top}\\
    &=2\fro{\bar\bSigma^{-1}\bar\bU^\top\bDelta_{\bgamma,\zeta}\bar\bU_{\perp}}^2\ge \frac{2\fro{\bar\bU^\top\bDelta_{\bgamma,\zeta}\bar\bU_{\perp}}^2}{\op{\bar\bSigma}^2}.
\end{align*}
We thus obtain that 
\begin{align*}
    \fro{\calS_{\overline\bM,1}\brac{\bDelta_{\bgamma,\zeta}} }\gtrsim\frac{\fro{\bar\bU^\top\bDelta_{\bgamma,\zeta}\bar\bU_{\perp}}}{\op{\bar\bSigma}}\gtrsim \frac{\gamma_0 T\sigma_{\rm C}^2\sigma_{\rm T}^2}{\op{\bM\bQ_\bgamma\bM^\top }},
\end{align*}
On the other hand, we have with probability at least $1-O\bbrac{n^{-10}}$,
\begin{align*}
    \fro{\calS_{\overline\bM,1}\brac{\bN_1}}&\le \fro{\bar\bU\bar\bSigma^{-1}\bar\bU^\top\bM\bQ_\bgamma\bE^{\natural\top }\bar\bU_{\perp}\bar\bU_{\perp}^\top}+\fro{\bar\bU_{\perp}\bar\bU_{\perp}^\top \bE^{\natural}\bQ_\bgamma\bM^\top \bar\bU\bar\bSigma^{-1}\bar\bU^\top}\\
    &\le 2\sqrt{r}\brac{p^{-1}\op{\bar\bU_{\perp}^\top \calP_{\Omega}\brac{\bE}\bQ_\bgamma\bM^\top \bar\bU\bar\bSigma^{-1}}+\op{\bar\bU_{\perp}^\top \bDelta_M\bQ_\bgamma\bM^\top \bar\bU\bar\bSigma^{-1}}}\\
    &\lesssim \frac{\sqrt{NTr}\lambda_1\brac{K\sigma_{\rm C}\sigma_{\rm T}\sqrt{\log n}\vee \op{\bM}_{\sf max}}}{\sigma_r\brac{\bM\bQ_\bgamma\bM^\top}}.
\end{align*}
where in the last inequality we've used \Cref{lem:GB-bound-mcar}. Moreover, we have
\begin{align*}
    \fro{\calS_{\overline\bM,1}\brac{\bar\bN_2}}&\le2\sqrt{r} \op{\bar\bU_{\perp}\bar\bU_{\perp}^\top \frakD\brac{{\sum_{k=1}^4\bN_{2,k}}}\bar\bU\bar\bSigma^{-1}\bar\bU^\top}\\
    &\lesssim \frac{\sqrt{NTr\log n}\brac{K\sigma_{\rm C}\sigma_{\rm T}\vee \op{\bM}_{\sf max}}^2}{\sigma_r\brac{\bM\bQ_\bgamma\bM^\top}}.
\end{align*}
where we've used \Cref{lem:quad-ineq-mcar}. Hence we conclude that with probability at least $1-O\bbrac{n^{-10}}$, 
\begin{align}\label{eq:first-order-lb}
    \fro{\calS_{\overline\bM,1}\brac{\bN}}\gtrsim \frac{\gamma_0 T\sigma_{\rm C}^2\sigma_{\rm T}^2}{\op{\bM\bQ_\bgamma\bM^\top }},
\end{align}
{Indeed, using $\bar\kappa_\bgamma=O(\kappa^2)$, the ratio between the displayed bound for $\bN_1$ and the deterministic term is at most
\[
C\bar\kappa_\bgamma\frac{K\lambda_1}{\sigma_{\rm C}\sigma_{\rm T}}
\sqrt{\frac{Nr\log n}{T}}\frac{1}{\gamma_0}
\lesssim\frac{\kappa^2\sqrt{r\log n}}
{(r+\log n)(B_1\wedge B_2)}=o(1).
\]
The corresponding ratio for $\bar\bN_2$ is smaller because $\lambda_1\ge\lambda_r$ and the stated SNR condition holds. Thus the deterministic term dominates for every candidate under consideration.}
It remains to show that $\sum_{p\ge 2}\fro{\calS_{\overline\bM,p}\brac{\bN}}$ can be bounded by $\frac{1}{2}\fro{\calS_{\overline\bM,1}\brac{\bN}}$.
Notice that 
\begin{align*}
    \op{\bN}\le  \op{\bN_1}+\op{\bar\bN_2}+\op{\bDelta_{\bgamma,\zeta}}\lesssim\vartheta_\bgamma,
\end{align*}
By the condition that {$\lambda_r\brac{\bM\bQ_\bgamma\bM^\top}\ge C\bar\kappa_\bgamma\sqrt r\,\vartheta_\bgamma$} for some large universal constant $C>0$, we conclude that $\fro{\wt\bU^\Omega_\bgamma\wt\bU^{\Omega\top}_\bgamma-\bU\bU^\top }\ge \fro{\calS_{\overline\bM,1}\brac{\bN}}-\sum_{p\ge 2}\fro{\calS_{\overline\bM,p}\brac{\bN}}$, and
\begin{align*}
    \sum_{p\ge 2}\fro{\calS_{\overline\bM,p}\brac{\bN}}
    &\le \sqrt{r}\sum_{p\ge 2}\brac{\frac{4\op{\bN}}{\lambda_r\brac{\bM\bQ_\bgamma\bM^\top}}}^p
    \lesssim\frac{\sqrt r\,\vartheta_\bgamma^2}{\lambda_r^2\brac{\bM\bQ_\bgamma\bM^\top}}\\
    &\lesssim\frac{\bar\kappa_\bgamma\sqrt r\,\vartheta_\bgamma}
    {\lambda_r\brac{\bM\bQ_\bgamma\bM^\top}}
    \frac{\vartheta_\bgamma}{\lambda_1\brac{\bM\bQ_\bgamma\bM^\top}}
    \le\frac{1}{2}\fro{\calS_{\overline\bM,1}\brac{\bN}}.
\end{align*}
{Here we used $\vartheta_\bgamma\lesssim\gamma_0T\sigma_{\rm C}^2\sigma_{\rm T}^2$ and the signal-gap condition.} Collecting above fact with \eqref{eq:Xi-gamma-lb-decomp} and \eqref{eq:first-order-lb}, we can conclude that with probability at least $1-O\bbrac{n^{-10}}$,
\begin{align*}
    \fro{\mXi_\bgamma }&\gtrsim \frac{\sigma_r\brac{\bM} \gamma_0 T\sigma_{\rm C}^2\sigma_{\rm T}^2 }{\lambda_1 {\brac{\bM\bQ_\bgamma\bM^\top}}}\gtrsim \frac{\gamma_0\sigma_{\rm C}^2\sigma_{\rm T}^2 \lambda_r  T^{3/2}}{\lambda_1{\brac{\bM\bQ_\bgamma\bM^\top}}}.
\end{align*} 

\subsection{Proof of \Cref{lem:Deltagam-fro-ub}}\label{pf-lem:Deltagam-fro-ub}
We start by noticing that 
\begin{align*}
    \fro{\mXi_\bgamma }&=\fro{\brac{\wt\bU^\Omega_\bgamma\wt\bU^{\Omega\top}_\bgamma-\bU\bU^\top }\brac{\bM+\bE^\natural}}\\
    &\le  \fro{\brac{\wt\bU^\Omega_\bgamma\wt\bU^{\Omega\top}_\bgamma-\bU\bU^\top}\bM }+\fro{\brac{\wt\bU^\Omega_\bgamma\wt\bU^{\Omega\top}_\bgamma-\bU\bU^\top}\bE^\natural}\\
    &\le   \fro{\wt\bU^\Omega_\bgamma\wt\bU^{\Omega\top}_\bgamma-\bU\bU^\top}\brac{\op{\bM}+\op{\bE^\natural}}.
\end{align*} 
{Note that $\sigma_r\brac{\bM}\gtrsim\lambda_r\sqrt T\gg\op{\bE^\natural}$ and hence $\op{\bM}+\op{\bE^\natural}\lesssim\lambda_1\sqrt T$. Moreover, the bounds above give $\op{\bN}\lesssim\vartheta_\bgamma\lesssim\gamma_0\sigma_{\rm C}^2\sigma_{\rm T}^2T$. Davis--Kahan's theorem and the signal gap therefore yield}
\begin{align*}
    {\fro{\wt\bU^\Omega_\bgamma\wt\bU^{\Omega\top}_\bgamma-\bU\bU^\top}
    \lesssim\frac{\sqrt r\,\vartheta_\bgamma}
    {\lambda_r\brac{\bM\bQ_\bgamma\bM^\top}}
    \lesssim\frac{\gamma_0\sigma_{\rm C}^2\sigma_{\rm T}^2T\sqrt r}
    {\lambda_r\brac{\bM\bQ_\bgamma\bM^\top}}.}
\end{align*}
{Substitution into the preceding display proves \Cref{lem:Deltagam-fro-ub}.}

\subsection{Proof of \Cref{lem:cv-Pdeltagam-lb}}\label{pf-lem:cv-Pdeltagam-lb}
For any $\bgamma\in\calG_{K_0,1}\cup\calG_{K_0,2}$, define 
\begin{align*}
    \alpha_{\sf sp}\brac{\mXi_\bgamma}:=\frac{\sqrt{NT}\op{\mXi_\bgamma}_{\sf max}}{\fro{\mXi_\bgamma}},\qquad \beta_{\sf ra}\brac{\mXi_\bgamma}:=\frac{\op{\mXi_\bgamma}_{*}}{\fro{\mXi_\bgamma}}.
\end{align*} 
Let $d:=(N+T)/2$. To apply Theorem B.4 in \cite{jin2021factor}, it suffices to verify that
\[
\alpha_{\sf sp}\brac{\mXi_\bgamma}\beta_{\sf ra}\brac{\mXi_\bgamma}
\le \frac{1}{c_0}\sqrt{\frac{NT}{d\log d}}\asymp\frac{1}{c_0}\sqrt{\frac{N}{\log n}},
\]
where the last relation uses $T\ge N$.
Notice that $\op{\mXi_\bgamma}_*\le\sqrt{2r}\fro{\mXi_\bgamma}$, then using \Cref{thm:U-two-infinty-mcar} we obtain with probability at least $1-O\brac{n^{-10}}$,
\begin{align}\label{eq:Xi-gamma-max-ub}
    \op{\mXi_\bgamma}_{\sf max}&\le \infn{\wt\bU^\Omega_\bgamma\wt\bU^{\Omega\top}_\bgamma-\bU\bU^\top}\brac{\infn{\bM^\top}+\infn{\bE^{\natural\top}}}\notag\\
    &\lesssim \frac{\vartheta_\bgamma\infn{\bU}\brac{\sqrt{T}\infn{\bV}\op{\bSigma}+\sigma_{\rm C}\sigma_{\rm T}\sqrt{N}\log n}}{\sigma_r\brac{\bM\bQ_\bgamma\bM^\top }}\notag\\
    &\lesssim \frac{\vartheta_\bgamma\infn{\bU}\sigma_{\rm C}\sigma_{\rm T}\sqrt{N}\log n}{\sigma_r\brac{\bM\bQ_\bgamma\bM^\top }}.
\end{align}
where in the last inequality we have used $\infn{\bV}=O\bbrac{\sqrt{\brac{r+\log n}/T}}$ and  $\lambda_1^2r/\brac{\sigma_{\rm C}\sigma_{\rm T}}^2=o\brac{N}$. 
By \Cref{lem:Deltagam-fro-lb}, for every candidate under consideration, with probability at least $1-O\bbrac{n^{-10}}$ we have 
\begin{align}\label{eq:Xi-gamma-F-lb}
    \fro{\mXi_\bgamma}\gtrsim   \frac{\gamma_0 \sigma_{\rm C}^2\sigma_{\rm T}^2\lambda_r T^{3/2}}{\lambda_1 {\brac{\bM\bQ_\bgamma\bM^\top}}}.
\end{align}
We thus arrive at  with probability at least $1-O\bbrac{n^{-10}}$,
\begin{align*}
    \alpha_{\sf sp}\brac{\mXi_\bgamma}\beta_{\sf ra}\brac{\mXi_\bgamma}&\le \frac{\sqrt{NTr}\op{\mXi_\bgamma}_{\sf max}}{\fro{\mXi_\bgamma}}
    \overset{(a)}{\lesssim}\frac{ \kappa^2\vartheta_\bgamma \sqrt{N}r\log n}{\gamma_0\sigma_{\rm C}\sigma_{\rm T}\lambda_r T}\overset{(b)}{\le }\frac{1}{c_0}\sqrt{\frac{N}{\log n}},
\end{align*}
where in (a) we've used \eqref{eq:Xi-gamma-max-ub} and \eqref{eq:Xi-gamma-F-lb}, in (b) we've used $\vartheta_\bgamma\lesssim \gamma_0 \sigma_{\rm C}^2\sigma_{\rm T}^2T$  and $\lambda_r/\brac{\sigma_{\rm C}\sigma_{\rm T}}\ge C\kappa^2r\log^{3/2} n$ 
for a sufficiently universal large constant $C>0$. 
Since $\beta_{\sf ra}(\mXi_\bgamma)\ge1$, the same bound also implies $\alpha_{\sf sp}(\mXi_\bgamma)/\sqrt{NT}=o(1)$. Theorem B.4 in \cite{jin2021factor} now gives, uniformly over its admissible class,
\begin{align*}
\frac{1}{\sqrt{1-p_*}}\fro{\calP_{\Omega_\perp}\brac{\mXi_\bgamma}}
\ge\frac18\fro{\mXi_\bgamma}
\brac{1-\frac{c_3\alpha_{\sf sp}\brac{\mXi_\bgamma}}{\sqrt{NT}}},
\end{align*}
with the probability stated in \Cref{lem:cv-Pdeltagam-lb}

    \subsection{Proof of \Cref{lem:high-order-bound-mcar}}\label{pf-lem:high-order-bound-mcar}
    The proof follows the same strategy as the proof of \Cref{lem:high-order-bound}.   Denote
    \begin{align}\label{eq:thetas-def-mcar}
        \vartheta_{\bgamma,1}&:=\sqrt{NT}\op{\bSigma}\brac{K\sigma_{\rm C}\sigma_{\rm T}\sqrt{\log n}+\op{\bM}_{\sf max}},\notag\\
        \vartheta_{\bgamma,2}&:=K^2\sigma_{\rm C}^2\sigma_{\rm T}^2\brac{\sqrt{NT\log n}+N}+\op{\bM}_{\sf max}^2\bbrac{\sqrt{NT}+N}+\op{\bDelta_{\bgamma,\zeta}}.
    \end{align}
    Define the following events for some universal constants $C_k>0, k\in[5]$:
    \begin{align}\label{eq:hp-event-mcar}
        \calE_1:=&\ebrac{\max\ebrac{\op{\bN_1},\op{\bar\bN_2}}\le C_1\vartheta_\bgamma},\notag\\
        \calE_2:=&\ebrac{\max\ebrac{\op{\bN_1^{(-i)}},\op{\bar\bN^{(-i)}_2}}\le C_2\vartheta_\bgamma\text{~for~}\forall i\in[N]},\notag\\
        \calE_3:=&\ebrac{\infn {\calP_{\Omega}\brac{\bSigma_{\rm C}^{1/2}\bG\bSigma_{\rm T}^{1/2}}\bQ_\bgamma\bM^\top \bar\bU}\le C_4K\sigma_{\rm C}\sigma_{\rm T}\sqrt{T}\op{\bSigma}r^{1/2}{\log n} },\notag\\
        \calE_4:=&\ebrac{\infn {\brac{\frakD\brac{\bOmega}\circ \bM }\bQ_\bgamma\bM^\top \bar\bU}\le C_5 T\op{\bSigma}^2\infn{\bar\bU}\infn{\bV}(r\log n)^{1/2}},
    \end{align}
    In addition, for any   ${q_0\ge 2}$ let 
    {\small
\begin{align*}
        \calE_{0,{q_0}}:=&\ebrac{\infn{\bar\bU_{\perp}\bar\bU_{\perp}^\top\brac{\bar\bU_{\perp}\bar\bU_{\perp}^\top\bN_2\bar\bU_{\perp}\bar\bU_{\perp}^\top }^{q-1}\bN\bar\bU}\le C_1\brac{C_0\vartheta_\bgamma }^{q}\infn{\bar\bU}\text{~for~} 1\le q\le {q_0-1}}.
    \end{align*}
}
    We introduce the following lemmas, whose proofs are deferred to \Cref{pf-lem:hp-event-mcar} and \Cref{pf-lem:first-order-bound-mcar}. 
    \begin{lemma}\label{lem:hp-event-mcar}
        For $\ebrac{\calE_j}_{j\in[4]}$ defined in \eqref{eq:hp-event-mcar},  we have $\PP\brac{\bigcap_{j\in[4]}\calE_j}\ge 1-O\brac{e^{-c_0N}+n^{-20}}$.
    \end{lemma}
    
    \begin{lemma}\label{lem:first-order-bound-mcar} 
        Consider the model defined in \eqref{eq:main-decomp-mcar}. We have with probability at least $1-O\bbrac{e^{-c_0N}+n ^{-10}}$ that 
        \begin{align*}
            \infn{\bar\bU_{\perp}\bar\bU_{\perp}^\top\bN\bar\bU}\le CC_0\vartheta_\bgamma\infn{\bar\bU},
        \end{align*}
        where $\vartheta_\bgamma$ is defined in \eqref{eq:thetas-def-mcar}.
    \end{lemma}
    By \Cref{lem:hp-event-mcar} and \Cref{lem:first-order-bound-mcar}, we have $\PP\brac{\calE_{0,{q_0}}}\ge 1-O\bbrac{{q_0} n ^{-10}}$ with ${q_0=2}$. Suppose that $\PP\brac{\calE_{0,{q_0}}}\ge 1-O\brac{{q_0} n ^{-10}}$ for ${q_0\ge 2}$. Notice that
    \begin{align}\label{eq:induction-bound-mcar}
        &\infn{\brac{\bar\bU_{\perp}\bar\bU_{\perp}^\top\bN_2\bar\bU_{\perp}\bar\bU_{\perp}^\top }^{{q_0-1}}\bN\bar\bU}= \infn{\bar\bU_{\perp}\bar\bU_{\perp}^\top\bN_2\bar\bU_{\perp}\bar\bU_{\perp}^\top\brac{\bar\bU_{\perp}\bar\bU_{\perp}^\top\bN_2\bar\bU_{\perp}\bar\bU_{\perp}^\top }^{{q_0-2}}\bN\bar\bU}\notag\\
        &\le \vartheta_\bgamma^{{q_0}}\infn{\bar\bU}+\infn{\bN_2\bar\bU_{\perp}\bar\bU_{\perp}^\top\brac{\bar\bU_{\perp}\bar\bU_{\perp}^\top\bN_2\bar\bU_{\perp}\bar\bU_{\perp}^\top }^{{q_0-2}} \bN\bar\bU}.
    \end{align}
    Fix $i\in[N]$, we are going to bound the $\ell_2$ norm of  the the $i$-th row of $\bN_2 \brac{\bar\bU_{\perp}\bar\bU_{\perp}^\top\bN_2\bar\bU_{\perp}\bar\bU_{\perp}^\top }^{{q_0-2}}\bN\bar\bU$.  Let $\bG^{(-i)}$ and $\bOmega^{(-i)}$ be the same matrices as $\bG$ and $\bOmega$ except zeroing the $i$-th row, i.e, $\bG^{(-i)}=\brac{\bI_N-\e_i\e_i^\top }\bG$ and $\bOmega^{(-i)}=\brac{\bI_N-\e_i\e_i^\top }\bOmega$, and define $\bE^{(-i)}:=\bSigma_{\rm C}^{1/2}\bG^{(-i)}\bSigma_{\rm T}^{1/2}$, $\bDelta_M^{(-i)}:=p^{-1}\bM\circ \brac{\bI_N-\e_i\e_i^\top }\bOmega-\brac{\bI_N-\e_i\e_i^\top }\bM=p^{-1}\frakD\brac{\bOmega^{(-i)}}\circ \bM$, and
    \begin{align*}
        \bN^{(-i)}_1&:=\bM\bQ_\bgamma\sqbrac{p^{-1}\calP_{\Omega}\bbrac{\bE^{(-i)}}+\bDelta_M^{(-i)} }^\top+\sqbrac{p^{-1}\calP_\Omega\bbrac{\bE^{(-i)}}+\bDelta_M^{(-i)}}\bQ_\bgamma\bM^\top,\\
        \bar \bN^{(-i)}_2&:=\underbrace{p^{-2}\calP_{\Omega}\bbrac{\bE^{(-i)}}\bQ_\bgamma\sqbrac{\calP_{\Omega}\bbrac{\bE^{(-i)}}}^\top}_{\bN_{2,1}^{(-i)}} -p^{-2}\EE\calP_{\Omega}\bbrac{\bE^{(-i)}}\bQ_\bgamma\sqbrac{\calP_{\Omega}\bbrac{\bE^{(-i)}}}^\top\\
        &+\underbrace{p^{-1}\calP_{\Omega}\bbrac{\bE^{(-i)}}\bQ_\bgamma\bDelta_M^{(-i)\top}}_{\bN_{2,2}^{(-i)}}+\underbrace{p^{-1}\bDelta_M^{(-i)}\bQ_\bgamma\sqbrac{\calP_{\Omega}\bbrac{\bE^{(-i)}}}^\top}_{\bN_{2,3}^{(-i)}}\\
        &+\underbrace{\bDelta_M^{(-i)}\bQ_\bgamma\bDelta_M^{(-i)\top}}_{\bN_{2,4}^{(-i)}}-\EE \bDelta_M^{(-i)}\bQ_\bgamma\bDelta_M^{(-i)\top}
    \end{align*}

    In addition, define $\bN^{(-i)}:=\bN^{(-i)}_1+\bN^{(-i)}_2$ where $\bN^{(-i)}_2:=\bar\bN^{(-i)}_2+\bDelta_{\bgamma,\zeta}$ and $\mPsi^{(-i)}:=\bN-\bN^{(-i)}$, and 
    \begin{align*}
        \bC:&=\bar\bU_{\perp}\bar\bU_{\perp}^\top\brac{\bar\bU_{\perp}\bar\bU_{\perp}^\top\bN_2\bar\bU_{\perp}\bar\bU_{\perp}^\top }^{{q_0-2}}\bN\bar\bU,\\
        \bC^{(-i)}:&=\bar\bU_{\perp}\bar\bU_{\perp}^\top\brac{\bar\bU_{\perp}\bar\bU_{\perp}^\top\bN^{(-i)}_2\bar\bU_{\perp}\bar\bU_{\perp}^\top }^{{q_0-2}}\bN^{(-i)}\bar\bU.
    \end{align*}
    Notice that $\bN_2 \bar\bU_{\perp}\bar\bU_{\perp}^\top\brac{\bar\bU_{\perp}\bar\bU_{\perp}^\top\bN_2\bar\bU_{\perp}\bar\bU_{\perp}^\top }^{{q_0-2}}\bN\bar\bU=\bN_2\bC=\bar\bN_2\bC+\bDelta_{\bgamma,\zeta}\bC$, hence 	 
    \begin{align}\label{eq:N2C-bound-mcar}
        \e_i^\top \bN_2\bC&=\e_i^\top \bar\bN_2\brac{\bC-\bC^{(-i)}}+\e_i^\top \bar\bN_2\bC^{(-i)}+\e_i^\top\bDelta_{\bgamma,\zeta}\bC.
    \end{align}
    
    \paragraph{Bound for $\op{\e_i^\top \bar\bN_2\brac{\bC-\bC^{(-i)}}}$.} 
    First   note that for ${q_0\ge 4}$,
    \begin{align*}
        &\brac{\bar\bU_{\perp}\bar\bU_{\perp}^\top\bN_2\bar\bU_{\perp}\bar\bU_{\perp}^\top }^{{q_0-2}}-\brac{\bar\bU_{\perp}\bar\bU_{\perp}^\top\bN^{(-i)}_2\bar\bU_{\perp}\bar\bU_{\perp}^\top }^{{q_0-2}}\\
        &=\sqbrac{\brac{\bar\bU_{\perp}\bar\bU_{\perp}^\top\bN_2\bar\bU_{\perp}\bar\bU_{\perp}^\top }^{{q_0-4}}-\brac{\bar\bU_{\perp}\bar\bU_{\perp}^\top\bN^{(-i)}_2\bar\bU_{\perp}\bar\bU_{\perp}^\top }^{{q_0-4}}}\brac{\bar\bU_{\perp}\bar\bU_{\perp}^\top\bN_2\bar\bU_{\perp}\bar\bU_{\perp}^\top }^2\\
        &+\brac{\bar\bU_{\perp}\bar\bU_{\perp}^\top\bN^{(-i)}_2\bar\bU_{\perp}\bar\bU_{\perp}^\top }^{{q_0-4}}\brac{\bar\bU_{\perp}\bar\bU_{\perp}^\top\mPsi^{(-i)}\bar\bU_{\perp}\bar\bU_{\perp}^\top }\brac{\bar\bU_{\perp}\bar\bU_{\perp}^\top\bN_2\bar\bU_{\perp}\bar\bU_{\perp}^\top }\\
        &+\brac{\bar\bU_{\perp}\bar\bU_{\perp}^\top\bN^{(-i)}_2\bar\bU_{\perp}\bar\bU_{\perp}^\top }^{{q_0-3}}\brac{\bar\bU_{\perp}\bar\bU_{\perp}^\top\mPsi^{(-i)}\bar\bU_{\perp}\bar\bU_{\perp}^\top },
    \end{align*}
    where we've used $\bar\bU_{\perp}\bar\bU_{\perp}^\top\mPsi^{(-i)}\bar\bU_{\perp}\bar\bU_{\perp}^\top =\bar\bU_{\perp}\bar\bU_{\perp}^\top\brac{\bN_2-\bN_2^{(-i)}}\bar\bU_{\perp}\bar\bU_{\perp}^\top$.
    Hence we get 
    \begin{align}\label{eq:A-A(-i)-decomp-MCAR}
        &\bC-\bC^{(-i)}
        =\sum_{m=1}^{{q_0-2}}\bar\bU_{\perp}\bar\bU_{\perp}^\top\brac{\bS^{(-i)} }^{m-1}\bS_{\Psi}^{(-i)}\bS^{{q_0-m-2}}\bN\bar\bU+\bar\bU_{\perp}\bar\bU_{\perp}^\top\brac{\bS^{(-i)} }^{{q_0-2}}\mPsi^{(-i)}\bar\bU,
    \end{align}
    where $\bS:=\bar\bU_{\perp}\bar\bU_{\perp}^\top\bN_2\bar\bU_{\perp}\bar\bU_{\perp}^\top$, $\bS^{(-i)}:=\bar\bU_{\perp}\bar\bU_{\perp}^\top\bN^{(-i)}_2\bar\bU_{\perp}\bar\bU_{\perp}^\top$, $\bS^{(-i)}_{\Psi}:=\bar\bU_{\perp}\bar\bU_{\perp}^\top\mPsi^{(-i)}\bar\bU_{\perp}\bar\bU_{\perp}^\top$, and the summation is empty if ${q_0=2}$. It is readily seen that the decomposition \eqref{eq:A-A(-i)-decomp-MCAR} still holds for ${q_0=3}$.  We start with bounding $\op{\bar\bU_{\perp}\bar\bU_{\perp}^\top\brac{\bS^{(-i)} }^{m-1}\bS_{\Psi}^{(-i)}\bS^{{q_0-m-2}}\bN\bar\bU}$ for ${q_0\ge 3}$, $m\ge 1$. We then have   
    \begin{align*}
        \bS^{(-i)}_{\Psi}
        &=\bar\bU_{\perp}\bar\bU_{\perp}^\top{\sum_{k=1}^4\frakD\brac{\bN_{2,k}-\bN_{2,k}^{(-i)}}}\bar\bU_{\perp}{\bar\bU_{\perp}^\top}.
    \end{align*}
    By construction, we have
    \begin{align*}
        p^{2}\brac{\bN_{2,1}-\bN_{2,1}^{(-i)}}&=\calP_{\Omega}\brac{\bSigma_{\rm C}^{1/2}\e_i\e_i^\top \bG\bSigma_{\rm T}^{1/2}}\bQ_\bgamma\sqbrac{\calP_{\Omega}\brac{\bSigma_{\rm C}^{1/2}\bG\bSigma_{\rm T}^{1/2}}}^\top\\
        &+\calP_{\Omega}\brac{\bSigma_{\rm C}^{1/2}\bG\bSigma_{\rm T}^{1/2}}\bQ_\bgamma\sqbrac{\calP_{\Omega}\brac{\bSigma_{\rm C}^{1/2}\e_i\e_i^\top \bG\bSigma_{\rm T}^{1/2}}}^\top\\
        &-\calP_{\Omega}\brac{\bSigma_{\rm C}^{1/2}\e_i\e_i^\top \bG\bSigma_{\rm T}^{1/2}}\bQ_\bgamma\sqbrac{\calP_{\Omega}\brac{\bSigma_{\rm C}^{1/2}\e_i\e_i^\top \bG\bSigma_{\rm T}^{1/2}}}^\top.
    \end{align*}
    Similarly, we have
    \begin{align*}
        &p^2\brac{\bN_{2,2}-\bN_{2,2}^{(-i)}}\\
        &=\calP_{\Omega}\brac{\bSigma_{\rm C}^{1/2}\e_i\e_i^\top \bG\bSigma_{\rm T}^{1/2}}\bQ_\bgamma\brac{ \frakD\brac{\bOmega}\circ \bM}^\top+\calP_{\Omega}\brac{\bSigma_{\rm C}^{1/2}\bG\bSigma_{\rm T}^{1/2}}\bQ_\bgamma{\brac{\e_i\e_i^\top \frakD\brac{\bOmega}\circ \bM}}^\top\\
        &-\calP_{\Omega}\brac{\bSigma_{\rm C}^{1/2}\e_i\e_i^\top \bG\bSigma_{\rm T}^{1/2}}\bQ_\bgamma{\brac{\e_i\e_i^\top \frakD\brac{\bOmega}\circ \bM}}^\top,
    \end{align*}
    and
    \begin{align*}
        &p^2\brac{\bN_{2,3}-\bN_{2,3}^{(-i)}}\\
        &=\brac{\e_i\e_i^\top \frakD\brac{\bOmega}\circ \bM}\bQ_\bgamma\sqbrac{\calP_{\Omega}\brac{\bSigma_{\rm C}^{1/2}\bG\bSigma_{\rm T}^{1/2}}}^\top+\brac{\frakD\brac{\bOmega}\circ \bM}\bQ_\bgamma\sqbrac{\calP_{\Omega}\brac{\bSigma_{\rm C}^{1/2}\e_i\e_i^\top \bG\bSigma_{\rm T}^{1/2}}}^\top\\
        &-\brac{\e_i\e_i^\top \frakD\brac{\bOmega}\circ \bM}\bQ_\bgamma\sqbrac{\calP_{\Omega}\brac{\bSigma_{\rm C}^{1/2}\e_i\e_i^\top \bG\bSigma_{\rm T}^{1/2}}}^\top.
    \end{align*}
    In addition, we have
    \begin{align*}
        &{p^2\brac{\bN_{2,4}-\bN_{2,4}^{(-i)}}}\\
        &=\brac{\e_i\e_i^\top \frakD\brac{\bOmega}\circ \bM}\bQ_\bgamma\brac{ \frakD\brac{\bOmega}\circ \bM}^\top+\brac{ \frakD\brac{\bOmega}\circ \bM}\bQ_\bgamma{\brac{\e_i\e_i^\top \frakD\brac{\bOmega}\circ \bM}}^\top\\
        &-\brac{\e_i\e_i^\top \frakD\brac{\bOmega}\circ \bM}\bQ_\bgamma{\brac{\e_i\e_i^\top \frakD\brac{\bOmega}\circ \bM}}^\top. 
    \end{align*}
    Since $\bSigma_{\rm C}^{1/2}$ is diagonal matrix, 
    $\calP_{\Omega}\brac{\bSigma_{\rm C}^{1/2}\e_i\e_i^\top \bG\bSigma_{\rm T}^{1/2}}=\e_i\e_i^\top \calP_{\Omega}\brac{\bSigma_{\rm C}^{1/2}\bG\bSigma_{\rm T}^{1/2}}$.
    We thus get
    \begin{align*}
        \sum_{k=1}^4\brac{\bN_{2,k}-\bN_{2,k}^{(-i)}}=\e_i\e_i^\top \brac{\sum_{k=1}^4\bN_{2,k}}+ \brac{\sum_{k=1}^4\bN_{2,k}}\e_i\e_i^\top-\e_i\e_i^\top\brac{\sum_{k=1}^4\bN_{2,k}}\e_i\e_i^\top.
    \end{align*}
    Moreover, since $\bar\bN_2=\frakD\brac{\sum_{k=1}^4\bN_{2,k}}$, we have
    \begin{align*}
        &\op{\bar\bU_{\perp}\bar\bU_{\perp}^\top\brac{\bS^{(-i)} }^{m-1}\bar\bU_{\perp}\bar\bU_{\perp}^\top\sum_{k=1}^4\frakD\brac{\bN_{2,k}-\bN_{2,k}^{(-i)}}\bar\bU_{\perp}\bar\bU_{\perp}^\top\bS^{{q_0-m-2}}\bN\bar\bU}\\
        &\le \op{\brac{\bS^{(-i)} }^{m-1}}\brac{\op{\e_i^\top \bar\bN_2\bar\bU_{\perp}\bar\bU_{\perp}^\top\bS^{{q_0-m-2}}\bN\bar\bU}+2\op{\bar\bN_2}\op {\e_i^\top \bar\bU_{\perp}\bar\bU_{\perp}^\top\bS^{{q_0-m-2}}\bN\bar\bU}}\\
        &\le \op{\brac{\bS^{(-i)} }^{m-1}}\infn{\bar\bU}\op{ \bar\bN_2\bar\bU_{\perp}\bar\bU_{\perp}^\top\bS^{{q_0-m-2}}\bN\bar\bU}\\
        &+\op{\brac{\bS^{(-i)} }^{m-1}}\brac{\op{\e_i^\top \bar\bU_{\perp}\bar\bU_{\perp}^\top\bN_2\bar\bU_{\perp}\bar\bU_{\perp}^\top\bS^{{q_0-m-2}}\bN\bar\bU}+\op{\e_i^\top \bar\bU_{\perp}\bar\bU_{\perp}^\top\bDelta_{\bgamma,\zeta}\bS^{{q_0-m-2}}\bN\bar\bU}}\\
        &+2\op{\brac{\bS^{(-i)} }^{m-1}}\op{\bar\bN_2}\infn {\bar\bU_{\perp}\bar\bU_{\perp}^\top\bS^{{q_0-m-2}}\bN\bar\bU}
    \end{align*}
    On $\calE_1$ and $\calE_2$, we obtain that 
    \begin{align*}
        \op{\brac{\bS^{(-i)} }^{m-1}}\lesssim \vartheta_\bgamma^{m-1},\quad \op{\bar\bN_2}\lesssim\vartheta_\bgamma,\quad \op{\bDelta_{\bgamma,\zeta}}_{1}\lesssim \vartheta_\bgamma.
    \end{align*}
    Hence by induction (i.e., on $\calE_{0,{q_0}}$) we can arrive at 
    {\small
\begin{align*}
        &\op{\bar\bU_{\perp}\bar\bU_{\perp}^\top\brac{\bS^{(-i)} }^{m-1}\bar\bU_{\perp}\bar\bU_{\perp}^\top\sum_{k=1}^4\frakD\brac{\bN_{2,k}-\bN_{2,k}^{(-i)}}\bar\bU_{\perp}\bar\bU_{\perp}^\top\bS^{{q_0-m-2}}\bN\bar\bU}\\
        &\qquad\lesssim  \vartheta_\bgamma^{m-1}\brac{C_0\vartheta_\bgamma}^{{q_0-m}}\infn{\bar\bU}.
    \end{align*}
}
    Next, it suffices to   bound the boundary term  $\op{\bar\bU_{\perp}\bar\bU_{\perp}^\top\brac{\bS^{(-i)} }^{{q_0-2}}\mPsi^{(-i)}\bar\bU}$.  Notice that
    \begin{align}\label{eq:boundary-term-decomp-mcar}
        &\op{\bar\bU_{\perp}\bar\bU_{\perp}^\top\brac{\bS^{(-i)} }^{{q_0-2}}\mPsi^{(-i)}\bar\bU}\notag\\
        &\le \op{\bar\bU_{\perp}\bar\bU_{\perp}^\top\brac{\bS^{(-i)} }^{{q_0-2}}\brac{\bN_1-\bN_1^{(-i)}}\bar\bU}+\op{\bar\bU_{\perp}\bar\bU_{\perp}^\top\brac{\bS^{(-i)} }^{{q_0-2}}\brac{\bar\bN_2-\bar\bN_2^{(-i)}}\bar\bU}.
    \end{align}
    The second term of \eqref{eq:boundary-term-decomp-mcar} can be bounded in the same way as 
\[
\op{ \bar\bU_{\perp}\bar\bU_{\perp}^\top\brac{\bS^{(-i)} }^{m-1}\bS_{\Psi}^{(-i)}\bS^{{q_0-m-2}}\bN\bar\bU}
\]
. We only bound the first term of \eqref{eq:boundary-term-decomp-mcar}. Notice that 
    \begin{align*}
        &\op{\bar\bU_{\perp}\bar\bU_{\perp}^\top\brac{\bS^{(-i)} }^{{q_0-2}}\brac{\bN_1-\bN_1^{(-i)}}\bar\bU}\\
        &\le p^{-1}\op{\brac{\bS^{(-i)} }^{{q_0-2}}\e_i\e_i^\top\sqbrac{\calP_{\Omega}\brac{\bSigma_{\rm C}^{1/2}\bG\bSigma_{\rm T}^{1/2}}+\frakD\brac{\bOmega}\circ \bM }\bQ_\bgamma\bM^\top\bU}\\
        &\le p^{-1}\op{\brac{\bS^{(-i)} }^{{q_0-2}}}\infn {\calP_{\Omega}\brac{\bSigma_{\rm C}^{1/2}\bG\bSigma_{\rm T}^{1/2}}\bQ_\bgamma\bM^\top \bar\bU}\\
        &+p^{-1} \op{\brac{\bS^{(-i)} }^{{q_0-2}}}\infn {\brac{\frakD\brac{\bOmega}\circ \bM }\bQ_\bgamma\bM^\top \bar\bU}\\
        &\lesssim\brac{C_0\vartheta_\bgamma}^{{q_0-1}}\infn{\bar\bU}
    \end{align*}
    where we've used the bounds in $\calE_3$ and $\calE_4$ and that $\infn{\bar\bU}\ge \sqrt{r/N}$ in the last inequality. We are ready to bound $\op{\e_i^\top\bar\bN_2\brac{\bC-\bC^{(-i)}}}$ on $\calE_{0,{q_0}}\bigcap\ebrac{\bigcap_{j=1}^4\calE_j}$: 
    \begin{align}\label{eq:loo-remain-bound-mcar}
        &\op{\e_i^\top \bar\bN_2\brac{\bC-\bC^{(-i)}}}\notag\\
        &\le \sum_{m=1}^{{q_0-2}}\op{\e_i^\top \bar\bN_2\bar\bU_{\perp}\bar\bU_{\perp}^\top\brac{\bS^{(-i)} }^{m-1}\bS_{\Psi}^{(-i)}\bS^{{q_0-m-2}}\bN\bar\bU}+\op{\e_i^\top \bar\bN_2\bar\bU_{\perp}\bar\bU_{\perp}^\top\brac{\bS^{(-i)} }^{{q_0-2}}\mPsi^{(-i)}\bar\bU}\notag\\
        &{\lesssim} \sum_{m=1}^{{q_0}}\brac{C_0\vartheta_\bgamma }^{{q_0-m}}\vartheta_\bgamma^{m}\infn{\bar\bU}+\brac{C_0\vartheta_\bgamma}^{{q_0}}\infn{\bar\bU}\lesssim  \brac{C_0\vartheta_\bgamma }^{{q_0}}\infn{\bar\bU}.
    \end{align}
    \paragraph{Bound for $\op{\e_i^\top\bar \bN_2\bC^{(-i)}}$.}
    \begin{align}\label{eq:eibarN2C}
        {\e_i^\top \bar\bN_2\bC^{(-i)}}&=p^{-2}\e_i^\top\frakD\brac{ \calP_{\Omega}\brac{\bE}\bQ_\bgamma\sqbrac{\calP_{\Omega}\brac{\bE}}^\top }\bC^{(-i)}	+\e_i^\top\frakD\brac{ \bDelta_M\bQ_\bgamma\bDelta_M^\top }\bC^{(-i)}\notag\\
        &+p^{-1}\e_i^\top \calP_{\Omega}\brac{\bE}\bQ_\bgamma\bDelta_M^\top \bC^{(-i)}+p^{-1}\e_i^\top \bDelta_M\bQ_\bgamma\sqbrac{\calP_{\Omega}\brac{\bE}}^\top \bC^{(-i)}.
    \end{align}
    We start with the bound for $\infn{\bC^{(-i)}}$. Using the expansion of $\bC-\bC^{(-i)}$ in \eqref{eq:A-A(-i)-decomp-MCAR} and bound for each term obtained before, we have
    \begin{align}
        &\op{\bC-\bC^{(-i)}}\notag\\
        &\overset{\eqref{eq:A-A(-i)-decomp-MCAR}}{\le} \sum_{m=1}^{{q_0-2}}\op{\bar\bU_{\perp}\bar\bU_{\perp}^\top\brac{\bS^{(-i)} }^{m-1}\bS_{\Psi}^{(-i)}\bS^{{q_0-m-2}}\bN\bar\bU}+\op{\bar\bU_{\perp}\bar\bU_{\perp}^\top\brac{\bS^{(-i)} }^{{q_0-2}}\mPsi^{(-i)}\bar\bU}\notag\\
        &\lesssim \sum_{m=1}^{{q_0}}\brac{C_0\vartheta_\bgamma }^{{q_0-m}}\vartheta_\bgamma^{m-1}\infn{\bar\bU}\le  \brac{C_0\vartheta_\bgamma }^{{q_0-1}}\infn{\bar\bU}
    \end{align}
    Hence on $\calE_{0,{q_0}}\bigcap\ebrac{\bigcap_{j=1}^4\calE_j}$, we get that 
    \begin{align*}
        \infn{\bC^{(-i)}}\le\infn{\bC}+\infn{\bC-\bC^{(-i)}}\lesssim \brac{C_0\vartheta_\bgamma }^{{q_0-1}}\infn{\bar\bU}.
    \end{align*}
    By \eqref{eq:eibarN2C} and \Cref{lem:GBGU-two-inf-bound-mcar} we can obtain that with probability at least $1-O\brac{e^{-c_0n}+n ^{-15}}$, 
    \begin{align}\label{eq:loo-term-bound-mcar}
        \op{\e_i^\top\bar\bN_2\bC^{(-i)}}&\lesssim\brac{C_0\vartheta_\bgamma }^{{q_0-1}}\infn{\bar\bU}\brac{K\sigma_{\rm C}\sigma_{\rm T}\vee \op{\bM}_{\sf max}}^2\sqrt{nNr}\log n .
    \end{align}
    \paragraph{Bound for $\infn{\brac{\bar\bU_{\perp}\bar\bU_{\perp}^\top\bN_2\bar\bU_{\perp}\bar\bU_{\perp}^\top }^{{q_0-1}}\bN\bar\bU}$.}
    By \eqref{eq:induction-bound-mcar}, \eqref{eq:N2C-bound-mcar}, \eqref{eq:loo-remain-bound-mcar} and \eqref{eq:loo-term-bound-mcar}, on $\calE_{0,{q_0}}\bigcap\ebrac{\bigcap_{j=1}^4\calE_j}$ and a union bound over $i\in[N]$,  we can obtain that with probability at least $1-O\brac{{q_0} n ^{-10}}$,
    \begin{align*}
        &\infn{\brac{\bar\bU_{\perp}\bar\bU_{\perp}^\top\bN_2\bar\bU_{\perp}\bar\bU_{\perp}^\top }^{{q_0-1}}\bN\bar\bU}\\
        &\le  \vartheta_\bgamma^{{q_0}}\infn{\bar\bU}+\max_{i\in[N]}\op{\e_i^\top \bN_2\brac{\bar\bU_{\perp}\bar\bU_{\perp}^\top\bN_2\bar\bU_{\perp}\bar\bU_{\perp}^\top }^{{q_0-2}} \bN\bar\bU}+\vartheta_\bgamma^{{q_0-1}}\infn{\bar\bU}\op{\bDelta_{\bgamma,\zeta}}_{1} \\
        &\lesssim \brac{C_0\vartheta_\bgamma }^{{q_0}}\infn{\bar\bU}
    \end{align*}
    We can get there exists some universal constant $C^\prime>0$ such that
    \begin{align*}
        \PP\brac{\infn{\brac{\bar\bU_{\perp}\bar\bU_{\perp}^\top\bN_2\bar\bU_{\perp}\bar\bU_{\perp}^\top }^{{q_0-1}}\bN\bar\bU}\ge C^\prime\brac{C_0\vartheta_\bgamma }^{{q_0}}\infn{\bar\bU}}\lesssim\brac{{q_0+1}} n ^{-10}.
    \end{align*}
    Hence we can arrive at   $\PP\brac{\calE_{0,{q_0+1}}}\ge 1-O\brac{\brac{{q_0+1}} n ^{-10}}$. The proof is completed by choosing sufficiently large $C_0$ such that $C^\prime/C$ is absorbed into $C_0$.

\subsection{Proof of \Cref{lem:V-first-order-bound}}\label{pf-lem:V-first-order-bound}
We work on $\calE_0$. For $l\in[T]$, let $\bG^{(-l)}:=\bG(\bI_T-\e_l\e_l^\top)$ and $\bX^{(-l)}:=\bM+\bSigma_{\rm C}^{1/2}\bG^{(-l)}\bSigma_{\rm T}^{1/2}$, and define $\bA^{(-l)}:=\bX^{(-l)}\bQ\bX^{(-l)\top}+\allowbreak[\bSigma_{\rm T}^{1/2}\bQ\bSigma_{\rm T}^{1/2}]_{l,l}\bSigma_{\rm C}$. Let $\wt\bU_\bQ^{(-l)}$ contain the leading $r$ eigenvectors of $\bA^{(-l)}$.  Note that such construction retains the population matrix:
\begin{align*}
 \EE[\bA^{(-l)}\mid\bF]
 =\bM\bQ\bM^\top+\textsf{Tr}(\bSigma_{\rm T}\bQ)\bSigma_{\rm C}
 =\EE[\bX\bQ\bX^\top\mid\bF].
\end{align*}
An exact expansion gives
\begin{align}\label{eq:loo-column-difference}
 \bX\bQ\bX^\top-\bA^{(-l)}
 ={}&\bM\bQ\bSigma_{\rm T}^{1/2}\e_l\e_l^\top\bG^\top\bSigma_{\rm C}^{1/2}
   +\bSigma_{\rm C}^{1/2}\bG\e_l\e_l^\top\bSigma_{\rm T}^{1/2}\bQ\bM^\top\notag\\
 &+\bSigma_{\rm C}^{1/2}\bG^{(-l)}\bSigma_{\rm T}^{1/2}\bQ\bSigma_{\rm T}^{1/2}
       \e_l\e_l^\top\bG^\top\bSigma_{\rm C}^{1/2}\notag\\
 &+\bSigma_{\rm C}^{1/2}\bG\e_l\e_l^\top
       \bSigma_{\rm T}^{1/2}\bQ\bSigma_{\rm T}^{1/2}\bG^{(-l)\top}\bSigma_{\rm C}^{1/2}\notag\\
 &+[\bSigma_{\rm T}^{1/2}\bQ\bSigma_{\rm T}^{1/2}]_{l,l}
       \bSigma_{\rm C}^{1/2}(\bG\e_l\e_l^\top\bG^\top-\bI_N)\bSigma_{\rm C}^{1/2}.
\end{align}
\paragraph{Spectral separation.}
The concentration arguments in \Cref{pf-thm:U-error-gen,pf-thm:U-error} apply to both the full and deleted matrices. For the latter, the linear coefficient is multiplied by $\bI_T-\e_l\e_l^\top$, and the centered quadratic coefficient is $(\bI_T-\e_l\e_l^\top)\bSigma_{\rm T}^{1/2}\bQ\bSigma_{\rm T}^{1/2}(\bI_T-\e_l\e_l^\top)$. Neither its operator norm nor its Frobenius norm increases, and the population term is unchanged. The signal condition of \Cref{thm:U-error} therefore gives a common scalar $\zeta\ge0$ such that, with  probability at least $1-O(Te^{-cN})$,
\begin{align}\label{eq:loo-spectral-event}
 \max\biggl\{&\op{\bX\bQ\bX^\top-\zeta\bI_N-\bM\bQ\bM^\top}, \max_{l\in[T]}\op{\bA^{(-l)}-\zeta\bI_N-\bM\bQ\bM^\top}\biggr\}
 \le\frac18\sigma_r(\bM\bQ\bM^\top).
\end{align}
Here $\sigma_r(\bM\bQ\bM^\top)\ge\lambda_r^2\sigma_r(\bF^\top\bQ\bF)$. Positive definiteness of $\bF^\top\bQ\bF$ ensures that these are the leading signal eigenspaces. Since $T=o(e^{c_0N})$ with $c_0$ sufficiently small, the failure probability is $O(e^{-c'N})$ for some $c'>0$.
\paragraph{Leave-one-out stability.}
Sub-Gaussian concentration with standard epsilon-net argument give, with probability at least $1-O(e^{-cN}+n^{-12})$, simultaneously for all $l\in[T]$,
\begin{align*}
 \op{\bSigma_{\rm C}^{1/2}\bG\e_l}
 &\lesssim K\op{\bSigma_{\rm C}^{1/2}}\sqrt N,\\
 \op{\e_l^\top\bG^\top\bSigma_{\rm C}^{1/2}\wt\bU_\bQ^{(-l)}}
 &\lesssim K\op{\bSigma_{\rm C}^{1/2}}\sqrt{r+\log n},\\
 \op{\bSigma_{\rm C}^{1/2}\bG^{(-l)}\bSigma_{\rm T}^{1/2}\bQ\bSigma_{\rm T}^{1/2}\e_l}
 &\lesssim K\op{\bSigma_{\rm C}^{1/2}}
       \op{\bSigma_{\rm T}^{1/2}\bQ\bSigma_{\rm T}^{1/2}}\sqrt N.
\end{align*}
The second bound uses independence of $\wt\bU_\bQ^{(-l)}$ and $\bG\e_l$, and the third uses a fixed linear combination of noise columns. We thus have
\begin{align}\label{eq:loo-full-basis-comparison}
 &\max_{j\in[T]}\op{\e_j^\top\bG^\top\bSigma_{\rm C}^{1/2}\wt\bU_\bQ} \notag\\
 &\quad\le \max_{j\in[T]}\brac{\op{\e_j^\top\bG^\top\bSigma_{\rm C}^{1/2}\wt\bU_\bQ^{(-j)}}+\op{\e_j^\top\bG^\top\bSigma_{\rm C}^{1/2}(\wt\bU_\bQ\wt\bU_\bQ^\top-\wt\bU_\bQ^{(-j)}\wt\bU_\bQ^{(-j)\top})}}\notag\\ 
&\quad\lesssim K\op{\bSigma_{\rm C}^{1/2}}\sqrt{r+\log n}
  +K\op{\bSigma_{\rm C}^{1/2}}\sqrt N\,
       \max_{j\in[T]}\op{\wt\bU_\bQ\wt\bU_\bQ^\top
                    -\wt\bU_\bQ^{(-j)}\wt\bU_\bQ^{(-j)\top}}.
\end{align}
This leads to 
\begin{align}\label{eq:loo-cross-projection}
 &\op{\e_l^\top\bSigma_{\rm T}^{1/2}\bQ\bSigma_{\rm T}^{1/2}
                   \bG^{(-l)\top}\bSigma_{\rm C}^{1/2}\wt\bU_\bQ^{(-l)}}\notag\\
 &\quad\le\sum_{j\ne l}|[\bSigma_{\rm T}^{1/2}\bQ\bSigma_{\rm T}^{1/2}]_{j,l}|
          \op{\e_j^\top\bG^\top\bSigma_{\rm C}^{1/2}\wt\bU_\bQ^{(-l)}}\notag\\
 &\quad\lesssim K\op{\bSigma_{\rm C}^{1/2}}
           \op{\bSigma_{\rm T}^{1/2}\bQ\bSigma_{\rm T}^{1/2}}_1
           \biggl[\sqrt{r+\log n}+\sqrt N\max_{j\in[T]}\op{\wt\bU_\bQ\wt\bU_\bQ^\top
                    -\wt\bU_\bQ^{(-j)}\wt\bU_\bQ^{(-j)\top}}\biggr].
\end{align}
On \eqref{eq:loo-spectral-event}, the  Davis--Kahan bound implies
\begin{align*}
 \op{\wt\bU_\bQ\wt\bU_\bQ^\top-\wt\bU_\bQ^{(-l)}\wt\bU_\bQ^{(-l)\top}}
 \lesssim\frac{\op{(\bX\bQ\bX^\top-\bA^{(-l)})\wt\bU_\bQ^{(-l)}}}
                  {\sigma_r(\bM\bQ\bM^\top)}.
\end{align*}
Substituting \eqref{eq:loo-column-difference} and \eqref{eq:loo-cross-projection}, and using $r+\log n\lesssim N$, gives
\begin{align*}
 &\max_{l\in[T]}\op{\wt\bU_\bQ\wt\bU_\bQ^\top
                    -\wt\bU_\bQ^{(-l)}\wt\bU_\bQ^{(-l)\top}}\\
 &\quad\lesssim\frac{K\op{\bSigma_{\rm C}^{1/2}}\sqrt N}
                  {\sigma_r(\bM\bQ\bM^\top)}
       {\max_{l\in[T]}\op{\bM\bQ\bSigma_{\rm T}^{1/2}\e_l}}\\
 &\qquad+\frac{K^2\op{\bSigma_{\rm C}^{1/2}}^2
                 \op{\bSigma_{\rm T}^{1/2}\bQ\bSigma_{\rm T}^{1/2}}_1
                 \sqrt{N(r+\log n)}}{\sigma_r(\bM\bQ\bM^\top)}\\
 &\qquad+\frac{K^2N\op{\bSigma_{\rm C}^{1/2}}^2
                 \op{\bSigma_{\rm T}^{1/2}\bQ\bSigma_{\rm T}^{1/2}}_1}
                  {\sigma_r(\bM\bQ\bM^\top)}
       \max_{j\in[T]}\op{\wt\bU_\bQ\wt\bU_\bQ^\top
                    -\wt\bU_\bQ^{(-j)}\wt\bU_\bQ^{(-j)\top}}.
\end{align*}
The last summand is absorbed by \eqref{eq:V-loo-separation}, choosing $C_0$ sufficiently large. Thus,
\begin{align}\label{eq:wtU-wtUl-bound}
 &\max_{l\in[T]}\op{\wt\bU_\bQ\wt\bU_\bQ^\top
                    -\wt\bU_\bQ^{(-l)}\wt\bU_\bQ^{(-l)\top}}\notag\\
 &\quad\lesssim\frac{K\op{\bSigma_{\rm C}^{1/2}}\sqrt N}
                  {\sigma_r(\bM\bQ\bM^\top)}
       \biggl[\max_{l\in[T]}\op{\bM\bQ\bSigma_{\rm T}^{1/2}\e_l}+K\op{\bSigma_{\rm C}^{1/2}}
              \op{\bSigma_{\rm T}^{1/2}\bQ\bSigma_{\rm T}^{1/2}}_1
              \sqrt{r+\log n}\biggr].
\end{align}
\paragraph{The noise--projector product.}
The difference $\wt\bU_\bQ^{(-l)}\wt\bU_\bQ^{(-l)\top}-\bU\bU^\top$ is independent of $\bG\e_l$ and has rank at most $2r$. A espilon-net argument therefore gives, simultaneously for all $l$,
\begin{align*}
 &\op{\e_l^\top\bG^\top\bSigma_{\rm C}^{1/2}
                  (\wt\bU_\bQ^{(-l)}\wt\bU_\bQ^{(-l)\top}-\bU\bU^\top)}\\
 &\quad\lesssim K\op{\bSigma_{\rm C}^{1/2}}\sqrt{r+\log n}\,
                  \op{\wt\bU_\bQ^{(-l)}\wt\bU_\bQ^{(-l)\top}-\bU\bU^\top}.
\end{align*}
For the difference between the full and deleted projectors, use the norm of $\bSigma_{\rm C}^{1/2}\bG\e_l$ instead of independence. The triangle inequality then yields
\begin{align*}
 &\infn{\bG^\top\bSigma_{\rm C}^{1/2}
                  (\wt\bU_\bQ\wt\bU_\bQ^\top-\bU\bU^\top)}\\
 &\quad\lesssim K\op{\bSigma_{\rm C}^{1/2}}\sqrt{r+\log n}\,
                  \op{\wt\bU_\bQ\wt\bU_\bQ^\top-\bU\bU^\top}\\
 &\qquad+K\op{\bSigma_{\rm C}^{1/2}}\sqrt N
       \max_{l\in[T]}\op{\wt\bU_\bQ\wt\bU_\bQ^\top
                    -\wt\bU_\bQ^{(-l)}\wt\bU_\bQ^{(-l)\top}}.
\end{align*}
Finally, $\bE^\top=\bSigma_{\rm T}^{1/2}\bG^\top\bSigma_{\rm C}^{1/2}$ and symmetry of $\bSigma_{\rm T}^{1/2}$ imply
\begin{align*}
 \infn{\bE^\top(\wt\bU_\bQ\wt\bU_\bQ^\top-\bU\bU^\top)}
 \le\op{\bSigma_{\rm T}^{1/2}}_1
       \infn{\bG^\top\bSigma_{\rm C}^{1/2}(\wt\bU_\bQ\wt\bU_\bQ^\top-\bU\bU^\top)}.
\end{align*}
Substitute \eqref{eq:wtU-wtUl-bound} and use
\begin{align*}
 \max_{l\in[T]}\op{\bM\bQ\bSigma_{\rm T}^{1/2}\e_l}
 &\le\sqrt T\op{\bSigma}\infn{\bSigma_{\rm T}^{1/2}\bQ\bV}\\
 &\le\sqrt T\op{\bSigma}
       \op{\bSigma_{\rm T}^{1/2}\bQ\bSigma_{\rm T}^{1/2}}_1
       \infn{\bSigma_{\rm T}^{-1/2}\bV}\\
 &\lesssim\lambda_1\sqrt T
       \op{\bSigma_{\rm T}^{1/2}\bQ\bSigma_{\rm T}^{1/2}}_1
       \op{\bSigma_{\rm T}^{-1/2}}_1\infn{\bV},\\
 \sigma_r(\bM\bQ\bM^\top)
 &\ge\lambda_r^2\sigma_r(\bF^\top\bQ\bF).
\end{align*}
Together with \Cref{lem:wtu-eq-hatu} and the definition of $\kappa_{\rm T,1}$, these inequalities prove \eqref{eq:loo-noise-projector}.

    \subsection{Proof of \Cref{lem:hp-event}}\label{pf-lem:hp-event}
    We first show $\calE_1$ holds with probability at least $1-O\brac{e^{-c_0N}}$. 
    By \Cref{lem:GB-bound}, we obtain  that  with probability at least $1-O\brac{e^{-c_0N}}$,
    \begin{align*}
        \op{\bN_1}\le 2 \op{\bSigma_{\rm C}^{1/2}\bG\bSigma_{\rm T}^{1/2}\bQ\bM^\top}\lesssim K\sqrt{NT} \op{\bSigma}\op{\bSigma_{\rm C}^{1/2}}\op{\bSigma^{1/2}_{\rm T}\bQ}\le \vartheta_{1}.
    \end{align*}
    On the other hand, by \Cref{lem:EQE-bound} we have with probability at least $1-O\brac{e^{-c_0N}}$, 
    \begin{align*}
        \op{\bar\bN_2}=\op{\bE\bQ\bE^\top-\EE\bE\bQ\bE^\top}&\lesssim K^2\sqrt{N}\op{\bSigma_{\rm C}^{1/2}}^2\brac{\fro{\bSigma^{1/2}_{\rm T}\bQ\bSigma^{1/2}_{\rm T}}+\sqrt{N}\op{\bSigma^{1/2}_{\rm T}\bQ\bSigma^{1/2}_{\rm T}}}\\
        &\lesssim K^2\sqrt{nN}\op{\bSigma_{\rm C}^{1/2}}^2\op{\bSigma^{1/2}_{\rm T}\bQ\bSigma^{1/2}_{\rm T}}.
    \end{align*}
    For $\calE_2$, we can similarly obtain the bounds by virtue of \Cref{lem:GBG-bound}. In addition, \Cref{lem:GBG-bound} also implies the bound in $\calE_3$. For $\calE_4$, using matrix Hoeffding inequality, we can obtain that 
    \begin{align*}
        \PP\brac{\infn {\bG\bSigma^{1/2}_{\rm T}\bQ\bM^\top \bar\bU}\ge CK\fro{\bSigma^{1/2}_{\rm T}\bQ\bM^\top \bar\bU}\log n }\lesssim n ^{-20}.
    \end{align*}
    The proof is completed by noticing that 
    \begin{align*}
        \fro{\bSigma^{1/2}_{\rm T}\bQ\bM^\top \bar\bU}\log n \le\sqrt{T}\op{\bQ}\op{\bSigma}\op{\bSigma^{1/2}_{\rm T}}r^{1/2}\log n.
    \end{align*}
    
    \subsection{Proof of \Cref{lem:first-order-bound}}\label{pf-lem:first-order-bound}
    We start with following decomposition:
    \begin{align*}
        \infn{\bar\bU_{\perp}\bar\bU_{\perp}^\top\bN\bar\bU}\le \infn{\bar\bU_{\perp}\bar\bU_{\perp}^\top\bN_1\bar\bU}+\infn{\bar\bU_{\perp}\bar\bU_{\perp}^\top\bDelta_{\zeta}\bar\bU}+\infn{\bar\bN_2\bar\bU}+\infn{\bar\bU}\op{\bar\bN_2}
    \end{align*}
    Notice  that $\op{\bar\bN_2}\le \vartheta_{2}^*\le \vartheta$ on $\calE_3$, and
    {\small
\begin{align*}
        \infn{\bar\bU_{\perp}\bar\bU_{\perp}^\top\bDelta_{\zeta}\bar\bU}\le \infn{\bDelta_{\zeta}\bar\bU}+\infn{\bar\bU\bar\bU^\top\bDelta_{\zeta}\bar\bU}\le \brac{\op{\bDelta_{\zeta}}_1+\op{\bDelta_{\zeta}}}\infn{\bar\bU}\le \vartheta\infn{\bar \bU}
    \end{align*}
}
    by definition of $\vartheta$. It suffices to bound $\infn{\bar\bU_{\perp}\bar\bU_{\perp}^\top\bN_1\bar\bU}$ and $\infn{\bar\bN_2\bar\bU}$. Observe that 
    \begin{align*}
        \infn{\bar\bU_{\perp}\bar\bU_{\perp}^\top\bN_1\bar\bU}&\le \infn{\bSigma_{\rm C}^{1/2}\bG\bSigma_{\rm T}^{1/2}\bQ\bM^\top \bar\bU}+\infn{\bar\bU}\op{\bSigma_{\rm C}^{1/2}\bG\bSigma_{\rm T}^{1/2}\bQ\bM^\top \bar\bU}.
    \end{align*}
    By \Cref{lem:GB-bound}, $\op{\bSigma_{\rm C}^{1/2}\bG\bSigma_{\rm T}^{1/2}\bQ\bM^\top \bar\bU}\lesssim K\sqrt{N}\op{\bSigma}\op{\bSigma_{\rm T}^{1/2}\bQ}\op{\bSigma_{\rm C}^{1/2}}\le \vartheta$ with probability at least $1-O\brac{e^{-c_0N}}$ .
    By \Cref{lem:GBGU-two-inf-bound}, we have with probability at least $1-O\bbrac{n^{-10}}$ that 
    \begin{align*}
        \infn{\bSigma_{\rm C}^{1/2}\bG\bSigma_{\rm T}^{1/2}\bQ\bM^\top \bar\bU}&\lesssim K\sqrt{r\log n}\infn{\bU^\top \bM\bQ\bSigma_{\rm T}^{1/2}}\infn{\bSigma_{\rm C}^{1/2}}\\
        &\lesssim K\sqrt{Tr\log n}\op{\bSigma}\op{\bSigma_{\rm T}^{1/2}\bQ}\infn{\bSigma_{\rm C}^{1/2}}\le \vartheta \infn{\bU},
    \end{align*}
    which completes the proof.

    \subsection{Proof of \Cref{lem:hp-event-mcar}}\label{pf-lem:hp-event-mcar}
    We first show $\calE_1$ holds with probability at least $1-O\brac{e^{-c_0N}}$.
    By \Cref{lem:GB-bound-mcar}, we obtain  that  with probability at least $1-O\brac{e^{-c_0N}+n^{-10}}$,
    \begin{align*}
        \op{\calP_{\Omega}\brac{\bSigma_{\rm C}^{1/2}\bG\bSigma_{\rm T}^{1/2}}\bQ_\bgamma\bM^\top}\lesssim K\sigma_{\rm C}\sigma_{\rm T}\op{\bSigma}\sqrt{NT\log n}. 
    \end{align*}
    Moreover, by \Cref{lem:GB-bound-mcar} we obtain that 
    $\op{\bDelta_{M}\bQ_\bgamma\bM^\top}\lesssim\op{\bM}_{\sf max}\op{\bSigma}\sqrt{NT}$ with probability at least $1-O\brac{e^{-c_0N}}$.
    Hence we get with probability at least $1-O\brac{e^{-c_0N}+n^{-10}}$,
    \begin{align*}
        \op{\bN_1}\le 2\op{\bE^\natural\bQ_\bgamma\bM^\top }\le &2\brac{p^{-1}\op{\calP_{\Omega}\brac{\bSigma_{\rm C}^{1/2}\bG\bSigma_{\rm T}^{1/2}}\bQ_\bgamma\bM^\top}+\op{\bDelta_{M}\bQ_\bgamma\bM^\top}}\\
        &\lesssim\sqrt{NT}\op{\bSigma}\brac{K\sigma_{\rm C}\sigma_{\rm T}\sqrt{\log n}+\op{\bM}_{\sf max}}.
    \end{align*}
    By \Cref{lem:quad-ineq-mcar}, we obtain that with probability at least $1-O\brac{e^{-c_0N}+n^{-10}}$,
    {\small
\begin{align*}
        \op{\bar\bN_2}&=\op{\bE^\natural\bQ_\bgamma\bE^{\natural\top}-\EE\bE^\natural\bQ_\bgamma\bE^{\natural\top}}\\
        &\le p^{-2}\op{\calP_{\Omega}\brac{\bE}\bQ_\bgamma\sqbrac{\calP_\Omega\brac{\bE}}^\top-\EE\calP_{\Omega}\brac{\bE}\bQ_\bgamma\sqbrac{\calP_\Omega\brac{\bE}}^\top}\\
        &+2p^{-1}\op{\calP_\Omega\brac{\bE}\bQ_{\bgamma}\bDelta_M^\top }+\op{\bDelta_M\bQ_{\bgamma}\bDelta_M^\top-\EE\bDelta_M\bQ_{\bgamma}\bDelta_M^\top}\\
        &\lesssim K^2\sigma_{\rm C}^2\sigma_{\rm T}^2\brac{\sqrt{NT\log n}+N}+{K\sigma_{\rm C}\sigma_{\rm T}\infn{\bM}\sqrt{N\log n}}+\op{\bM}_{\sf max}^2\bbrac{\sqrt{NT}+N}
        {\lesssim\vartheta_\bgamma}.
    \end{align*}
}
    {Here the mixed term is absorbed by the first term in $\vartheta_\bgamma$, since $\infn{\bM}\le\sqrt{T}\op{\bM}_{\sf max}$, $n\ge T$, and $r\ge1$.}
    The proof for  $\calE_2$ is almost the same and hence omitted.  The proofs for $\calE_3$ and $\calE_4$ directly follows from \Cref{lem:first-order-two-inf-bound-mcar} and the following bounds:
    \begin{align*}
        &\infn{\bQ_\bgamma\bM^\top \bar\bU}\le \op{\bQ_\bgamma}_1\infn{\bV}\sqrt{T}\op{\bSigma}\lesssim \infn{\bV}\sqrt{T}\op{\bSigma},\quad \infn{\bar\bU^\top \bM\bQ_\bgamma }\le \sqrt{T}\op{\bSigma}.
    \end{align*}

    \subsection{Proof of \Cref{lem:first-order-bound-mcar} }\label{pf-lem:first-order-bound-mcar}
    We start with following decomposition:
    \begin{align*}
        \infn{\bar\bU_{\perp}\bar\bU_{\perp}^\top\bN\bar\bU}\le \infn{\bar\bU_{\perp}\bar\bU_{\perp}^\top\bN_1\bar\bU}+\infn{\bar\bU_{\perp}\bar\bU_{\perp}^\top\bDelta_{\bgamma,\zeta}\bar\bU}+\infn{\bar\bN_2\bar\bU}+\infn{\bar\bU}\op{\bar\bN_2}
    \end{align*}
    Notice  that $\op{\bar\bN_2}\le \vartheta_\bgamma$ on $\calE_1$, and $\infn{\bar\bU_{\perp}\bar\bU_{\perp}^\top\bDelta_{\zeta}\bar\bU}\le \vartheta_\bgamma\infn{\bar \bU}$ by defintion of $\vartheta_\bgamma$, 
    it suffices to bound $\infn{\bar\bU_{\perp}\bar\bU_{\perp}^\top\bN_1\bar\bU}$ and $\infn{\bar\bN_2\bar\bU}$. {By the definitions of $\calE_3,\calE_4$ and \Cref{lem:GB-bound-mcar}, with probability at least $1-O(e^{-c_0N}+n^{-10})$ we have}
    \begin{align*}
        \infn{\bar\bU_{\perp}\bar\bU_{\perp}^\top\bN_1\bar\bU}&{\le p^{-1}\infn{\bbrac{\calP_{\Omega}\brac{\bSigma_{\rm C}^{1/2}\bG\bSigma_{\rm T}^{1/2}}+\frakD\brac{\bOmega}\circ\bM}\bQ_\bgamma\bM^\top \bar\bU}}\notag\\
        &\quad{+p^{-1}\infn{\bar\bU}\op{\bbrac{\calP_{\Omega}\brac{\bSigma_{\rm C}^{1/2}\bG\bSigma_{\rm T}^{1/2}}+\frakD\brac{\bOmega}\circ\bM}\bQ_\bgamma\bM^\top \bar\bU}}\notag\\
        &\lesssim \vartheta_\bgamma\infn{\bar\bU},
    \end{align*}
    {where we also used $p\in[c,1-c]$.}
    By \Cref{lem:GBGU-two-inf-bound-mcar}, we have with probability at least $1-O\bbrac{e^{-c_0n}+n^{-10}}$ that 
    \begin{align*}
        \infn{\bar\bN_2\bar\bU}&\lesssim\infn{\bU}{\brac{K\sigma_{\rm C}\sigma_{\rm T}\vee\op{\bM}_{\sf max}}^2}\sqrt{nNr}\log n\lesssim \vartheta_\bgamma \infn{\bU},
    \end{align*}
    which completes the proof.

    \subsection{Proof of \Cref{lem:GBG-bound}}\label{pf-lem:GBG-bound}
    Fix $\u,\v\in \SS^{N-1}$, we have
    \begin{align}\label{eq:EQEt-decomp}
        \u^T\bG\bB \bG^\top \v&=\u^T\brac{\sum_{t,t^\prime\in[T]}B_{t,t^\prime}\bG_{\cdot,t}\bG^\top_{\cdot,t^\prime}} \v=\sum_{i,j\in[N]}\sum_{t,t^\prime\in[T]}u_iv_jB_{t,t^\prime}G_{i,t}G_{j,t^\prime}.
    \end{align}
    Define $\g^*\in\RR^{NT}$ and $\bB^* \in\RR^{NT\times NT}$ such that  
    \begin{align*}
        g^*_l=&G_{i,t},\quad l=T(i-1)+t, \quad i\in[N],\quad  t\in[T],\\
        B_{ll^\prime}^*=&u_iv_jB_{t,t^\prime},\quad l=T(i-1)+t, \quad l^\prime=T(j-1)+t^\prime, \quad i,j\in[N],\quad  t,t^\prime\in[T].
    \end{align*}
    It is readily seen that $\bB^* =\brac{uv^\top }\otimes \bB$, and from \eqref{eq:EQEt-decomp} we have that
    \begin{align*}
        \u^T\bG\bB \bG^\top \v=\sum_{l,l^\prime\in[N T]} B^*_{ll^\prime}g^*_lg^*_{l^\prime}=\g^{*\top }\bB^* \g^*.
    \end{align*}
    {Let $\widetilde\bB^*:=\brac{\bB^*+\bB^{*\top}}/2$. Since $\g^{*\top}\bB^*\g^*=\g^{*\top}\widetilde\bB^*\g^*$, Hanson-Wright inequality (e.g., Theorem 1.1. in \cite{rudelson2013hanson}) gives, for any $x\ge 0$,}
    \begin{align*}
        \PP\brac{\ab{\u^T\brac{\bG\bB \bG^\top-\EE \bG\bB \bG^\top} \v}\ge x}
        &{\le 2\exp\brac{-c\min\sqbrac{\frac{x^2}{K^4\fro{\widetilde\bB^*  }^2},\frac{x}{K^2\op{\widetilde\bB^*  }}}}.}
    \end{align*}
    It is readily seen that  $\fro{\bB^*  }^2=\fro{\bB}^2$ and $\op{\bB^*  }=\op{\bB}$, {while $\fro{\widetilde\bB^*}\le\fro{\bB^*}$ and $\op{\widetilde\bB^*}\le\op{\bB^*}$}. By setting $x=CK^2\brac{\fro{\bB}\sqrt{N}+\op{\bB}N}$ and a standard $\epsilon$-net argument, we obtain that with probability exceeding $1-O\brac{e^{-cN}}$,
    \begin{align*}
        \op{\bG\bB \bG^\top -\EE \bG\bB \bG^\top}\lesssim K^2\brac{\fro{\bB}\sqrt{N}+\op{\bB}N}.
    \end{align*}

    \section{Proofs of Technical Lemmas}\label{sec:pf-pftlem}
    
    \subsection{Proof of \Cref{lem:GB-bound}}\label{pf-lem:GB-bound}
Denote $m=T\wedge T_0$.  Let $(\bU_B,\bSigma_B,\bV_B)$ be the compact SVD of $\bB$, where $\bU_B\in\OO_{T,m}$, $\bV_B\in\OO_{T_0,m}$, $\bSigma_B\in \RR^{m\times m}$.  We thus have
$\op{\bG\bB}\le \|\bG\bU_B\| \op{\bSigma_B}$. It suffices to bound $\op{\bG\bU_B}$.  By definition of operator norm, we have
    \begin{align*}
        \op{\bG\bU_B}&=\sup_{\u\in\SS^{N-1}, \v\in\SS^{m-1}}\u^\top \bG\bU_B \v=\sup_{\u\in\SS^{N-1}, \v^\prime\in\SS_{B}}\u^\top \bG\v^\prime,
    \end{align*}
where $\SS_B:=\ebrac{\bU_B\v:\v\in\SS^{m-1}}\subseteq\SS^{T-1}$ is an $m$-dimensional sphere embedded in $\RR^T$.  For any fixed $\u\in\SS^{N-1}$ and $\v^\prime\in\SS_B$, we have that $\u^\top \bG\v^\prime$ is a sum of independent mean-zero subgaussian random variables with subgaussian norm at most $K$. By a standard union bound argument over an $\epsilon$-net of $\SS^{N-1}\times \SS_B$, we can obtain that with probability at least $1-O\brac{e^{-cN}}$,
    \begin{align*}
        \op{\bG\bU_B}=\sup_{\u\in\SS^{N-1}, \v^\prime\in\SS_{B}}\u^\top \bG\v^\prime\lesssim K\sqrt{N+m}\lesssim K\sqrt{NT}.
    \end{align*}

    \subsection{Proof of \Cref{lem:EQE-bound}}\label{pf-lem:EQE-bound}
Let $\bar\bQ:=\bSigma^{1/2}_{\rm T}\bQ\bSigma^{1/2}_{\rm T}$, then we get that
    \begin{align}\label{eq:EQE-decomp}
        \bE\bQ\bE^\top&{=}\bSigma^{1/2}_{\rm C}\bG \bSigma^{1/2}_{\rm T}\bQ\bSigma^{1/2}_{\rm T}\bG^{\top}\bSigma^{1/2}_{\rm C}=\bSigma_{\rm C}^{1/2}\bG\bar\bQ\bG^{\top }\bSigma_{\rm C}^{1/2}.
    \end{align}
    In view of \eqref{eq:EQE-decomp}, we are going to  bound $\op{\bSigma_{\rm C}^{1/2}\bG\bar\bQ\bG^{\top }\bSigma_{\rm C}^{1/2}}$.  The following lemma is needed, whose proof is deferred to \Cref{pf-lem:GBG-bound}.
    \begin{lemma}\label{lem:GBG-bound}
        Let $\bG$ be a $N\times T$ random matrix with independent, {not necessarily identically distributed} entries which satisfy $\EE G_{ij}=0$ and $\op{G_{ij}}_{\psi_2}\le K$. There exists some universal constants $c_0,C_0,C_1>0$ such that for any fixed $\bB\in\RR^{T\times T}$, we have
        \begin{align*}
            \PP\brac{\op{\bG\bB \bG^\top -\EE \bG\bB \bG^\top}\ge C_0K^2\brac{\fro{\bB}\sqrt{N}+\op{\bB}N}}\le C_1{e^{-c_0N}}.
        \end{align*}
    \end{lemma}
    By \Cref{lem:GBG-bound}, we obtain that with probability exceeding $1-O\brac{e^{-c_0N}}$,
    \begin{align}\label{eq:GQG-hp-bound}
        \op{\bG\bar\bQ \bG^\top -\textsf{Tr}\brac{\bar\bQ}\bI_N}\lesssim K^2\brac{\fro{\bar\bQ}\sqrt{N}+\op{\bar\bQ}N}.
    \end{align}
    where we've used  that $\EE \bG\bar\bQ\bG^\top=\textsf{Tr}\brac{\bar\bQ}\bI_N$. Combined with \eqref{eq:EQE-decomp}, we can obtain
    {\small
\begin{align*}
        \op{\bE\bQ\bE^\top-\EE \bE\bQ\bE^\top}=\op{\bSigma_{\rm C}^{1/2}\brac{\bG\bar\bQ\bG^{\top }-\textsf{Tr}\brac{\bar\bQ}\bI_N}\bSigma_{\rm C}^{1/2}}\le \op{\bSigma^{1/2}_{\rm C}}^2\op{\bG\bar\bQ\bG^{\top }-\textsf{Tr}\brac{\bar\bQ}\bI_N}.
    \end{align*}
} 
    The proof is completed by using the above inequality and \eqref{eq:GQG-hp-bound}.

    \subsection{Proof of \Cref{lem:GBGU-two-inf-bound}}\label{pf-lem:GBGU-two-inf-bound}
    \paragraph{Proof of claim (1)} Fix $i\in[N]$ and $k\in[r]$, we then have
    \begin{align*}
        \op{[\bB\bG\bV]_{i,k}}_{\psi_2}=\op{\sum_{i^\prime\in[N]}\sum_{l\in[T]}G_{i^\prime,l} B_{i,i^\prime}V_{l,k}}_{\psi_2} \lesssim K\op{\bB_{i,\cdot}}\op{\bV_{\cdot,k}}.
    \end{align*}
    Hence there exists some constant $C>0$ such that  for any $x\ge 0$,
    {\small
\begin{align*}
        &\PP\brac{\frac{1}{\sqrt{r}}\infn{ \bB\bG \bV}\ge x}\le \PP\brac{\max_{i\in[N],k\in[r]}\ab{[ \bB\bG \bV]_{i,k}}\ge x}\lesssim Nr\exp\brac{-\frac{Cx^2}{K^2\infn{ \bB}^2\infn{\bV^\top }^2}}.
    \end{align*}
}
    By choosing $x=C_0K\infn{ \bB}\infn{\bV^\top }\sqrt{\log n}$ for some large constant $C_0>0$, we could obtain the first claim.
    \paragraph{Proof of claim (2)}
    Fix $k\in[r]$, we have
    \begin{align*}
        [\bG\bB\bG^\top \bU^{(-i)}]_{i,k}&=\sum_{l_1\in[T]}G_{i,l_1}\sum_{j\in[N]}\sum_{l_2\in[T]}G_{j,l_2}B_{l_1,l_2}U^{(-i)}_{j,k}\\
        &=\sum_{l_1\in[T]}\sum_{l_2\in[T]}G_{i,l_1}G_{i,l_2}B_{l_1,l_2}U^{(-i)}_{i,k}+\sum_{j\in[N]\backslash\ebrac{i}}\sum_{l_1,l_2\in[T]}G_{i,l_1}G_{j,l_2}B_{l_1,l_2}U^{(-i)}_{j,k}
    \end{align*}
    Notice that $\sum_{l_1\in[T]}\sum_{l_2\in[T]}G_{i,l_2}G_{i,l_1}B_{l_1,l_2}=\bG_{i,\cdot} \bB\bG_{i,:}^\top$. By Hanson-wright inequality in \cite{rudelson2013hanson}, we obtain that with probability at least $1-O(n^{-12})$,
    \begin{align*}
        \ab{\sum_{l_1\in[T]}\sum_{l_2\in[T]}G_{i,l_1}G_{i,l_2}B_{l_1,l_2}-\textsf{Tr}\brac{\bB}}\lesssim K^2\brac{\op{\bB}\log n+\fro{\bB}\sqrt{\log n}}.
    \end{align*}
    On the other hand, note that 
    \begin{align*}
        \brac{\sum_{j\in[N]\backslash\ebrac{i}}\sum_{l_2\in[T]}G_{j,l_2}B_{l_1,l_2}U^{(-i)}_{j,k}}^2=\sqbrac{\bB\bG^{(-i)\top} \bU^{(-i)}}_{l_1,k}^2.
    \end{align*}
    Hence we have with probability at least $1-O\brac{e^{-c_0n}}$,
    \begin{align*}
        \ab{\sum_{j\in[N]\backslash\ebrac{i}}\sum_{l_1,l_2\in[T]}G_{i,l_1}G_{j,l_2}B_{l_1,l_2}U^{(-i)}_{j,k}}&\lesssim K\sqrt{\sum_{l_1\in[T]}\sqbrac{\bB\bG^{(-i)\top} \bU^{(-i)}}_{l_1,k}^2\log n}\\
        &\le K\infn{\bU^{(-i)\top}\bG^{(-i)}\bB^\top  }\sqrt{\log n}\\
        &\lesssim K^2\infn{\bU^{(-i)}}\op{\bB}\sqrt{nN\log n},
    \end{align*}
    where the last inequality holds due to \Cref{lem:GB-bound}. Collecting the above bounds, we have conditioning on $\bU^{(-i)}$ with probability at least $1-O(e^{-c_0n}+n^{-12})$,
    \begin{align*}
        \ab{[\bG\bB\bG^\top \bU^{(-i)}]_{i,k}-\EE[\bG\bB\bG^\top \bU^{(-i)}]_{i,k}}&\lesssim K^2\infn{\bU^{(-i)}}\op{\bB}\sqrt{nN\log n}.
    \end{align*}
    The proof is completed by noticing
    {\small
\begin{align*}
        &\PP\brac{\frac{1}{\sqrt{r}}\op{\e_i^\top \bG\bB \bG^\top\bU^{(-i)}-\EE \e_i^\top \bG\bB \bG^\top\bU^{(-i)}}\ge K^2\infn{\bU^{(-i)}}\op{\bB}\sqrt{nN\log n}\mid \bU^{(-i)}}\\
        &\le \PP\brac{\max_{k\in[r]}\ab{[\bG\bB\bG^\top \bU^{(-i)}]_{i,k}-\EE[\bG\bB\bG^\top \bU^{(-i)}]_{i,k}}\ge K^2\infn{\bU^{(-i)}}\op{\bB}\sqrt{nNr\log n}\mid \bU^{(-i)}}\\
        &\lesssim e^{-cn}+n^{-11}.
    \end{align*}
}

\subsection{Proof of \Cref{lem:subg_clt_weighted}}\label{pf-lem:subg_clt_weighted}
Fix any unit vector $\a\in\mathbb S^{r-1}$ and consider the scalar projection
\begin{align*}
S_m:=\a^\top \bar\bSigma^{-1/2}\bZ=\sum_{k=1}^{m} g_k b_k,
\qquad b_k:=\a^\top \bar\bSigma^{-1/2}\w_k.
\end{align*}
By construction, $\sum_{k=1}^{m} b_k^2 = \a^\top \bar\bSigma^{-1/2}\bar\bSigma \bar\bSigma^{-1/2}\a=\|\a\|_2^2=1$.
We verify Lindeberg's condition.  For any $\varepsilon>0$,
\begin{align*}
\sum_{k=1}^{m}\mathbb E\Big[(g_k b_k)^2\mathbf 1\{|g_k b_k|>\varepsilon\}\Big]
\le \sum_{k=1}^{m} b_k^2 \,\mathbb E\Big[g_k^2\mathbf 1\{|g_k|>\varepsilon/|b_k|\}\Big].
\end{align*}
Since $\|g_k\|_{\psi_2}\le K$, there exist constants $c,C>0$ such that
$\mathbb P(|g_k|>u)\le 2\exp(-c u^2/K^2)$, hence by integrating the tail,
$\mathbb E[g_k^2\mathbf 1\{|g_k|>u\}] \le C K^2 \exp(-c u^2/K^2)$ for all $u>0$.
Therefore, since $\delta_m \rightarrow 0$, we have
\[
\sum_{k=1}^{m}\mathbb E\Big[(g_k b_k)^2\mathbf 1\{|g_k b_k|>\varepsilon\}\Big]
\le C K^2 \sum_{k=1}^{m} b_k^2 \exp\!\Big(-c \frac{\varepsilon^2}{K^2 b_k^2}\Big)
\le C K^2 \exp\!\Big(-c \frac{\varepsilon^2}{K^2\delta_m^2}\Big)\ \rightarrow\ 0,
\]
where we have used $|b_k|\le \delta_m$ for all $k$ by definition. Hence Lindeberg--Feller yields
$S_m\Rightarrow \mathcal N(0,1)$.
By the Cram\'er--Wold device, $\bar\bSigma^{-1/2}\bZ \overset{d}{\rightarrow} \mathcal N(0,\bI_r)$.

    \subsection{Proof of \Cref{lem:GB-bound-mcar}}\label{pf-lem:GB-bound-mcar}
    Without loss of generality, we assume $T_0\le T$. Otherwise we can consider a compact SVD of $\bB=\bU_B\bSigma_B\bV_B^\top $, let $\wt\bB:=\bU_B\bSigma_B\in\RR^{T\times T}$, and bound $\op{\calP_{\Omega}\brac{\bE}\wt\bB}$ and $\op{\bDelta_M\wt\bB}$.
    
    We start with the proof of the first claim. Notice that 
    $$\op{\bOmega\circ\brac{\bSigma_{\rm C}^{1/2}\bG\bSigma_{\rm T}^{1/2}}\bB}\le p\op{\brac{\bSigma_{\rm C}^{1/2}\bG\bSigma_{\rm T}^{1/2}}\bB}+\op{\frakD\brac{\bOmega}\circ\brac{\bSigma_{\rm C}^{1/2}\bG\bSigma_{\rm T}^{1/2}}\bB}.$$ 
    The first term can be bounded as $\op{\brac{\bSigma_{\rm C}^{1/2}\bG\bSigma_{\rm T}^{1/2}}\bB}\le \sigma_{\rm C}\op {\bG\bSigma_{\rm T}^{1/2}\bB}\lesssim \sigma_{\rm C}\sigma_{\rm T}K\sqrt{N+ T_0}\op{\bB}$ with probability at least $1-O(\exp\brac{-c\brac{N+ T_0}})$, where we've used \Cref{lem:GB-bound}. It suffices to bound $\op{\frakD\brac{\bOmega}\circ\brac{\bSigma_{\rm C}^{1/2}\bG\bSigma_{\rm T}^{1/2}}\bB}$. 
    Fix $\u\in\SS^{N-1}$ and $\v\in\SS^{T_0-1} $, we have 
    \begin{align*}
        \u^\top\frakD\brac{\bOmega}\circ\brac{\bSigma_{\rm C}^{1/2}\bG\bSigma_{\rm T}^{1/2}}\bB\v
        &=\sum_{i\in[N]}\sum_{l_2\in[T]}G_{i,l_2}u_i\sigma_{{\rm C},i}\sum_{t\in[T]}\frakD\brac{\Omega_{i,t}} \sqbrac{\bSigma_{\rm T}^{1/2}}_{t,l_2}\sum_{j\in[T_0]}B_{t,j}v_j.
    \end{align*}
    We have with probability at least $1-O(n^{-10})$  that for $i\in[N]$,
    \begin{align}\label{eq:GB-var-bound}
        \ab{\sum_{t\in[T]}\frakD\brac{\Omega_{i,t}}\sqbrac{\bSigma_{\rm T}^{1/2}}_{t,l_2}\sum_{j\in[T_0]}B_{t,j}v_j}\lesssim\sqrt{\sum_{t\in[T]}\sqbrac{\bSigma_{\rm T}^{1/2}}_{t,l_2}^2\brac{\sum_{j\in[T_0]}B_{t,j}v_j}^2\log n}.
    \end{align}
    We thus have 
    \begin{align*}
        &\PP\brac{\ab{\sum_{i\in[N]}\sum_{l_2\in[T]}G_{i,l_2}u_i\sigma_{{\rm C},i}\sum_{t\in[T]}\frakD\brac{\Omega_{i,t}} \sqbrac{\bSigma_{\rm T}^{1/2}}_{t,l_2}\sum_{j\in[T_0]}B_{t,j}v_j}\ge x\mid \bOmega}\\
        &\le \exp\brac{-\frac{x^2}{K^2\sum_{i\in[N]}\sum_{l_2\in[T]}u_i^2\sigma_{{\rm C},i}^2\brac{\sum_{t\in[T]}\frakD\brac{\Omega_{i,t}} \sqbrac{\bSigma_{\rm T}^{1/2}}_{t,l_2}\sum_{j\in[T_0]}B_{t,j}v_j}^2}}.
    \end{align*}
    Take $x=CK\sqrt{\brac{N+T_0}\sum_{i\in[N]}\sum_{l_2\in[T]}u_i^2\sigma_{{\rm C},i}^2\brac{\sum_{t\in[T]}\frakD\brac{\Omega_{i,t}} \sqbrac{\bSigma_{\rm T}^{1/2}}_{t,l_2}\sum_{j\in[T_0]}B_{t,j}v_j}^2 }$ and apply a standard $\varepsilon$-net argument, we obtain that (conditioning on $\bOmega$) with probability at least $1-O\bbrac{e^{-c\brac{N+T_0}}}$,
    \begin{align*}
        &\op{\frakD\brac{\bOmega}\circ\brac{\bSigma_{\rm C}^{1/2}\bG\bSigma_{\rm T}^{1/2}}\bB} \lesssim K\sqrt{N\sum_{i\in[N]}\sum_{l_2\in[T]}u_i^2\sigma_{{\rm C},i}^2\brac{\sum_{t\in[T]}\frakD\brac{\Omega_{i,t}} \sqbrac{\bSigma_{\rm T}^{1/2}}_{t,l_2}\sum_{j\in[T_0]}B_{t,j}v_j}^2 }.
    \end{align*}
    Combining \eqref{eq:GB-var-bound} and the above bound, we get with probability at least $1-O(e^{-c\brac{N+ T_0}}+n^{-10})$,
    \begin{align*}
        \op{\frakD\brac{\bOmega}\circ\brac{\bSigma_{\rm C}^{1/2}\bG\bSigma_{\rm T}^{1/2}}\bB} &\lesssim K\sqrt{N\sum_{i\in[N]}u_i^2\sigma_{{\rm C},i}^2\sum_{t\in[T]}\sum_{l_2\in[T]}\sqbrac{\bSigma_{\rm T}^{1/2}}_{t,l_2}^2\brac{\sum_{j\in[T_0]}B_{t,j}v_j}^2\log n }\\
        &\lesssim \sigma_{\rm C}\sigma_{\rm T}K \op{\bB}\sqrt{\brac{N+ T_0}\log n}.
    \end{align*}
    where we've used the fact that $\op{\bB}^2\ge \sum_{t\in[T]}\brac{\sum_{j\in[T_0]}B_{t,j}v_j}^2$ by definition. This concludes the proof.
    
    It remains to show the second claim.
    Fix $\u\in\SS^{N-1}$ and $\v\in\SS^{T_0-1} $, we have 
    \begin{align*}
        \u^\top\brac{\frakD\brac{\bOmega}\circ\bM}\bB\v
        &=\sum_{i\in[N]}\sum_{t\in[T]}\frakD\brac{\Omega_{i,t}}u_i M_{i,t} \sum_{j\in[T_0]}B_{t,j}v_j.
    \end{align*}
    Thus we have with probability at least $1-O(e^{-c_0\brac{N+T_0}})$  that,
    {\small
\begin{align*}
        \ab{\u^\top\brac{\frakD\brac{\bOmega}\circ\bM}\bB\v}\lesssim \sqrt{\sum_{i\in[N]}\sum_{t\in[T]}u_i ^2M^2_{i,t} \brac{\sum_{j\in[T_0]}B_{t,j}v_j}^2 \brac{N+ T_0}}\lesssim\op{\bM}_{\sf max}\op{\bB}\sqrt{N+T_0}.
    \end{align*}
}
    The proof is completed by applying a standard $\varepsilon$-net argument.

    \subsection{Proof of \Cref{lem:quad-ineq-mcar}}\label{pf-lem:quad-ineq-mcar}
    \paragraph{Proof of claim (1).}
    By definition, we have that
    \begin{align}\label{eq:EQE-decomp-mcar}
        \calP_{\Omega}\brac{\bE}\bQ_\bgamma\brac{\calP_{\Omega}\brac{\bE}}^\top&=\bOmega\circ\brac{\bSigma^{1/2}_{\rm C}\bG \bSigma^{1/2}_{\rm T}}\bQ_\bgamma\brac{\bSigma^{1/2}_{\rm T}\bG^{\top}\bSigma^{1/2}_{\rm C}}\circ\bOmega^\top.
    \end{align}
    Denote $\bar\bG:=\bOmega\circ\brac{\bSigma^{1/2}_{\rm C}\bG \bSigma^{1/2}_{\rm T}}$, then we get
    $$\bar G_{i,t}=\Omega_{i,t}\sum_{l_1\in[N]}\sum_{l_2\in[T]}G_{l_1,l_2}\sqbrac{\bSigma^{1/2}_{\rm C}}_{i,l_1}\sqbrac{\bSigma^{1/2}_{\rm T}}_{t,l_2},$$ for any $i\in[N]$ and $t\in[T]$.
    Fix $\u\in \SS^{N-1}$, we have
    \begin{align}\label{eq:EQEt-decomp-mcar}
        &\u^T\bar\bG\bQ_\bgamma \bar\bG^\top \u\notag\\
        &=\sum_{i,j\in[N]}\sum_{t,t^\prime\in[T]}u_iu_jQ_{\bgamma,tt^\prime}\Omega_{i,t}\Omega_{j,t^\prime}\sum_{\substack{l_1\in[N]\\l_2\in[T]}}\sum_{\substack{l_1^\prime\in[N]\\l_2^\prime\in[T]}}G_{l_1,l_2}G_{l^\prime_1,l^\prime_2}\sqbrac{\bSigma^{1/2}_{\rm C}}_{i,l_1}\sqbrac{\bSigma^{1/2}_{\rm T}}_{t,l_2}\sqbrac{\bSigma^{1/2}_{\rm C}}_{j,l_1^\prime}\sqbrac{\bSigma^{1/2}_{\rm T}}_{t^\prime, l_2^\prime}.
    \end{align}
    Define $\g^*\in\RR^{NT}$ and $\bB^* \in\RR^{NT\times NT}$ such that  
    \begin{align*}
        g^*_l&=G_{l_1,l_2},\quad l=T(l_1-1)+l_2, \quad l_1\in[N],\quad  l_2\in[T],\\
        B_{ll^\prime}^*&=\sum_{i,j\in[N]}\sum_{t,t^\prime\in[T]}u_iu_jQ_{\bgamma,tt^\prime}\Omega_{i,t}\Omega_{j,t^\prime}\sqbrac{\bSigma^{1/2}_{\rm C}}_{i,l_1}\sqbrac{\bSigma^{1/2}_{\rm C}}_{j,l_1^\prime}\sqbrac{\bSigma^{1/2}_{\rm T}}_{t,l_2}\sqbrac{\bSigma^{1/2}_{\rm T}}_{t^\prime, l_2^\prime},\\
        l&= T(l_1-1)+l_2, \quad  l^\prime =T(l_1^\prime-1)+l_2^\prime, \quad l_1,l_1^\prime\in[N],\quad  l_2,l_2^\prime\in[T].
    \end{align*}
    First notice that $\u^T\bar\bG\bQ_\bgamma \bar\bG^\top \u=\g^{*\top }\bB^*\g^* $.
    By definition of $\bQ_\bgamma$ and $\bSigma_{\rm C}$, we have that 
    \begin{align*}
	    {B^*_{ll^\prime,0}}&{:=\gamma_0} u_{l_1}u_{l^\prime_1}\sigma_{{\rm C},l_1}\sigma_{{\rm C},l_1^\prime}\sum_{t\in[T]}\Omega_{l_1,t}\Omega_{l^\prime_1,t}\sqbrac{\bSigma^{1/2}_{\rm T}}_{t,l_2}\sqbrac{\bSigma^{1/2}_{\rm T}}_{t ,l_2^\prime},\\
	    {B^*_{ll^\prime,k}}&{:=}\gamma_ku_{l_1}u_{l^\prime_1}\sigma_{{\rm C},l_1}\sigma_{{\rm C},l_1^\prime}\sum_{t\in[T-k]}\Omega_{l_1,t}\Omega_{l_1^\prime,t+k}\sqbrac{\bSigma^{1/2}_{\rm T}}_{t,l_2}\sqbrac{\bSigma^{1/2}_{\rm T}}_{t+k, l_2^\prime}{,\quad k\in[K_0]}.
    \end{align*}
    Denote $\bB^*_{k}=\bbrac{B^*_{ll^\prime,k}}\in\RR^{NT\times NT}$ for $k=0,1,\cdots,{K_0}$. {The exact matrix decomposition is $\bB^*=\bB_0^*+\sum_{k=1}^{K_0}\brac{\bB_k^*+\bB_k^{*\top}}$. The factor $2$ arises only after forming the scalar quadratic form.} We are going to bound $\fro{\bB^*_{k}}$ and $\op{\bB^*_{k}}$ for each $k$.  We start with the bound for $\fro{\bB^*_{{0}}}$. Notice that with probability at least $1-O\brac{n^{-10}}$ such that for any $l_1=l_1^\prime \in[N]$,
    \begin{align*}
        &\ab{\sum_{t\in[T]}\Omega_{l_1,t}\Omega_{l^\prime_1,t}\sqbrac{\bSigma^{1/2}_{\rm T}}_{t,l_2}\sqbrac{\bSigma^{1/2}_{\rm T}}_{t ,l_2^\prime}-p\bSigma_{{\rm T},l_2l_2^\prime}}\lesssim \sqrt{\sum_{t\in[T]}\sqbrac{\bSigma^{1/2}_{\rm T}}^2_{t,l_2}\sqbrac{\bSigma^{1/2}_{\rm T}}^2_{t ,l_2^\prime}\log n},
    \end{align*}
    and for any $l_1\ne l_1^\prime \in[N]$,
    \begin{align}
        &\ab{\sum_{t\in[T]}\Omega_{l_1,t}\Omega_{l^\prime_1,t}\sqbrac{\bSigma^{1/2}_{\rm T}}_{t,l_2}\sqbrac{\bSigma^{1/2}_{\rm T}}_{t ,l_2^\prime}-p^2\bSigma_{{\rm T},l_2l_2^\prime}}\lesssim \sqrt{\sum_{t\in[T]}\sqbrac{\bSigma^{1/2}_{\rm T}}^2_{t,l_2}\sqbrac{\bSigma^{1/2}_{\rm T}}^2_{t ,l_2^\prime}\log n}.
    \end{align}
    Notice that $\op{\e_t^\top\bSigma_{\rm T}^{1/2}}\le\op{\e_t^\top\bSigma_{\rm T}}\op{\bSigma_{\rm T}^{-1/2}}\lesssim \sigma^2_{\rm T} \op{\bSigma_{\rm T}^{-1/2}}\lesssim \sigma_{\rm T}$, thus we have 
    \begin{align*}
        \sum_{l_2,l_2^\prime\in[T]}\sum_{t\in[T]}\sqbrac{\bSigma^{1/2}_{\rm T}}^2_{t,l_2}\sqbrac{\bSigma^{1/2}_{\rm T}}^2_{t ,l_2^\prime}\log n\le \brac{\log n}\sum_{t\in[T]}\op{\e_t^\top\bSigma^{1/2}_{\rm T}}^4\le \sigma_{\rm T}^4T\log n.
    \end{align*}
    Hence we arrive at with probability at least $1-O\brac{n^{-10}}$,
    \begin{align}\label{eq:B1-Fro-bound}
        \fro{\bB^*_{0}}^2&
        =\gamma_0^2\sum_{\substack{l_1\in[N]\\l_2\in[T]}}\sum_{\substack{l_1^\prime\in[N]\\l_2^\prime\in[T]}}u^2_{l_1}u^2_{l^\prime_1}\sigma^2_{{\rm C},l_1}\sigma^2_{{\rm C},l_1^\prime}\brac{\sum_{t\in[T]}\Omega_{l_1,t}\Omega_{l^\prime_1,t}\sqbrac{\bSigma^{1/2}_{\rm T}}_{t,l_2}\sqbrac{\bSigma^{1/2}_{\rm T}}_{t ,l_2^\prime}}^2\notag\\
        &\lesssim \gamma_0^2\sigma_{\rm C}^4\brac{\sum_{l_2,l_2^\prime\in[T]}p^2\bSigma^2_{{\rm T},l_2l_2^\prime}+\sigma_{\rm T}^4T{\log n}}
        \lesssim \gamma_0^2\sigma_{\rm C}^4 \sigma_{\rm T}^4T\log n.
    \end{align}
    To bound $\fro{\bB^*_{k}}$ with $k\ge 1$, we first fix  any $l_1\ne l_1^\prime \in[N]$ and $l_2,l_2^\prime \in{[T]}$. 
    Define $\bH_k\in\RR^{T\times T}$ such that $[\bH_k]_{i,j}=1$ if $\ab{i-j}=k$ and $[\bH_k]_{i,j}=0$ otherwise. 
    By definition we have ${\bQ_\bgamma=\gamma_0\bI_T+\sum_{k=1}^{K_0}\gamma_k\bH_k}$, 
    thus with probability at least $1-O\brac{n^{-10}}$,
    \begin{align*}
        &\ab{\sum_{t\in[T-k]}\Omega_{l_1,t}\Omega_{l^\prime_1,t+k}\sqbrac{\bSigma^{1/2}_{\rm T}}_{t,l_2}\sqbrac{\bSigma^{1/2}_{\rm T}}_{t+k, l_2^\prime}-p^2[\bSigma^{1/2}_{{\rm T}}\bH_k\bSigma^{1/2}_{{\rm T}}]_{l_2,l_2^\prime}}\\
        &\hspace{4cm}\lesssim \sqrt{\sum_{t\in[T-k]}\sqbrac{\bSigma^{1/2}_{\rm T}}^2_{t,l_2}\sqbrac{\bSigma^{1/2}_{\rm T}}^2_{t+k,l_2^\prime}\log n}.
    \end{align*}
    For $l_1=l_1^\prime\in{[N]}$, we can obtain similar results by decoupling the sum into two parts. In particular, define 
    \begin{align*}
\calI_{k,0}&:=\ebrac{t\in[T-k]:\lfloor(t-1)/k\rfloor\equiv 0~(\text{mod}~2)},\\
\calI_{k,1}&:=\ebrac{t\in[T-k]:\lfloor(t-1)/k\rfloor\equiv 1~(\text{mod}~2)}.  
    \end{align*} 
   Then we can decompose 
    \begin{align*}
        &\sum_{t\in[T-k]}\Omega_{l_1,t}\Omega_{l^\prime_1,t+k}\sqbrac{\bSigma^{1/2}_{\rm T}}_{t,l_2}\sqbrac{\bSigma^{1/2}_{\rm T}}_{t+k, l_2^\prime}\\
        &=\sum_{t\in\calI_{k,0}}\Omega_{l_1,t}\Omega_{l^\prime_1,t+k}\sqbrac{\bSigma^{1/2}_{\rm T}}_{t,l_2}\sqbrac{\bSigma^{1/2}_{\rm T}}_{t+k, l_2^\prime}+\sum_{t\in\calI_{k,1}}\Omega_{l_1,t}\Omega_{l^\prime_1,t+k}\sqbrac{\bSigma^{1/2}_{\rm T}}_{t,l_2}\sqbrac{\bSigma^{1/2}_{\rm T}}_{t+k, l_2^\prime}.
    \end{align*}
    where each sum consists of independent terms. Thus we arrive at with probability  at least $1-O\brac{n^{-10}}$,
    \begin{align}\label{eq:B2-Fro-bound}
        \fro{\bB^*_k}^2=\sum_{l,l^\prime\in[NT]}B^{*2}_{ll^\prime,k}&\lesssim \gamma_k^2\sigma_{\rm C}^4\brac{p^2\fro{\bSigma^{1/2}_{{\rm T}}\bH_k\bSigma^{1/2}_{{\rm T}}}^2+\sigma_{\rm T}^4T{\log n}}
        \lesssim \gamma_k^2\sigma_{\rm C}^4 \sigma_{\rm T}^4T\log n.
    \end{align}
    Next we bound $\op{\bB^*_{0}}$. For any fixed $\x=(x_{l_1 l_2}), \y=(y_{l^\prime_1 l^\prime_2})\in \SS^{NT-1}$, it suffices to  bound the following quantity: 
    \begin{align}
        &\ab{\sum_{\substack{l_1\in[N]\\ l_2\in[T]}}\sum_{\substack{l_1^\prime\in[N]\\ l_2^\prime\in[T]}} x_{l_1,l_2}y_{l^\prime_1 l^\prime_2}{\gamma_0} u_{l_1}u_{l^\prime_1}\sigma_{{\rm C},l_1}\sigma_{{\rm C},l_1^\prime}\sum_{t\in[T]}\Omega_{l_1,t}\Omega_{l^\prime_1,t}\sqbrac{\bSigma^{1/2}_{\rm T}}_{t,l_2}\sqbrac{\bSigma^{1/2}_{\rm T}}_{t ,l_2^\prime}}.
    \end{align}
    Let $\bD_{\Omega,l_1l_1^\prime}:=\textsf{diag}\brac{\Omega_{l_11}\Omega_{l_1^\prime 1},\cdots,\Omega_{l_1 T}\Omega_{l_1^\prime T}}\in\RR^{T\times T}$,  $\x_{l,\cdot}:=\brac{\x_{l,1},\cdots, x_{l,t}}^\top $ and $\y_{l,\cdot}:=\brac{y_{l,1},\cdots, y_{l,t}}^\top $,  using bilinear form of operator norm we can obtain:   
    \begin{align*}
        \brac{\sum_{l_2,l^\prime_2\in[T]}x_{l_1,l_2}y_{l_1^\prime l^\prime_2} \sum_{t\in[T]}\Omega_{l_1,t}\Omega_{l^\prime_1,t}\sqbrac{\bSigma^{1/2}_{\rm T}}_{t,l_2}\sqbrac{\bSigma^{1/2}_{\rm T}}_{t ,l_2^\prime}}^2\le \op{\x_{l,\cdot}}^2\op{\y_{l,\cdot}}^2\op{\bSigma_{\rm T}^{1/2}\bD_{\Omega,l_1l_1^\prime}\bSigma_{\rm T}^{1/2}}^2
    \end{align*}
    This leads to 
    \begin{align}\label{eq:op-norm-term-bound}
        &\ab{\sum_{l_2,l^\prime_2\in[T]}\sum_{l_1, l_1^\prime\in[N]} x_{l_1,l_2}y_{l^\prime_1 l^\prime_2}{\gamma_0} u_{l_1}u_{l^\prime_1}\sigma_{{\rm C},l_1}\sigma_{{\rm C},l_1^\prime}\sum_{t\in[T]}\Omega_{l_1,t}\Omega_{l^\prime_1,t}\sqbrac{\bSigma^{1/2}_{\rm T}}_{t,l_2}\sqbrac{\bSigma^{1/2}_{\rm T}}_{t ,l_2^\prime}}\notag\\
        &\le \gamma_0\sum_{l_1,l_1^\prime\in[N]} \ab{u_{l_1}u_{l^\prime_1}\sigma_{{\rm C},l_1}\sigma_{{\rm C},l_1^\prime}}\op{\x_{l_1,\cdot}}\op{\y_{l_1^\prime,\cdot}}\sigma_{\rm T}^2\le \gamma_0 \sigma_{\rm C}^2\sigma_{\rm T}^2.
    \end{align}
    Since the bound \eqref{eq:op-norm-term-bound} holds for any $\x,\y\in \SS^{NT-1}$,  we can conclude that $\op{\bB^*_0}\lesssim \gamma_0 \sigma_{\rm C}^2\sigma_{\rm T}^2$. The bound for $\op{\bB^*_{k}}$ can be obtained similarly by replacing $\bD_{\Omega,l_1l_1^\prime}$ with $${\bD_{\Omega,l_1l_1^\prime}^{(k)}:=\sum_{t\in[T-k]}\Omega_{l_1,t}\Omega_{l_1^\prime, t+k}\e_t \e^\top _{t+k}}\in\RR^{T\times T}.$$ 
    We can arrive at with probability at least $1-O(n^{-10})$,
    \begin{align}\label{eq:Bstar-hp-bound}
        \fro{\bB^*}\le {\fro{\bB_0^*}+2\sum_{k=1}^{K_0}\fro{\bB_k^*}}\lesssim \sigma_{\rm C}^2 \sigma_{\rm T}^2\sqrt{T\log n},\quad \op{\bB^*}\le {\op{\bB_0^*}+2\sum_{k=1}^{K_0}\op{\bB_k^*}}\lesssim \sigma_{\rm C}^2\sigma_{\rm T}^2.
    \end{align}
    {Let $\bA_\Omega:=\EE\brac{\bar\bG\bQ_\bgamma\bar\bG^\top\mid\bOmega}$. Since $\bSigma_{\rm C}$ is diagonal, $\bA_\Omega$ is diagonal. Its lag-zero part is a sum over $\Omega_{i,t}\sqbrac{\bSigma_{\rm T}}_{t,t}$, while its lag-$k$ part is a sum over $\Omega_{i,t}\Omega_{i,t+k}\sqbrac{\bSigma_{\rm T}}_{t,t+k}$. Partitioning the latter into $2k$ classes gives sums of independent bounded variables. Bernstein's inequality, the diagonal-dominance condition, and a union bound over $i\in[N]$ therefore give, with probability at least $1-O(n^{-10})$,
    \begin{align}\label{eq:conditional-mean-mcar}
        \op{\bA_\Omega-\EE\bA_\Omega}\lesssim\sigma_{\rm C}^2\sigma_{\rm T}^2\sqrt{T\log n}.
    \end{align}}
    By Hanson-Wright inequality in \cite{rudelson2013hanson}, the fact that $\g^*$ is independent of $\bB^*$, 
    and a standard $\epsilon$-net argument, we obtain that
    \begin{align*}
        {\PP\brac{\op{\bar\bG\bQ_\bgamma \bar\bG^\top-\bA_\Omega}\ge CK^2\brac{\fro{\bB^*}\sqrt{N}+\op{\bB^*}N}\mid \bOmega}\lesssim e^{-cN}.}
    \end{align*}
    {The proof is completed by combining \eqref{eq:Bstar-hp-bound} and \eqref{eq:conditional-mean-mcar} with the triangle inequality.}
    \paragraph{Proof of claim (2).}
    By definition we have $[\bDelta_M]_{i,t}=p^{-1}\frakD\brac{\Omega_{i,t}}M_{i,t}$, notice that 
    \begin{align*}
        \op{\sqbrac{\calP_\Omega\brac{\bE}}\bQ_{\bgamma}\bDelta_M^\top }\le  \gamma_0\op{\sqbrac{\calP_\Omega\brac{\bE}}\bDelta_M^\top }+\sum_{k=1}^{{K_0}}\gamma_k\op{\sqbrac{\calP_\Omega\brac{\bE}}\bH_k\bDelta_M^\top }.
    \end{align*}
    We first bound the term $\op{\sqbrac{\calP_\Omega\brac{\bE}}\bDelta_M^\top }$. Fix $\u,\v \in\SS^{N-1}$, we have 
    \begin{align}\label{eq:EQDelta-1}
        \u^\top&\calP_{\Omega}\brac{\bSigma_{\rm C}^{1/2}\bG\bSigma_{\rm T}^{1/2}}{\bDelta_M^\top}\v\notag\\
        &=p^{-1}\sum_{i\in[N]}\sum_{l_2\in[T]}G_{i,l_2}u_i\sigma_{{\rm C},i}\sum_{t\in[T]}{\Omega_{i,t}} \sqbrac{\bSigma_{\rm T}^{1/2}}_{t,l_2}\sum_{j\in[N]}\frakD\brac{\Omega_{j,t}}M_{j,t}v_j.
    \end{align}
    {Next we bound $\ab{\sum_{t\in[T]}\Omega_{i,t} \sqbrac{\bSigma_{\rm T}^{1/2}}_{t,l_2}\sum_{j\in[N]}\frakD\brac{\Omega_{j,t}}M_{j,t}v_j}$. For the contribution from $j=i$, use $\Omega_{i,t}\frakD\brac{\Omega_{i,t}}=(1-p)\Omega_{i,t}=p(1-p)+(1-p)\frakD\brac{\Omega_{i,t}}$. Thus, with probability at least $1-O(n^{-10})$, for all $i\in[N]$,}
    \begin{align*}
        {\ab{\sum_{t\in[T]}\Omega_{i,t}\frakD\brac{\Omega_{i,t}} \sqbrac{\bSigma_{\rm T}^{1/2}}_{t,l_2}M_{i,t}v_i}}
        &\lesssim \ab{\sum_{t\in[T]}\sqbrac{\bSigma_{\rm T}^{1/2}}_{t,l_2}M_{i,t}v_i}+ \sqrt{\sum_{t\in[T]} \sqbrac{\bSigma_{\rm T}^{1/2}}_{t,l_2}^2M^2_{i,t}v^2_i\log n}.
    \end{align*}
    {For the off-diagonal contribution, $\Omega_{i,t}$ is independent of the centered sub-Gaussian variable $\sum_{j\ne i}\frakD\brac{\Omega_{j,t}}M_{j,t}v_j$, and these products are independent across $t$. Standard sub-Gaussian concentration therefore gives, with probability at least $1-O(n^{-10})$,}
    \begin{align*}
        {\ab{\sum_{t\in[T]}\Omega_{i,t} \sqbrac{\bSigma_{\rm T}^{1/2}}_{t,l_2}\sum_{j\in[N]\backslash\ebrac{i}}\frakD\brac{\Omega_{j,t}}M_{j,t}v_j}}
        &\lesssim\sqrt{\sum_{t\in[T]} \sqbrac{\bSigma_{\rm T}^{1/2}}^2_{t,l_2}\sum_{j\in[N]\backslash\ebrac{i}}M^2_{j,t}v^2_j\log n}.
    \end{align*}
    We thus arrive at with probability at least $1-O(n^{-10})$  that for $i\in[N]$,
    \begin{align}\label{eq:EQDelta-var-bound}
        &{\ab{\sum_{t\in[T]}\Omega_{i,t} \sqbrac{\bSigma_{\rm T}^{1/2}}_{t,l_2}\sum_{j\in[N]}\frakD\brac{\Omega_{j,t}}M_{j,t}v_j}}\notag\\
        &\lesssim \ab{\sum_{t\in[T]}\sqbrac{\bSigma_{\rm T}^{1/2}}_{t,l_2}M_{i,t}v_i}+ \sqrt{\sum_{t\in[T]} \sqbrac{\bSigma_{\rm T}^{1/2}}^2_{t,l_2}\sum_{j\in[N]}M^2_{j,t}v^2_j\log n}.
    \end{align}
    We thus can 
    apply a standard $\varepsilon$-net argument to get that (conditioning on $\bOmega$) with probability at least $1-O(e^{-cN})$,
    \begin{align*}
        &\op{\calP_{\Omega}\brac{\bSigma_{\rm C}^{1/2}\bG\bSigma_{\rm T}^{1/2}}{\bDelta_M^\top}} \lesssim K\sqrt{N\sum_{i\in[N]}\sum_{l_2\in[T]}u_i^2\sigma_{{\rm C},i}^2{\brac{\sum_{t\in[T]}\Omega_{i,t} \sqbrac{\bSigma_{\rm T}^{1/2}}_{t,l_2}\sum_{j\in[N]}\frakD\brac{\Omega_{j,t}}M_{j,t}v_j}^2} }.
    \end{align*}
    Combining \eqref{eq:EQDelta-var-bound} and the above bound, we get with probability at least $1-O(e^{-cN}+n^{-10})$ that
    \begin{align*}
        \op{\calP_{\Omega}\brac{\bSigma_{\rm C}^{1/2}\bG\bSigma_{\rm T}^{1/2}}{\bDelta_M^\top}} 
        &\lesssim\sigma_{\rm C}\sigma_{\rm T} K\infn{\bM}\sqrt{N\log n}.
    \end{align*}
    It remains to  bound $\sum_{k=1}^{{K_0}}\gamma_k\op{\sqbrac{\calP_\Omega\brac{\bE}}\bH_k\bDelta_M^\top }$. Notice that for each $k\in[K_0]$,
    \begin{align*}
        &\u^\top\calP_{\Omega}\brac{\bSigma_{\rm C}^{1/2}\bG\bSigma_{\rm T}^{1/2}}\bH_k{\bDelta_M^\top}\v\\
        &=p^{-1}\sum_{i\in[N]}\sum_{l_2\in[T]}G_{i,l_2}u_i\sigma_{{\rm C},i}\sum_{t\in[T-k]}{\Omega_{i,t}} \sqbrac{\bSigma_{\rm T}^{1/2}}_{t,l_2}\sum_{j\in[N]}\frakD\brac{\Omega_{j,t+k}}M_{j,t+k}v_j.
    \end{align*}
    {This admits the same structure as \eqref{eq:EQDelta-1}. For $k\ge1$, $\Omega_{i,t}$ and $\frakD(\Omega_{i,t+k})$ are independent and their product is centered. Splitting the lagged sum into the two alternating blocks used above gives the same bound. We thus omit the repeated details.}
    \paragraph{Proof of claim (3)}   
    By definition we have $[\bDelta_M]_{i,t}=p^{-1}\frakD\brac{\Omega_{i,t}}M_{i,t}$, $\EE [\bDelta_M]_{i,t}=0$ and $\op{[\bDelta_M]_{i,t}}_{\psi_2}\lesssim \op{\bM}_{\sf max}$ for $i\in[N], t\in[T]$.  By \Cref{lem:GBG-bound}, we conclude that with probability at least $1-O\bbrac{e^{-cN}}$,
    \begin{align*}
        \op{\bDelta_M\bQ_\bgamma\bDelta_M^\top -\EE\bDelta_M\bQ_\bgamma\bDelta_M^\top }\lesssim {\op{\bM}_{\sf max}^2}\brac{\fro{\bQ_\bgamma}\sqrt{N}+\op{\bQ_\bgamma}N}.
    \end{align*}
    The proof of this claim is completed by using $\fro{\bQ_\bgamma}\lesssim \sqrt{T}$ and $\op{\bQ_\bgamma}\lesssim 1$. 

    \subsection{Proof of \Cref{lem:first-order-two-inf-bound-mcar}} \label{pf-lem:first-order-two-inf-bound-mcar} 
    \paragraph{Proof of claim (1)}
     Fix $i\in{[N]}$ and $k\in[r]$, we then have
    \begin{align}\label{eq:BGU-inf-term}
        &\sqbrac{\bOmega \circ \brac{\bSigma_{\rm C}^{1/2}\bG \bSigma_{\rm T}^{1/2}} \bV}_{i,k}=
        p\sqbrac{\bSigma_{\rm C}^{1/2}\bG \bSigma_{\rm T}^{1/2}{\bV}}_{i,k}+\sum_{l\in[T]}\sigma_{{\rm C},i}\frakD\brac{\Omega_{i,l}}V_{l,k}\sum_{l^\prime\in[T]}G_{i,l^\prime}\sqbrac{\bSigma_{\rm T}^{1/2}}_{l^\prime,l}.
    \end{align}
    For the first term in \eqref{eq:BGU-inf-term}, by \Cref{lem:GBGU-two-inf-bound} we have with probability at least $1-O(n^{-10})$ that
    \begin{align*}
        \ab{\sqbrac{\bSigma_{\rm C}^{1/2}\bG \bSigma_{\rm T}^{1/2}\bV}_{i,k}}\lesssim \sigma_{{\rm C}} K\infn{\bV^\top \bSigma_{\rm T}^{1/2}}\sqrt{\log n}. 
    \end{align*}
    For the second term in \eqref{eq:BGU-inf-term}, notice that $\op{\sum_{l^\prime\in[T]}G_{i,l^\prime}\sqbrac{\bSigma_{\rm T}^{1/2}}_{l^\prime,l}}_{\psi_2}\lesssim K\op{\sqbrac{\bSigma_{\rm T}^{1/2}}_{l,\cdot }}$, hence we have with probability at least $1-O(n^{-10})$ that 
    \begin{align*}
        \ab{\sum_{l^\prime\in[T]}G_{i,l^\prime}\sqbrac{\bSigma_{\rm T}^{1/2}}_{l^\prime,l}}\lesssim K\op{\sqbrac{\bSigma_{\rm T}^{1/2}}_{l,\cdot }}\sqrt{\log n}.
    \end{align*}
    Thus with probability at least $1-O(n^{-10})$, we get
    \begin{align*}
        \ab{\sum_{l\in[T]}\sigma_{{\rm C},i}\frakD\brac{\Omega_{i,l}}V_{l,k}\sum_{l^\prime\in[T]}G_{i,l^\prime}\sqbrac{\bSigma_{\rm T}^{1/2}}_{l^\prime,l}}&\lesssim \sigma_{{\rm C}}K\sqrt{\sum_{l\in[T]}\op{\sqbrac{\bSigma_{\rm T}^{1/2}}_{l,\cdot }}^2V_{l,k}^2}\log n\\
        &\le \sigma_{\rm C}K\infn{\bV^\top }\infn{\bSigma_{\rm T}^{1/2}}\log n.
    \end{align*}
    We then obtain the claim by applying a union bound over $(i,k)\in[N]\times[r]$.
    \paragraph{Proof of claim (2)}  Fix $i\in[N]$ and $k\in[r]$, by definition we have $[\bDelta_M]_{i,t}=p^{-1}\frakD\brac{\Omega_{i,t}}M_{i,t}$. We then have with probability at least $1-O(n^{-20})$,
    \begin{align*}
        &\ab{\sqbrac{\bDelta_M  \bV}_{i,k}}=\ab{\sum_{t\in[T]}p^{-1}\frakD\brac{\Omega_{i,t}}M_{i,t}V_{t,k}}\lesssim\sqrt{\sum_{t\in[T]}M^2_{i,t}V^2_{t,k}\log n}\le \infn{\bV}\infn{\bM }\sqrt{\log n}.
    \end{align*}
    We obtain the claim by applying a union bound over $(i,k)\in[N]\times[r]$.

    \subsection{Proof of \Cref{lem:GBGU-two-inf-bound-mcar}}\label{pf-lem:GBGU-two-inf-bound-mcar} 
    \paragraph{Proof of claim (1)}
    We are going to bound $\op{\e_i^\top \frakD\brac{\calP_{\Omega}\brac{\bE}\sqbrac{\calP_{\Omega}\brac{\bE}}^\top }\bU^{(-i)}}$ and $\op{\e_i^\top\allowbreak~~~~~~~~~~~~~~~~~~~~~ \allowbreak\frakD\brac{\calP_{\Omega}\brac{\bE}\bH_\tau\sqbrac{\calP_{\Omega}\brac{\bE}}^\top }\bU^{(-i)}}$ for $\tau\in[K_0]$ separately.
    Fix $k\in[r]$,  we have    
    \begin{align}\label{eq:EEU-decomp-PC-mcar}
        &{\sqbrac{\calP_{\Omega}\brac{\bE}\sqbrac{\calP_{\Omega}\brac{\bE}}^\top \bU^{(-i)}}_{i,k}}\notag\\
        &=\sum_{t\in[T]}\sum_{l\in[N]}\sigma_{{\rm C},i}\sigma_{{\rm C},l}U^{(-i)}_{l,k}\Omega_{i,t}\Omega_{l,t}\sum_{l_1\in[T]}G_{i,l_1} \sqbrac{\bSigma_{\rm T}^{1/2}}_{l_1,t}\sum_{l_2\in[T]}G_{l,l_2}\sqbrac{\bSigma_{\rm T}^{1/2}}_{l_2,t}.
    \end{align}
    For the summand in \eqref{eq:EEU-decomp-PC-mcar} with $l=i$ , we have 
    \begin{align*}
        &\sum_{t\in[T]}\sigma_{{\rm C},i}^2U^{(-i)}_{i,k}\Omega_{i,t}\sum_{l_1\in[T]}\sum_{l_2\in[T]}G_{i,l_1} G_{i,l_2}\sqbrac{\bSigma_{\rm T}^{1/2}}_{l_1,t}\sqbrac{\bSigma_{\rm T}^{1/2}}_{l_2,t}=\sigma_{{\rm C},i}^2U^{(-i)}_{i,k}\bG_{i,:}\bB_{\Omega,i}\bG^\top _{i,:},
    \end{align*}
    where we define $\bB_{\Omega,i}:=\bSigma_{\rm T}^{1/2}\bD_{\Omega,i}\bSigma_{\rm T}^{1/2}$ and $\bD_{\Omega,i}:=\textsf{diag}\brac{\Omega_{i 1},\cdots,\Omega_{i T}}\in\RR^{T\times T}$. By Hanson-wright inequality in \cite{rudelson2013hanson}, we obtain that conditioning on $\bOmega$, we have with probability at least $1-O(n^{-20})$,
    {\small
\begin{align}\label{eq:EEU-decomp-1-PC-mcar}
        &\ab{\frakD\brac{\sum_{t\in[T]}\sigma_{{\rm C},i}^2U^{(-i)}_{i,k}\Omega_{i,t}\sum_{l_1\in[T]}\sum_{l_2\in[T]}G_{i,l_1} G_{i,l_2}\sqbrac{\bSigma_{\rm T}^{1/2}}_{l_1,t}\sqbrac{\bSigma_{\rm T}^{1/2}}_{l_2,t}}}\notag\\
        &\lesssim \sigma_{\rm C}^2K^2\infn{\bU^{(-i)}}\brac{\fro{\bB_{\Omega,i}}\sqrt{\log n}+\op{\bB_{\Omega,i}}\log n}\lesssim \sigma_{\rm C}^2\sigma_{\rm T}^2K^2\infn{\bU^{(-i)}}\brac{\sqrt{T\log n}+\log n}.
    \end{align}
}
    For the summand in \eqref{eq:EEU-decomp-PC-mcar}  with $l\ne i\in[N]$,   we have 
    \begin{align}\label{eq:EEU-decomp-2-PC-mcar}
        &\ab{\sum_{t\in[T]}\sum_{l\ne i\in[N]}\sigma_{{\rm C},i}\sigma_{{\rm C},l}U^{(-i)}_{l,k}\Omega_{i,t}\Omega_{l,t}\sum_{l_1\in[T]}G_{i,l_1} \sqbrac{\bSigma_{\rm T}^{1/2}}_{l_1,t}\sum_{l_2\in[T]}G_{l,l_2}\sqbrac{\bSigma_{\rm T}^{1/2}}_{l_2,t}}\notag\\
        &\lesssim \sigma_{\rm C} \sigma_{\rm T} K\infn{\bU^{(-i)}}\sqrt{N}\op{\brac{\bSigma_{\rm C}^{1/2}\bG^{(-i)}\bSigma_{\rm T}^{1/2}}\circ \bOmega^{(-i)} }\notag\\
        &\lesssim \sigma_{\rm C}^2 \sigma_{\rm T}^2K^2\infn{\bU^{(-i)}}\sqrt{nN\log n}.
    \end{align}
    with probability at least $1-O\brac{e^{-c_0n}+n^{-20}}$, where we have applied \Cref{lem:GB-bound-mcar} with $\bB=\bI_T$. Collecting \eqref{eq:EEU-decomp-1-PC-mcar}, \eqref{eq:EEU-decomp-2-PC-mcar} and applying a union bound over $k\in[r]$, we obtain the desired bound for $\op{\e_i^\top\frakD\brac{ \calP_{\Omega}\brac{\bE}\sqbrac{\calP_{\Omega}\brac{\bE}}^\top }\bU^{(-i)}}$. It remains to bound $\op{\frakD\brac{\e_i^\top\calP_{\Omega}\brac{\bE}\bH_\tau\sqbrac{\calP_{\Omega}\brac{\bE}}^\top} \bU^{(-i)}}$, which further reduces to bound $\op{\frakD(\e_i^\top \calP_{\Omega}\brac{\bE}\brac{\sum_{t=1}^{{T-\tau}}\e_{t}\e^\top _{t+\tau}}\sqbrac{\calP_{\Omega}\brac{\bE}}^\top )\bU^{(-i)}}$. Fix $k\in[r]$, 
    \begin{align*}
        &{\sqbrac{\calP_{\Omega}\brac{\bE}\brac{\sum_{t=1}^{{T-\tau}}\e_{t}\e^\top _{t+\tau}}\sqbrac{\calP_{\Omega}\brac{\bE}}^\top \bU^{(-i)}}_{i,k}}\notag\\
        &=\sum_{t\in[T-\tau]}\sum_{l\in[N]}\sigma_{{\rm C},i}\sigma_{{\rm C},l}U^{(-i)}_{l,k}\Omega_{i,t}\Omega_{l,t+\tau}\sum_{l_1\in[T]}G_{i,l_1} \sqbrac{\bSigma_{\rm T}^{1/2}}_{l_1,t}\sum_{l_2\in[T]}G_{l,l_2}\sqbrac{\bSigma_{\rm T}^{1/2}}_{l_2,t+\tau},
    \end{align*}
    which admits a similar form to \eqref{eq:EEU-decomp-PC-mcar}. The same bound can be obtained by repeating the above argument and we omit the details for brevity. The proof of this claim is completed by applying a union bound over $k\in[r]$ and using the fact that $\bgamma\in\Delta_{T,K_0}$. 
    
    \paragraph{Proof of claim (2)}
    It boils down to bound $\op{\e_i^\top \calP_{\Omega}\brac{\bE}\bDelta_M^\top \bU^{(-i)}}$ and $\op{\e_i^\top \calP_{\Omega}\brac{\bE}\bH_{\tau}\bDelta_M^\top \bU^{(-i)}}$ for $\tau\in[K_0]$. We start with the bound for  $\op{\e_i^\top \calP_{\Omega}\brac{\bE}\bDelta_M^\top \bU^{(-i)}}$.
    Fix $k\in[r]$,  by definition we have $[\bDelta_M]_{i,t}=p^{-1}\frakD\brac{\Omega_{i,t}}M_{i,t}$. We then have
    \begin{align}\label{eq:claim4-term-decomp}
        {\sqbrac{\calP_{\Omega}\brac{\bE}\bDelta_M^\top \bU^{(-i)}}_{i,k}}
        &=p^{-1}\sigma_{{\rm C},i}\sum_{l_1\in[T]}G_{i,l_1} \sum_{t\in[T]}\sqbrac{\bSigma_{\rm T}^{1/2}}_{l_1,t}\Omega_{i,t}U_{i,k}^{(-i)}\frakD\brac{\Omega_{i,t}}M_{i,t}\notag\\
        &+p^{-1}\sigma_{{\rm C},i}\sum_{l_1\in[T]}G_{i,l_1} \sum_{t\in[T]}\sqbrac{\bSigma_{\rm T}^{1/2}}_{l_1,t}\Omega_{i,t}\sum_{l\in[N]\backslash\ebrac{i}}U_{l,k}^{(-i)}\frakD\brac{\Omega_{l,t}}M_{l,t}.
    \end{align}
    We first bound $\ab{p^{-1}\sigma_{{\rm C},i}\sum_{l_1\in[T]}G_{i,l_1} \sum_{t\in[T]}\sqbrac{\bSigma_{\rm T}^{1/2}}_{l_1,t}\Omega_{i,t}U^{(-i)}_{i,k}\frakD\brac{\Omega_{i,t}}M_{i,t}}$. Notice that $\Omega_{i,t}\frakD\brac{\Omega_{i,t}}=\Omega_{i,t}\brac{1-p}$, we thus have 
    \begin{align*}
        &\sum_{t\in[T]}\sqbrac{\bSigma_{\rm T}^{1/2}}_{l_1,t}\Omega_{i,t}\frakD\brac{\Omega_{i,t}}U_{i,k}^{(-i)}M_{i,t}\\
        &=p\brac{1-p}U_{i,k}^{(-i)}\sqbrac{\bSigma_{\rm T}^{1/2}\bM^\top }_{l_1,i}+\brac{1-p}U^{(-i)}_{i,k} \sum_{t\in[T]}\frakD\brac{\Omega_{i,t}}\sqbrac{\bSigma_{\rm T}^{1/2}}_{l_1,t}M_{i,t}.
    \end{align*}
    With probability at least $1-O(n^{-20})$ for $i\in[N]$, we have
    \begin{align*}
        \ab{\sum_{t\in[T]}\frakD\brac{\Omega_{i,t}}\sqbrac{\bSigma_{\rm T}^{1/2}}_{l_1,t}M_{i,t}}\lesssim\sqrt{\sum_{t\in[T]}\sqbrac{\bSigma_{\rm T}^{1/2}}^2_{l_1,t}M^2_{i,t}\log n}.
    \end{align*}
    Hence we have with probability at least $1-O(n^{-20})$ for $i\in[N]$,
    \begin{align*}
        &\ab{p^{-1}\sigma_{{\rm C},i}\sum_{l_1\in[T]}G_{i,l_1} \sum_{t\in[T]}\sqbrac{\bSigma_{\rm T}^{1/2}}_{l_1,t}\Omega_{i,t}U^{(-i)}_{i,k}\frakD\brac{\Omega_{i,t}}M_{i,t}}\\
        &\lesssim \sigma_{{\rm C},i} K\sqrt{\brac{U^{(-i)}_{i,k}}^2\sum_{l_1\in[T]}\sqbrac{\bSigma_{\rm T}^{1/2}\bM^\top }^2_{l_1,i}+\brac{U^{(-i)}_{i,k}}^2\sum_{l_1\in[T]}\sum_{t\in[T]}\sqbrac{\bSigma_{\rm T}^{1/2}}^2_{l_1,t}M^2_{i,t}\log n}\\
        &\lesssim \sigma_{\rm C}\sigma_{\rm T}K\infn{\bU^{(-i)}}\brac{\op{\bM}+\infn{\bM}\sqrt{\log n }}.
    \end{align*}
    We then bound $\ab{p^{-1}\sigma_{{\rm C},i}\sum_{l_1\in[T]}G_{i,l_1} \sum_{t\in[T]}\sqbrac{\bSigma_{\rm T}^{1/2}}_{l_1,t}\Omega_{i,t}\sum_{l\in[N]\backslash\ebrac{i}}U^{(-i)}_{l,k}\frakD\brac{\Omega_{l,t}}M_{l,t}}$. Note that
    \begin{align*}
        \sum_{t\in[T]}\sqbrac{\bSigma_{\rm T}^{1/2}}_{l_1,t}\Omega_{i,t}\sum_{l\in[N]\backslash\ebrac{i}}U^{(-i)}_{l,k}\frakD\brac{\Omega_{l,t}}M_{l,t}=\sqbrac{\bSigma_{\rm T}^{1/2}\bD_{\Omega,i}\brac{\frakD\brac{\bOmega^{(-i)}} \circ \bM}^\top\bU^{(-i)}}_{l_1,k}
    \end{align*}
    is independent of $\ebrac{G_{i,l_1},l_1\in[T]}$, we thus have
    \begin{align*}
        &\ab{p^{-1}\sigma_{{\rm C},i}\sum_{l_1\in[T]}G_{i,l_1} \sum_{t\in[T]}\sqbrac{\bSigma_{\rm T}^{1/2}}_{l_1,t}\Omega_{i,t}\sum_{l\in[N]\backslash\ebrac{i}}U^{(-i)}_{l,k}\frakD\brac{\Omega_{l,t}}M_{l,t}}\\
        &\lesssim \sigma_{{\rm C},i}K\op{\sqbrac{\bSigma_{\rm T}^{1/2}\bD_{\Omega,i}\brac{\frakD\brac{\bOmega^{(-i)}} \circ \bM}^\top\bU^{(-i)}}_{\cdot,k}}\sqrt{\log n}\\
        &\lesssim\sigma_{{\rm C}}\sigma_{\rm T}K\infn{\bU^{(-i)}}\sqrt{nN\log n}\op{\bM}_{\sf max},
    \end{align*}
    with probability at least $1-O\brac{e^{-c_0n}+n^{-20}}$, where the last inequality holds since entries of  $\frakD\brac{\bOmega^{(-i)}} \circ \bM$ are independent, mean zero, sub-gaussian random variables with subgaussian norm bounded by $C\op{\bM}_{\sf max}$.
    
    The bound for $\op{\e_i^\top \calP_{\Omega}\brac{\bE}\bH_\tau\bDelta_M^\top \bU^{(-i)}}$ for $\tau\in[K_0]$ can be obtained similarly. In particular, observe that
    \begin{align*}
        &{\sqbrac{\calP_{\Omega}\brac{\bE}\bH_\tau\bDelta_M^\top \bU^{(-i)}}_{i,k}}\notag\\
        &=p^{-1}\sigma_{{\rm C},i}\sum_{l_1\in[T]}G_{i,l_1} \sum_{t\in[T-\tau]}\sqbrac{\bSigma_{\rm T}^{1/2}}_{l_1,t}\Omega_{i,t}U^{(-i)}_{i,k}\frakD\brac{\Omega_{i,t+\tau}}M_{i,t+\tau}\\
        &+p^{-1}\sigma_{{\rm C},i}\sum_{l_1\in[T]}G_{i,l_1} \sum_{t\in[T-\tau]}\sqbrac{\bSigma_{\rm T}^{1/2}}_{l_1,t}\Omega_{i,t}\sum_{l\in[N]\backslash\ebrac{i}}U^{(-i)}_{l,k}\frakD\brac{\Omega_{l,t+\tau}}M_{l,t+\tau}.
    \end{align*}
    which admits the same structure as \eqref{eq:claim4-term-decomp}. The remaining proof follows by repeating the argument for bounding $\op{\e_i^\top \calP_{\Omega}\brac{\bE}\bDelta_M^\top \bU^{(-i)}}$. We omit the details for brevity. The proof of this claim is completed by applying a union bound over $k\in[r]$ and using the fact that $\bgamma\in\Delta_{T,K_0}$. 
    \paragraph{Proof of claim (3)}
    Similar to the proof of claim (2), it suffices to bound $\op{\e_i^\top \bDelta_M\sqbrac{\calP_{\Omega}\brac{\bE}}^\top \bU^{(-i)}}$ and $\op{\e_i^\top \bDelta_M\bH_\tau \sqbrac{\calP_{\Omega}\brac{\bE}}^\top \bU^{(-i)}}$ for $\tau\in[K_0]$.  Fix $k\in[r]$, we then have
    \begin{align}\label{eq:claim5-term-decomp}
        {\sqbrac{\bDelta_M\sqbrac{\calP_{\Omega}\brac{\bE}}^\top {\bU^{(-i)}}}_{i,k}}
        &=p^{-1}\sigma_{{\rm C},i}\sum_{l_1\in[T]}G_{i,l_1} \sum_{t\in[T]}\sqbrac{\bSigma_{\rm T}^{1/2}}_{l_1,t}\Omega_{i,t}U^{(-i)}_{i,k}\frakD\brac{\Omega_{i,t}}M_{i,t}\notag\\
        &+p^{-1}\sum_{l_1\in[T]}\sum_{l\in[N]\backslash\ebrac{i}}G_{l,l_1} \sigma_{{\rm C},l}U^{(-i)}_{l,k}\sum_{t\in[T]}\sqbrac{\bSigma_{\rm T}^{1/2}}_{l_1,t}\Omega_{l,t}M_{i,t}\frakD\brac{\Omega_{i,t}}.
    \end{align}
    The first term in \eqref{eq:claim5-term-decomp} is exactly the same as \eqref{eq:claim4-term-decomp}, and we only need to bound the second term. Note that 
    \begin{align*}
        \sum_{l_1\in[T]}\sum_{l\in[N]\backslash\ebrac{i}}G_{l,l_1} \sigma_{{\rm C},l}U^{(-i)}_{l,k}\sqbrac{\bSigma_{\rm T}^{1/2}}_{l_1,t}\Omega_{l,t}=\sqbrac{\brac{\bOmega^{(-i)}\circ\bSigma_{\rm C}^{1/2}\bG^{(-i)}\bSigma_{\rm T}^{1/2}}^\top \bU^{(-i)}}_{t,k}
    \end{align*}
    Hence we get with probability at least $1-O(n^{-20})$ that 
    \begin{align*}
        &\ab{p^{-1}\sum_{l_1\in[T]}\sum_{l\in[N]\backslash\ebrac{i}}G_{l,l_1} \sigma_{{\rm C},l}U^{(-i)}_{l,k}\sum_{t\in[T]}\sqbrac{\bSigma_{\rm T}^{1/2}}_{l_1,t}\Omega_{l,t}M_{i,t}\frakD\brac{\Omega_{i,t}}}\\
        &\lesssim\op{\bM}_{\sf max}\op{\sqbrac{\brac{\bOmega^{(-i)}\circ\bSigma_{\rm C}^{1/2}\bG^{(-i)}\bSigma_{\rm T}^{1/2}}^\top \bU^{(-i)}}_{\cdot,k}}\sqrt{\log n}\\
        &\lesssim \sigma_{\rm C}\sigma_{\rm T}K\infn{\bU^{(-i)}}\sqrt{nN}\log n\op{\bM}_{\sf max},
    \end{align*}
    where the last inequality we've used \Cref{lem:GB-bound-mcar}. For $\op{\e_i^\top \bDelta_M\bH_\tau \sqbrac{\calP_{\Omega}\brac{\bE}}^\top {\bU^{(-i)}}}$, notice that 
    \begin{align*}
        &{\sqbrac{\bDelta_M\bH_\tau\sqbrac{\calP_{\Omega}\brac{\bE}}^\top {\bU^{(-i)}}}_{i,k}}\notag\\
        &=p^{-1}\sigma_{{\rm C},i}\sum_{l_1\in[T]}G_{i,l_1} \sum_{t\in[T-\tau]}\sqbrac{\bSigma_{\rm T}^{1/2}}_{l_1,t+\tau}\Omega_{i,t+\tau}{U^{(-i)}_{i,k}}\frakD\brac{\Omega_{i,t}}M_{i,t}\notag\\
        &+p^{-1}\sum_{l_1\in[T]}\sum_{l\in[N]\backslash\ebrac{i}}G_{l,l_1} \sigma_{{\rm C},l}{U^{(-i)}_{l,k}}\sum_{t\in[T-\tau]}\sqbrac{\bSigma_{\rm T}^{1/2}}_{l_1,t+\tau}\Omega_{l,t+\tau}M_{i,t}\frakD\brac{\Omega_{i,t}}.
    \end{align*}
    which admits the same structure as \eqref{eq:claim5-term-decomp} and hence we omit the details.  The proof of this claim is completed by applying a union bound over $k\in[r]$ and using the fact that $\bgamma\in\Delta_{T,K_0}$. 
    \paragraph{Proof of claim (4)} The proof of this claim is almost identical to (but simpler than) that of claim (1), by viewing $\bDelta_M$ as a random matrix with independent entries whose sub-gaussian norm is uniformly bounded by $C\op{\bM}_{\sf max}$. We omit the details for brevity. 

\end{document}